# Epsilon-Near-Zero Materials based Photonic Architectures for Absorption and Emission Control

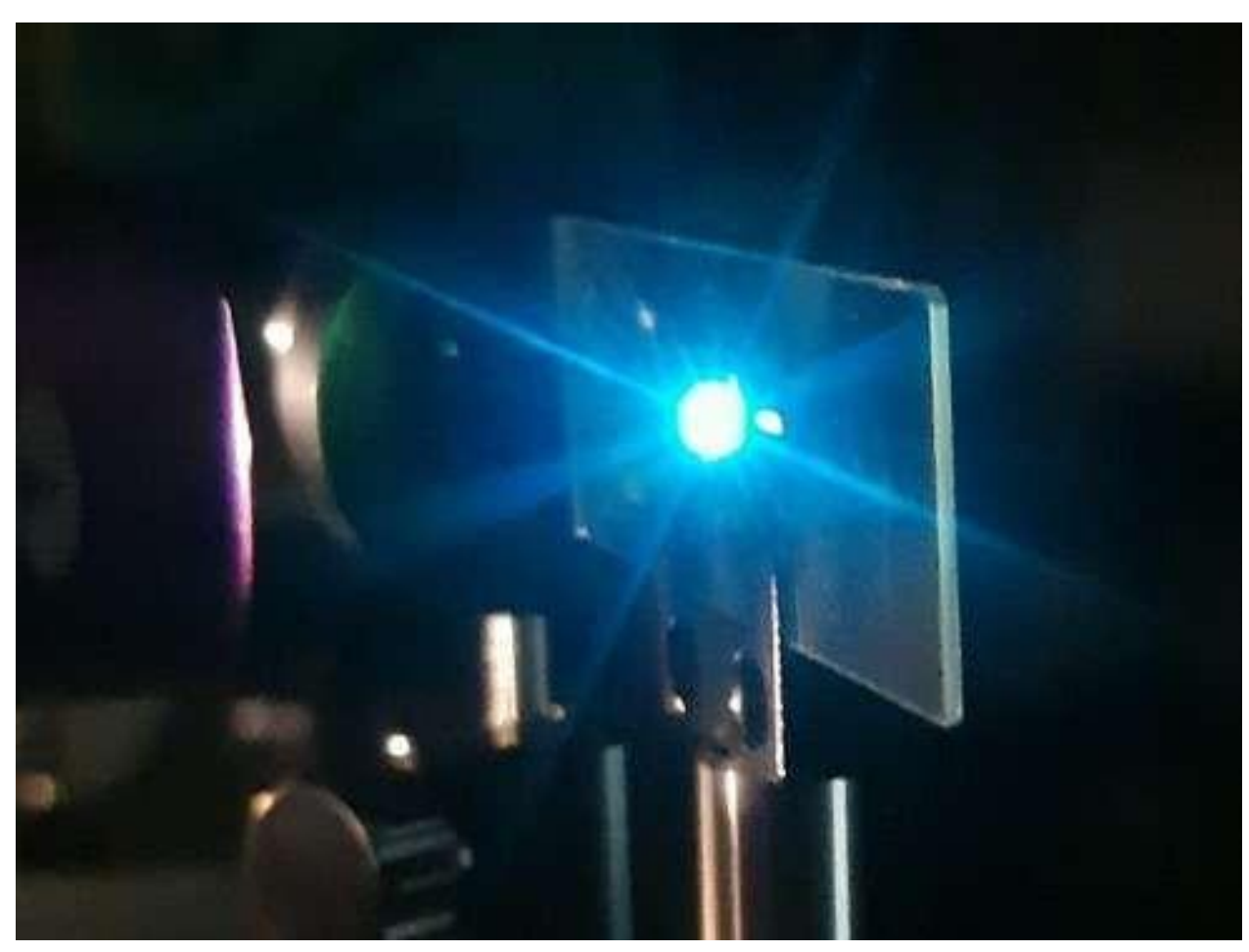

**Sraboni Dey**

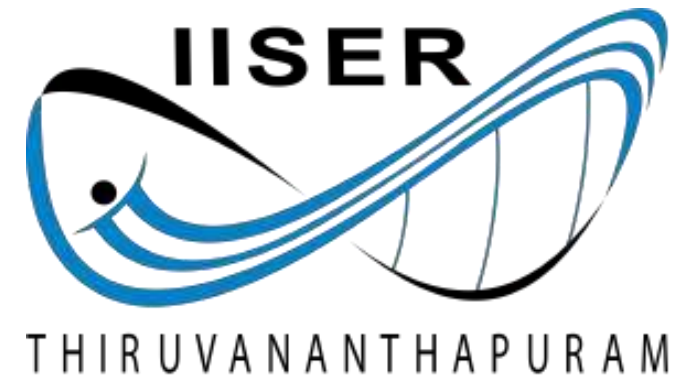

SCHOOL OF PHYSICS
INDIAN INSTITUTE OF SCIENCE EDUCATION AND RESEARCH
THIRUVANANTHAPURAM
KERALA- 695551, INDIA

August 2025

# Epsilon-Near-Zero Materials based Photonic Architectures for Absorption and Emission Control

**Sraboni Dey**

**PHD 201046**

*A thesis submitted for the degree of*

**Doctor of Philosophy (Ph.D.)**

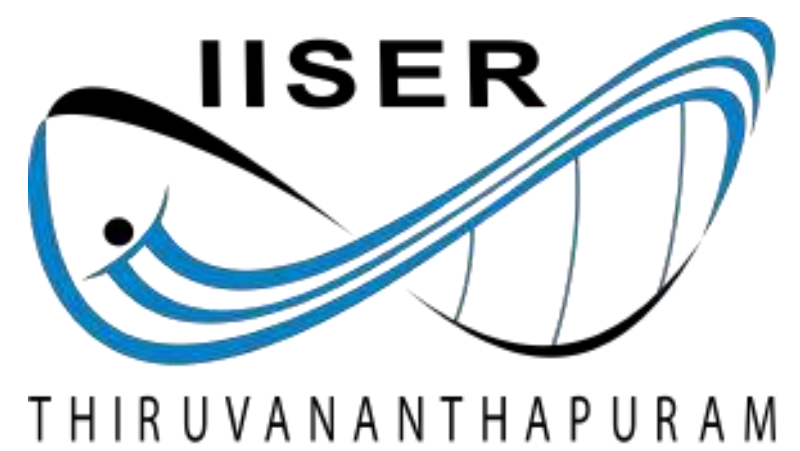

SCHOOL OF PHYSICS
INDIAN INSTITUTE OF SCIENCE EDUCATION AND RESEARCH
THIRUVANANTHAPURAM
KERALA- 695551, INDIA

August 2025

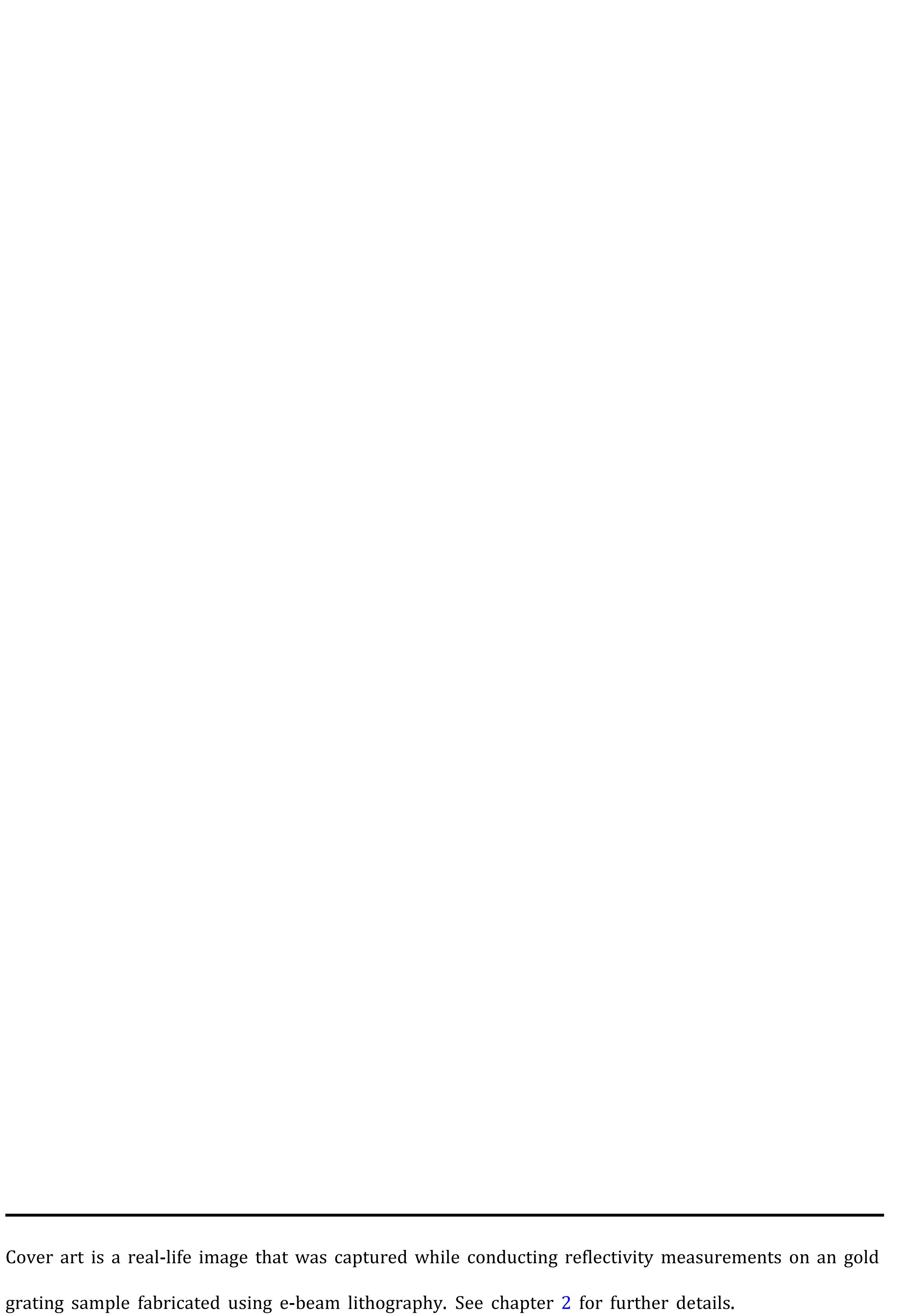

Cover art is a real-life image that was captured while conducting reflectivity measurements on an gold grating sample fabricated using e-beam lithography. See chapter 2 for further details.

*Dedicated to*

*Baba, Maa and Dada*

# AUTHOR'S DECLARATION

I hereby declare that the work reported in this dissertation is original and has been carried out by me during my tenure as a PhD student at the **School of Physics, Indian Institute of Science Education and Research Thiruvananthapuram** under the supervision of Prof Joy Mitra. This work has not been submitted earlier as a whole or in part for a degree/diploma at this or any other Institution/University. Material obtained from other sources has been duly acknowledged in the dissertation.

Sraboni Dey.

Thiruvananthapuram

Sraboni Dey

Date: 21-10-2025

## CERTIFICATE

This is to certify that the work contained in this thesis entitled **"Epsilon-Near-Zero Materials based Photonic Architectures for Absorption and Emission Control "** submitted by **Sraboni Dey** (**Roll No: PHD201046**) to **School of Physics, Indian Institute of Science Education and Research Thiruvananthapuram (IISER - TVM)** towards the requirement of **Doctor of Philosophy** in **Physics** has been carried out by her under my supervision and that it has not been submitted elsewhere for the award of any degree.

Thiruvananthapuram

Date: 21-10-2025

Prof Joy Mitra

Doctoral Advisor

## ACKNOWLEDGMENTS

Completing this journey would not have been possible without the support and guidance of numerous individuals who helped me at different stages of my PhD. First and foremost, I would like to express my deepest gratitude to my doctoral supervisor, Prof. Joy Mitra, for giving me the opportunity to work under his supervision. The constant support and the freedom he offered to explore diverse research directions and topics greatly shaped my academic development. From him, I learned numerous invaluable skills, including thinking in the right direction, systematic problem-solving, and scientific writing. I am also extremely grateful to my doctoral committee members, Dr. Ravi Pant and Dr. Madhu Thalakulam, for dedicating their valuable time to openly discussing and providing constructive suggestions, especially during moments when I was stuck or needed direction. My sincere thanks go to Dr. Deepshikha Jaiswal Nagar for allowing me to access and operate the sputtering system in her lab, which was essential for timely thin-film depositions. I am particularly thankful to Mr. Kiran for training me in the sputtering technique and for our long-standing collaboration on ITO and dielectric depositions. I want to acknowledge Dr. Somu Kumaragubaran for granting me access to the rapid thermal annealing (RTA) facility in his lab and Ms. Anjana for training me in RTA. I am also thankful to Prof. M.M. Shaijumon for collaborating on 2D material samples and to Mr. Renjith N for consistently providing $MoS_2$ so that I can perform the device fabrications. I would like to acknowledge Prof Hema Somanathan from School of Biology for giving me access to the thermal imaging facility. Special thanks to Prof. Satheesh Krishnamurthy and Prof. Ravi Silva for being gracious and supportive hosts during my stay at the University of Surrey and for engaging in valuable discussions regarding the ion implantation results. I am immensely grateful to all my past and present lab members- Krishna, Ben, Arijit, Hari, Kritika, Shashwata, Arun, Soumyadip, Narendra, and Satyam for contributing to my learning experience throughout my PhD.

A few deserve special mention: Dr. Arijit, for training me in AFM and introducing me to 2D materials; Dr. Ben, for his insightful discussions and expertise in ENZ-related studies; and Dr. Krishna and Dr. Hari, for their guidance on various lab equipment and their continued cooperation. Beyond the lab, I am very thankful to Anusha for training me in e-beam lithography at IISER, without which I could not have completed a major portion of my thesis work. I would also like to acknowledge Prof. Sanatan Chattopadhyay from the University of Calcutta, who first introduced me to lithography techniques during my Master's project. I would like to thank a few friends I met at different points in this journey: Vijay Bhaiyya and Sayani di at IISER, Praveen from CENSe IISc, and Pramod from Surrey, for their continuous support. I would also like to acknowledge my teachers, starting from school to master's, who have taught me at various stages and improved my understanding, and a few childhood friends, Soumya, Riddhi, Bishal, and Pritishna, who have always been in touch and given me company. I am also grateful to various funding agencies- DST INSPIRE, DST-SERB, UKIERI-SPARC and European Physical Society (EPS) for their financial support during my PhD tenure. I am also thankful to the Indian Nanoelectronics Users Programme (INUP), CENSE, IISc Bangalore for providing me the opportunity to utilize their research facilities during certain projects. Most importantly, I would like to convey my gratitude to IISER Thiruvananthapuram for providing various facilities throughout my PhD.

Last but certainly not least, I am deeply grateful to my family-my parents, my brother Somnath, and Shipra di- for their unwavering love and support, without which I could not have reached this point in my life.

# Preface

The study of light-matter interaction with engineered surfaces has grown into a pivotal area of research, driven by its extensive potential in diverse technological domains. From energy harvesting and optical sensing to smart windows and displays, controlling the spectral optical response of materials remains central to advancing these applications. In recent years, global concerns such as overpopulation, excessive reliance on non-renewable energy sources, and the inability to efficiently recycle energy have intensified the search for sustainable alternatives. Among these, solar energy stands out as a promising candidate, offering clean, abundant power. However, its effective utilization depends on the ability to efficiently capture and manage optical response, particularly through the design and integration of advanced optical materials and coatings.

This thesis is motivated by the need to tailor surface optical responses-reflection, absorption, transmission, and scattering for optimized performance across the electromagnetic spectrum. Achieving this requires controlled engineering of surface structures, refractive indices, and geometries such as gratings or nanoantennas. One of the primary aims of this work is to explore and exploit the capabilities of epsilon-near-zero (ENZ) materials, namely indium tin oxide (ITO) and titanium nitride (TiN), for producing spectrally selective optical responses as well as to enhance light absorption and emission, especially in two-dimensional (2D) materials such as monolayer $MoS_2$, which, despite their compelling electronic and optical properties, suffer from inherently weak light interaction due to their atomic-scale thickness. The research presented in this thesis is structured across six main chapters and two appendices, providing a comprehensive study through theory, simulation, fabrication, and experimental validation.

Chapter 1 provides a broad overview of the motivation behind optical response engineering, particularly for energy and thermal management applications. It introduces ENZ materials as promising candidates for spectral selectivity and outlines their fundamental optical characteristics and relevance in enhancing the performance of 2D materials like $MoS_2$.

Chapter 2 details the numerical, simulation, and experimental techniques used in the thesis. Finite Element Method simulations using COMSOL Multiphysics enabled precise modeling of the multilayer optical coatings based on ENZ and 2D systems, while a range of nanofabrication and characterization tools including electron-beam lithography, sputtering, SEM, reflectance and photoluminescence spectroscopy were employed for experimental studies.

Chapter 3 presents the development of a multilayer optical coating using ITO as the ENZ material. This design exhibits a step-function-like reflectivity profile, enabling low reflectance in the visible to near-infrared and high reflectance beyond a cut-in wavelength ($\lambda_0$). Tunable and thermally stable, the coating is promising for energy harvesting applications.

Chapter 4 explores a band-selective absorber using an ITO-based grating structure layered over a dielectric coated metal substrate. This coating demonstrates high absorption across a broad angular range in the near infrared, making it suitable for thermal emission applications. Both simulation and thermal imaging results underline the efficacy of ENZ-assisted light absorption.

Chapter 5 investigates a novel approach for enhancing absorption and emission in monolayer $MoS_2$ using TiN thin films, without complex nanostructuring. This work highlights the potential of TiN in optical engineering and its compatibility with 2D materials for optoelectronic applications such as photodetectors and photovoltaics.

Chapter 6 concludes the thesis by summarizing the key findings and outlining directions for future research, particularly in scaling the demonstrated technologies and extending their functionalities.

Two appendices complement the main chapters with additional explorations: Appendix A discusses how periodically patterned substrates can significantly alter the local electronic and optical properties of $MoS_2$ by introducing strain, thereby improving carrier mobility and enhancing hydrogen evolution performance. Appendix B examines the impact of defect engineering via ion irradiation on the optical behavior of $MoS_2$, opening up further possibilities for material property tuning. Altogether, this thesis seeks to bridge the gap between material science and optical engineering through the strategic use of ENZ materials and nanostructures. It contributes to the broader goal of developing efficient, tunable, and scalable solutions for energy, thermal, and optical applications.

# Contents

# List of Figures

# List of Tables

# Chapter 1

# Introduction

## 1.1 Engineering optical response of surfaces

Light matter interactions are the cornerstone of the existence of life on earth. Its singular importance in the physical, chemical and biological aspects of our existence has driven investigations over centuries, leading to many scientific discoveries and inventions.[1]. Starting with experiments utilizing various optical components such as mirrors, lenses and prisms, researchers have gained insights on how to engineer materials to deliver a wide range of desired optical properties [2, 3], with effects which are not realizable with natural materials[4, 5]. Engineering the optical response of surfaces has been a crucial area of research for its wide range of applications. Global energy crisis due to several factors like overpopulation, excessive consumption of non-renewable energy sources, inability to efficiently recycling of energy has encouraged the search and utilization of renewable energy sources[6]. Solar energy is one of the major renewable energy sources; however, its effective utilization has still been a challenge [8]. The right assembly of materials, structuring, and engineering them to the right architectures, for example, in the form of optical coatings, has been conducted extensively to efficiently

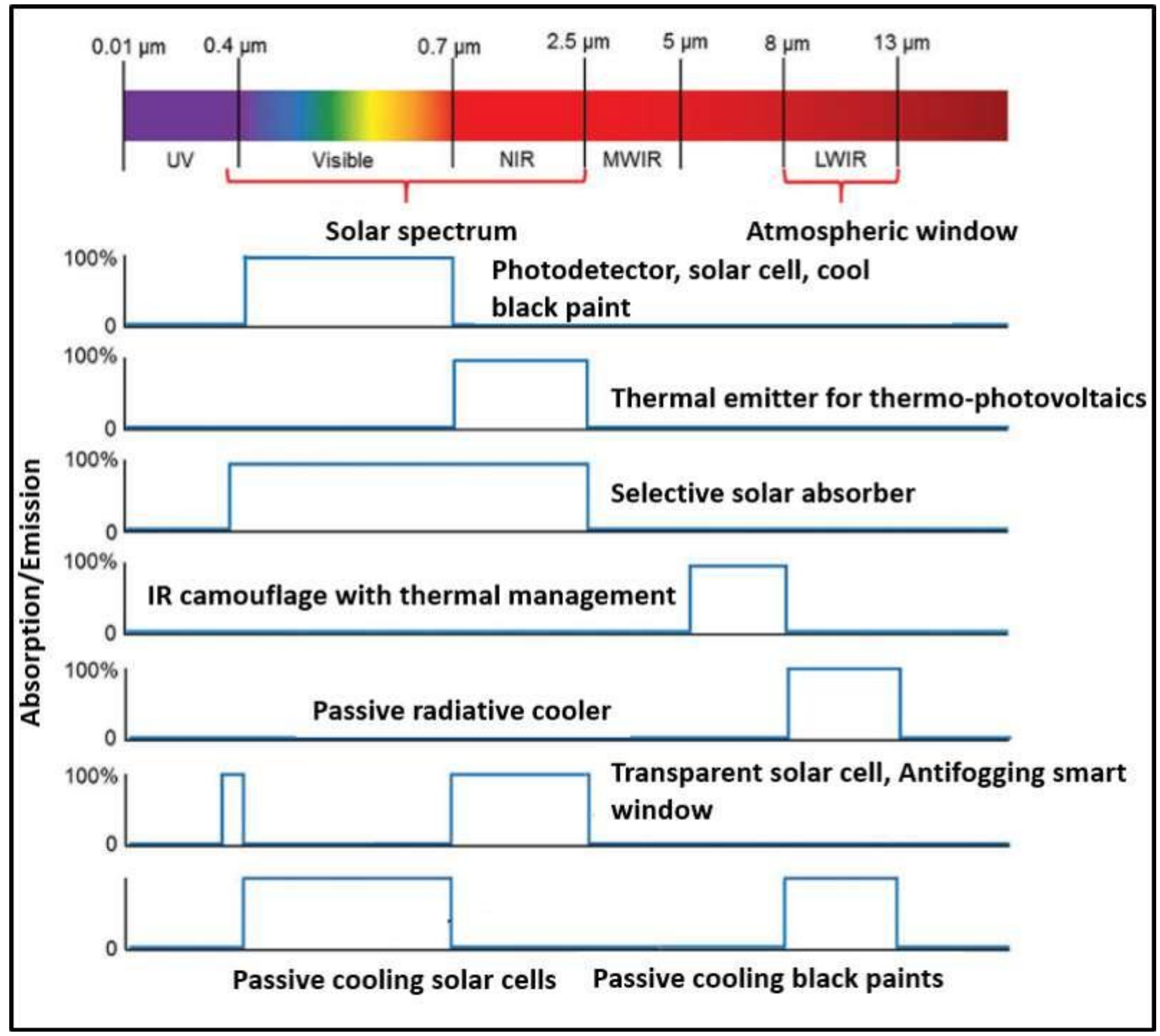

**Figure 1.1:** Engineering spectrally selective absorbers/emitters with enhanced absorption/emission over specific spectral windows in the UV–vis–IR range, and their relevant applications. The image has been adapted from [7]

harness solar energy [9, 10]. They provide the opportunity to improve light-matter interaction and absorb light in different spectral regimes of the electromagnetic spectrum. Efficient absorption of light can be achieved by modifying surface properties like roughness [11, 12], refractive index [13, 14], and designing structures like gratings [15, 16] or nanoantennas [17–19]. Fig.1.1 showcases the relevant applications that can result from the enhancement of light absorption/emission over particular spectral regimes from UV to IR [7]. The objective is to tailor the surface's response, in terms of reflection, transmission, absorption, or light scattering, catering to specific applications including sensing [20, 21], displays [22], filters[23] or energy harvesting[24, 25] leading to technological advancement.

## 1.2 Optical response modulation: Challenges

Optical engineering is used to design or modify the optical response in materials or surfaces in terms of intensity, directionality, and polarization with a broad range of applications [26, 27] via various approaches as shown in Fig. 1.2. For each application, as shown in Fig. 1.1, the ideal spectral profiles of the optical devices must be carefully engineered in order to meet the necessary requirements such that they absorb within a particular spectral regime and reflect or transmit beyond that. For example,

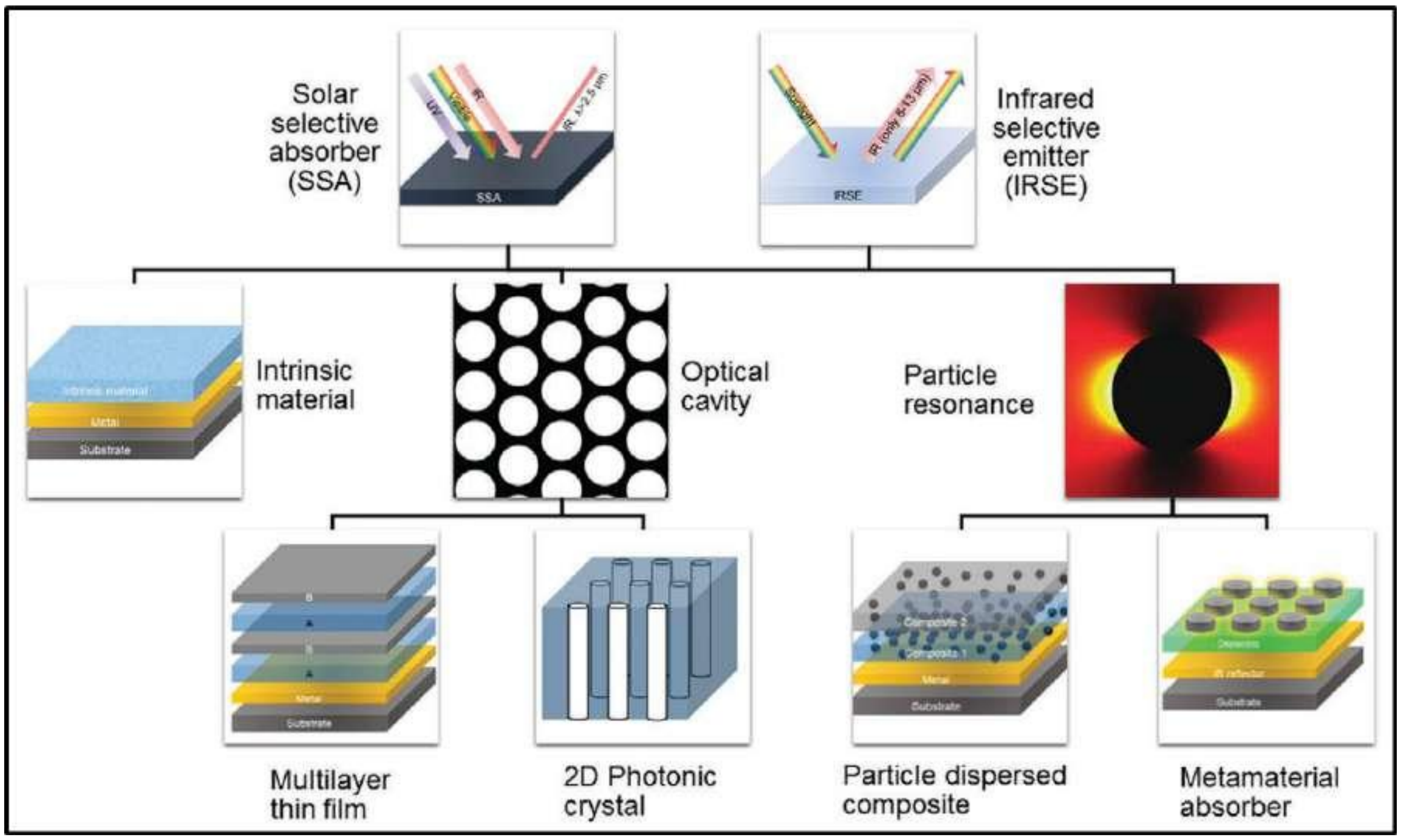

**Figure 1.2:** Classification of spectrally selective absorbers/emitterson the basis of the mechanism of absorption: intrinsic materials, optical cavities, and particle-based resonance.. The image is adapted from [7]

it is essential to design band-selective absorption/emission to exhibit near-unity ab-sorptivity/emissivity in the transparent atmospheric window (0.4$\mu$m-1.3$\mu$m and 8$\mu$m-14$\mu$m) and near-unity reflectivity at non-atmospheric-window wavelengths for passive cooling applications [28, 29]. On the other hand, a broadband absorber/emitter with high emissivity over the entire mid-infrared regime is important for daytime above-ambient cooling [30, 31] and for extraterrestrial applications[32]. As a result, the design of efficient spectrally selective reflectors/absorbers, band-selective absorbers/emitters becomes important for the optimization of absorption and emission characteristics in terms of emission bandwidths, band positions, and numbers with synchronous control over the entire electromagnetic spectrum from the ultraviolet to infrared[33, 34].

Various issues that often pose significant challenges in efficient optical engineering are discussed as follows:

#### 1.2.0.1 Choice of material and environmental factors

The right selection of material is highly important for achieving the desired optical performance. This depends on the specific application to be met, considering various factors like wavelength, temperature, and mechanical properties. Different materials transmit/reflect/absorb different wavelengths of light. For instance, fused silica is ideal for applications dealing with the ultraviolet regime[35], while fluoride and chalcogenide glasses are suitable for the infrared[36]. In addition, for applications dealing with significant variation in temperature, materials with a low expansion coefficient are often preferred. Fused silica is often chosen for its high transmission and thermal stability[37]. Moreover, environmental factors significantly impact the performance and longevity of optical systems. Fluctuations in temperature, presence of humidity [38], and local dielectric environment can affect the physical properties and performance of optical devices significantly [39]. Hence, the choices of the materials should be such that they have high thermal, chemical, and mechanical stability. Hence, careful selection of materials that can effectively meet various requirements of light modulation without involving much complexity is a significant challenge in optical engineering. Therefore, engineering the right architectures with efficient material platforms is needed as a solution for desired light modulation, which forms the core of this thesis.

#### 1.2.0.2 Complexity in design

Complexity in design is another major existing challenge that is frequently faced in designing a specific optical response. This involves the use of several optical elements, with complex designs, in realizing specific spectral responses with characteristic spectral features. In order to meet these spectral requirements (e.g., for certain wavelengths or bandwidths), specialized materials or multilayer coatings of complex composites [40] are often used, adding to fabrication challenges. This thesis partly tries to address this issue by proposing straightforward coating designs for absorption and emission control.

#### 1.2.0.3 Cost of fabrication

The cost of fabrication is an integral factor in optical engineering that varies widely depending on the design complexity, usage of different materials, the process of manufacturing and the quantity produced [41]. In addition, the type of optical element that is to be designed like optical windows, lenses, the different fabrication processes involved and the required precision of the different components to develop the final device, influence the total cost massively. Hence, leveraging the use of materials that would be economical compared to the frequently used conventional metals, such as gold (Au), silver (Ag), etc. has been explored in this thesis.

## 1.3 Optical response modulation: Strategies

This section discusses various strategies that have been adopted in spectral modulation of light, which controls the variation of the intensity or distribution of light across different wavelength regimes.

### 1.3.1 Cavity resonances

A cavity resonator is a structure that is formed by an enclosure surrounded by conducting walls, with dimensions comparable to the wavelength of electromagnetic wave in free space[42]. It can trap electromagnetic waves by making them reflect back and forth from the cavity walls with a large dependence on the refractive index of the cavity media in confining light. A Fabry-Perot cavity is a kind of cavity resonator made from two relatively parallel mirrors with partial transmission and partial reflectivity, as shown in Fig. 1.3. The light incident into a Fabry-Perot resonator undergoes reflections multiple times inside the cavity. The transmission of the resonator is maximum if the resonance condition is satisfied. With a relatively high reflectivity of either mirror, the reflectivity is high throughout the spectrum except for the wavelengths at resonance, where the light is absorbed[43]. In this thesis, a Fabry-Perot cavity was formed with a stainless steel (SS) and thin reflecting chromium (Cr) layer with chromium oxide ($Cr_2O_3$) layer of optimal thickness in between to enhance light absorption over the entire visible regime as discussed in Chapter 3.

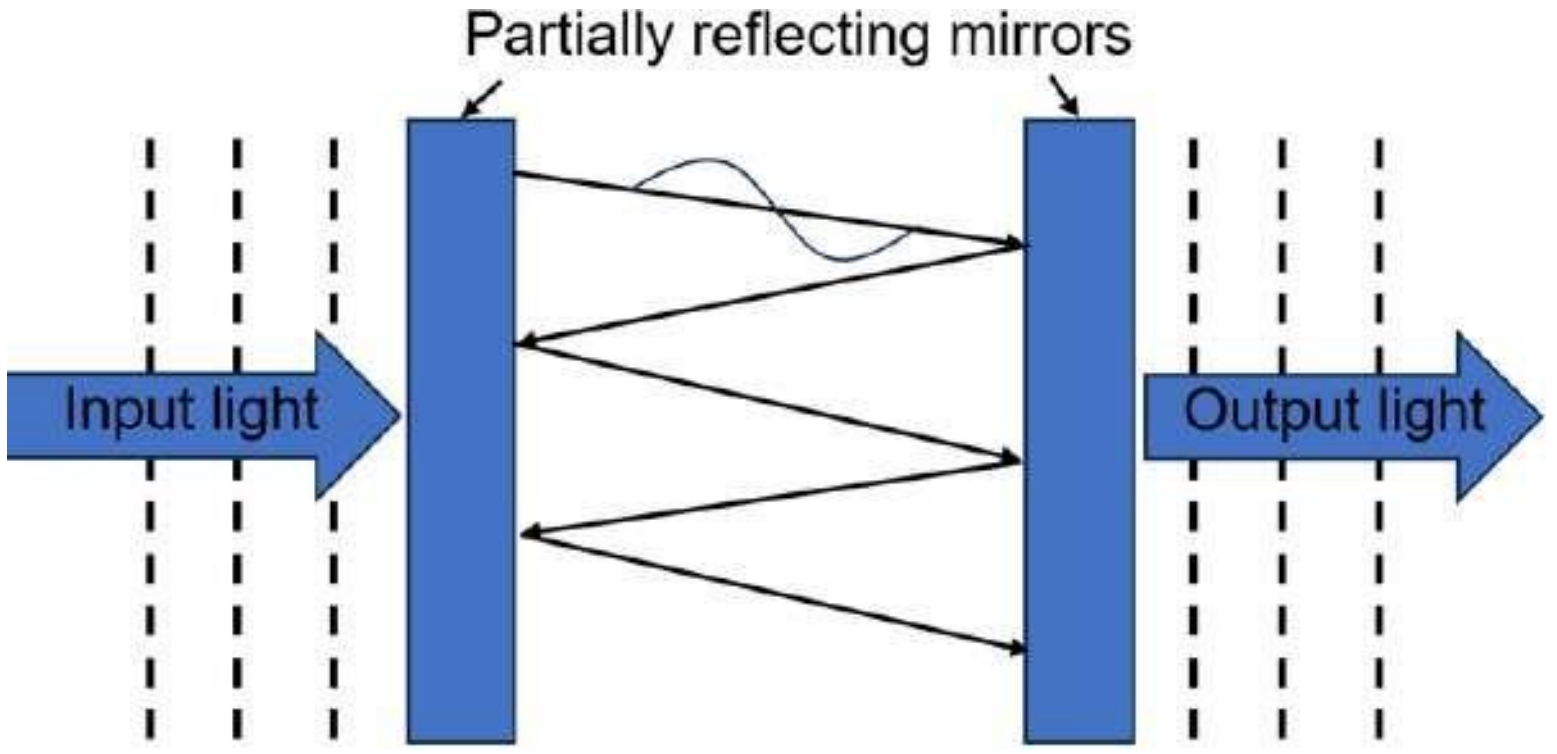

**Figure 1.3:** Schematic of a Fabry-Perot cavity

### 1.3.2 Meta-surfaces

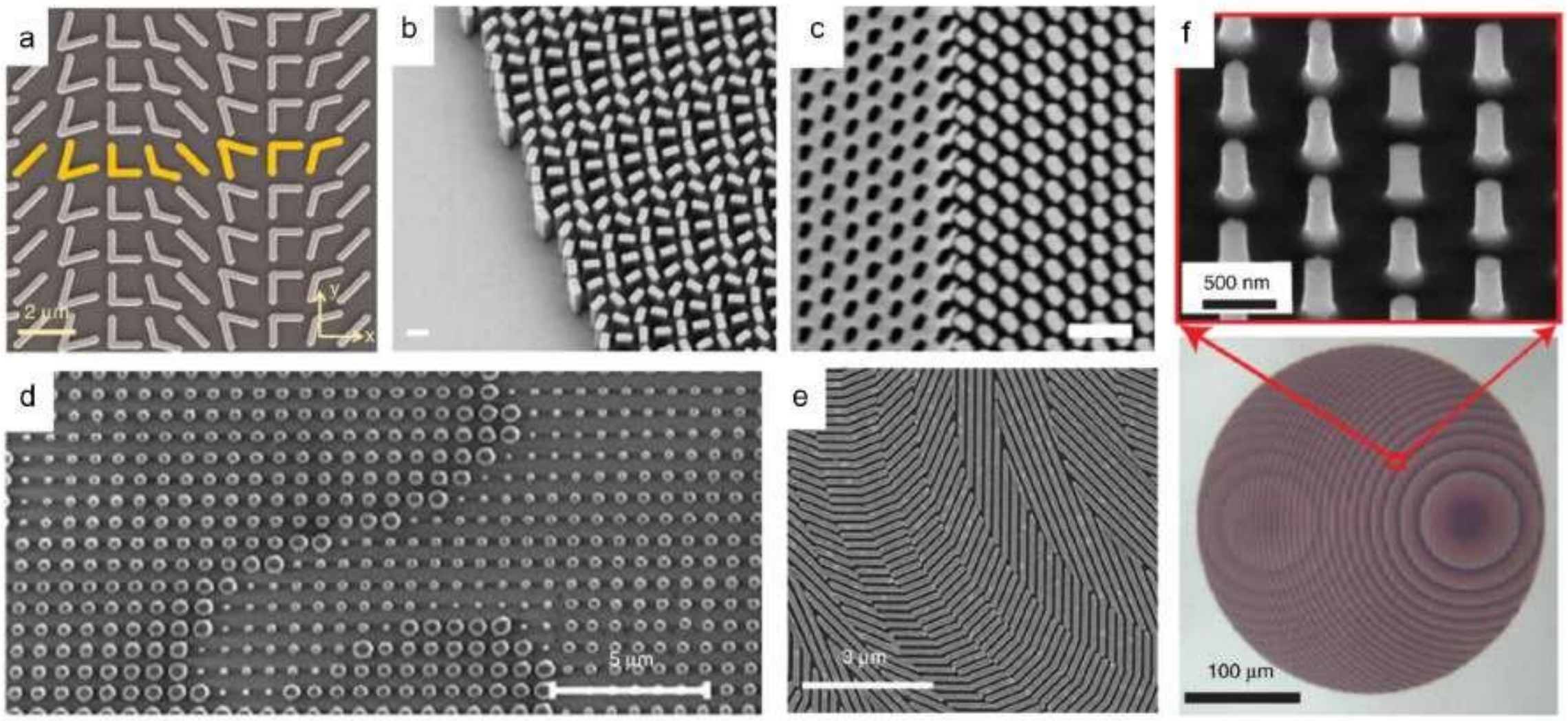

**Figure 1.4:** (a) Optical metasurface composed of a gold antenna array. The unit cell of the plasmonic interface (yellow) comprises eight gold V-antennas. (b) A metalens operating at 660 nm and consisting of $TiO_2$ nanofins on a glass substrate. Scale bar is 300 nm. (c) Achromatic metalens with NA ∼0.1. Scale bar is 500nm. The vertical boundary of nanopillars and Babinet structures is visible.(d) Fabricated meta-hologram that produces 5mm large images at a distance of 10mm. The posts are silicon on SiO. (e) SEM image of a dielectric metasurface lens based on Si nanobeams that results in a local Bessel spot focal length of 100 mm at $\lambda$ =550nm.(f) Dielectric metasurface made from amorphous silicon pillars on a $SiO_2$ that separates x- and y-polarized light and focuses them to two different points. The image is adapted from [44]

Metasurfaces are engineered surfaces composed of sub-wavelength structures that can manipulate electromagnetic waves, often unattainable via natural materials. They have recently emerged as a key platform in flat optics, a branch of metaphotonics focused on manipulating electromagnetic fields in nanostructures to control light propagation [45, 46]. A few recent examples of experimentally explored metasurfaces are shown in Fig.1.4. Applications of metasurfaces range from optical communication and solar energy harvesting [46] to sensing technologies and biophotonics [47, 48]. In particular, nonlinear metaphotonics, which involves light interacting with materials exhibiting second- and third-order nonlinearities, offers exciting possibilities for dynamic light control, though its exploration remains in early stages[49]. Metasurfaces enable precise control over transmission, reflection, refraction, dispersion, interference, and beam steering, especially over small space and time scales[50]. Advanced variants, such as chiral metasurfaces, are being explored for use in broadband communication, beam shaping, cloaking, and hyperlens imaging[51–53]. Of special interest are epsilon-near-zero (ENZ) metasurfaces, which offer remarkable light manipulation capabilities. This thesis draws motivation from such systems, in fabricating 1D gratings of indium-tin-oxide (ITO) as the ENZ material to develop spectrally selective reflector and band-selective absorber coatings as detailed in Chapters 3 and 4.

### 1.3.3 Thin film interferences

Thin-film interference is the phenomenon that takes place when a light wave enters a medium and is reflected off two surfaces that are at a distance comparable to its wavelength. When light waves that reflect from the top and bottom surfaces interfere with one another, different coloured patterns are produced. In this scenario, the light reaches the boundary between two media and part of it gets reflected, while some part gets transmitted and therefore leads to selective absorption of light [54]. Hence, this also provides a simple yet effective strategy for light modulation.

### 1.3.4 Plasmonic resonances

Plasmonics deals with the optical response of collective free electron oscillations at the surface of materials with large free carrier densities. Surface plasmons, often referred to as a charge density wave, are coherent collective oscillations of electrons that exist at the interface of two materials, one with negative and the other with positive permittivity [55]. Metals with high free electron density have negative permittivity, whereas dielectrics have positive permittivity, and this is an essential condition for surface plasmons to exist at the metal-dielectric interface. Surface plasmons exist in two forms: as surface plasmon polaritons (SPP), the propagating mode along the metal-dielectric interface, while the other is the localized surface plasmons (LSP) that is localized in particles of nanoscale dimensions with high free electron density [56] as shown in Fig.1.5. Individual nanostructures of sub-wavelength dimensions exhibit localized surface plas-mon resonances (LSPRs), where electromagnetic fields couple to the surface plasmons at a metal–dielectric interface. Depending on its shape, dimension, an individual nanostructure may be polarized by an incident light beam, which leads to resonances due to amplification of the near-field, leading to absorption of light [58].

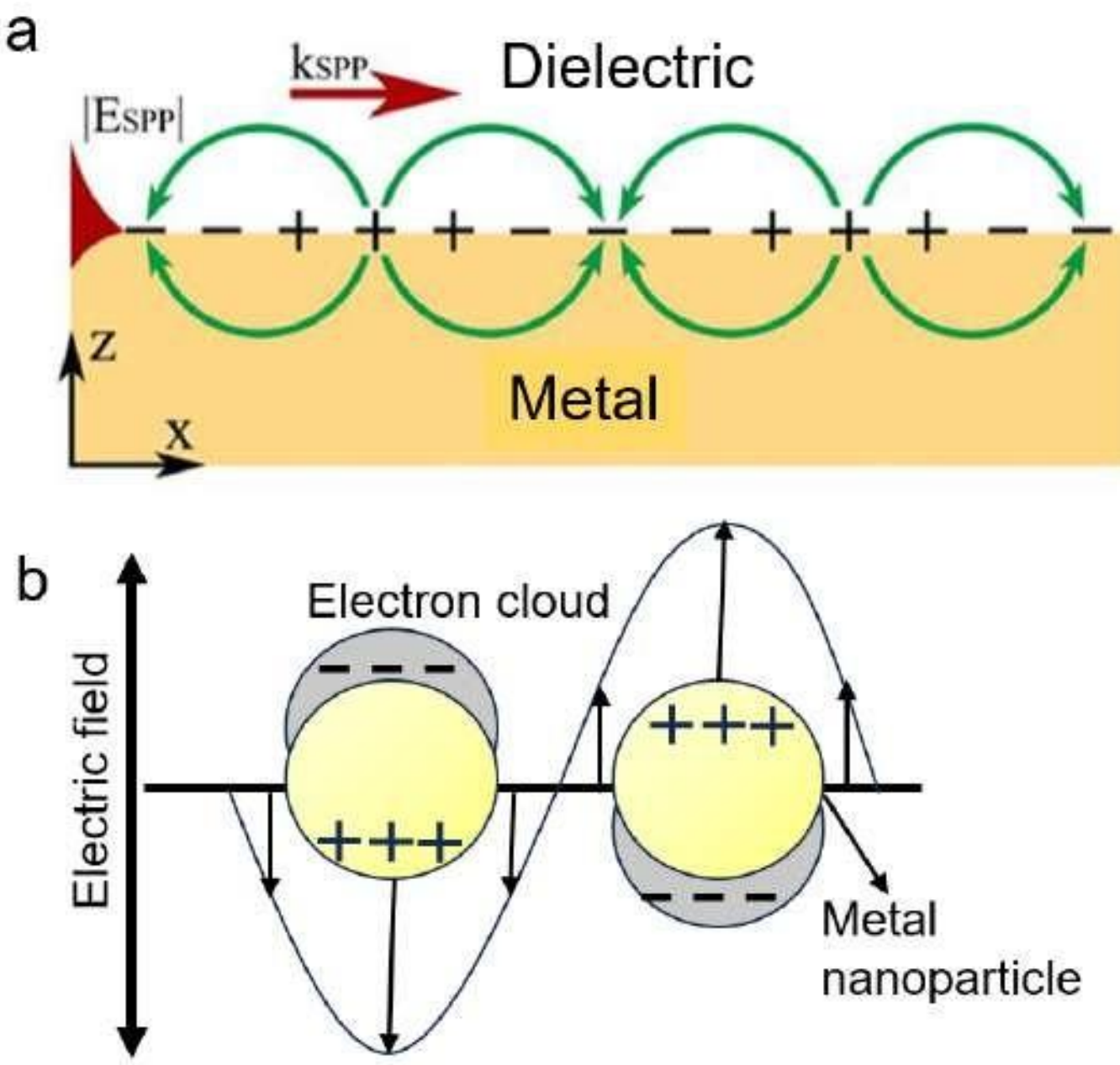

**Figure 1.5:** Schematic of (a) propagating surface plasmon polaritons, the image has been adapted from [57] (b) Localized surface plasmon polaritons at the interface between metal nanoparticles and dielectric environment.

Surface plasmon resonances (SPRs) have a wide range of applicability. Various kinds of sensors [59], such as refractive index sensors for biomolecule detection, use SPRs as they provide high sensitivity as well as selectivity. Spectral filters and color displays are also developed based on SPRs [60]. However, the biggest problem encountered in plasmonic applications with metals is the existence of high inherent losses in metals, which significantly affect their performance in terms of the quality factors of SPPs and LSPRs. To overcome this, low-loss transparent conducting oxides have emerged as the near-infrared (NIR) plasmonic materials such as indium-tin-oxide (ITO), doped zinc-oxide (AZO, GZO), doped cadmium oxide (CdO), etc [61]. In addition, transition metal nitrides like titanium nitride (TiN) have been researched for visible plasmonic properties and as a potential substitute for gold [62, 63]. The Drude and Lorentz models are used to model the optical properties in metals[64]. Similarly, due to high free carrier density, the same model can be applied to model ITO and TiN's optical properties. This thesis explores LSPRs in ITO nanostructures in the near-infrared, arising from their high carrier density aspect in developing spectrally selective optical responses as discussed later.

### 1.3.5 Epsilon-near-zero (ENZ) phenomena

Epsilon-near-zero (ENZ) materials are an exclusive category of near-zero-index (NZI) materials [65, 66] that have provided a novel platform for the development of integrated photonics and nanophotonic devices [67]. In continuous ENZ systems where the ENZ effect arises from free carrier response, the real part of the relative permittivity goes to zero at particular wavelengths known as the ENZ wavelength ($\lambda_{ENZ}$), demarcating an optical dielectric to metal transition (Fig.1.6). The ENZ regime is defined as the wavelength range where the real part of the relative permittivity varies from +1 to -1. Also, the real and imaginary parts of the refractive index exhibit the same value at the ENZ wavelength[68]. Several interesting phenomena are realized in this regime, like strong field enhancement (Fig.1.7a) and confinement [69–71], perfect absorption (Fig.1.7b) due to the excitation of ENZ modes[72], enhanced non-linear optical response[73], which makes them ideal candidates in exploring light-matter interactions. The electric field (E) enhancement in the material happens due to the electromagnetic boundary conditions[75], where the normal component of the electric field across a boundary of two materials follows: $\varepsilon_1 E_{1n} = \varepsilon_2 E_{2n}$. As the $\varepsilon_2$ value goes below 1 and approaches zero, the normal component of the field in the ENZ material $E_{2n}$ is enhanced by a factor of $\varepsilon_1/\varepsilon_2$. In addition, as $\varepsilon_2$ goes to zero, an effective field confinement is also realized inside the ENZ layer.

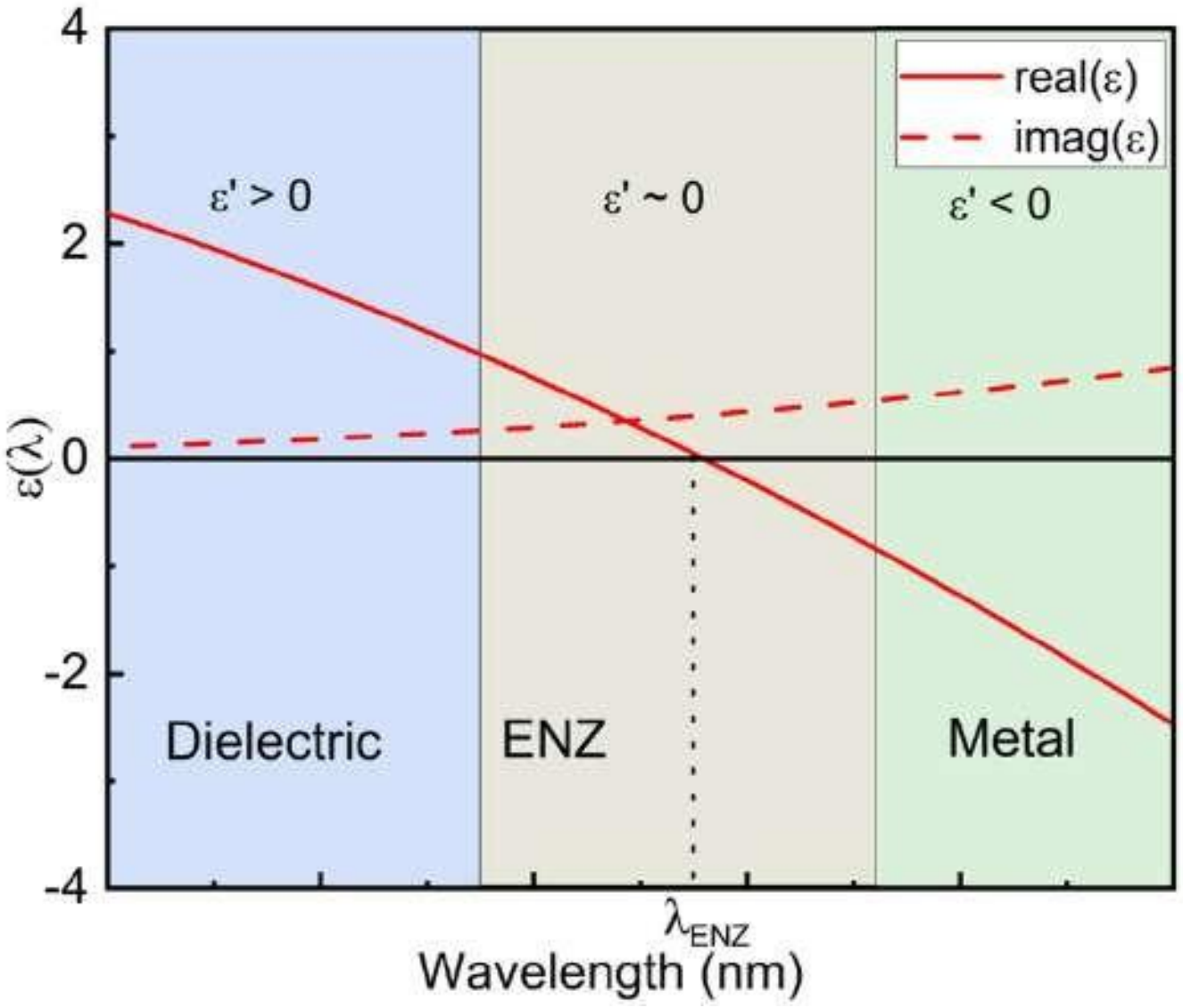

**Figure 1.6:** Plot of the dielectric function showing the real ($\varepsilon^{'}$) and imaginary ($\varepsilon^{''}$) components, highlighting the optical dielectric-to-metal transition marked by the ENZ (epsilon-near-zero) wavelength. The ENZ regime, indicated by the light brown box, corresponds to the wavelength range where the real part of the permittivity ($\varepsilon^{'}$) transitions from +1 to -1.

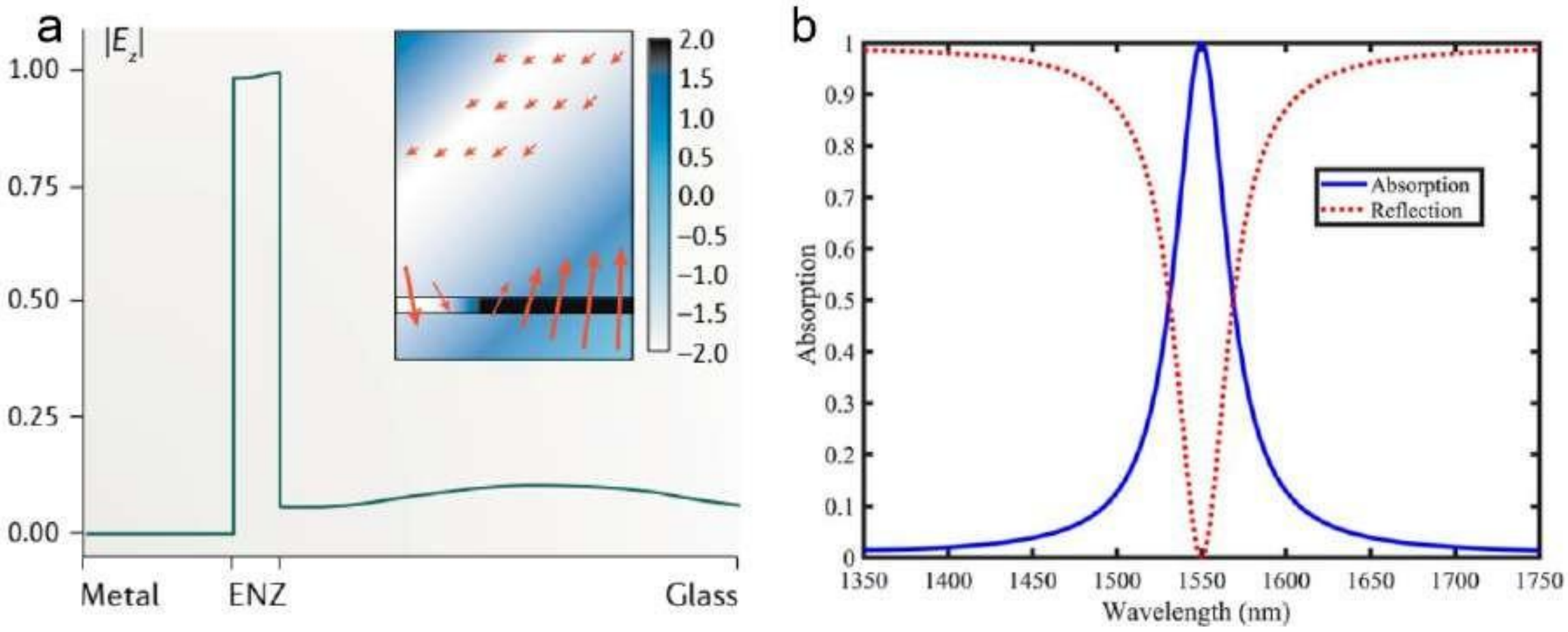

**Figure 1.7:** (a) Electric field enhancement across an ENZ film between a metal and a glass layer. The inset displays the electric field vector map under oblique plane wave excitation (as shown by the white shaded region), which couples into the ENZ layer, (b) Perfect absorption based on ENZ multilayer structure. The image (a) has been adapted from [74] and image (b) from [72]

Different mechanisms have been utilized to realize ENZ behaviour in different spectral regimes. First, as already mentioned, utilizing the collective motion of free carriers at the bulk plasma frequency in semiconductors[61]. Second, through the engineering of photonic multilayers of alternating layers of dielectric and metals, thereby creating an effective medium [76]. Third, using a metallic shell such that electromagnetic (EM) energy is forcefully confined in a waveguide below cutoff frequency to obtain a near-zero effect[77, 78]; and finally, exploiting the photonic Dirac cone in the Brillouin zone. Among these, the first mechanism is widely used in realizing ENZ phenomena because mere adjustment of the free carrier concentration can be easily realized in heavily doped silicon-based semiconductors [80], III-V group semiconductors, metals [67], and transparent conducting oxides (TCOs) [81]. Most of the ENZ materials operate in a wide range of the wavelengths, from ultraviolet (UV), visible, near-infrared (NIR), to the mid-infrared (MIR) and far-infrared (FIR), making them perfect candidates to be used in applications of visible light communications (VLC) and the IR telecommunica-

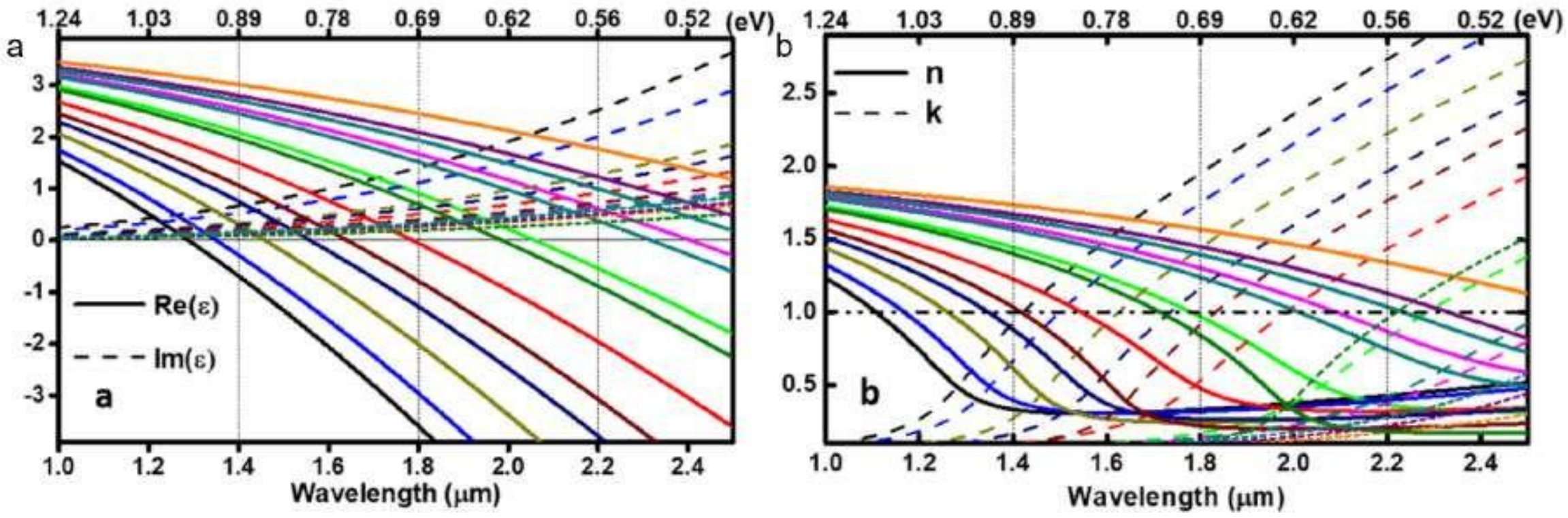

**Figure 1.8:** (a) Real and imaginary parts of the dielectric function of ITO thin film samples showing tunability on annealing. (b) Corresponding real and imaginary parts of their refractive index. The image has been adapted from [79]

.

tion windows of 1.3, 1.55, and 2 $\mu$m [82]. In metals and doped semiconductors, the ENZ property is realized due to their free carrier response. The collective oscillation of free carriers in such material systems creates the bulk plasma effect, where ENZ phenomena is the result of the carriers' response to a certain frequency range of the electric field oscillation in the EM waves. Their behaviour can be well modeled by the Drude model as discussed in section 1.5.

Inspite of having high intrinsic losses in free carrier ENZ materials, most of them are complementary metal oxide semiconductor (CMOS) compatible and therefore can be realized in on-chip integration and are suitable for nanoscale photonic devices[67]. Also, several fabrication techniques, such as magnetron sputtering, chemical vapor deposition, and atomic layer deposition (ALD) of these materials have evolved over time, which have substantially reduced their cost of fabrication. The most popular materials in this category are TCOs in which both transparency and conductivity can be realized. The dielectric permittivity and conductivity in these materials help to realize and tune the intended optical properties, such as EM field strength, phase, velocity, nonlinearity, and loss. In ENZ-related studies, the commonly used TCOs include ITO, doped

ZnO, doped CdO, which have their ENZ response mostly in NIR and IR, depending on the carrier concentration, effective mass, and loss, which determines the plasma frequency[83, 84].

Understanding the relevance of these materials and the utilization of the ENZ property and its implementation in optical devices can be done in various forms, for example, thin film, two- and three-dimensional (2D and 3D) structures, and metasurfaces [85]. Hence, recent advances in free carrier ENZ materials have focused more on the optical science and applications. Moreover, the tunability aspect via varying the carrier density (Fig.1.8) in these materials can be beneficial for the development of spectrally selective optical systems in different spectral regimes. For free carrier continuous ENZ systems, the tunability can be realized either through static methods like annealing or by dynamic methods like electrostatic gating, which effectively tunes the number density and consequently the $\lambda_{ENZ}$.
In this thesis, we have exploited the dielectric and metallic aspects of ITO as a manifestation of its ENZ behaviour in developing tunable, spectrally selective reflector and absorber coatings as discussed in Chapters 3 and 4. While a visible ENZ material, TiN has been used in boosting emission in monolayer $MoS_2$ as discussed in Chapter 5.

## 1.4 ENZ Material Systems explored in the Thesis

### 1.4.1 Indium-tin-oxide (ITO)

Indium tin oxide is an n-type semiconductor with a wide bandgap of around 3.2 eV. ITO is one of the most widely used transparent conducting oxides (TCOs) with high electrical conductivity and optical transparency well known for various industrial applications (Fig.1.9). It has a low electrical resistivity of $\sim 10^{-4}$ Ω cm, and a thin film has a high optical transmittance of greater than 80% in the visible regime[86]. These properties are utilized to great advantage in touch-screen applications such as mobile phones[87]. Moreover, ITO thin films can be prepared relatively easily with physical vapour deposition methods. Additionally, it is a chemically stable material with very little susceptibility to environmental conditions like temperature and humidity. Material properties can only be modified when annealed at elevated temperatures ($> 300^{\circ}C$)[79, 89]. However, with all transparent conducting films, a compromise is often made between conductivity and transparency if the thickness is increased, which increases the concentration of charge carriers which resulting in increased film conductivity and decreasing the transparency.

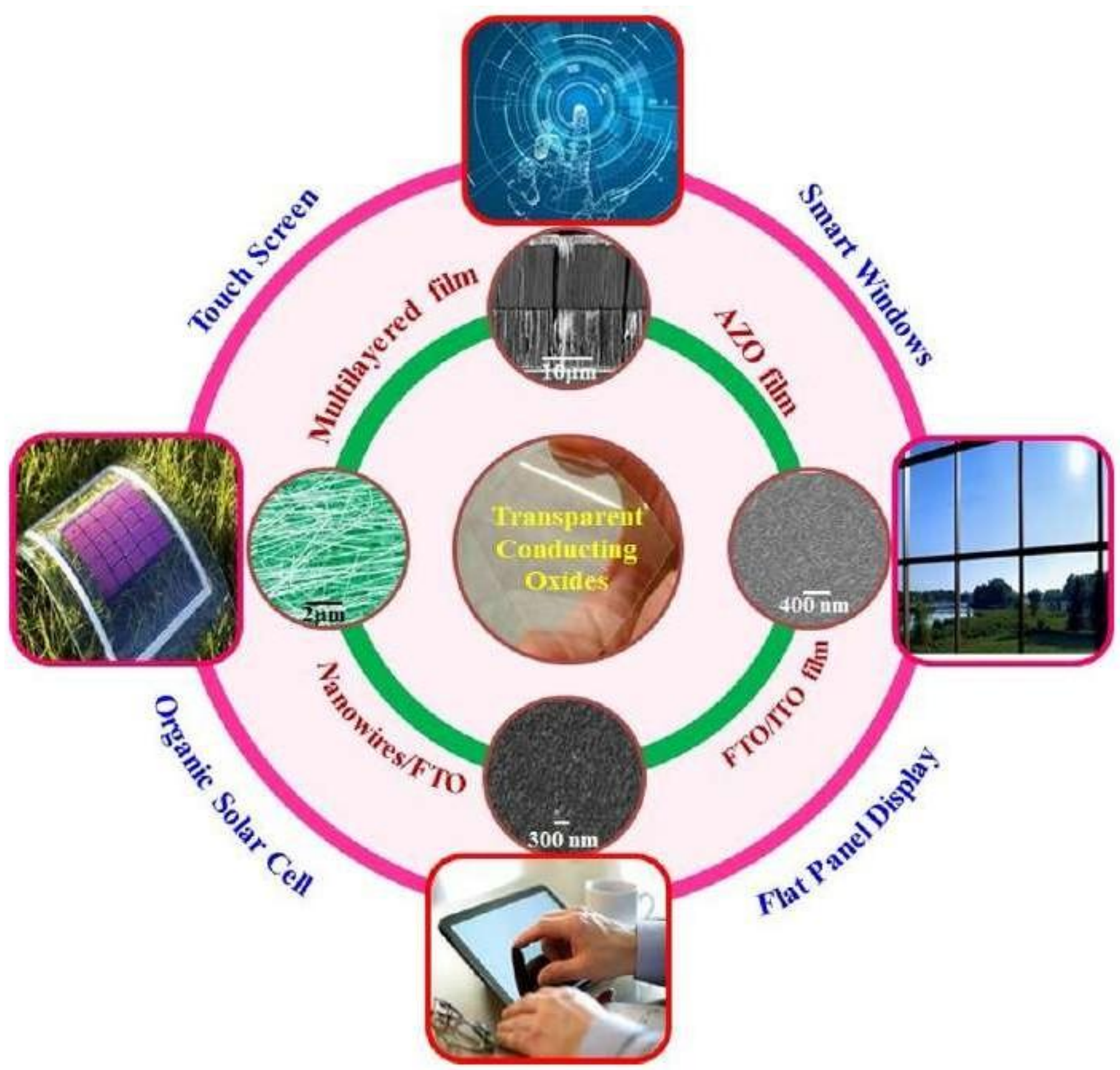

**Figure 1.9:** Schematic diagram of various transparent conducting oxides and applications. The image has been adapted from [88].

Apart from these versatile properties, ITO is a NIR plasmonic material due to high free carrier density[90]. Besides being a plasmonic material, it also exhibits the ENZ phenomena as discussed before, which originates from its high free carrier concentration ($\sim 10^{21}$/cc), making it feasible to model its optical properties via Drude model [79]. Thin films of ITO exhibit strong absorption near its ENZ wavelength ($\lambda_{ENZ}$) due to extreme field amplification and confinement under p-polarized excitation due to the excitation of ENZ and Berreman modes[91, 92]. When nanostructured, it exhibits SPPs and LSPRS in NIR because of high free electron concentration. The spectral position of the localized resonances is largely governed by the dimensions and shape of the nanostructures as well as the environment surrounding them. Moreover, the ability to tailor its optical properties via modulating its free carrier concentration through annealing or electrostatic gating serves as another important aspect in its utilization in photonics, especially spectrally selective coatings. Despite its interesting optical properties, it often becomes non-trivial to lithographically pattern these oxides into nanostructures because of their low adhesion with the substrates[93]. In this thesis, both thin film as well as nanostructured ITO have been fabricated successfully using RF sputtering as discussed in Chapters 2 and 3. ITO's ENZ and plasmonic properties are exploited extensively to develop tunable, spectrally selective coatings as described in Chapters 3 and 4.

### 1.4.2 Titanium nitride

Titanium nitride (TiN) belongs to the family of transition metal nitrides and is also classified as a refractory compound. It offers several advantages for industrial applications, including cost-effectiveness and CMOS compatibility for fabrication processes. It is chemically stable with high mechanical strength and has a melting point around $\sim 2900^{\circ}C$, which makes it an ideal candidate for high temperature applications[62, 94]. Recently, titanium nitride (TiN) has gained significant attention for its plasmonic prop-

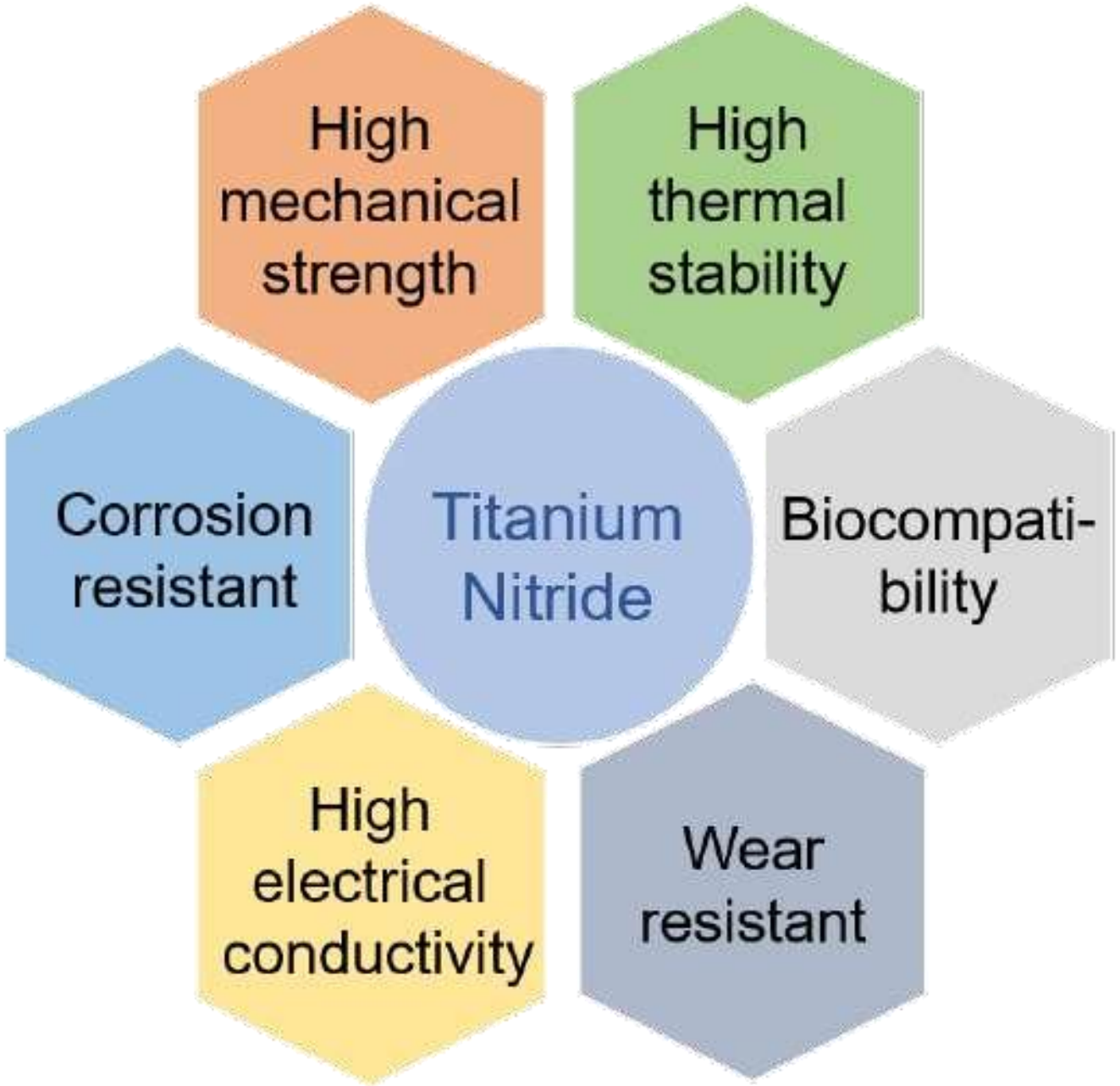

**Figure 1.10:** Various properties of TiN elucidating the relevance of the material for industrial applications. The image has been reproduced from [62].

.

erties in the visible spectrum. It has also been widely studied as a promising alternative to traditional metals due to its refractory characteristics. One notable application is in heat-assisted magnetic recording (HAMR), where gold (Au) is currently used. However, Au often experiences shape distortion under high-intensity laser illumination, highlighting the need for more thermally stable materials like TiN. In such cases, TiN performs better due to its very high melting point, which can improve the performance compared to conventional materials like Au and silver (Ag)[62]. Although TiN has a wide band gap (~3.45 eV), it has very high carrier density ($\sim 10^{22}$/cc) comparable to metals, which permits it to exhibit plasmonic behaviour in the visible [95]. Moreover, it can potentially replace metals and significantly improve the performance of devices as it exhibits lower loss[96]. In addition to the various characteristics illustrated in Fig. 1.10, TiN exhibits epsilon-near-zero (ENZ) behavior due to its high free carrier concentration. This property enhances its potential for practical applications, especially since ENZ materials are rare in the visible regime. However, while the high free carrier density is advantageous, it also leads to optical losses which increase progressively from the visible to the near-infrared (NIR) range, primarily due to interband transitions associated with these carriers. Moreover, the ENZ wavelength ($\lambda_{ENZ}$) of TiN is highly dependent on the fabrication process. Many studies have reported epitaxial growth of TiN thin films[97, 98] such that they have improved quality and hence encourage utilization in absorbers, waveguides, etc [99]. However, epitaxial growth often needs substrates that should have lattice matching with TiN. The problem of lattice mismatch leads to a search for techniques that can grow TiN without having significant dependence on the substrate material. One such method is reactive sputtering, using which TiN can be deposited onto different substrates. Even though it is not epitaxial, the thin films grown via reactive sputtering are of good quality in terms of surface roughness. In this thesis, thin films of TiN were grown on silicon (Si) and glass substrates via reactive sputtering with an average roughness of around 4 nm quantified through AFM. The IR reflectivity is around 60% as discussed in Chapter 5 and we obtained an $\lambda_{ENZ}$ $\sim$690 nm. The prepared TiN films were used to enhance the absorption and emission in a CVD-grown monolayer $MoS_2$. While monolayer $MoS_2$ exhibits very low absorption in the visible. However, a monolayer $MoS_2$/TiN on Si shows a wide-angle (0-60°) enhancement in absorption and emission as described in Chapter 5.

## 1.5 Modelling of the dielectric function of free electron ENZ media

#### 1.5.0.1 Drude Model

The Drude model was proposed in 1900 by Paul Drude to explain the transport properties of electrons in materials with high free carrier density. It is one of the fundamental theories used for expressing the complex dielectric function of metals or materials exhibiting high free carrier density. This model assumes that there are no electron-electron interactions as well as no electron-ion interactions[64]. The damping arises from collisions of the electrons with the lattice ions or other imperfections, which leads to a loss in momentum and scattering. Under the influence of a time-varying externally applied ac field ($E = E_0 e^{-i\omega t}$), the current density in a free electron system can be written as

$$\mathbf{J}(\omega) = \sigma(\omega)\mathbf{E}(\omega) \tag{1.1}$$

Under the presence of the external field, the AC dielectric function can be written as:

$$\varepsilon(\omega) = 1 - \frac{\omega_P^2}{\omega^2 - i\gamma\omega} \tag{1.2}$$

where $\omega_P$ is the plasma frequency and $\gamma$ is the scattering paramater. Plasma frequency can be expressed as

$$\omega_P = \sqrt{\frac{Ne^2}{m^*\varepsilon_0}} \tag{1.3}$$

where $N$ is the carrier concentration, $m^*$ is the effective mass, $e$ is the electronic charge and $\varepsilon_0$ is the free space permittivity. In noble metals and TCOs, further, the contribution of the core electrons is considered through the introduction of a term $\varepsilon_\infty$ which is called the high-frequency permittivity. Hence, the modified dielectric function can be written

as

$$\varepsilon = \varepsilon_\infty - \frac{\omega_P^2}{\omega^2 - i\gamma\omega} \tag{1.4}$$

where the real part of the dielectric function is

$$\varepsilon' = \varepsilon_\infty - \frac{\omega_P^2}{\omega^2 + \gamma^2} \tag{1.5}$$

and the imaginary part of the dielectric function is

$$\varepsilon'' = \frac{\omega_P^2\gamma}{\omega(\omega^2 + \gamma^2)} \tag{1.6}$$

The real part of permittivity signifies the material's polarization response to an applied field, whereas the imaginary part signifies the loss in a material when subjected to an external field. The epsilon-zero wavelength ($\lambda_{ZE}$) where the $\varepsilon'$ goes to zero is expressed as:

$$\lambda_{ZE} = \frac{2\pi c}{\sqrt{\omega_p^2/\varepsilon_\infty - \gamma^2}} \tag{1.7}$$

In this thesis, the Drude model has been used to model the material properties in ITO and TiN as they have high free electron density ($\sim 10^{21} cm^{-3}$) [79]and ($\sim 10^{22} cm^{-3}$) [95] respectively. The various other parameters have been discussed in Chapters 2, 3, 4 and 5.

## 1.6 Thermal radiation and Kirchhoff's law

Thermal radiation is defined as the emission of electromagnetic waves from all objects with a temperature greater than absolute zero [100]. It signifies the conversion of thermal energy into electromagnetic energy. The characteristics of thermal radiation depend on various properties of the surface from which it is radiating, including its tempera-ture and its spectral emissivity, as expressed by Kirchhoff's law which states at thermal equilibrium, the ratio of emissive power to absorptive power is equal for all bodies at a given temperature and is equal to the emissive power of a blackbody at that temperature. Emissivity of a material is the ratio of the energy radiated from the surface of a material to that radiated from a perfect emitter, also known as a blackbody, for the same values of temperature and wavelength and under the same viewing conditions. Emissivity does not have a dimension, and its value varies from 0 to 1. Emissivity value '0' for a material signifies that it is a perfect reflector, while that of '1' signifies a perfect emitter/absorber i.e. blackbody [101]. Absorptivity, reflectivity, and emissivity of all objects have direct dependence on the wavelength of the radiation. Absorptivity and emissivity for any particular wavelength are equal at equilibrium, and the temperature determines the wavelength distribution of the electromagnetic radiation. Moreover, for an opaque object, the emissivity ($E$) can be directly written as

$$E = 1 - R \tag{1.8}$$

where 'R' is the reflectivity of the surface. In this thesis, we will look into the emission from an ENZ grating in Chapter 4.

## 1.7 Light absorption and emission in 2D materials

Since Graphene has been discovered [102], 1-atomically thin 2D-layered materials have been in the spotlight for applications in optoelectronic devices due to their unique optical and electronic properties. Beyond graphene, various 2D materials such as hexagonal boron nitride (hBN), transition metal dichalcogenides (TMDCs), such as $MoS_2$, $WS_2$, $WSe_2$, $MoSe_2$, etc., black phosphorus (BP) have been widely investigated for their ver-satile properties [103]. Various fabrication approaches via bottom-up and top-down methods have been developed over time to obtain high-quality samples of these materials. Moreover, large quantum confinement effects in these materials have made them strong contenders as quantum materials[104].

TMDCs exhibit excitonic properties in the visible regime as a result of their unique electronic band structure and reduced dimensionality. This often leads to the formation of tightly bound electron-hole pairs (excitons) with large binding energies [105]. Most TMDCs, when they are monolayers, exhibit a direct bandgap, which boosts light absorption and emission in them. However, significant challenges often discourage their integration in practical opto-electronic, photonic, and quantum devices because of their ultrathin dimension. The inherent light absorption in these materials becomes very weak due to their limited light-matter interaction length[106]. Another significant challenge is the presence of defects introduced during the growth process, which can have considerable impact on the material's properties. These defects can substantially influence excitonic behavior, photoluminescence (PL) intensity, and optical absorption. For instance, sulfur vacancies in $MoS_2$ can act as electron donors, leading to increased electron concentration and potential quenching of PL [107]. The effects of defects are not only limited to their optical properties but also have substantial effects on the transport properties. They can affect the conductivity, carrier mobility in the samples by acting as scattering centres or charge traps[108].

Hence, to mitigate the problem of inefficient light absorption in these systems, strategies for scalable and efficient device fabrication or integration with material platforms or structures are required, which can enhance their optical and electrical properties. So far, various approaches have been undertaken to improve light absorption and emission in 2D materials. Among these, the use of localized surface plasmon resonances sustained by metal nanoparticles or patterned substrates has appeared to be an effective approach to enhance the PL in 2D semiconductors[109]. However, the rigorous sophisticated fabrication process involved increases the cost and discourages their practical applications. The TMDC explored in this thesis is molybdenum disulphide ($MoS_2$) which has been integrated with a visible ENZ material to improve its light absorption and emission characteristics.

### 1.7.1 Molybdenum di-sulphide ($MoS_2$)

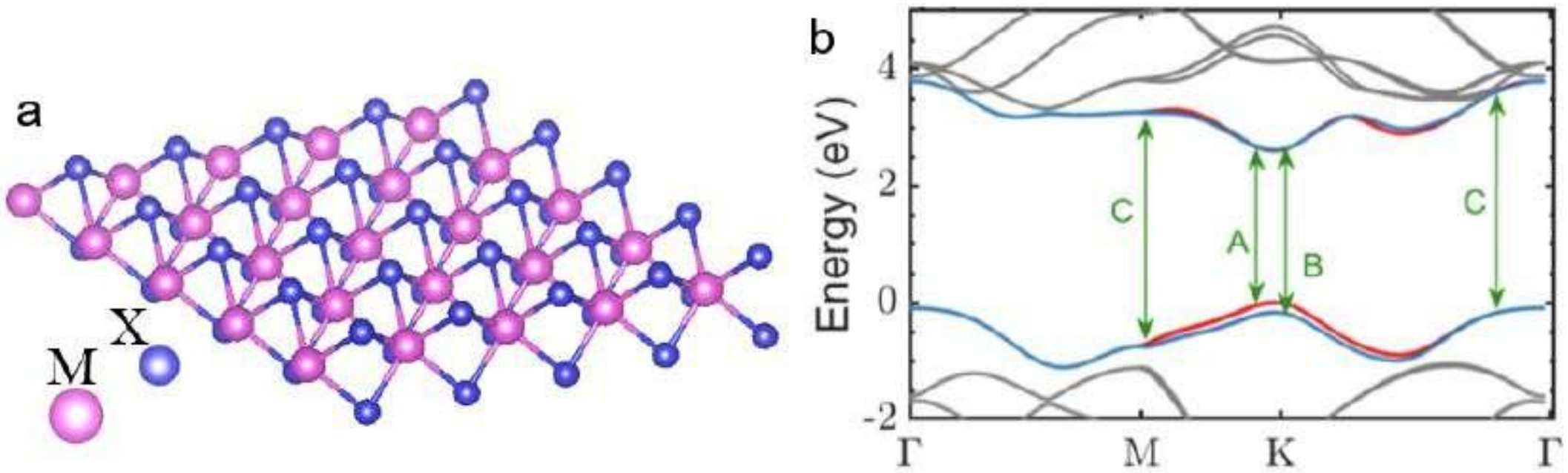

**Figure 1.11:** (a) Crystal structure of TMDCs with 'M' denoting the transition metal and 'X' denoting the chalcogen atom, (b) Band structure of $MoS_2$ showing the three excitonic (A, B and C) transitions. Image (b) has been adapted from[110]

As previously mentioned, molybdenum disulfide ($MoS_2$) is one of the most extensively studied two-dimensional (2D) transition metal dichalcogenides (TMDCs), an inorganic material analogous to graphene. It consists of layered structures held together by weak van der Waals forces, while the molybdenum (Mo) and sulfur (S) atoms within each layer are bonded covalently. Each individual $MoS_2$ layer is approximately 0.7 nm thick. However, graphene is a zero-bandgap semiconductor, but bulk, multi-layer, and few-layer $MoS_2$ are indirect bandgap semiconductors. Its band gap essentially becomes direct through monolayer growth or exfoliation from bulk [111]. A single layer of $MoS_2$ consists of a Mo atom sandwiched between two S atoms. The coordination and stacking order decide the polymorphic crystalline forms of $MoS_2$, which are 1T (tetragonal), 2H (hexagonal), and 3R (rhombohedral). The 2H phase is the thermodynamically stable phase of $MoS_2$, while the 3R and 1T phases are metastable and often found in synthetic $MoS_2$ [112]. The typical crystal structure of a monolayer $MoS_2$ is shown in Fig. 1.11a. $MoS_2$ shows unique layer-dependent optical, electronic and mechanical properties and is influenced by factors like layer thickness[113], presence of defects induced as a result of the growth process and the environment[114]. In addition, monolayer $MoS_2$ shows strong excitonic properties because of its ultrathin nature and strong Coulomb interactions, resulting in sharp three absorption peaks- A, B and C excitons (Fig. 1.11b) positioned at 1.85 eV (670 nm), 1.98 eV (627 nm) and 2.8 eV (427 nm), respectively and relative positions matches with the emission peaks[115]. Apart from excitons, phonons also play a significant role in the optoelectronic response in these systems. $MoS_2$ has two active Raman modes: $E^1_{2g}(\Gamma)$ positioned at 384 cm$^{-1}$ (in-plane mode) and $A_{1g}(\Gamma)$403 cm$^{-1}$ (out-of-plane mode). However, their absorption is often compromised due to their ultralow thickness, as discussed earlier and thereby affecting emission. This thesis addresses this challenge by enhancing the absorption and emission properties of $MoS_2$ through the use of an ENZ thin film of TiN, as detailed in Chapter 5.

### 1.7.2 Modelling the dielectric function of $MoS_2$

#### 1.7.2.1 Lorentz oscillator model

Unlike the Drude model, the Lorentz oscillator model is used to model the optical response of bound charges and vibrational resonances [64]. It involves modeling an electron as a driven, damped harmonic oscillator where the driving force is the externally applied oscillating electric field. The dielectric function according to the Lorentz model is given as

$$\varepsilon = \varepsilon_\infty + \frac{N_o e^2}{\varepsilon_0 m^*} \frac{1}{\omega_0^2 - \omega^2 - i\gamma\omega} \tag{1.9}$$

where $N_o$ is the oscillator number density, $\omega_o$ is the resonance frequency, $\gamma$ is the scattering paramater. In this thesis, we have used a superposition of Lorentzian oscillators to define a function as given in equation 1.10 to fit the three excitonic absorption peaks in $MoS_2$. Hence, we observed that the modified model can adequately fit the dielectric function of materials exhibiting excitonic resonances, such as transition metal-dichalcogenides (TMDCs) having sharp peaks.

$$\varepsilon(\omega) = 1 + \Sigma \frac{f_k}{(\omega_k^2 - \omega^2) - i\omega\gamma_k} \tag{1.10}$$

where we have defined $f_k$ to be

$$f_k = a_k \frac{Ne^2}{m^* \varepsilon_0} \tag{1.11}$$

where k is 1, 2 and 3 for the three excitonic peaks in $MoS_2$ and $a_k$ is the fitting parameter, $N$ is the number density and $m^*$ is the effective mass, $\omega_k$ and $\gamma_k$ are the frequency and linewidth of the $k$th oscillator. The parameter details are given in Chapters 2 and 5.

## 1.8 Motivation and objectives

As elucidated from the discussions in the previous sections, there is a significant attention to improve light-matter interaction, which is central to countless applications starting with optical control via designing spectrally selective coatings to optoelectronics, sensing, and even quantum technologies. To enhance effective interaction of light with a medium, various novel material platforms and designs have been explored, which can control and manipulate light at nanoscale dimensions, including confining light within specific materials. In such cases, phenomena such as the plasmonic resonances, metasurfaces, metamaterials, photonic crystals, and cavity resonances have been extensively explored to intensify interaction of light with a medium. However, all the approaches have often involved numerous material systems, which can be metallic or dielectric in behaviour, often adding to the cost of engineering these devices. In addition, huge optical losses in the conventional metals like Au, Ag, have resulted in deteriorated performances with low quality factors of LSPR etc. In this scenario, search for optical platforms where a single material system can exhibit metallic/ dielectric response across wavelengths with lower optical loss can serve as an advantage, which will not only limit the usage of numerous materials but also improve the performance. In this thesis, we have explored such materials that behave as a dielectric at shorter wavelengths and a metal at longer wavelengths, referred to as epsilon-near-zero materials. The dielectric to metallic behaviour transition is demarcated by the wavelength where the real part of the dielectric constant proceeds to zero, termed as $\lambda_{ENZ}$. As discussed, degenerately doped TCOs like ITO, doped ZnO, etc., are NIR plasmonic materials due to their high free carrier density, and further, their ($\varepsilon < 0$) also results in the ENZ response. Most importantly, they have an order of magnitude lower loss than metals. In addition, the tunability in the optical properties via simple annealing and gating protocols, together with field enhancement, perfect absorption and strong non-linearity features, makes them ideal candidates for real-world applications and their utilization in photonic devices offering a cost-effective solution. Thus, the main objectives and scope of the work emphasized in this thesis are discussed as follows:

- Investigating the epsilon-near-zero (ENZ) and plasmonic properties of indium tin oxide (ITO) in both thin-film and periodic grating configurations to develop a spectrally selective reflector coating, supported by underlayers of materials such as chromium and chromium oxide, and fabricated on commercially relevant substrates including opaque stainless steel (SS) and transparent BK7 glass. The objective was to maximize absorption in the visible ($\sim$ 90%) and reflection ($\sim$ 90%) in the IR through the multilayer developed. The emphasis was also made on the fact to obtain a sharp transition (a "step-function"-like reflection coefficient with a cut-in wavelength denoted as $\lambda_o$) from low to high reflectivity through the structure optimization of the different material layers alongwith dimensions of the ENZ grating as described in Chapter 3 of the thesis. To take it a step forward towards practical applications, the tunability aspect of the reflector coating was also demonstrated via tuning the carrier density in the ENZ ITO film and grating, and thereby $\lambda_{ENZ}$ through annealing at elevated temperatures (>300°C), validating its thermal stability.

- As discussed in the first case, the ENZ and plasmonic property of ITO was exploited to develop a tunable spectrally selective reflector coating. Next, the idea was to employ the same ENZ ITO grating to develop a tunable, band-selective absorber integrated with two simple underlayers of a dielectric ($SiO_2$) and metal (Au), the combination of which results in a sharp high absorption window ($\sim$ 90%) in IR as discussed in Chapter 4. The motivation for developing band-selective absorber coating also stems from the applications as discussed in section 1.1. Further, checking the high emission from the coating was also one of the objectives, as that would be a direct validation of Kirchhoff's law.

- After successful modulation of the optical response of surfaces, the final objective was to check the credibility of ENZ materials in boosting the absorption and emission properties of ultrathin 2D materials such as $MoS_2$ as elucidated in Chapter 5. The focus shifted to the visible regime and a material exhibiting ENZ response in the visible regime i.e. TiN was investigated as $MoS_2$ shows excitonic properties in the visible regime and therefore, how would the optical response of $MoS_2$ be modified in proximity to an ENZ material.

In all the mentioned studies, the fabrication and optimization of different structures to obtain the desired optical responses become extremely vital. Numerical investigations set the stage before delving into experiments; otherwise, the experiments become extremely tedious. The finite element modelling as well as the various experimental techniques, which have played an essential role towards the completion of this thesis, are detailed in Chapter 2. Chapter 6 ends the thesis with a summary of key findings of the various studies and gives an overview of the future direction of work.

Appendix A and B outline the related investigations that have been carried out extensively apart from the main works reported in this thesis. Appendix A investigates the strain-induced modification of local optical and electronic properties in $MoS_2$ using periodically nano-patterned substrates, while Appendix B discusses the controlled introduction of defects in $MoS_2$ via ion irradiation and thereby modification in its optical properties.

# Chapter 2

# Numerical methods and Experimental Techniques

## 2.1 Introduction

This chapter discusses the numerical and simulation methods along with the experimental fabrication and characterization techniques that have been used during the course of this thesis work. The Finite Element Method (FEM) modelling has been used in order to simulate the optical responses of the multilayer coatings, the optimization of the layer dimensions in COMSOL Multiphysics 5.3a software, as discussed in section 2.2. The experimental techniques are broadly divided into two sections: fabrication (section 2.3) and characterization (section 2.4). The major experimental tool that has been used all throughout is electron beam (e-beam) lithography for nano-patterning of surfaces and nanoscale device fabrication, elaborately discussed in section 2.3.1. Other lithographic processes are also discussed. Various physical vapor deposition methods, such as thermal evaporation and sputtering, for thin film coating are discussed in section 2.3.5. Various surface characterization techniques, such as optical microscopy, scan-ning electron microscopy (SEM), and tapping mode atomic force microscopy (AFM), utilized in this thesis are discussed in section 2.4, followed by material characterization tools such as energy dispersive x-ray spectroscopy (EDS), X-ray diffraction (XRD). Optical characterizations such as angle-resolved reflectance spectroscopy, spatially resolved photoluminescence and Raman spectroscopy as discussed in various sub-sections of section 2.4.

## 2.2 Finite Element modeling

The description of the laws of physics for space- and time-dependent problems is usually expressed in terms of partial differential equations (PDEs). These PDEs for various geometries or problems cannot always be solved by analytical methods. Instead, an approximation of the equations can be constructed on the basis of different types of discretizations. These discretization methods approximate the PDEs with numerical model equations, which can be solved using numerical methods[116]. The solution to the numerical model equations is, in turn, an approximation of the real solution to the PDEs. The finite element method (FEM) is used to compute such approximations. In this thesis, FEM calculations are performed in a software known as COMSOL Multiphysics. In this software, different modules have been developed for carrying out different numerical calculations related to various branches of physics. Starting from wave optics, semiconductors, acoustics to structural mechanics, several modules deal with models concerning various complex problems[117] that cannot be solved analytically.

### 2.2.1 Optical response calculation

In this thesis, the wave optics module in COMSOL Multiphysics 5.3a has been used explicitly to calculate the reflectivity, transmissivity, and absorptivity from the multi-layer coatings as discussed in Chapters 3 and 4. Maxwell's equations forms the heart of all the calculations performed in the wave optics module. A schematic of the model geometry mainly explored in this thesis is shown in Fig.2.1. A plane wave($E_0 e^{ikr}$) is incident at the input port normally or obliquely and after interaction with the structure (different layers and nanostructures),it exits from the output port as shown in Fig.2.1. All the calculations are done with periodic boundary conditions (Floquet periodicity) to mimic the real-life structure for an array of nanostructures. To solve for reflectivity and transmission, the main parameters that are fed to the model are the refractive index ($n$) and extinction coefficient ($k$) for each material. Each material is assigned as individual domain in the model and they are considered isotropic. Also, each domain are divided into discrete units via creation of "mesh". The meshing is vital to simulate the model correctly as it is largely dependent on the dimensions of each layer. Most of the models solved in this thesis were '2D', hence a triangular mesh was used as shown in Fig.2.2. But when 3D calculations were conducted, both mapped and triangular meshing were used.

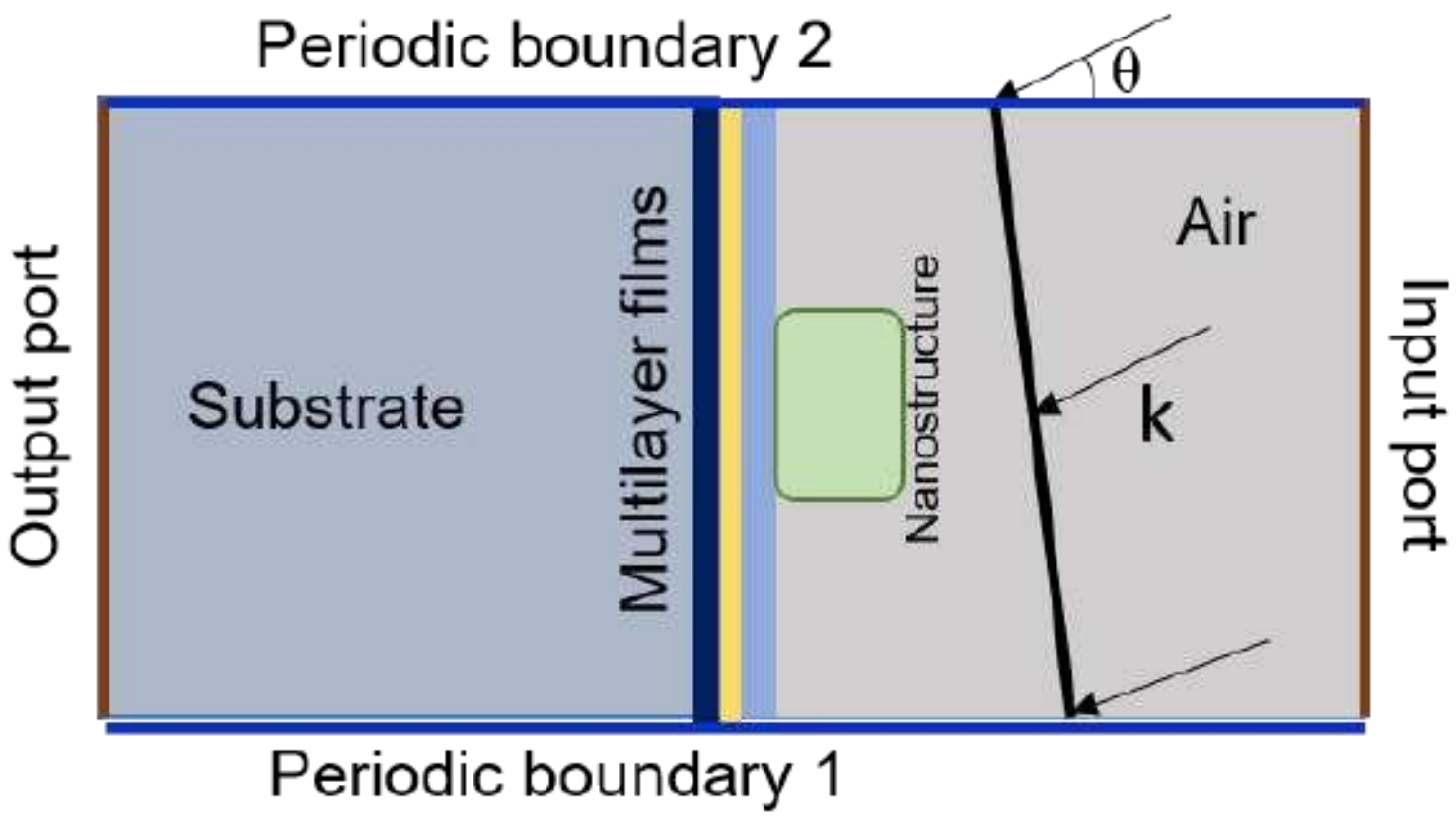

**Figure 2.1:** Schematic of a typical model investigated in the COMSOL Multiphysics to calculate reflectivity, transmissivity, and absorption from the layered structure.

Reflectivity calculations are performed for different polarizations of the input field. In this thesis reflectivity is calculated for both s-polarized and p-polarized incident light as discussed in Chapters 3 and 4 and 5. The reflectivity is calculated from the S parameters of the input port which is the ratio of the field propagating to the direction of the input port due to reflection from the structure ($E_{ref}$) and the incident field ($E_{in}$) given as

$$S_{11} = \frac{E_{ref}}{E_{in}} \tag{2.1}$$

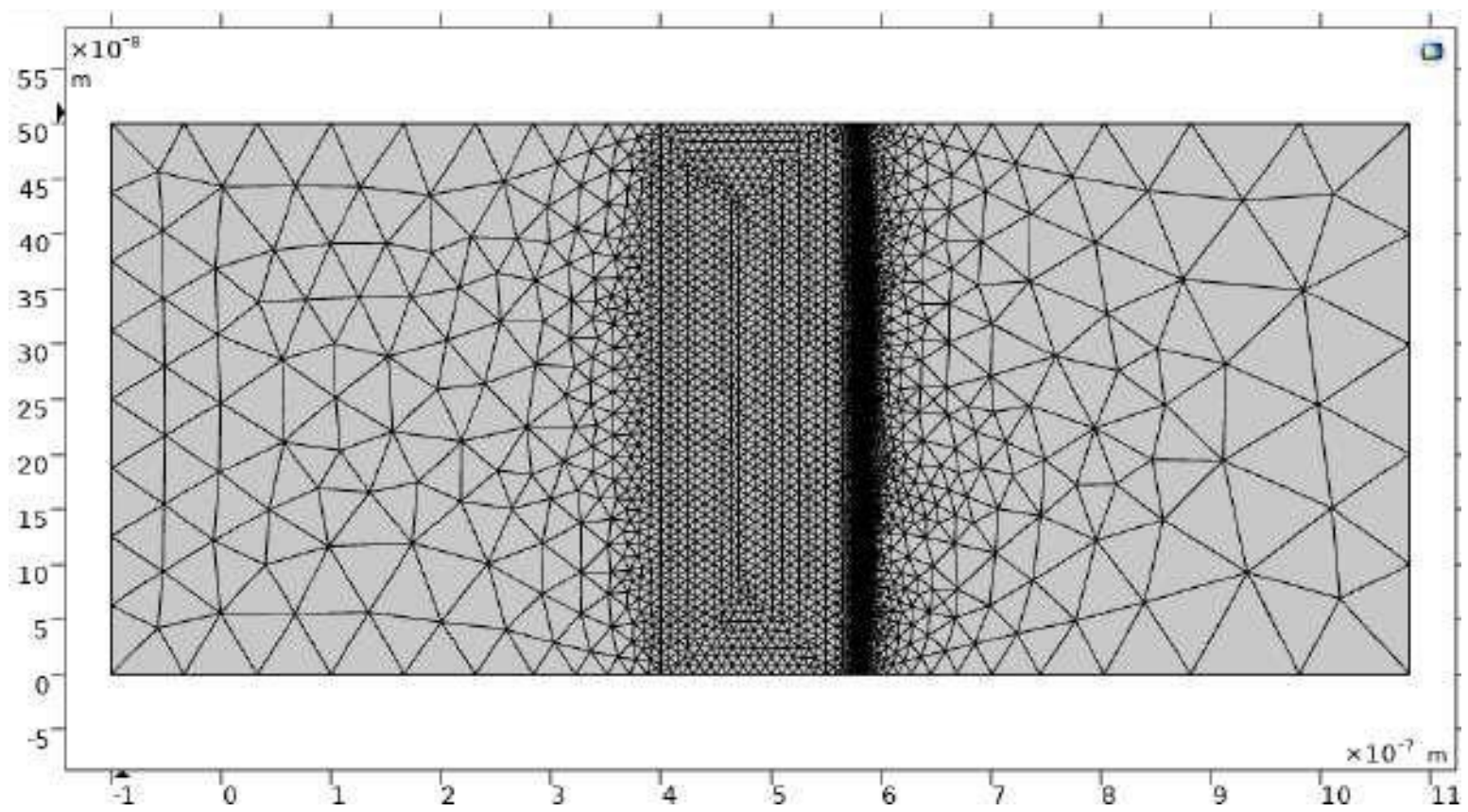

**Figure 2.2:** Cross-sectional view of a model investigated in the Wave Optics module in COMSOL Multiphysics showing the typical mesh used. The mesh density was varied according to the domain dimension.

The reflectivity (R) is calculated to be $R = |S_{11}|^2$. The transmissivity is also calculated from the S parameter of the output port as the ratio of the field exiting after interaction with the structure ($E_t$) to the input field ($E_{in}$) given as

$$S_{21} = \frac{E_t}{E_{in}} \tag{2.2}$$

The transmissivity (T) is calculated to be $T = |S_{21}|^2$. The absorptivity is calculated to be $A = 1 - R - T$.

### 2.2.2 Electric field distribution and Power dissipation calculations

While performing the reflectivity calculations it solves for the electric field. Hence, it returns the magnitude of the electric field distribution (ewfd.norm**E**) across the structure. It also gives the individual electric field components for x, y and z directions. The electric field distribution solved for a multilayer coating with a grating is shown in Fig.2.3a. The strong electric field near the nanostructure helps to understand that there are resonances in the vicinity of the nanostructure, as discussed in chapter 3. It also generates the power dissipation (ewfd.Qh) distribution as shown in Fig.2.3b for the different multilayer structures to comprehensively understand the role of each layer in developing the particular optical response. The spectral power dissipation density ($W/m^3$) is given by the equation

$$P_D = \frac{1}{2}\omega\varepsilon''|\mathbf{E}|^2 \tag{2.3}$$

where "$\omega$" is the frequency, $\varepsilon''$ is the imaginary part of relative permittivity of each system used in the model and the **E** is the magnitude of the electric field. Apart from visualizing the distributions in electric field and power dissipation, we can also obtain the integrated/maximum/minimum values of electric field or power dissipated over particular domains in the model as discussed in chapters 3 and 4.

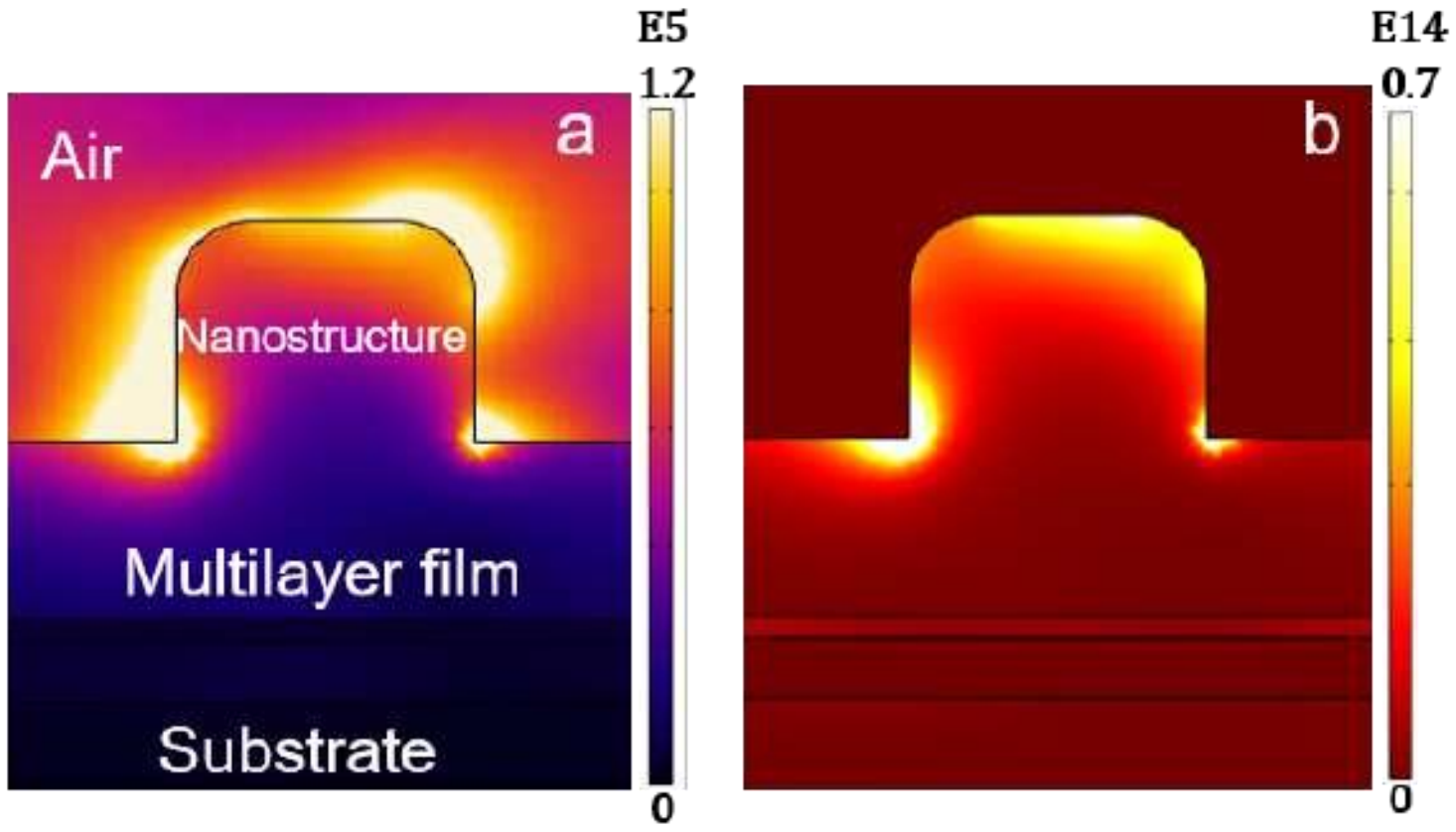

**Figure 2.3:** (a) Electric field (V/m) distribution and (b) Power dissipation ($W/m^3$) across a multilayer structure investigated.

### 2.2.3 Calculation of extinction cross section

In chapter 4, the extinction cross-section ($\sigma_{ex}$ of the ITO grating/$SiO_2$ and ITO grating/air was calculated through a 3D model in the same wave optics module. It basically calculates the scattering cross-section($\sigma_s$) and the absorption cross-section and takes a sum of the two to give the extinction cross-section($\sigma_{abs}$). The scattering cross section is given as

$$\sigma_s = 1/I_0 \iint n \cdot S_s dS \tag{2.4}$$

where **n** is the normal vector pointing outward from the grating which is the scatterer, $S_s$ is the scattered intensity (Poynting) vector,and $I_0$ is the incident intensity. The closed surface integral over the scatterer yields the scattering cross-section. On the other hand, the absorption cross section is given as

$$\sigma_{abs} = 1/I_0 \iiint Q dV \tag{2.5}$$

where Q is the power dissipation/loss density in the grating and a volume integral of that. Hence, the extinction cross section is

$$\sigma_{ex} = \sigma_s + \sigma_{abs} \tag{2.6}$$

### 2.2.4 Materials used and modelling parameters

The Drude model was used to model the dielectric function of ITO and TiN as discussed in Chapter 1 to solve the model in COMSOL, while the superposition of Lorentz oscilla-tors was used to define the dielectric function of $MoS_2$. The various parameters such as carrier concentration $N_e$, scattering parameter $\gamma$, and background permittivity $\varepsilon_\infty$, effective mass $m*$ used for the different materials are tabulated in Table 2.1. The origin of the various parameters and the corresponding equations are discussed in sections 1.5 and 1.7 of Chapter 1.

| Material parameters | ITO ($\lambda_{ENZ}$=1220 nm) | TiN ($\lambda_{ENZ}$=690 nm) | $MoS_2$ |
|---|---|---|---|
| $N_e$ ($cm^{-3}$) | 1.04 x $10^{21}$ | 1.3 x $10^{22}$ | 4 x $10^{12}$ |
| $\gamma$ (eV) | 0.1 | 0.6 | 0.059, 0.12, 0.37 [118] |
| $m^*$ | 0.35 [119] | 1.09 | 0.54 [120] |
| $\varepsilon_\infty$ | 3.9 [121] | 4.564 [122] | - |
| $\omega_k$ (eV) | - | - | 1.85, 1.98, 2.877 [118] |

**Table 2.1:** The material parameters used for ITO, TiN, and $MoS_2$ to model their dielectric function.

## 2.3 Experimental Fabrication Techniques

### 2.3.1 Electron beam lithography

Electron beam (E-beam) lithography is one of the most effective technique used for designing structures at nanoscale dimensions. It has gained wide significance due to its ability to fabricate devices with scalability, especially in semiconductors, optoelectronics, or quantum technologies [123]. This technique has huge utility for global micro- and nano-scale device fabrication, serving as the backbone of complex technological processes like in photonic integrated circuits, sensors, quantum devices, etc. It is a powerful technique for patterning nanostructures on a wide variety of substrates[124]. It is possible to create nanostructures having dimensions even around 10 nm by this technique. This method was developed in the late 1960s by modifying the design of scanning electron microscope (SEM). In this process, both SEM and light microscopes (LM) are used for controlling and investigating the developed nanostructures on the sample. It is a high-resolution technique in which electrons with high energy (10 – 100) KeV focused into a narrow beam and exposed on the wafer uniformly coated with electron-sensitive resist [125]. Fig. 2.4 shows the image of the e-beam lithography instrument (Raith Pioneer 2) used for all the nano-fabrication in this thesis.

#### 2.3.1.1 Scanning electron microscope

A scanning electron microscope uses a focused beam of electrons to scan the surface of a sample to create an image. The electrons interact with the sample, producing various signals which can be utilised to obtain the surface topography and composition. Optical microscopes cannot serve the purpose of giving the required resolution to image the micron/nano-dimensional particles. That is why electrons are needed for imaging, as it has a shorter wavelength ($\sim$ 10 nm), which gives higher resolution. Electrons are generated from an electron gun (thermionic /field emission) which is maintained at $\sim 10^{-10}$ mbar pressure, and are accelerated downwards. The electron beam passes through a combination of apertures and electromagnetic lenses and a focused electron beam is produced, which hits the surface of the sample as shown in Fig. 2.5. The sample is mounted on a stage called stub in the sample chamber maintained at $\sim 10^{-6}$ mbar.A combination of pumps (scroll, turbo and ion pump) is used to evacuate both the column and the chamber. The position of the electron beam on the sample is controlled by scan coils located above the objective lens. The beam is scanned over the surface of the sample with the help of these coils. This scanned beam gives us the information about a defined area on the sample and the image is collected. When an electron beam interacts with the sample surface, signals like secondary electrons are produced, back-scattered electrons and characteristic x-rays are generated.

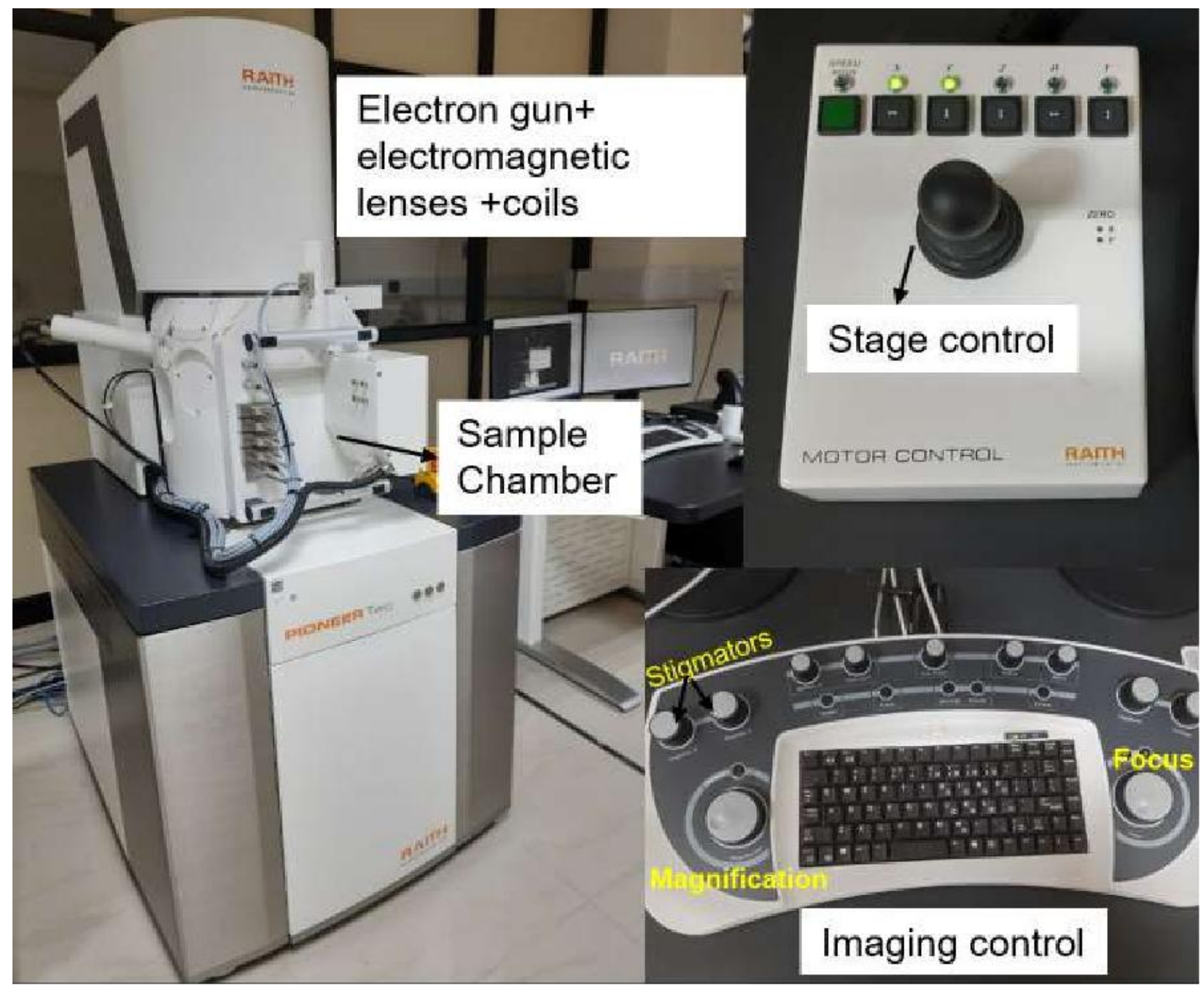

**Figure 2.4:** Image of the e-beam lithography instrument (Raith Pioneer 2) with magnified view of the operational units

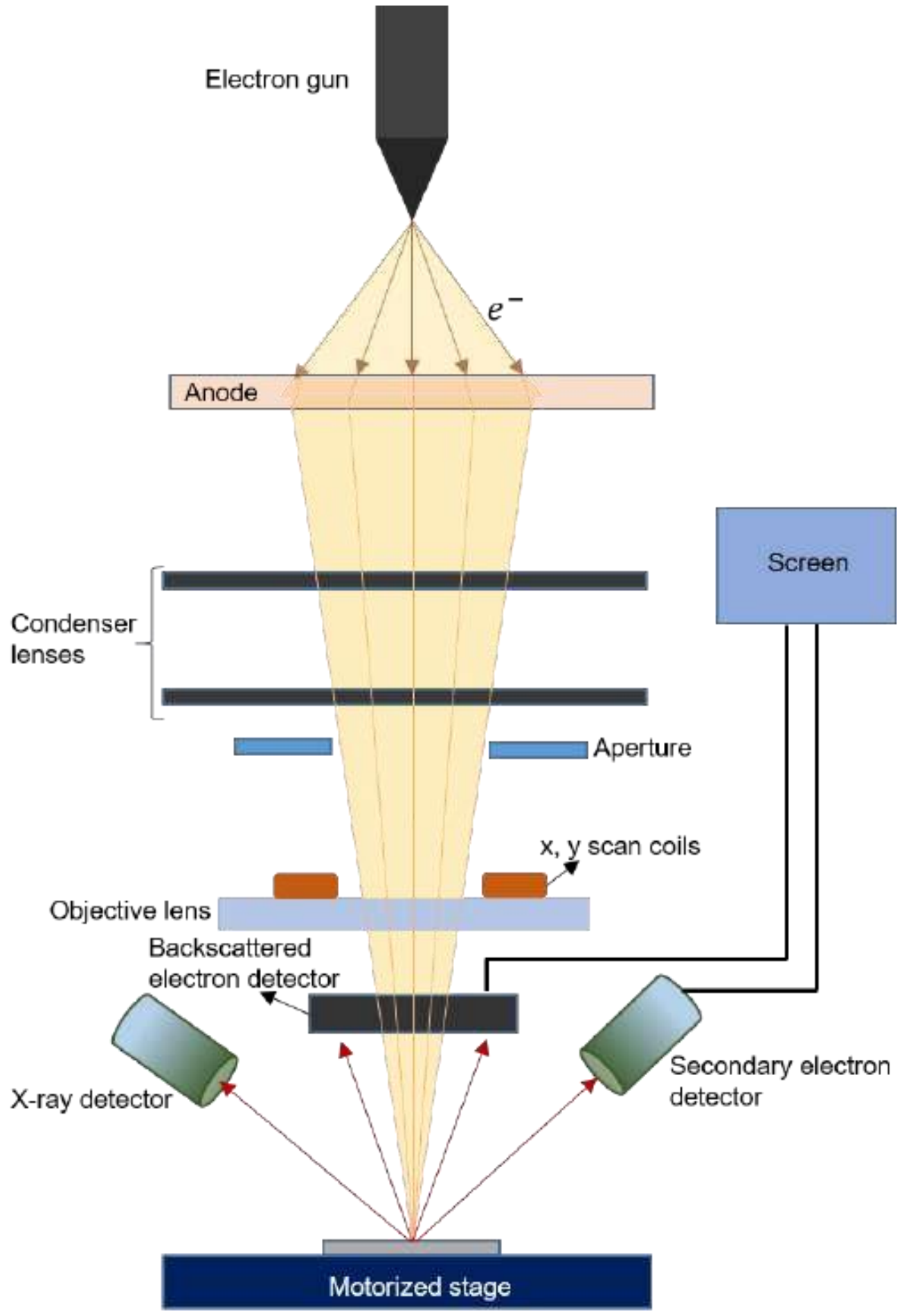

**Figure 2.5:** Schematic of a typical scanning electron microscope showing the various operational units.

Specific detectors like Everhart-Thornley detector are used to detect the secondary electrons. Back-scattered electrons and the characteristic x-rays are also separately detected to form secondary or back-scattered images, which are then displayed on the computer screen. When an electron beam is incident on the surface of the sample, a penetration into the sample occurs to a depth of a few microns. It depends on the accelerating voltage (EHT) and density of the sample and also the dose (charge per unit area) of the incident electron beam. The resolution power depends on multiple factors such as the electron spot size, electron beam dose, etc., incident on the sample. The typical resolution that can be achieved with an SEM is 1-20 nm. Besides being a sophisticated imaging tool, it also performs the task of nano-patterning when integrated with an E-beam lithography module. In this case, a focused beam of electrons exposes a substrate coated with electron-sensitive material, thereby rupturing the bonds, which become soluble in a developer. Further metal coating and lift-off or etching is done to obtain the desired structure.

#### 2.3.1.2 E-beam lithography sample preparation

In this thesis, fabrication of periodic structures of Au, ITO and TiN is done using electron beam lithography. Field-effect transistor (FET) devices on strained $MoS_2$ were also fabricated using this technique. Fig.2.6 shows the schematic of the entire lithography process. Prior to patterning, the substrates are coated with a custom-made e-beam resist. We used poly (methyl methacrylate) (PMMA 120k) powder dissolved in anisole as the positive e-beam resist. For the preparation of different concentrations of the resist, we used the following equation:

$$m_{solute} = \frac{Cm_{solvent}}{100 - C} \tag{2.7}$$

where $m_{solute}$ is the mass of the PMMA powder to be dissolved in a particular mass of anisole ($m_{solvent}$) for a given concentration $C$. Most of the e-beam patterning was conducted with 4 wt%, 8 wt% and 10 wt% PMMA resist concentrations. With 4% PMMA, spin-coated at 4000 rpm for 30 seconds, a thickness of 110 nm was obtained, and it helped to obtain structures with dimensions 100 nm- 500 nm and height around 50 nm post-exposure, deposition and lift off. For 8 wt% and 10 wt%, the thicknesses with the same spin coating recipe, were obtained to be around 280 nm and 390 nm, respectively as shown in Fig.2.14. These two thicknesses helped to obtain structures with a thick-ness ~200 nm after deposition and lift-off. Following spin coating, the samples were also baked at 140 $^{\circ}C$ for 10 minutes before e-beam exposure.

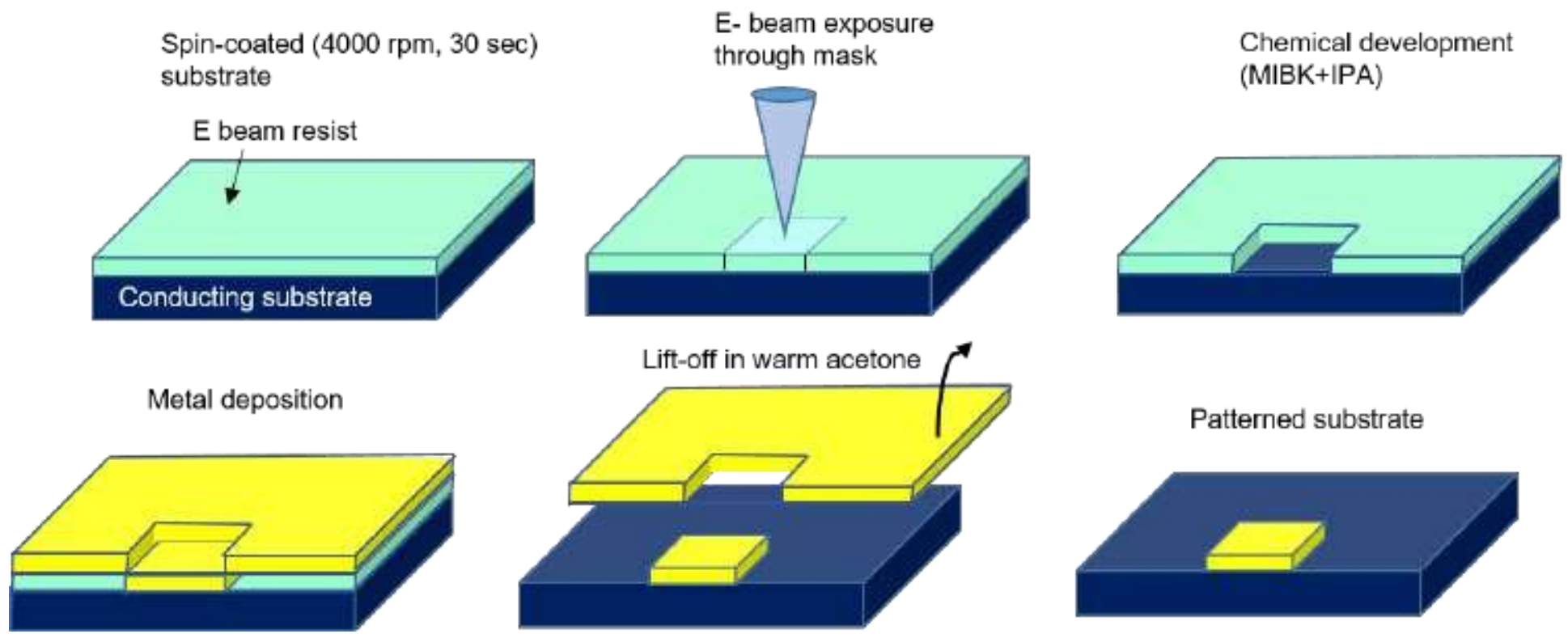

**Figure 2.6:** Schematic of the steps followed during electron beam lithography.

#### 2.3.1.3 E-beam lithography parameter Optimization

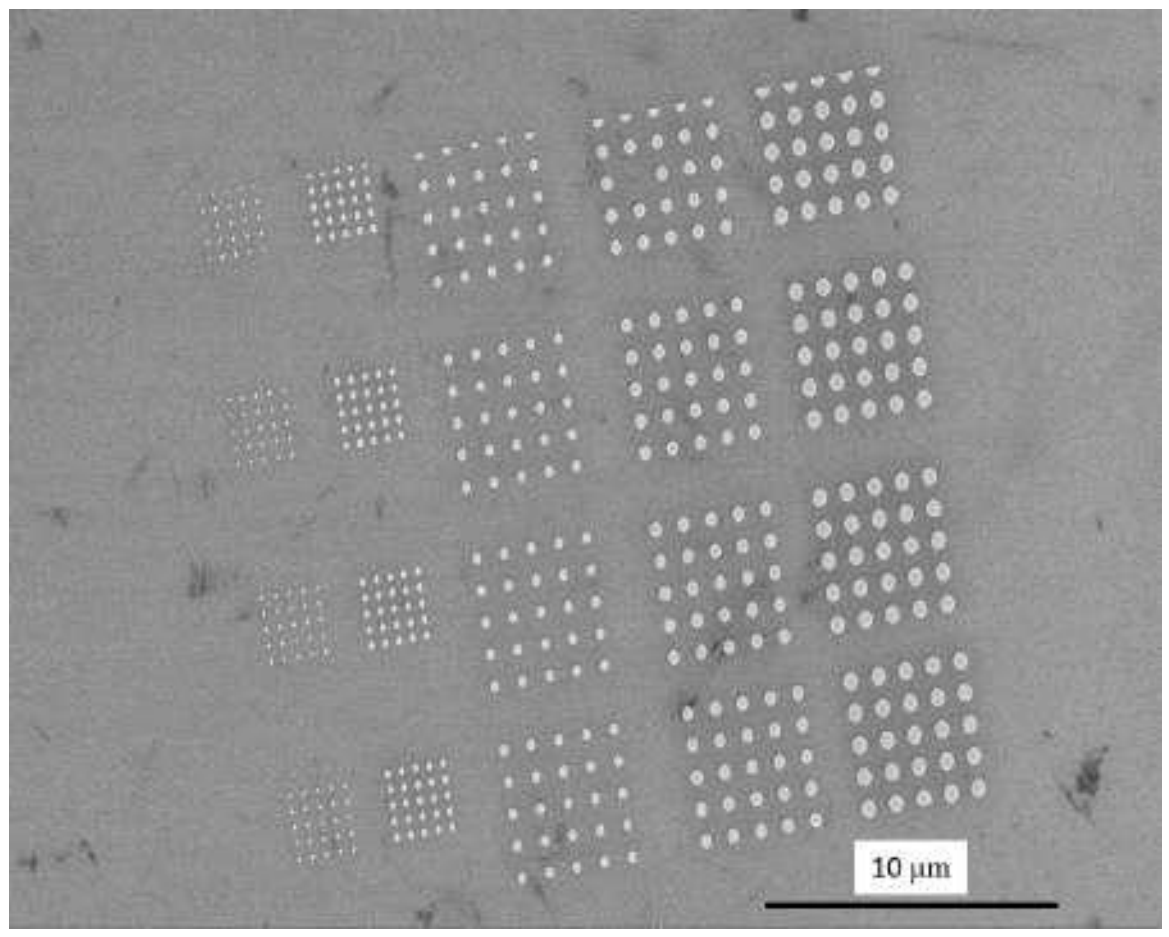

**Figure 2.7:** SEM images of various dimensions of periodic nanopatterns fabricated via e-beam lithography with electron dose varying from 110 $\mu C/cm^2$ to 140 $\mu C/cm^2$ (top row to bottom row)

After coating with PMMA, the samples are fed to the e-beam chamber for patterning. The patterns are created through e-beam exposure via a mask which are designed by software such as CLEWIN Layout and K Layout. The various parameters that were optimized to obtain the desired dimensions were:**electron dose; write-field; PMMA**

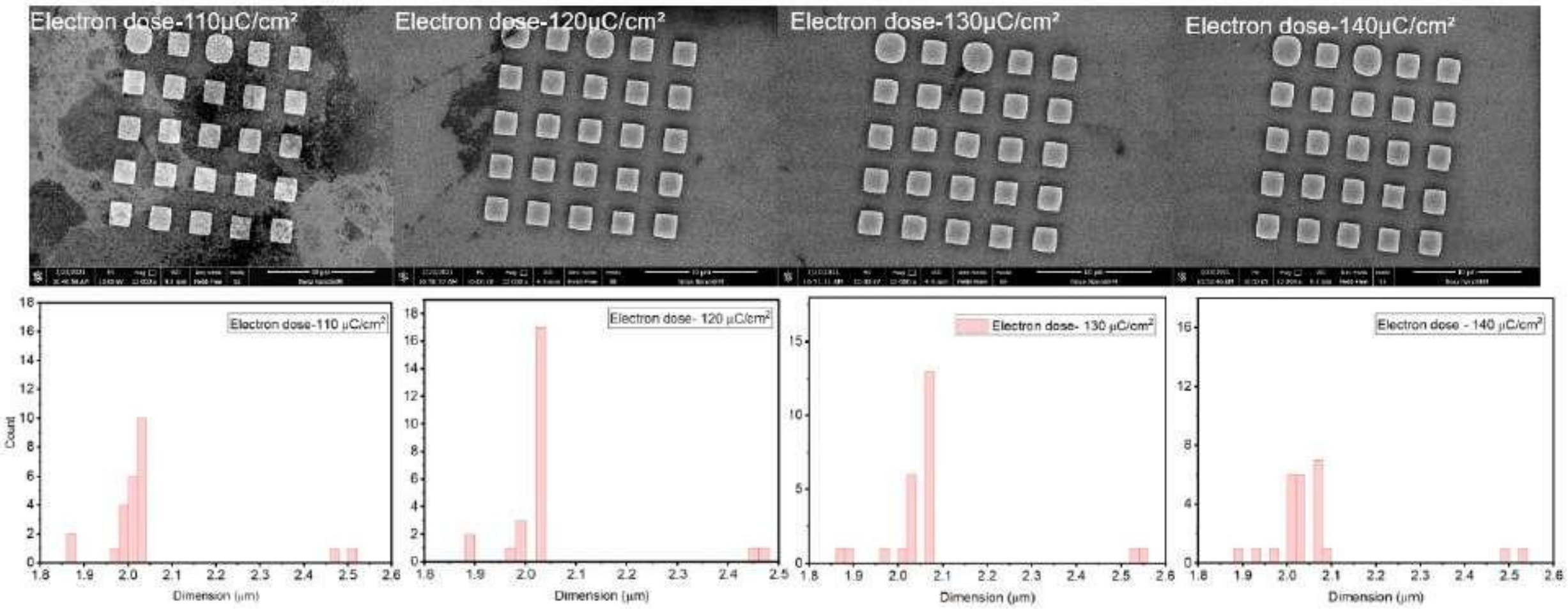

**Figure 2.8:** SEM images of repeating square patterns of intended dimension of 2$\mu$m written with variation in electron dose with statistical analysis of the dimension to understand the optimized dose.

**thickness and electron high tension (EHT) or the accelerating voltage.** Now, two sets of optimizations were conducted. One for the very low dimension structures (50-500 nm), the other for the higher dimension (> 1 $\mu m$). The electron dose was varied from 100 $\mu C/cm^2$ to 140 $\mu C/cm^2$ with a step size of 10 $\mu C/cm^2$, keeping the EHT fixed at 10 KV and write field at 25 $\mu m$ to get sub-100 nm structures. Fig.2.7 shows an SEM image of repeating patterns with different dimensions (100-500 nm) written with different electron doses for optimization to obtain the desired dimension. We obtained 120 $\mu C/cm^2$ to be the optimum dose for the lower dimension structures with periodicity (center to center distance) being 300 nm. A typical example of optimization is shown in Fig.2.8 where we see the maximum count of the intended pattern dimension is obtained with 120 $\mu C/cm^2$. For the higher dimension structures we obtained 135 $\mu C/cm^2$ to be the optimum dose with write field varying from 50 $\mu m$ to 1 mm. Note that the effective electron dose has an error of around $\pm 5$ $\mu C/cm^2$ after the software calculates the area dwell time with the input dose and area step-size of writing. Post exposure, the samples were developed in a solution of methyl isobutyl ketone (MIBK) and isopropyl alcohol (IPA) in a volumetric ratio of(1:3) for 10 seconds and washed in IPA for 10 seconds to stop the development process. Post-development, the substrates were coated with gold or indium-tin-oxide using thermal evaporation or sputtering (techniques are discussed later) followed by lift-off in warm acetone. Various geometries of patterns were fabricated during the initial optimization process as shown in Fig.2.9 in order to check the flexibility of the parameters that are optimized. The smallest dimension achieved was 70 nm and periodicity/separation of 250 nm.

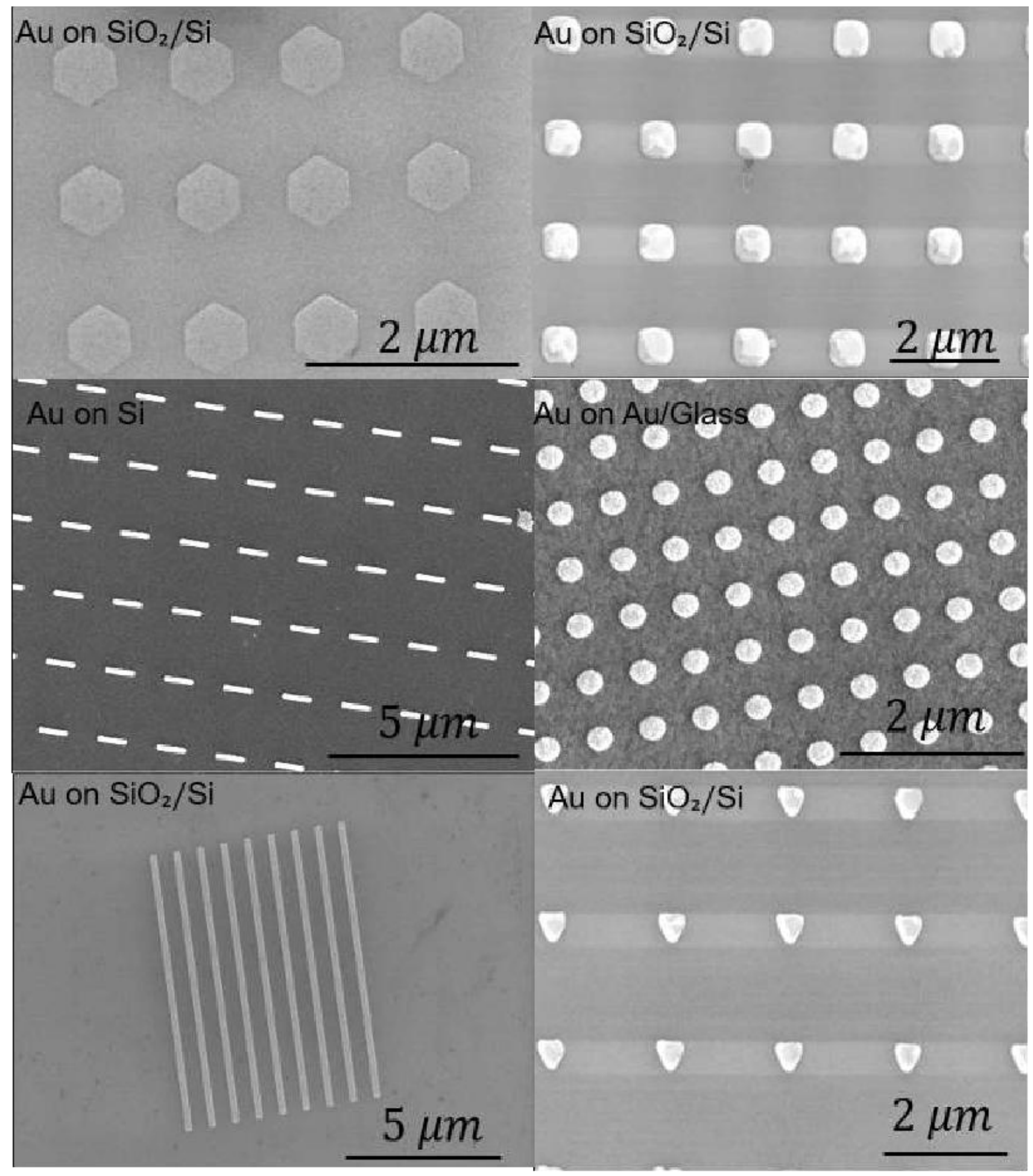

**Figure 2.9:** SEM images of various geometries (hexagonal, square, rectangles, circular, ridges and triangular) of nano patterns with varying dimensions to check the robustness of the optimization.

### 2.3.2 Photolithography

Photo-lithography is also a method of surface patterning. Mostly, a UV source is used to expose a surface through a photosensitive resist to draw a particular pattern. It has a lower resolution compared to electron beam lithography and is effective for sub-micron structures. Mostly, large electrodes are patterned through photolithography as it is a faster process and is economical. In this thesis, large contact pads of gold for the field-effect-transistor devices on strained $MoS_2$ were fabricated using a positive photo resist (SU 1813). The pohotoresist was spin coated on the substrates at 5000 rpm for 30 sec followed by baking at 140°C for 1 min. The resist-coated substrates were exposed to UV LIGHT (385 nm) through a photo-mask. Further, the exposed samples were developed using Tetramethylammonium Hydroxide (TMAH) and washed with DI water. Au was deposited on the developed patterns and finally, lift-off was performed in warm acetone to get the final structure. The FET devices are shown in Appendix A.

### 2.3.3 Nanosphere lithography

Nanosphere lithography (NSL) is a cost-effective, simple nano-fabrication technique that is used for surface patterning over large areas (hundreds of microns). We utilized polystyrene latex spheres of diameter 200 nm to create a self-assembly on Si wafer as shown in Fig.2.10a. Later 100 nm gold was evaporated using thermal evaporation and post-deposition the latex spheres were lifted-off using Scotch tape leaving the gold in the interstitial gaps as shown in Fig.2.10b. This method was explored as an alternative for the creation of plasmonic meta-surface.

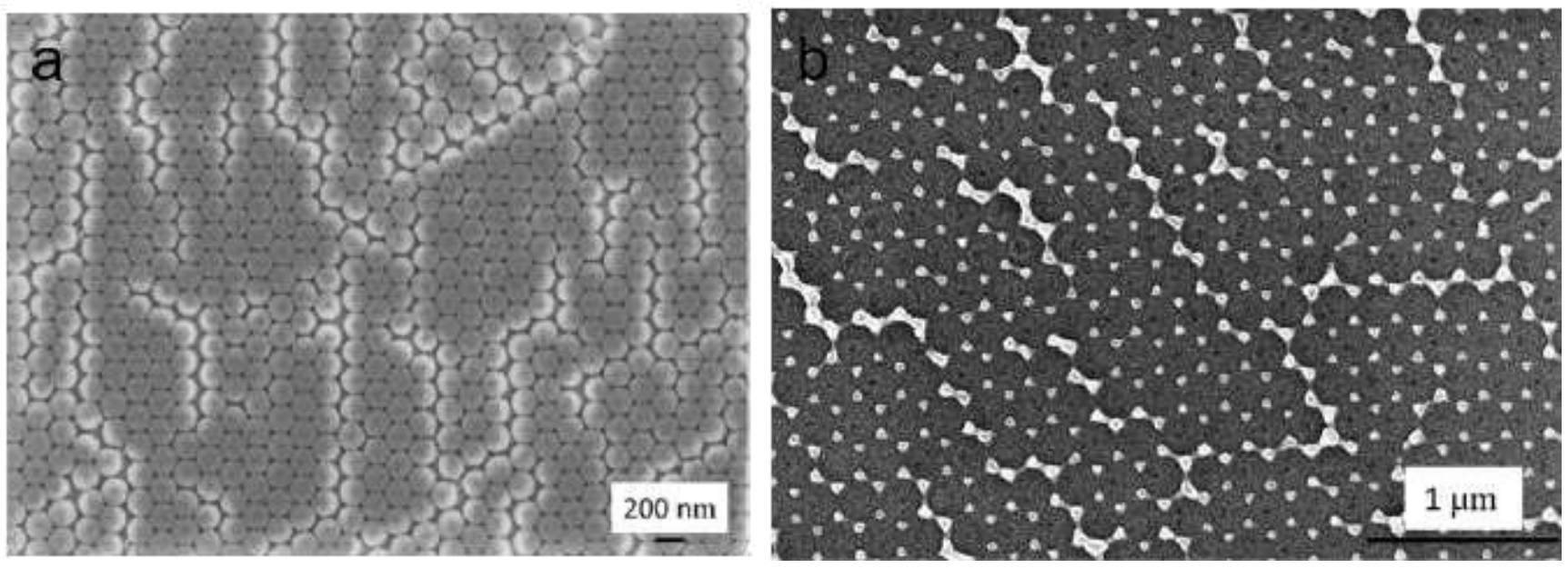

**Figure 2.10:** SEM image of (a) self-assembly of polystyrene latex spheres of diameter 200 nm on $SiO_2$/Si substrate and (b) Nano-star pattern obtained after metal (Au) deposition and lift-off by scotch-tape.

### 2.3.4 Transfer of $MoS_2$ on patterned substrates

CVD-grown monolayer $MoS_2$ flakes were transferred on TiN/Si substrates using the PMMA-based wet etch transfer method as discussed in Chapter 5. In this method, CVD substrates ($MoS_2$ on $SiO_2$/Si) were spin-coated with 10 wt% PMMA. The PMMA-coated substrates were put in a 2M NaOH solution and heated at 60°C until the $SiO_2$ layer was completely etched, and the PMMA film detached along the flakes. Further, the PMMA film with the flakes was transferred to a DI water bath to get rid of NaOH. Finally, the film was transferred onto the TiN substrate with Si windows such that the flakes fall partially on TiN and Si. The PMMA film was dissolved in warm acetone and flakes remained on the TiN. Also, $MoS_2$ was transferred onto different geometries of nanostructures to study strain-induced modification of optical and electronic properties. Also, the formation of various strain features like wrinkles and bubbles is studied with variation of the geometry of patterns as shown in Fig. 2.11 where we found that wrinkles originate from the corners of nanostructures as discussed in Appendix A.

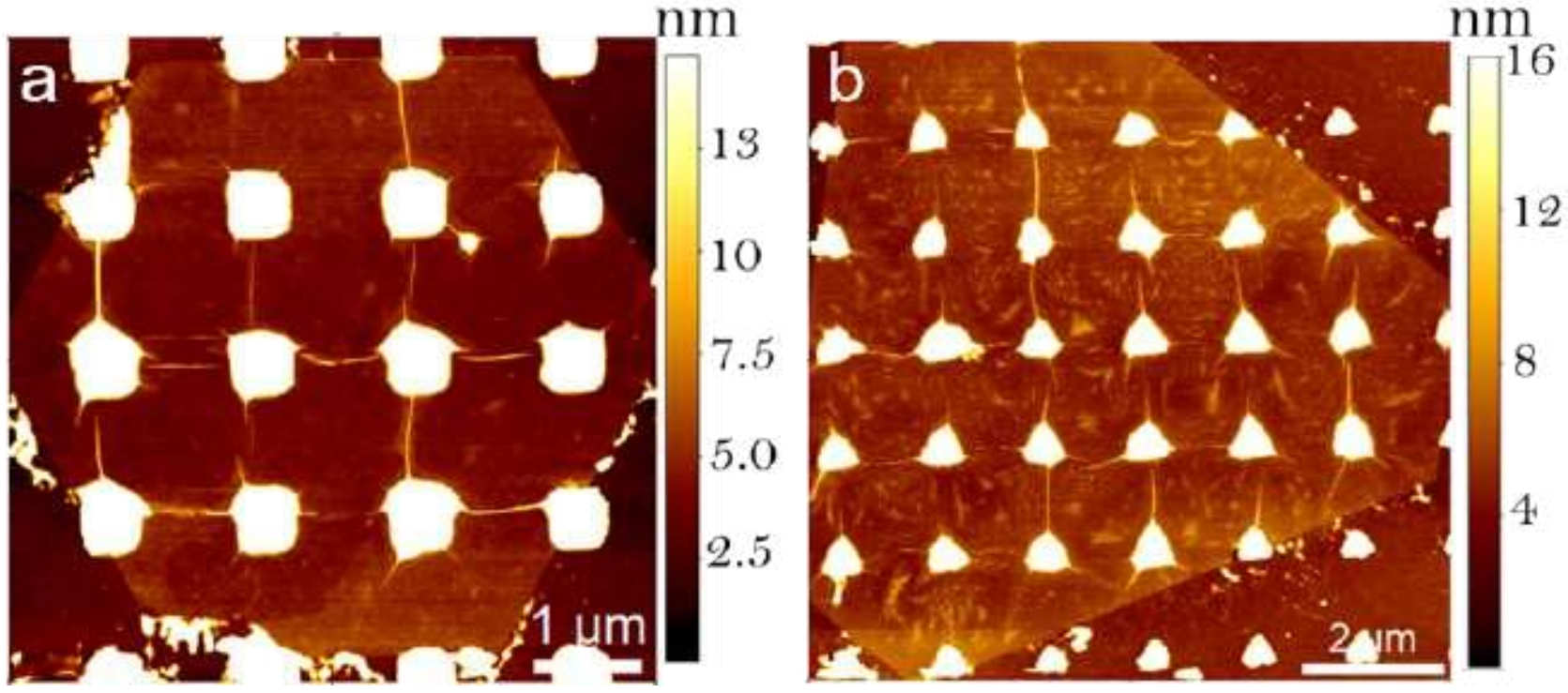

**Figure 2.11:** AFM topography image of monolayer $MoS_2$ transferred on (a) square and (b) Triangular nano-pattern showing strain features: wrinkles from corners of the nanostructures and occasional nanobubbles.

## 2.3.5 Physical vapour deposition methods

### 2.3.5.1 RF Magnetron sputtering

RF sputtering is one of the main fabrication tool that has been used extensively in this thesis. Generally, sputtering of materials is conducted in two ways depending on the conductivity of the material to be deposited: direct current (DC) and radio frequency(RF) sputtering. RF sputtering technique is one of the widely used thin film deposition techniques, especially for dielectrics, as shown in the schematic in Fig.2.12a. In this method, an inert gas, mainly argon, is introduced into the sputtering chamber evacuated to $\sim 10^{-6}$ mbar pressure and is ionised to generate Ar ions plasma. The plasma is created using a radio frequency power source. The plasma of Ar ions is accelerated by an RF electric field, which bombards a target of the material to be deposited on the substrates. In addition, a magnetic field is applied to confine the plasma near the surface of the target, thereby increasing the density of ions and electrons. When the plasma hits the target surface, it knocks out atoms from the target, which get deposited on the substrates. The deposition rate is mainly decided by the RF power and Argon gas flow rate. The main difference between RF and DC sputtering is that DC sputtering requires much higher voltage than RF to deliver the same deposition rate. This is mainly because DC sputtering involves direct ion bombardment of the gas plasma by electrons, while RF sputtering uses kinetic energy to remove the electrons from the outer shells of the gas atoms. Another difference between DC and RF sputtering is that RF sputtering can maintain the gas plasma at a significantly lower chamber pressure of under $10^{-3}$ mbar, compared to the $10^{-2}$ mbar required for DC sputtering which reduces the number of collisions between the charged plasma particles and the target material thereby creating a more direct pathway to the sputter target. Moreover, RF power prevents the charge build-up on the target material, unlike DC sputtering, due to the large number of energetic ions in the chamber, which allows relatively easy sputtering of non-conducting materials. In this thesis, thin films of indium-tin-oxide (ITO) (120-350 nm) and silicon dioxide ($SiO_2$) (200 nm) are deposited using RF sputtering. The deposition chamber was evacuated to 2 x $\sim 10^{-6}$ mbar to remove all the atmospheric gases. ITO depositions were mainly conducted at 90 W RF power with a 30 sccm Ar flow rate which resulted in a deposition pressure 1 x $\sim 10^{-2}$ mbar. While $SiO_2$ depositions were mainly conducted at 60 W at the same Ar flow rate. The time of deposition was varied according to the thickness needed.

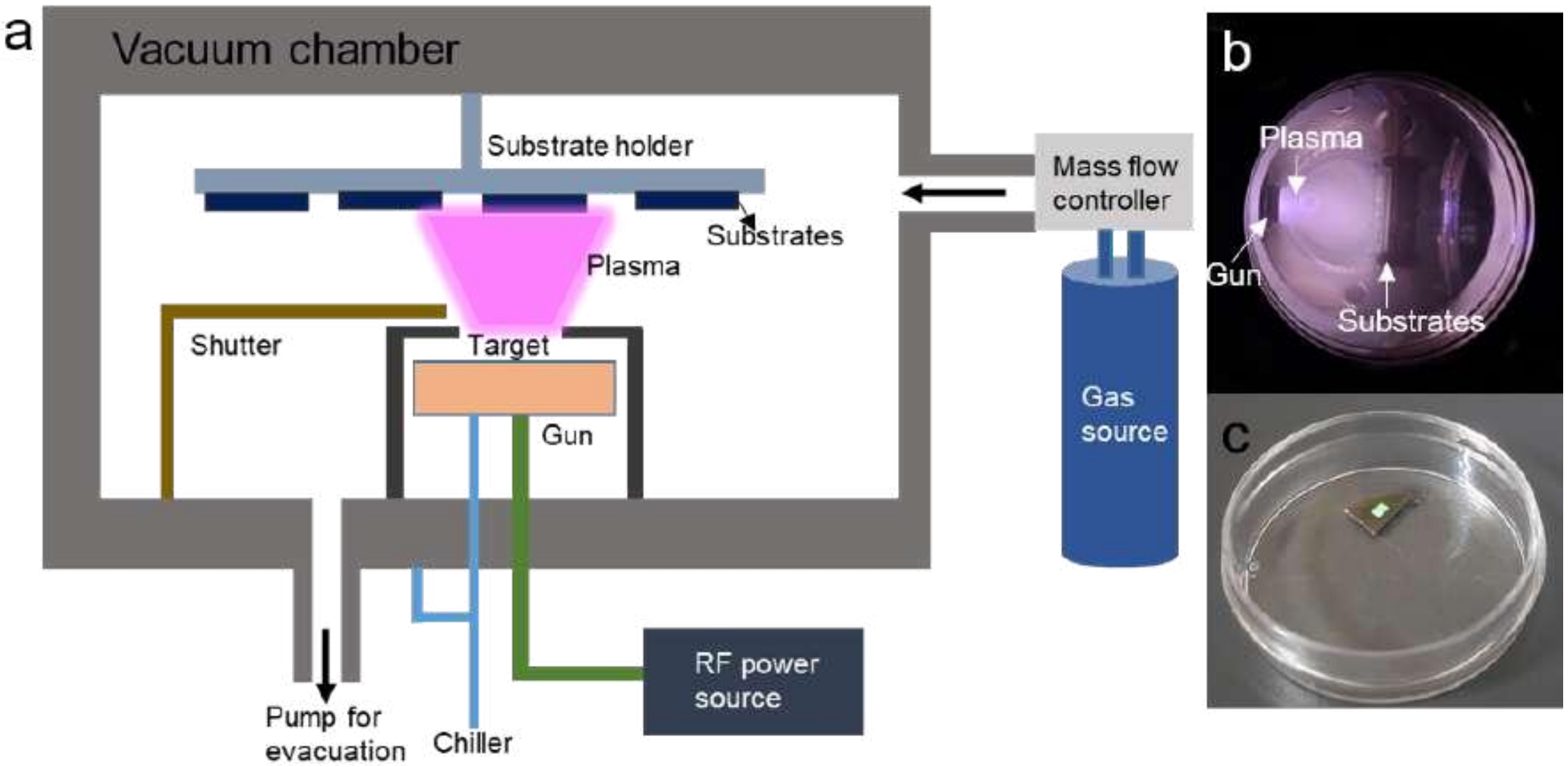

**Figure 2.12:** (a) Schematic of an RF sputtering system with its various units, (b) Image captured while deposition of a thin film by RF sputtering showing the plasma hitting the substrate, (c) Optical image of a substrate with nanopatterns after sputtering.

#### 2.3.5.2 Reactive sputtering

Reactive sputtering is technically a modified version of RF sputtering in which thin films are deposited by introducing a reactive gas (typically oxygen or nitrogen) into the inert gas plasma, commonly argon. In this process, the reactive gas "activated" by the plasma reacts chemically with the target material to form the desired material thin films on the substrates. The relative amounts of the inert and reactive gases decide the composition of the resultant film. Generally, thin films of oxides, nitrides, and carbides are formed through reactive sputtering. In this thesis, thin films of titanium nitride (150 nm) are deposited using reactive sputtering using a Titanium target in Ar and $N_2$ partial pressure.

#### 2.3.5.3 Thermal Evaporation

Thermal evaporation is of the most common thin film deposition methods especially metal films. The material to be evaporated is kept in a crucible/ boat generally made of tungsten as it has a higher melting point (3422 °C) compared to that of the material to be evaporated. The material is evaporated by passing a high electric current ($\sim$ 100A typically for gold). As the heat generation is due to the electrical resistance of the evaporation source, this technique is also known as resistive evaporation. The substrates to be coated are generally mounted upside down in a substrate holder so that the material evaporated from the bottom gets deposited directly onto the substrate surface. The deposition chamber is generally evacuated to $\sim 10^{-6}$ mbar pressure to minimise the collision of the material that is evaporated with the air molecules. The deposition rate is determined by the current that is passed. Keeping a low deposition rate is preferred for uniform deposition of the films with a substrate holder having a slow rotation rate ($\sim$ 8 rpm ) while the material is being deposited. In this thesis, thin films of metal like gold (Au) (50-100 nm) on glass and lithographically patterned substrates are deposited. Chromium (Cr) of thickness $\sim$3 nm has been deposited as the adhesive layer for Au on silicon substrates. Thin films of 15-40 nm of Cr have been deposited on stainless steel and glass substrates for spectrally selective reflector coating as discussed in chapter 3. Around 70A of current was passed through a chromium-plated tungsten rod, which maintained a deposition rate of 0.3 Å/s.

## 2.4 Experimental charcterization techniques

### 2.4.1 Optical microscopy

Optical microscopy is a technique that uses visible light to magnify small objects so that they can be observed and analyzed. In an optical microscope, light from a source is focused onto the specimen through a series of lenses, and the light that is transmitted through or reflected by the sample is collected by another set of lenses and directed to the observer's eye or a camera. However, the resolution of an optical microscope is diffraction-limited. This limits the ability of optical microscopy to observe structures at the nanoscale. Also, it has a limited depth of field. We used a standard, upright optical microscope (OLYMPUS BX53M) to investigate the quality of patterns obtained lithographically. It was also used to identify CVD-grown monolayer $MoS_2$ to be transferred to patterned substrates as well as to image the flakes post transfer, as shown in Fig.2.13. It has also been used to image the FET devices fabricated on strained $MoS_2$ as discussed in Appendix A.

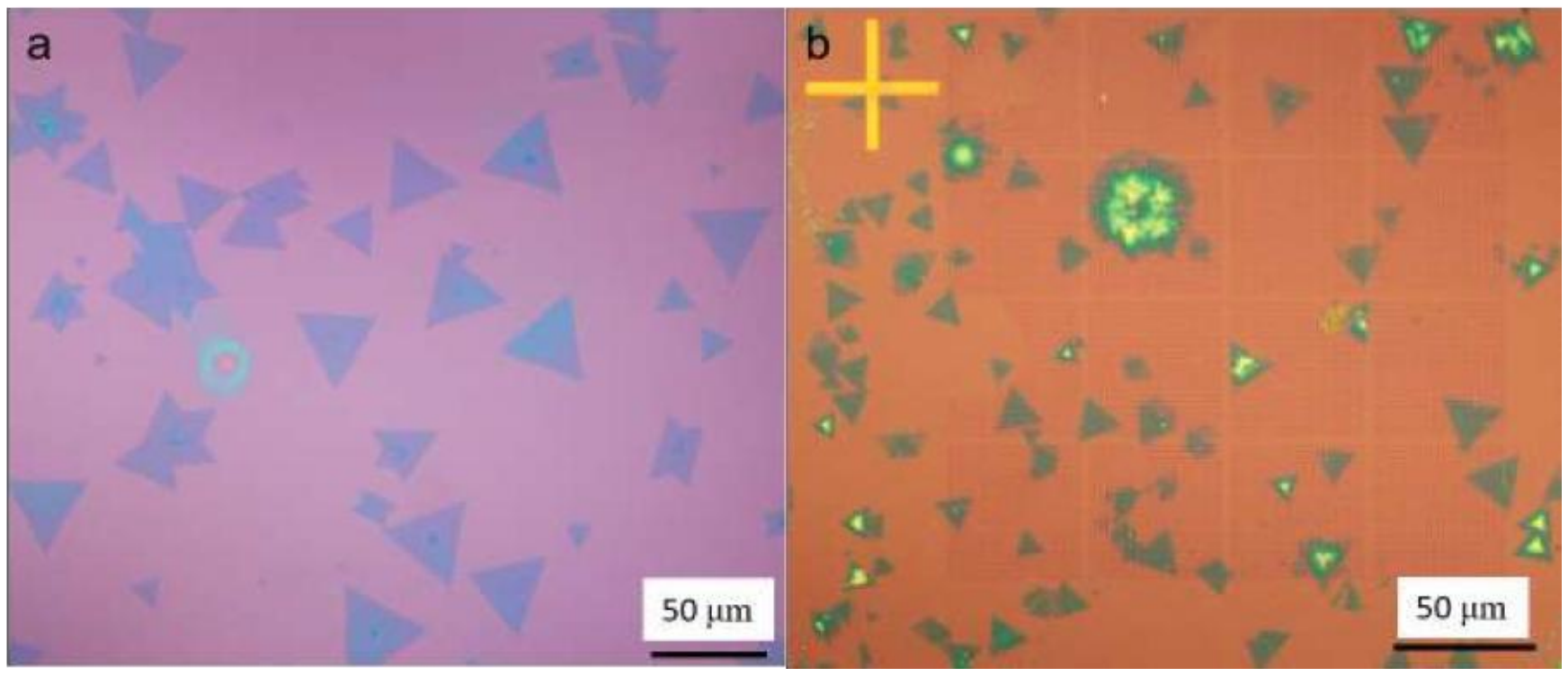

**Figure 2.13:** Optical microscope images of (a) CVD grown $MoS_2$ (monolayers to few layers) on $SiO_2$/Si substrate and(b) CVD grown $MoS_2$ transferred on patterned substrate

### 2.4.2 Scanning electron microscopy (SEM) and energy dispersive x-ray spectroscopy (EDS)

The operation of scanning electron microscope has been discussed under section 2.3.1. In this thesis, Nova Nano SEM 450 field emission scanning electron microscope coupled with an Apollo X EDS system has been extensively used to image the quality of nanoscale lithography patterns after metallization and lift-off. Also, it has been used to image the morphology of the various thin films that have been deposited during the fabrication of the multilayer coatings at each developmental stage, as discussed in chapters 3, 4 and 5.

Energy dispersive X-ray spectroscopy (EDX) is used for quantitative and/or qualitative identification and analysis of chemical composition of a sample. High energy electron beam bombards the sample, characteristic x-rays are emitted and are detected by the specific detectors which gives the information about the specific element present and the concentration. In this thesis, EDS was used as one of the tools in confirming the formation of chromium oxide on annealing chromium-coated glass substrates. It was also used to check the formation of indium-tin-oxide and titanium nitride after sputtering.

### 2.4.3 X-ray diffraction

X-ray diffraction (XRD) is a non-destructive technique that is used to determine chemical composition, crystal structure, crystal orientation, crystallite size, lattice strain, preferred orientation and layer thickness of a sample. It can be used for a wide variety of samples, from powders to thin films[126]. In this thesis, XRD studies were performed with Cu K=1.540  at 45 kV (Empyrean, PANanalytical). It was used to confirm the formation of thin films of $Cr_2O_3$ on glass, ITO and TiN on different substrates as discussed in Chapters 3 and 5.

### 2.4.4 Atomic Force microscopy

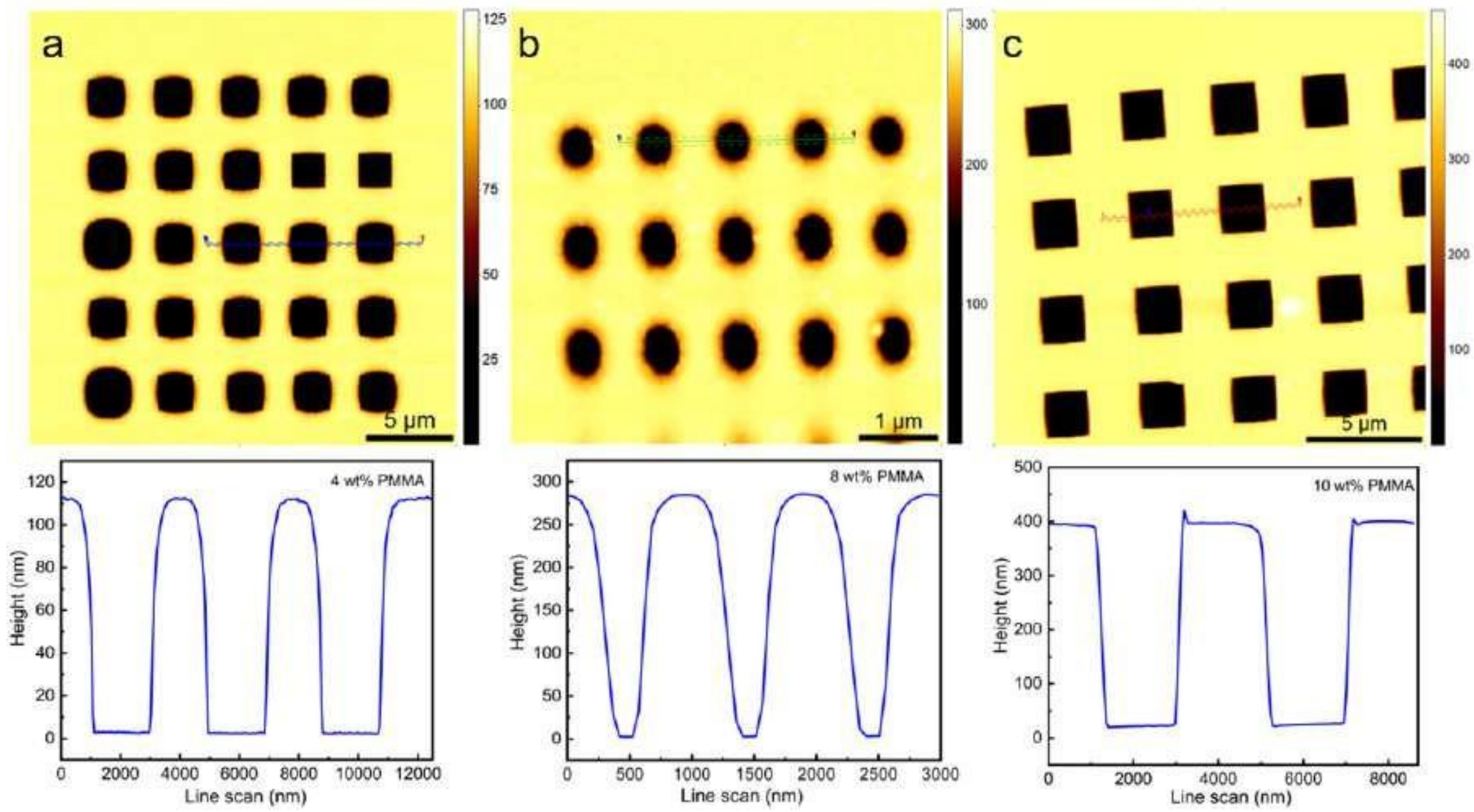

**Figure 2.14:** Tapping mode AFM images of developed e-beam patterns (after exposure) for PMMA concentration(a) 4 wt%(b) 8 wt% (c) 10wt%. The bottom panel shows the line scans showing the thickness of resist obtained for each concentration.

Atomic force microscopy (AFM) is a high-resolution scanning probe microscopy technique that is used for imaging diverse surface: conducting or non-conducting. Apart from topography information (variation in height and surface roughness), it also provides information on adhesion strength, magnetic forces and mechanical properties, and electronic properties of a sample. AFM is operated in two basic modes: contact and tapping modes. In the contact mode, the AFM tip is in continuous contact with the surface. Whereas, in tapping mode, the AFM cantilever is vibrated above the sample surface such that the tip is only in intermittent contact with the surface and hence less damage is caused to the sample surface and is ideal for imaging. AFM tip is approximately 10 to 20 nm in diameter, which is attached to a cantilever. The tip moves in response to tip–surface interactions, and this movement is measured by focusing a laser beam on a photodiode. In this thesis, AFM (Bruker Multimode 8) in tapping mode is used to measure the thicknesses of different concentration PMMA resist used for e-beam lithography optimization, as shown in Fig.2.14. It has also been used to quantify the average roughness for the various thin films deposited, like ITO and TiN thin films as well as bare stainless steel (SS) substrates.

### 2.4.5 Reflectance spectroscopy

Reflectance spectroscopy is a technique that provides spectral information about a sample in terms of the light that is reflected from the sample as a function of wavelength. Reflectivity is defined as the ratio of the reflected light intensity to the incident light intensity. The experimental arrangement, as shown in Fig. 2.15a, generally consists of a broadband light source which passes through apertures and is collimated into a beam which is incident on the sample through an objective lens. The reflected light from the sample is collected through another set of lenses and directed to a spectrometer that is interfaced with software that displays the reflectance spectra. A polarizer can also be introduced into the incident light path for polarization-dependent reflectivity measurements. Moreover, the reflection is also measured with variation in angle of incidence to understand the angle-dependent optical response of the sample. It provides information regarding different optical phenomena, like resonances observed from dips in the reflectivity spectra. There are two cases in reflection. First is specular reflection from an optically smooth surface in which the light reflects at an angle same as that of the incident angle and second is diffuse reflectance, in which scatters in multiple directions after hitting a surface, mostly observed in real-life scenarios. In this thesis, majority of the reflectivity measurements are conducted in a commercial UV-vis-NIR spectrometer setup (Perkin Elmer 950). It consists of two light sources, a deuterium lamp for the UV regime and a tungsten-halogen lamp for the visible and NIR regimes. It has different modules for reflectivity and transmission measurements. The specular reflectivity measurements in chapters 3 and 4 were conducted using the Universal Reflectance Accessory module, while the diffuse reflectance measurements were conducted using the Integrating Sphere module. Few standard measurements were done with a custom-made setup as shown in the schematic (Fig.2.15a) with a broadband source- tungsten halogen lamp (Thorlabs SLS301) and a visible and NIR spectrometer from Ocean Optics as the detector. Fig.2.15b shows the image of an illuminated periodic grating sample fabricated by EBL through an infinity corrected objective lens ( Plan Apo Mitutoyo 10X, NA=0.28, focal length=2 cm). Fig. 2.15c shows the images of the different colours captured due to the diffraction from the grating through angular variation in detection.

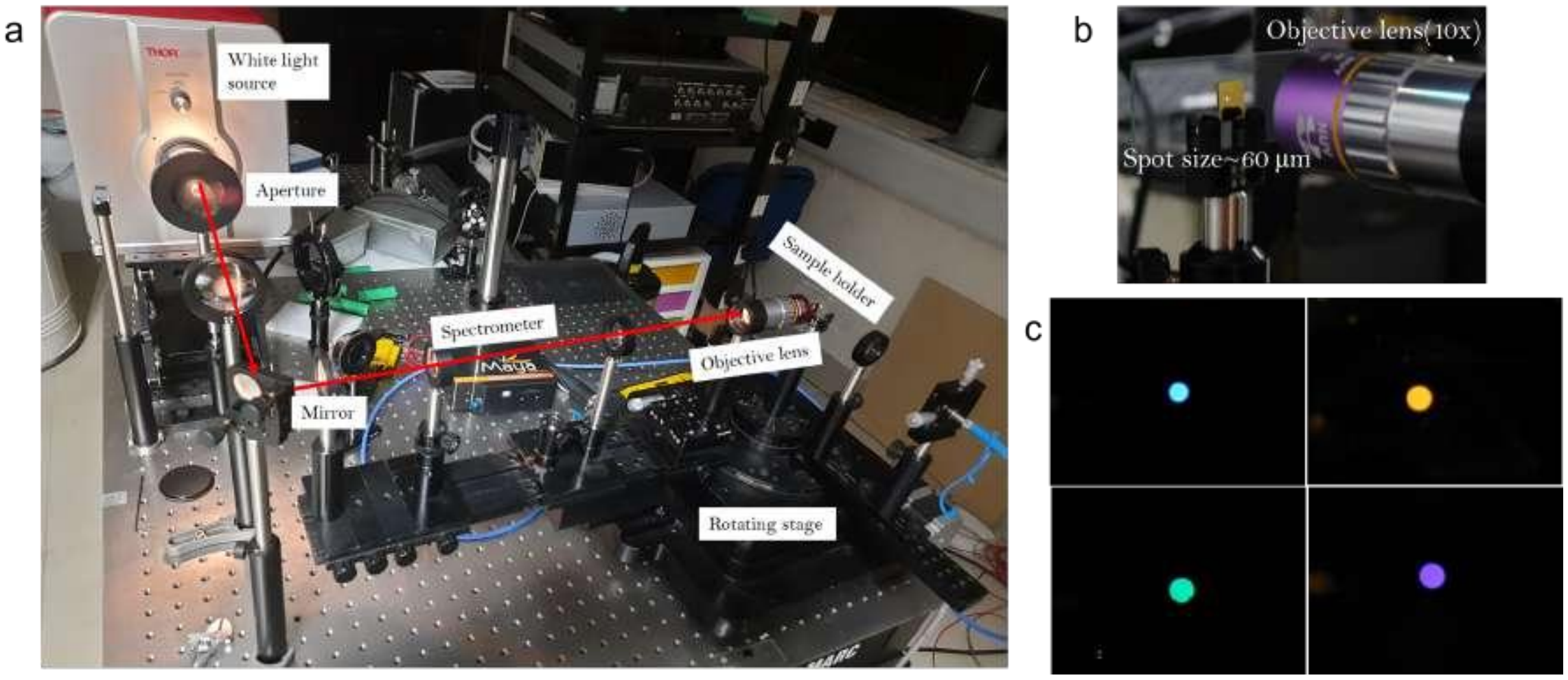

**Figure 2.15:** (a) Custom-made angle-resolved reflectivity cum transmittance measurement setup, (b) Optical image showing an illuminated grating sample with a light spot size $\sim$ 60 $\mu$m, (c) Optical images of the different colours captured due to diffraction from the grating via angular variation in detection

### 2.4.6 Raman spectroscopy

Raman spectroscopy is a spectroscopic technique typically used to determine vibrational modes of molecules. It also gives information about the crystallinity, lattice structure and phase of the sample. It mostly depends on the inelastic scattering of photons known as Raman scattering [127]. The sample is excited with a laser and the light scattered from is collected with a lens. Elastic scattered light at the wavelength corresponding to the laser line (Rayleigh scattering) is filtered out by either a notch filter, edge pass filter, or a band pass filter, while the rest of the collected light is dispersed onto a detector. In this thesis, a confocal Raman microscope (Horiba Xplora Plus) has been used to conduct spatially resolved measurements on CVD-grown monolayer $MoS_2$ transferred on dielectric and metal patterns as well as transferred on TiN-coated Si substrates. The sample is mounted on a motorized stage, which is movable with sub-micron precision via a controller. A green laser (532 nm) was used for excitation through a 100X objective lens (NA =0.9) with 2400 gr/mm grating. The signals were detected using a thermoelectric cooled (-60°C) camera. Fig.2.16 shows the Raman intensity maps for $E^{1}_{2g}(\Gamma)$ mode and $A_{1g}(\Gamma)$ mode of a CVD-grown monolayer $MoS_2$ on $SiO_2$/Si substrate and the corresponding spectra.

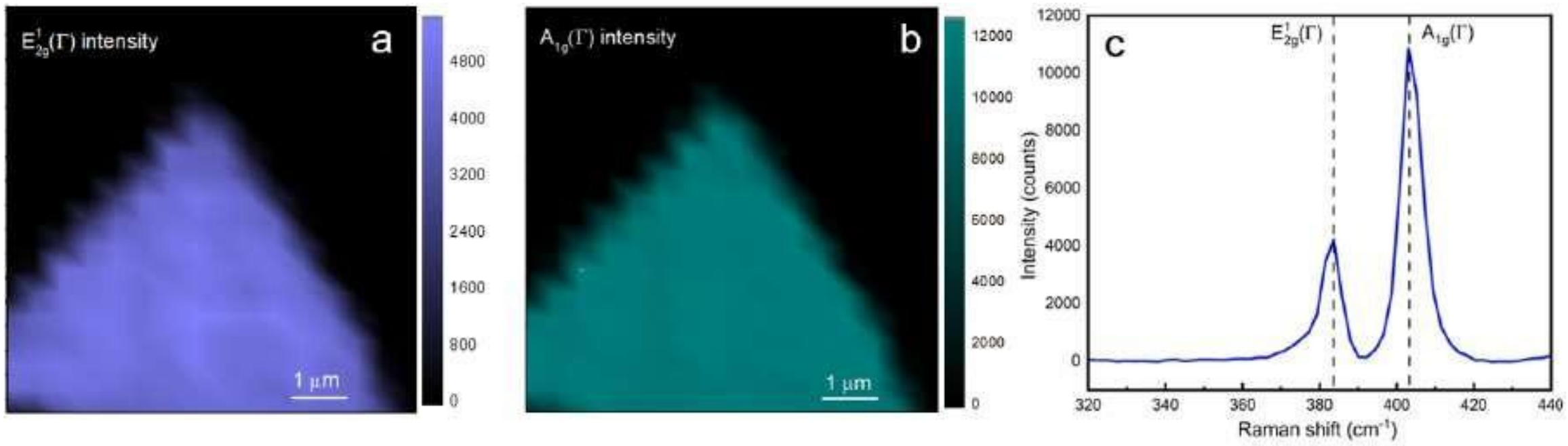

**Figure 2.16:** Raman intensity maps for (a)$E^1_{2g}$($\Gamma$) mode, (b)$A_{1g}$($\Gamma$) mode and (c) Raman spectra of CVD-grown monolayer $MoS_2$ on $SiO_2$/Si

### 2.4.7 Photoluminescence spectroscopy

Photoluminescence (PL) spectroscopy is a technique typically used to study the electronic and optical properties of materials. This technique also uses a laser where the incident photons excite electrons to higher energy states. The subsequent relaxation of the excited electrons back to lower-energy states results in the emission of photons, which is analyzed to understand the material's band structure, defect states, and composition.

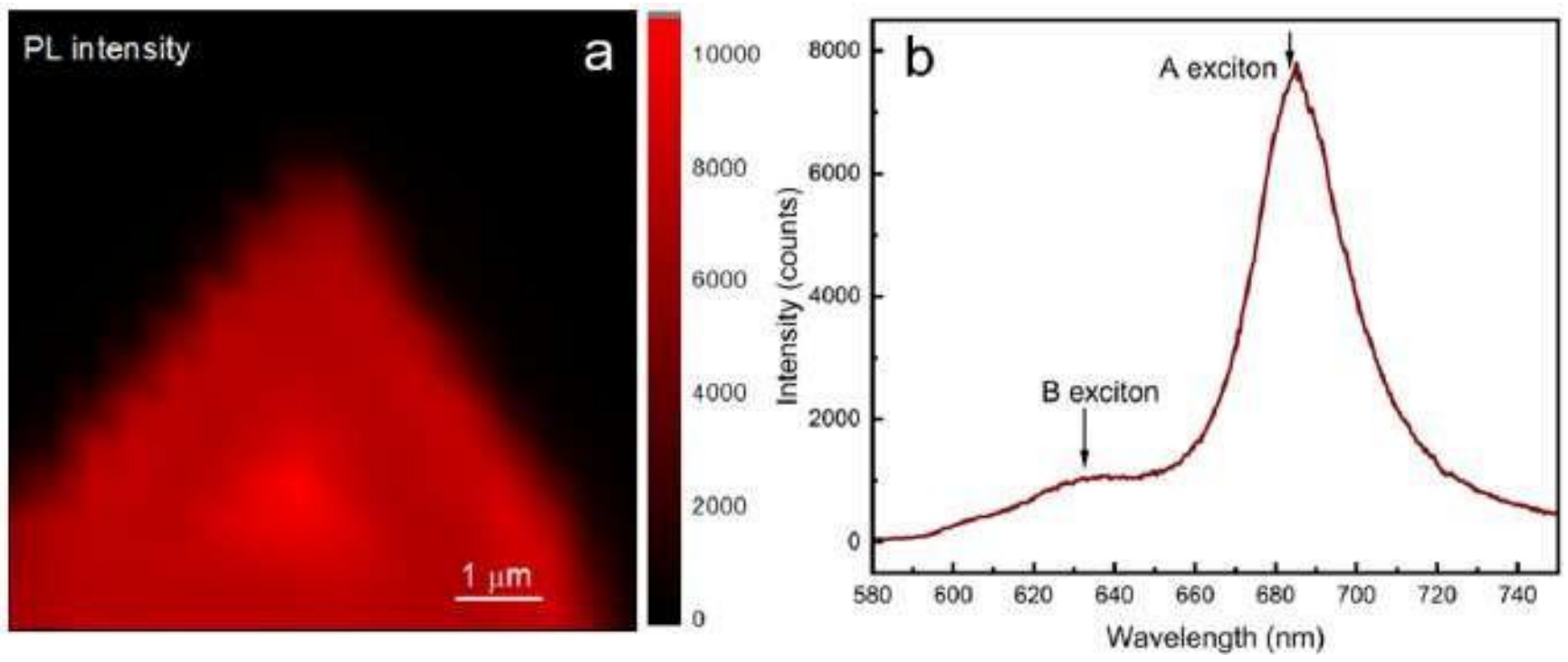

**Figure 2.17:** (a) PL intensity map and (b) the spectra of CVD-grown monolayer $MoS_2$ on $SiO_2$/Si.

In this thesis, spatially resolved PL has been used to study the PL response of $MoS_2$ transferred on an ENZ substrate (TiN). It is also used to detect the strain induced in $MoS_2$ transferred on metal and dielectric nano-patterns, as shown in Appendix A. The spatial resolution is important for these systems in order to understand the strain dis-tribution when the flakes are resting on periodically patterned substrates. PL measurements were also done with a Horiba Xplora Plus Raman microscope with 532 nm laser excitation, 100X objective lens (NA =0.9) with 600 gr/mm grating. Fig. 2.17 shows the PL intensity map for the A exciton wavelength of a CVD-grown monolayer $MoS_2$ on $SiO_2$/Si substrate and the corresponding spectra.

### 2.4.8 Thermal imaging

Thermal imaging is a sophisticated and non-invasive technique that uses infrared technology to detect thermal emissions from various objects. This tool converts the infrared energy, which is not visible to the naked human eye, into a visible light display. Above absolute zero temperature, all bodies emit thermal radiation and the variations in the emissions attributes to thermal imaging. A thermal imaging device or a camera detects the infrared radiation or heat emitted by objects and generates the corresponding temperature map of the area under focus [128]. It follows the Stefan-Boltzmann law, which is given as

$$P = e\sigma T^4 \tag{2.8}$$

where $P$ is the power radiated per unit area by the surface, $'e'$ is the emissivity of the object, $\sigma$ is the Stefan-Boltzmann constant and $T$ is the temperature. In this thesis, IR emission from ITO grating/$SiO_2$/Au coating was captured using a thermal IR camera(Fluke Ti480 Pro) as discussed in Chapter 4.

## 2.5 Summary

The numerical and experimental techniques used to conduct various studies in this thesis are discussed in this chapter. Finite element method calculations in optimizing opti-cal responses, along with the experimental fabrication and characterization techniques, are thoroughly explained.

# Chapter 3

# Epsilon-near-zero metal oxide based spectrally selective reflectors

Epsilon-near-zero (ENZ) materials can contribute significantly to the advancement of spectrally selective coatings aimed at enhancing efficient use of solar radiation and thermal energy management. Here, we demonstrate a subwavelength thick, multilayer optical coating that imparts a spectrally "step function" like reflectivity onto diverse surfaces, from stainless steel to glass, employing indium tin oxide as the key ENZ material. The coating, harnessing the ENZ and plasmonic properties of nominally nanostructured ITO along with ultrathin layers of Cr and $Cr_2O_3$ show 15% reflectivity over the visible to near-infrared and 80% reflectivity (and low emissivity) beyond a cut-in wavelength around 1500 nm, which is tunable in the infrared. A combination of simulations and experimental results are used to optimize the coating architecture and gain insights into the relevance of the components. The straightforward design with high thermal stability will find applications in diverse areas, from passive cooling to energy harvesting.

## 3.1 Introduction

Solar radiation management, a key component of sustainable energy generation, is an area of wide topical interest wherein spectrally selective coatings form an important domain of research [129, 130]. Engineering spectral selectivity of surfaces through coatings offers an opportune way to maximize effective utilization of incident and radiated energies. This discussion focuses on realizing a coating exhibiting idealized step-function-like reflectivity (R), characterized by high absorption in the visible and high reflectivity and low emissivity in the infrared, separated by a cut-in wavelength $\lambda_o$. Design and fabrication of such coatings that are durable, cost-effective and scalable,ensuring compatibility with diverse surfaces remain open challenges. Strategies for inducing such selectivity have broadly centered on assembling the right materials (semiconductors, metals and dielectrics) in conducive form factors like multilayers or composites with two and three-dimensional texturing [131, 132]. The increasing ability to engineer hybrid materials with sub-wavelength scale layering [133] and structuring [134] has broadened the scope of tailoring spectral selectivity. Fundamentally, the above schemes rely on the physics of light-matter interactions and engineering radiative vs. non-radiative modes, as realized in photonic crystals [135, 136] and optical metamaterials [137, 138], including hyperbolic metamaterials [139–141]. Interestingly, non-radiative regimes are also realized in continuous media systems such as the epsilon-near-zero (ENZ) materials, near their ENZ wavelength ($\lambda_{ENZ}$). Conducting oxides like Al-doped ZnO, indium tin oxide (ITO), doped CdO etc. [83, 84] harbor ENZ regimes that originate from their free electron response. The real part of their relative permittivity ($\varepsilon(\omega) = \varepsilon^{'}+i\varepsilon^{''}$), $\varepsilon^{'} = 0$ at $\lambda_{ENZ}$ (typically beyond 1200 nm), which demarcates a dielectric to metal transition in terms of their optical properties (see Fig.3.1). Consequently, ENZ systems are transparent for $\lambda < \lambda_{ENZ}$, and become reflecting progressively at longer wavelengths. Further, they sustain the non-radiative ENZ modes around $\lambda_{ENZ}$, and ultra-thin layers (thickness ~ 10 nm) evidence perfect absorption, under p-polarized illumination [92]. Spectrally, perfect absorption is realized at wavelengths $\lambda_{PA} \lessapprox \lambda_{ENZ}$ and is coupled with excitation of the ENZ modes accompanied by extreme field amplification and confinement [91]. Here, we show that the high reflectivity ($R \rightarrow 1$) for $\lambda > \lambda_{ENZ}$, together with perfect absorption ($R \sim 0$) around $\lambda_{ENZ}$ and plasmonic resonances in ITO nanostructures enable a niche application of ENZ media in developing spectrally selective coatings of sub-wavelength dimensions. Coatings that impart a step-function-like reflectivity in the 1500 nm – 2000 nm range are of vital interest in solar energy conversion [149, 157], thermal management [158, 159], smart windows [160] and related areas. The optical properties of commercially relevant surfaces, like stainless steel (SS), Si, glass, Cu etc. have been variously modified to impart step-function-like reflectivity, as summarized in Table 1, typically characterized by a cut-in wavelength ($\lambda_o$) and spectral transition width $\Delta\lambda$ from the low to high reflectivity state. The coatings referenced in Table 1 are typically microns thick and have a wide transition, $\Delta\lambda \geq 2\mu m$, involve multiple materials with significant complexity in growth, optimization and incorporation into a coating format. Hence, thinner coatings with simplified architecture, with narrow $\Delta\lambda$ for lower overall emissivity, higher thermal stability and spectral tunability of $\lambda_o$ are some of the current challenges that are being researched. Utilizing the unique optical properties of ENZ materials like ITO, our main objective has been to develop a sub-wavelength thick, spectrally selective coating for opaque surfaces like stainless steel, that works over wide angles and unpolarized light. For comparison, a similar coating has been evaluated on a transparent substrate like glass. Design of ITO-based meta-surfaces, broadband absorbers, and spectrally selective absorbers and reflectors [156, 161–164] have gained traction recently owing to material and processing advantages [165]. Interestingly, the $\sim$500 nm thick multilayer coating of ITO/Cr/$Cr_2O_3$ investigated here imparts the same spectral selectivity to both SS and glass, leveraging the ENZ properties of ITO. Optical and thermal response of the coated surfaces and their simulated response are used to comprehend the physics of optical engineering and demonstrate that a final elementary ITO grating (Fig.3.5) red-shifts $\lambda_o$ beyond the $\lambda_{ENZ}$ of ITO. The wide angle (0° - 60°), "step function" like response is demonstrated for both flat and nanostructured coatings with $\lambda_o \sim$ 1500 nm and a narrow $\Delta\lambda \sim$ 400 nm, for the latter. Finally, control of $\lambda_o$ via changing the free electron density of ITO is demonstrated. The coating structure is readily adapted as a traditional solar absorber for $\lambda_o \sim$ 2500 nm and the concept is indeed functional in other spectral ranges with relevant ENZ materials.

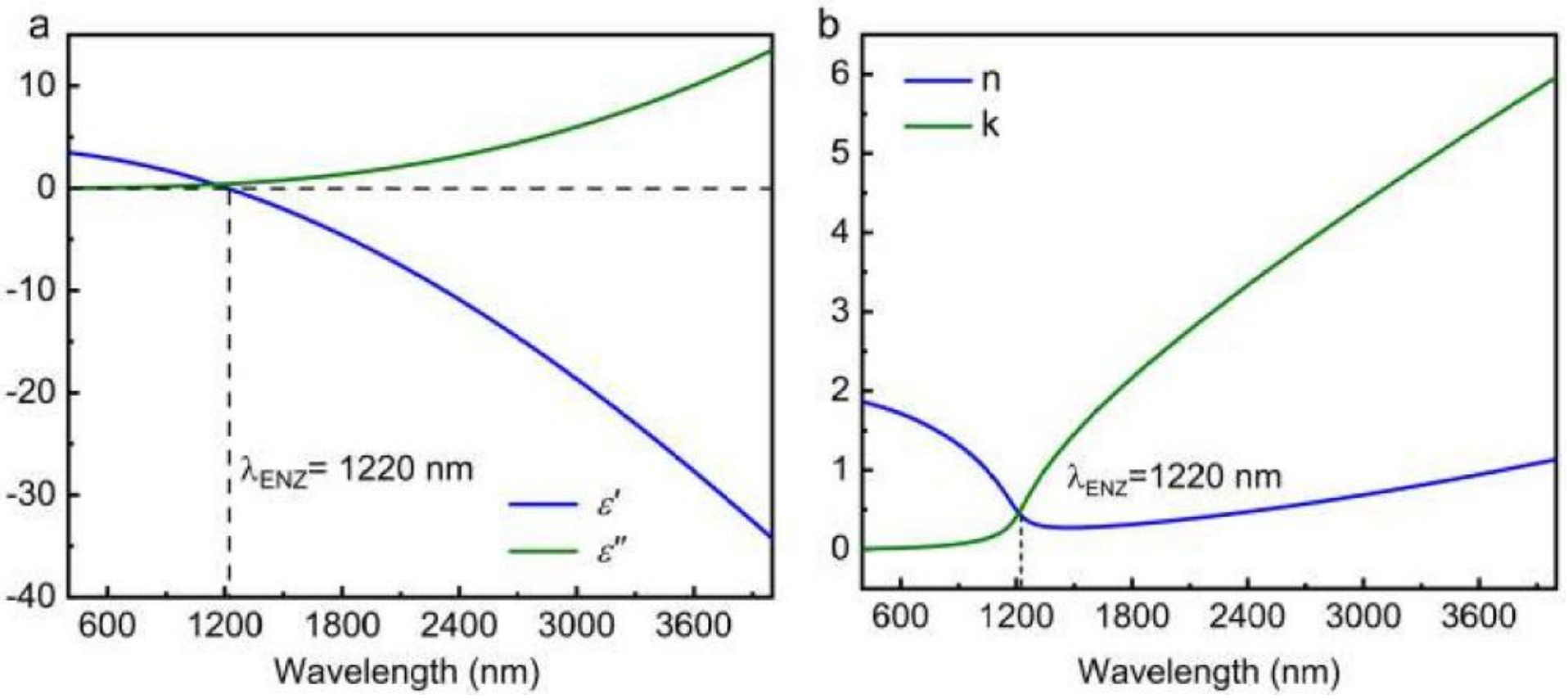

**Figure 3.1:** (a) Real ($\varepsilon^{'}$) and Imaginary ($\varepsilon^{''}$) plots of permittivity variation with wavelength, (b) Refractive index and extinction coefficient of ITO for $\lambda_{ENZ} = 1220nm$

| **Coating and substrate** | **Absorptivity (vis/NIR)** | **Emissivity (IR)** | **Transition width ($\Delta\lambda$) (nm)** | **Ref.** |
|---|---|---|---|---|
| Graphene +$SiO_2$ metamaterial* | - | 0.01 | 400 | [140] |
| Au/ $TiO_2$ based metamaterial | 0.9 | 0.1 | 250 | [141] |
| Cr/$SiO_2$/Cr on Si | 0.99 | - | - | [142] |
| W/WAlN/WAlON/$Al_2O_3$ on SS | 0.90 | 0.15 | 2000 | [143, 144] |
| TiAlC/TiAlCN/TiAlSiCN/TiAlSiCO/ TiAlSiO on SS | 0.961 | 0.07 | 1000 | [145] |
| 3 SiC-W nanocomposite layers on Si | 0.95 | 0.05 | 2000 | [146] |
| Structured graphene metamaterial on Cu* | 0.90 | 0.04 | 1800 | [147] |
| ITO NS/ITO /Cr/ $Cr_2O_3$ on SS and glass | 0.87 (SS) | 0.18 (SS) | 400 | This work |
| Cu/$Al_2O_3$/Cr/$SiO_2$/Cr/$SiO_2$ on Glass and Si | 0.954 | 0.196 | 2000 | [148] |
| Au/NiCr–$MgF_2$ (HMVF)/ NiCr–$MgF_2$ (LMVF)/$MgF_2$ on SS | 0.976 | 0.045 | 3000 | [149] |
| Multilayer graphene + TiN based meta-material * | 0.88 | 0.03 | 2000 | [150] |
| $SiO_2$/ $Al_2O_3$/ $ZrB_2$/ $Al_2O_3$/ $ZrB_2$/ $ZrB_2$/ $Al_2O_3$ / $ZrB_2$ on SS and Si | 0.96 | 0.16 | 2000 | [151] |
| W/WAlSiN/SiON/$SiO_2$ on SS and Si | 0.955 | 0.10 | 2000 | [152] |
| Multilayer reduced graphene oxide on Al* | 0.92 | 0.04 | 1250 | [153] |
| CrN(H)/CrN(L)/CrON/$Al_2O_3$ on SS | 0.93 | 0.14 | 2000 | [154] |
| Pt/ $TiO_2$/ Al on SS * | 0.90 | - | - | [155] |
| Black chrome/ITO/$SiO_2$ on SS | 0.90 | 0.40 | 1200 | [156] |

**Table 3.1:** Coatings that impart step function like reflectivity to various surfaces.
*Tunability demonstrated

## 3.2 Material and methods

### 3.2.1 Simulation

The multilayer structures were optimized in terms of the material properties and physical dimensions using finite element method modelling with 3D simulations conducted in the Wave Optics Module of COMSOL® Multiphysics 5.3a. The optimization scheme yielded the best optical response of the system considering feasibility of fabrication and experimental verification of the samples. It minimizes reflectance over the visible and maximizes reflectance above the cut-in wavelength ($\lambda_o$) with a straightforward fabrication protocol. The simulations were carried out in the range 400 – 4000 nm though the experimental verification is limited to 2500 nm. The optical properties of the materials were taken from the literature [166–168], with the Drude model used to determine that of ITO as a function of three parameters, carrier concentration $N_e$, scattering parameter $\gamma$, and background permittivity $\varepsilon_\infty$. The typical values of the parameters ($N_e$, $\gamma$ and $\varepsilon_\infty$) follow those from a previous publication [169] from the group. Fig.3.2 shows the model geometry designed in COMSOL with the optimized grating dimension and periodicity simulated using periodic boundary conditions. The geometric dimensions of periodicity (centre-to-centre distance) helps to achieve the typical "step function" like reflectivity.

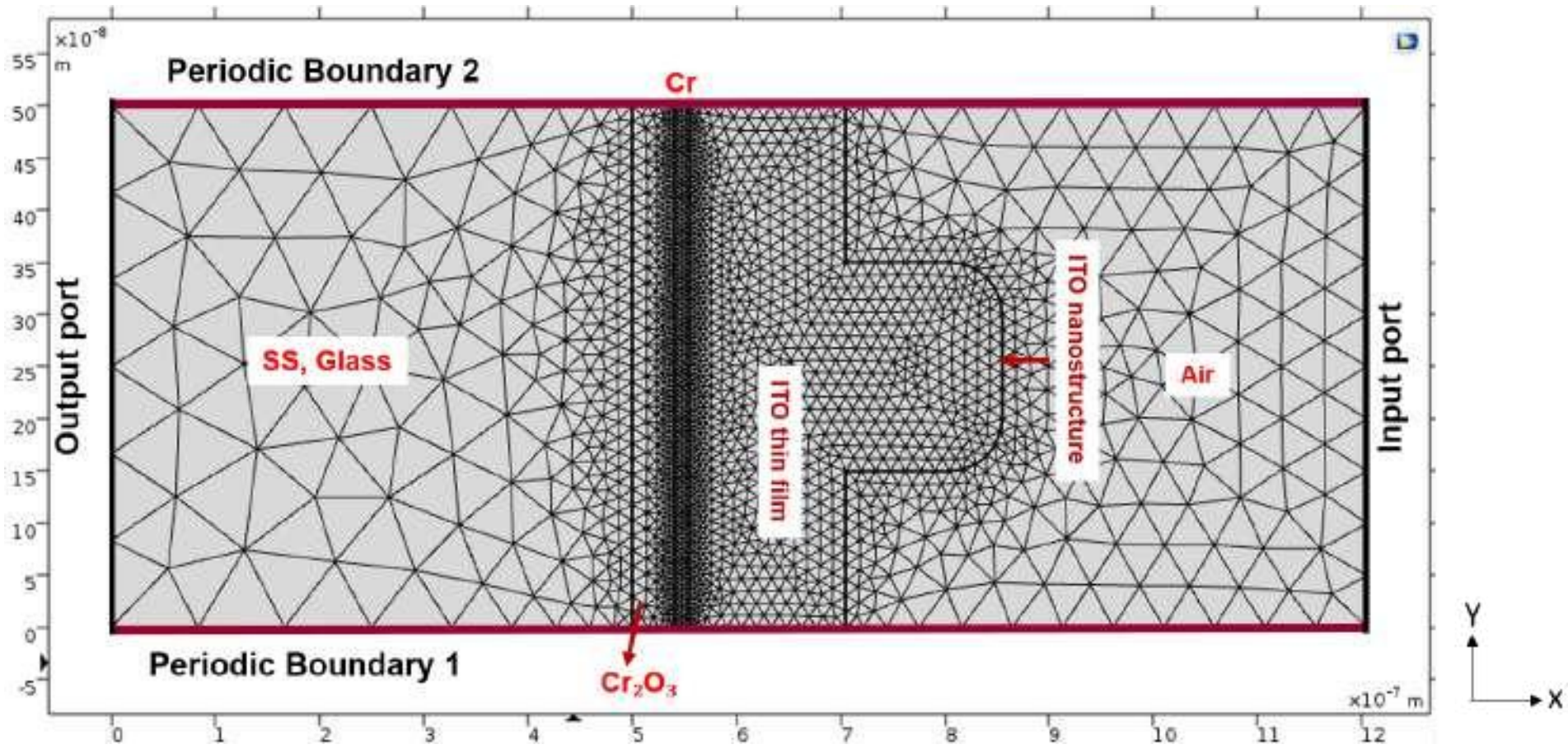

**Figure 3.2:** Cross-sectional view of the unit cell designed in the Wave optics module of COMSOL Multiphysics 5.3a the nanostructures (NS) (width, height), as well as the periodicity of the array, were optimized systematically in COMSOL while developing the spectrally selective reflector. The width was varied keeping the height fixed, the height was varied keeping the width fixed and finally, the periodicity was varied keeping the width and height fixed. Fig.3.10 shows that the nanostructure with ∼200 nm width and ∼150 nm height with ∼500 nm

### 3.2.2 Sample fabrication

Polished stainless steel (SS) substrates (MTI, USA) and NBK7 glass substrates were cleaned thoroughly using acetone, isopropyl alcohol and finally with DI water. 40 nm Cr was thermally evaporated onto the substrates at ∼0.5 Å/s. The films were annealed at 600°C for 2 hours in the ambient to form $Cr_2O_3$, verified by XRD, EDS and Raman spectroscopy, as shown in Fig.3.3. Subsequently, the samples were coated with 15 nm Cr (thermal evaporation) and vacuum annealed at 500°C for 1 hour, followed by ITO deposition (RF sputtering) of variable thickness between 120 nm - 350 nm. Chamber pressure for all depositions were maintained at lower than $2\times10^{-6}$ mbar. ITO nanostructures of ∼ 200 nm width and ∼ 500 nm periodicity were patterned using electron beam lithography (Raith Pioneer 2) using 10 wt% PMMA resist. Lithographed samples were developed in a solution of methyl isobutyl ketone (MIBK) and isopropyl alcohol (IPA) in 1:3 volumetric ratio for 10 s followed by IPA wash for 10 s. Finally, 150 nm ITO was sputter coated onto the patterned substrates with lift-off using warm acetone to yield the grating structure. Finally, the nanostructures samples were annealed using rapid thermal annealer (Mini Lamp Annealer-MILA 5000) for 15 minutes at 450 °C in $O_2$ lean atmosphere (∼ $10^{-6}$ mbar pressure), which yielded ITO with $\lambda_{ENZ}$ ∼ 1220 nm. The XRD pattern of the ITO/Cr/$Cr_2O_3$ coating on the substrates before and after the final annealing step are shown in Fig 3.3d. The plots show improved crystallinity of the various ITO facets along with signature of $Cr_2O_3$.

### 3.2.3 Material and Optical Characterization

Morphological characterization was conducted using Nova Nano SEM 450 field emission scanning electron microscope (SEM) coupled with an Apollo X energy dispersive x-ray analysis (EDS) system. X-ray diffraction (XRD) studies were performed with CuK$\alpha$ = 1.540  at 45 kV (Empyrean, PANanalytical).

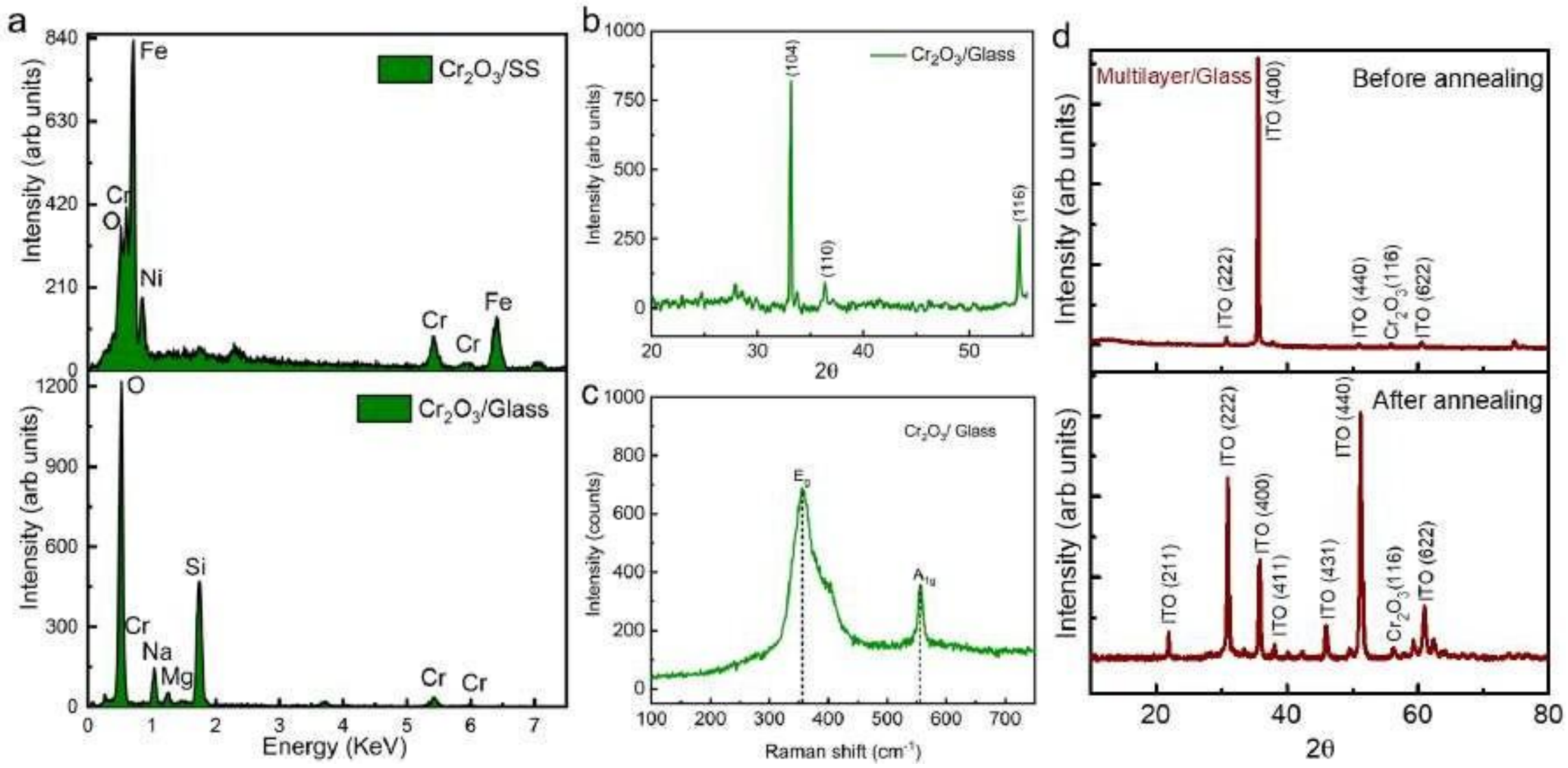

**Figure 3.3:** (a) EDS spectra of $Cr_2O_3$ on SS and glass, (b) XRD of $Cr_2O_3$ on glass, (c) Raman spectra of $Cr_2O_3$ on glass, (d) XRD of multilayer coating on glass before and after annealing.

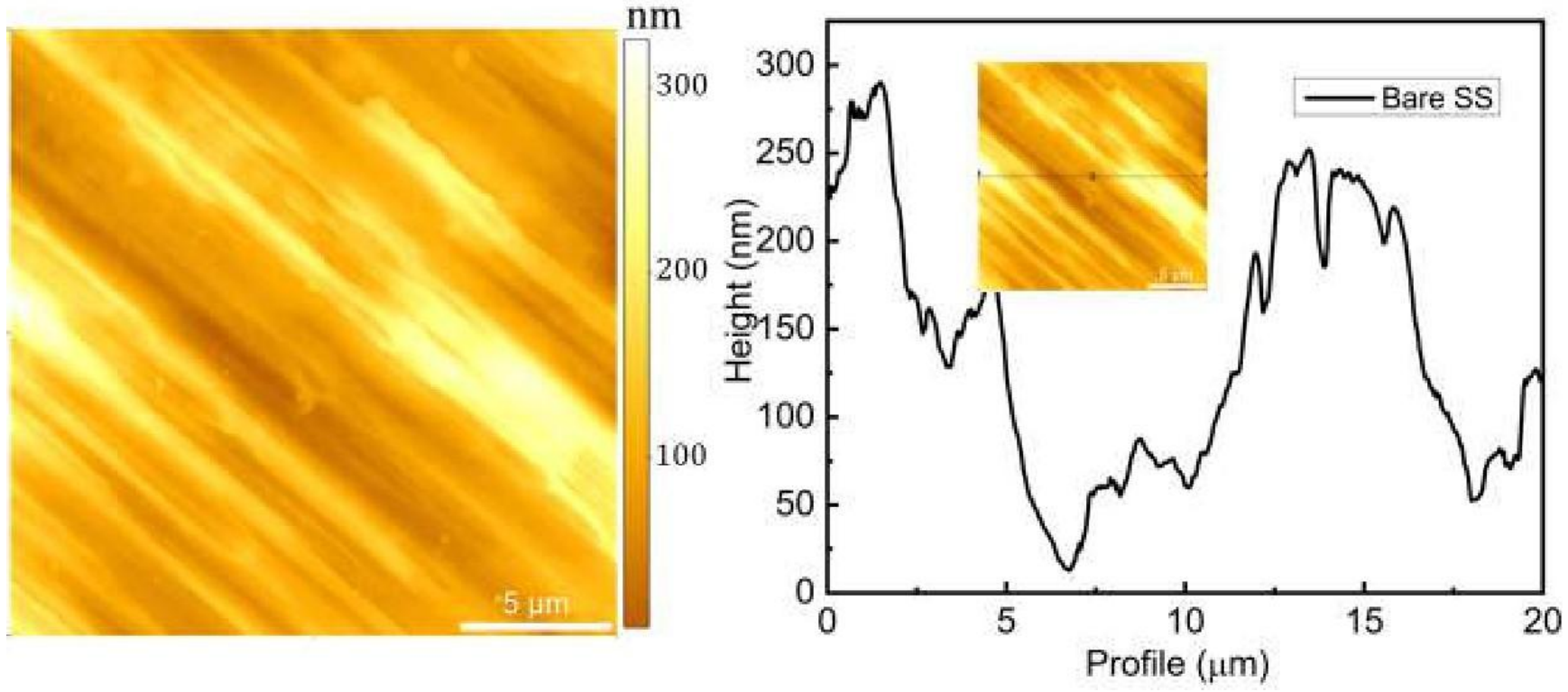

**Figure 3.4:** AFM topography of bare SS with the line-scan showing the height variation

Raman spectra for $Cr_2O_3$ on glass were recorded using a Raman setup (HORIBA Xplora Plus) with 532 nm laser excitation (Fig.3.3c). The atomic force microscopy topography images of bare SS to quantify the roughness were taken using a Bruker Multimode 8 AFM as shown in Fig.5.1. The polished SS substrate shows significantly higher rms roughness ($\sim$ 120 nm) than the BK7 glass (roughness < 3 nm), estimated over 20$\times$20 $\mu m^2$ areas. The average roughness of bare SS was calculated to be 118.44 nm. Reflectance spectra were recorded in a spectrometer (Perkin Elmer Lambda 950), in the wavelength range 400-2500 nm with unpolarized light using the universal reflectance accessory module and a 60 mm integrating sphere module.

## 3.3 Results and Discussion

### 3.3.1 Development of the spectrally selective reflector and its optical aspects

Fig. 3.5 shows the simulated reflectivity ($R$) for stainless-steel and glass substrates with nanostructured ITO coating with $\lambda_{ENZ}$ = 1220 nm under p-polarized excitation and averaged over incident angles from 0° - 60°. It also plots an ideal step-function like reflectivity (black dashed line) along with R($\lambda$ ) for bare stainless steel and glass. The inset shows the schematic of the multilayer coating with nanostructure (NS). The NS coating radically changes the reflectivity of both the opaque and transparent substrates with their coated response R($\lambda$ ) displaying step-function like reflectivity response. Fig. 3.6a shows the secondary electron images of the coated substrates, evidencing the grating NS. Morphological features of the bare substrates and after deposition of each layer are shown in Figs.3.7 and 3.8, with the latter demonstrating large area patterning of the substrates.

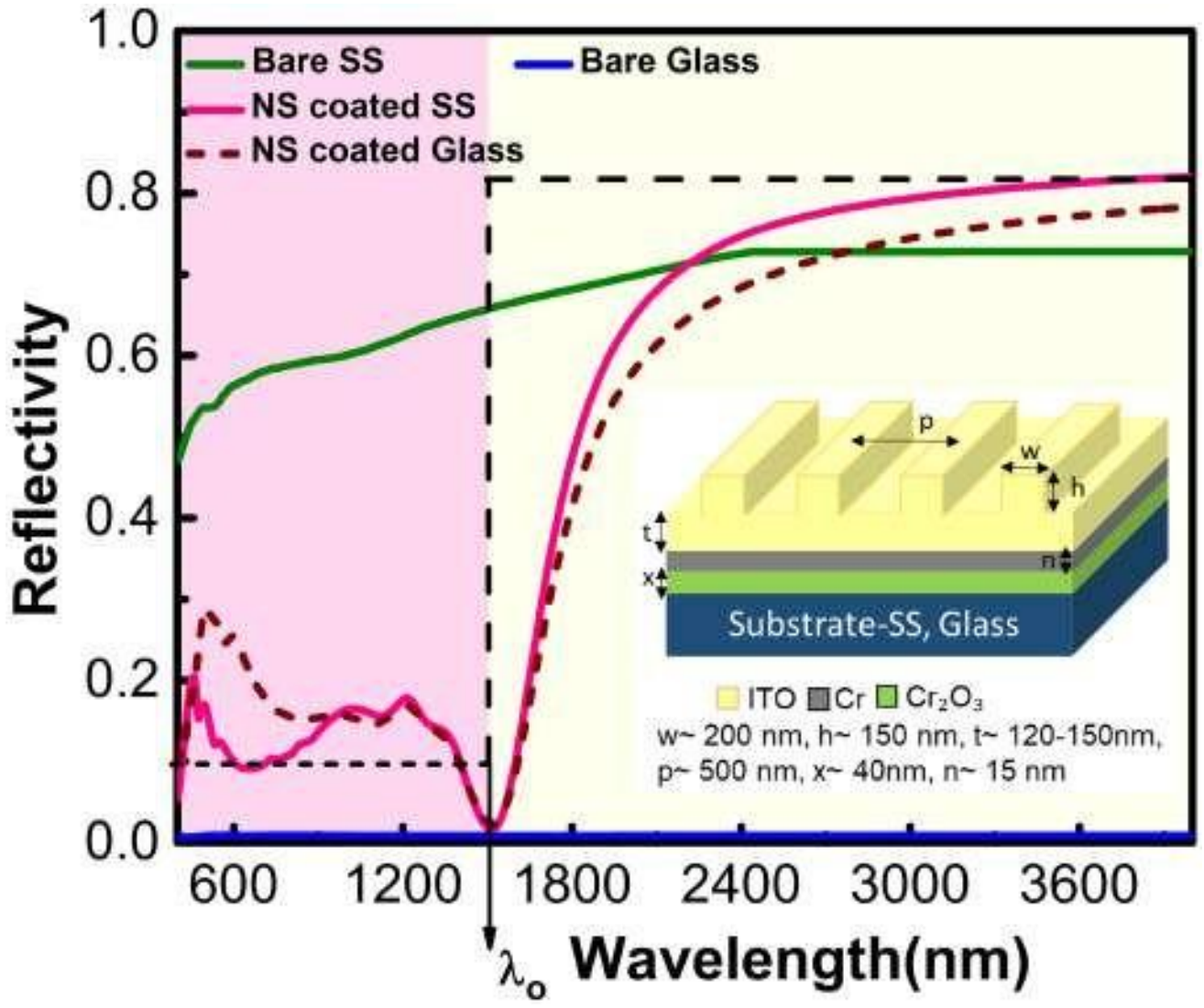

**Figure 3.5:** Simulated reflectivity for glass and stainless-steel substrates with nanostructured coating. The black dashed line shows the ideal "step function" like response (black), and the green and blue lines show simulated reflectivity for bare SS and glass. The inset shows a schematic of the nanostructured coating.

Figs. 3.6(c-d) show the simulated (with p-polarized light) and experimentally detected (with unpolarized light) reflectivity spectra for SS as it is coated by the various components of the multilayer, and Figs.3.6 (e-f) show the same for glass. The plots record evolution of reflectivity with quantifiable agreement between the simulated and experimental results and elucidating the role of individual layers in realizing the final response. Coatings with flat ITO and NS ITO, both show step-function like reflectivity response with different $\lambda_o$ at $\sim$ 1160 nm and $\sim$ 1500 nm and $\Delta\lambda$ of $\sim$250 nm and $\sim$ 400 nm, respectively. $\lambda_o$ is the wavelength at which the reflectivity shows a minima just before the transition from low to high reflectivity, as denoted by arrows in Fig.3.6(c-f). The width $\Delta\lambda$ is defined such that the reflectivity reaches 70% of the saturation (high reflectivity) value at the wavelength $\lambda_o + \Delta\lambda$ as shown in Fig.3.9. ITO is a low-loss dielectric for $\lambda < \lambda_{ENZ}$, thus the optical response at shorter $\lambda$ arise primarily from the properties of the underlying layers Cr/$Cr_2O_3$/substrate. On SS, this structure realizes a Fabry Perot cavity for p-polarized light at wavelengths below 500 nm, with the re-

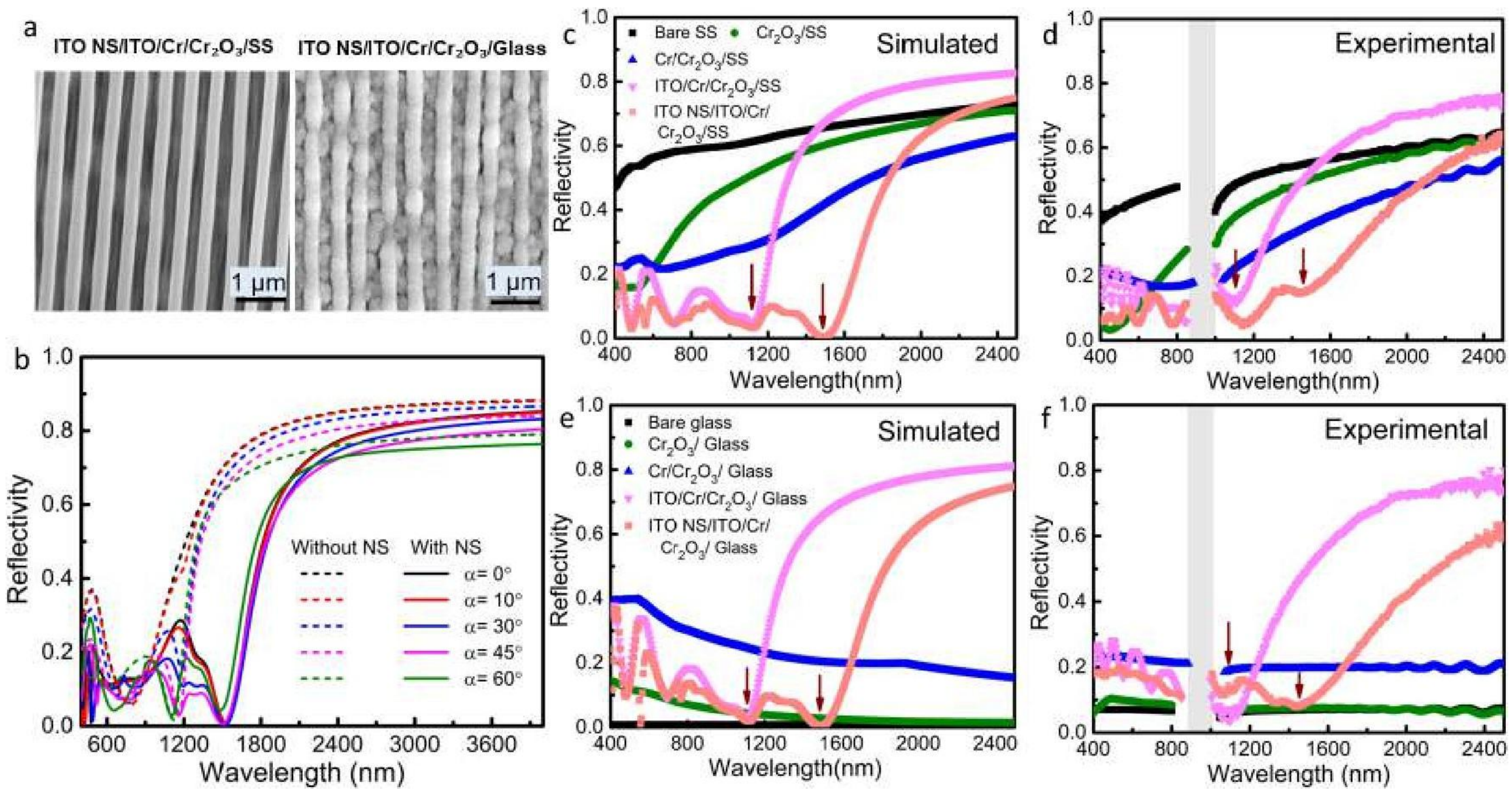

**Figure 3.6:** (a) SEM images of ITO NS/ITO/Cr/$Cr_2O_3$ on SS and glass; (b) Simulated reflectivity spectra for SS coated with flat ITO and NS ITO coating at selected angles ($\lambda_{ENZ} = 1220nm$). Simulated and Experimental reflectivity spectra of (c, d) Bare SS, $Cr_2O_3$/SS, Cr/$Cr_2O_3$/SS, ITO/Cr/$Cr_2O_3$/SS, ITO NS/ITO/Cr/$Cr_2O_3$/SS and (e, f) Bare BK7 glass, $Cr_2O_3$/Glass, Cr/$Cr_2O_3$ /Glass, ITO/Cr/$Cr_2O_3$/Glass, ITO NS/ITO/Cr/$Cr_2O_3$/Glass. Down arrows denote the $\lambda_o$ in each case. The shaded regions in (d, f) are excluded to mask spectrometer noise.

flective Cr and SS increasing the effective interaction of light with these lossy materials that increase absorption and decrease $R$. The low loss, non-dispersive nature of $Cr_2O_3$ from 400 - 4000 nm [166] allows absorption tuning by optimizing the thickness of the Cr layer. The optimal 15 nm Cr layer is critically responsible for balancing the trade-off between light penetration and absorption while maintaining its optical integrity. ITO undergoes a dielectric to metal transition at $\lambda_{ENZ}$ and progressively becomes more reflective for $\lambda > \lambda_{ENZ}$. For wavelengths just below $\lambda_{ENZ}$, in the ENZ regime (where $\varepsilon' > 0$ and $n < 1$) ITO thin films may also exhibit angle-dependent perfect absorption for p-polarized light, thereby reducing $R \sim 0$. Thus, reflectivity for the coating with flat ITO (Fig.3.6b) starts increasing for $\lambda > \lambda_{ENZ}$ with $\lambda_o \approx \lambda_{ENZ}$. The relevance of the NS ITO grating is evidenced in the simulated plots of $R(\lambda)$, shown in Fig. 3.6b for which the $\lambda_o \sim 1500$ nm is red-shifted from $\lambda_{ENZ}$, and is determined by the grating dimensions. The final dimensions of each element are optimized via simulations to deliver the desired spectral response, as shown in Fig.3.10. In spite of the correlations between the experimentally measured reflectivity and the simulations in Figs.3.6(c-f) major differences are apparent. The measured $R$ around $\lambda_o$ are higher than the simulated values though the spectral position of the minima ($\lambda_o$) are reproduced. At longer wavelengths, the measured reflectivity are lower than the simulated values across each layer, and the $\Delta\lambda$ obtained experimentally is higher than predicted in the simulations. Note that the simulations were conducted with p-polarized light to elucidate ENZ response, while the experiments were recorded with unpolarized light. Fig. 3.11 plots the simulated $R(\lambda)$ for the NS ITO coating on SS under s- and p-polarized light, along with the experimental data. Evidently, the higher reflectivity around $\lambda_o$ arises due to the unpolarized nature of the excitation, though the actual response is better than the simulated averaged value. The lower reflectivity at longer wavelengths primarily originates from scattering due to the surface and interfacial roughness of the layers [156] and, finally, the non-uniformity of the nanostructures over the large areas over which they were patterned using e-beam lithography. Importantly, though the predicted and observed values of $\lambda_o$ for coatings with and without NS are in agreement with each other. It is also interesting that the coatings render comparable reflectivity to diverse substrates like glass and SS, which highlights the determining role played by the multilayer. The coated glass substrate shows low reflectivity and transmittivity ($< 20\%$, as shown in Fig.3.12) in the visible for $\theta_{in} = 0^\circ - 90^\circ$, which benefits from the refractive index contrast of 1.5 (glass) and 2.2 ($Cr_2O_3$). It also enables total internal reflection for $\theta_{in} > 42^\circ$ thus, increases interaction with the lossy Cr layer leading to absorption and suppressing reflectivity in the visible regime over a wide angular regime. Inspite of having low reflectivity in the visible, the nanostructured portions show structural colours as evident in the optical images around 45° angles in Fig.3.13.

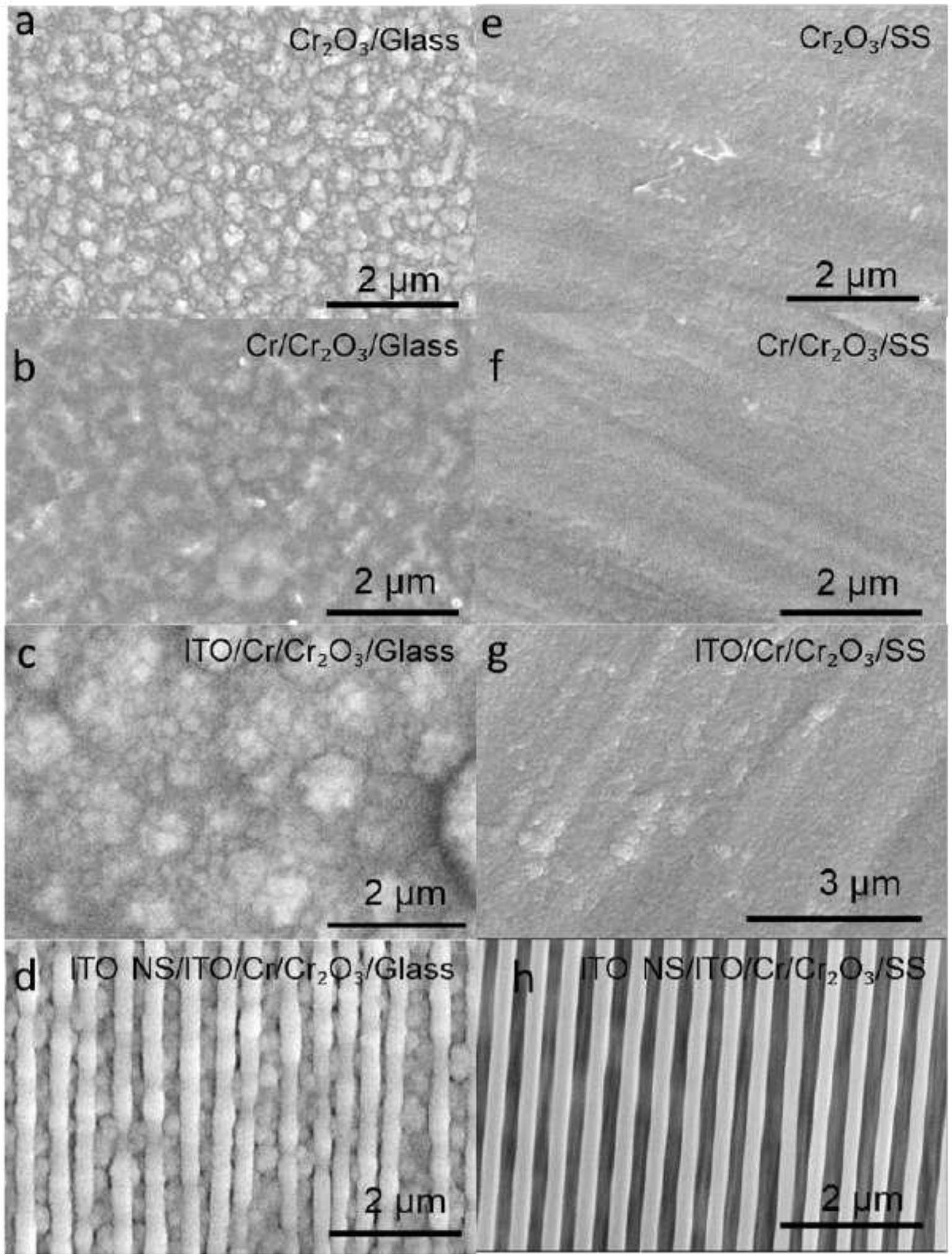

**Figure 3.7:** SEM images of the substrates after deposition of each layer, (a) $Cr_2O_3$, (b) $Cr/Cr_2O_3$ (c) $ITO/Cr/Cr_2O_3$ (d) ITO NS/ $ITO/Cr/Cr_2O_3$ on glass, (e) $Cr_2O_3$, (f) $Cr/Cr_2O_3$, (g) $ITO/Cr/Cr_2O_3$, (h) ITO NS/ $ITO/Cr/Cr_2O_3$ on SS.

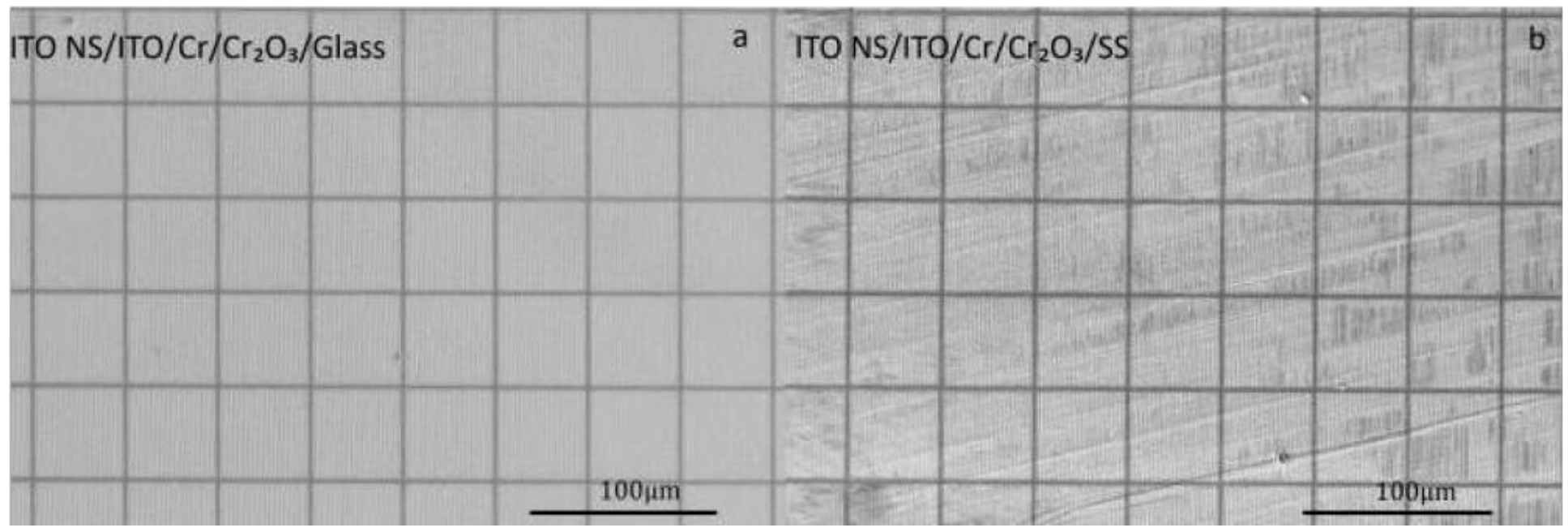

**Figure 3.8:** Large area SEM Images of the final multi-layered sample on (a) Glass substrate, (b) SS substrate

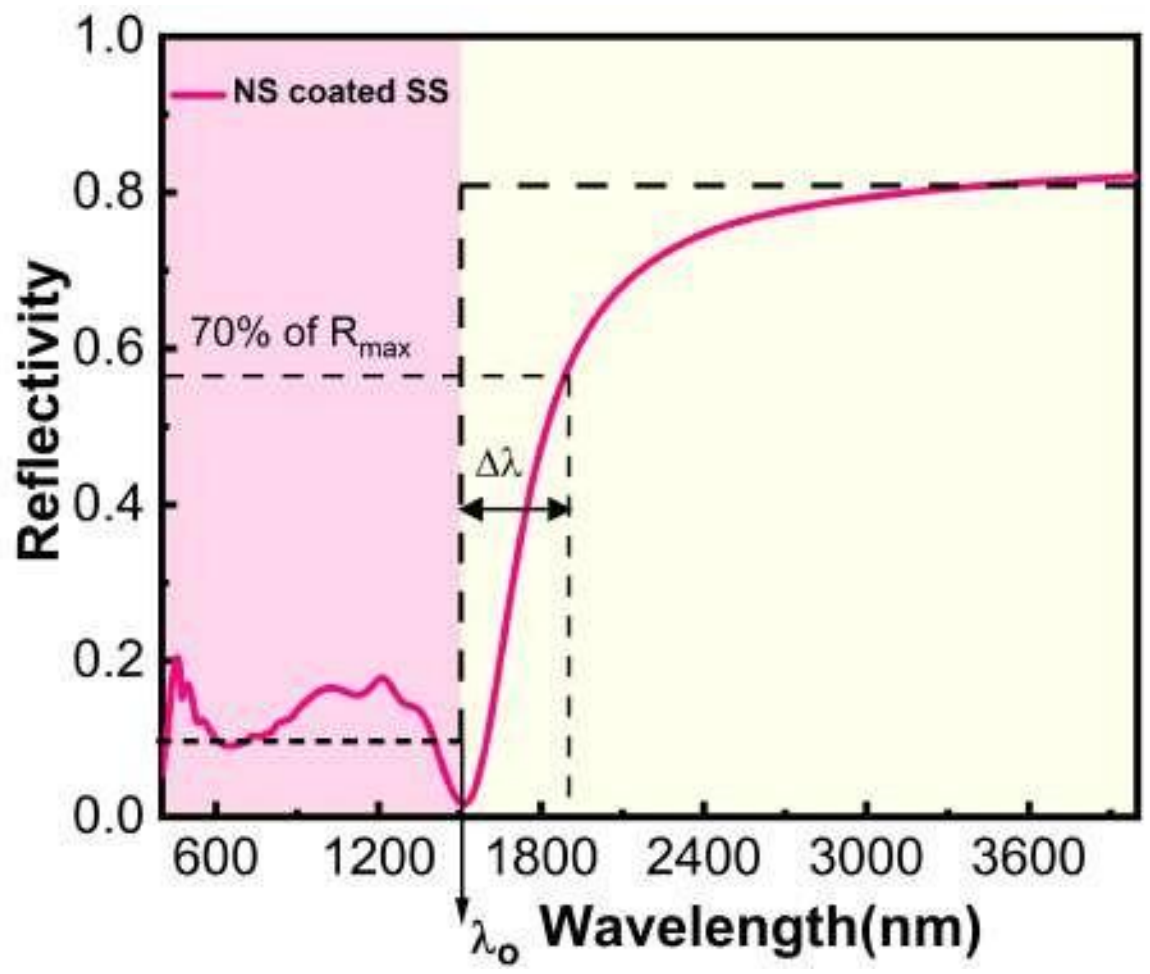

**Figure 3.9:** Step-function-like reflectivity showing the width of transition $\Delta\lambda$

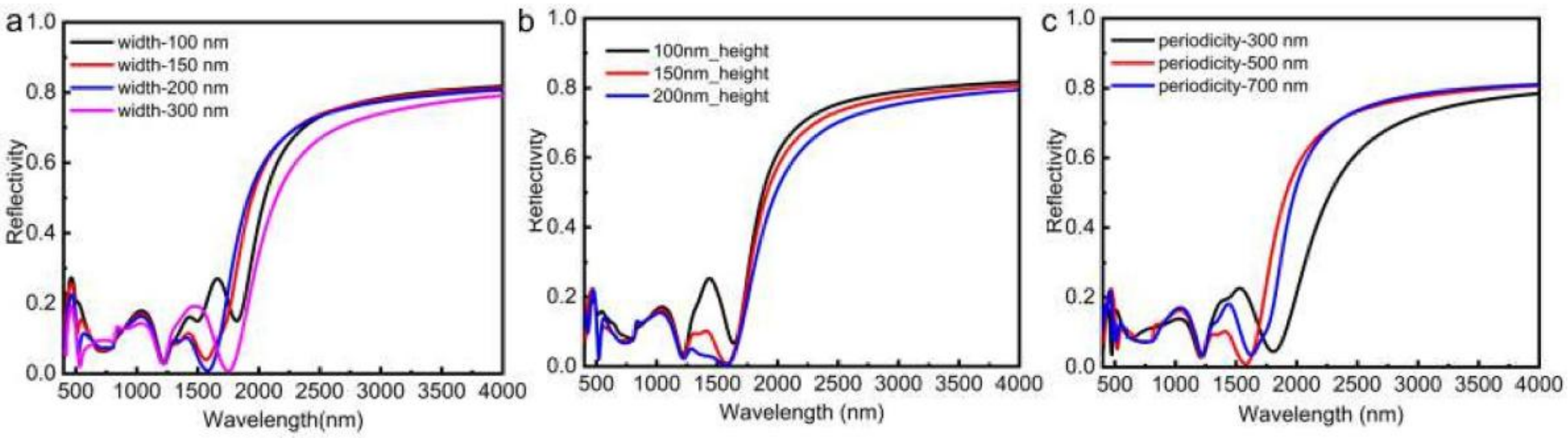

**Figure 3.10:** (a) Simulated reflectivity plots for different widths (100 nm, 150 nm, 200 nm, 300nm) of the NS, (b) for different heights (100 nm, 150 nm, 200 nm) of the NS, (c) for different periodicity (300 nm, 500 nm, 700 nm) of the NS array.

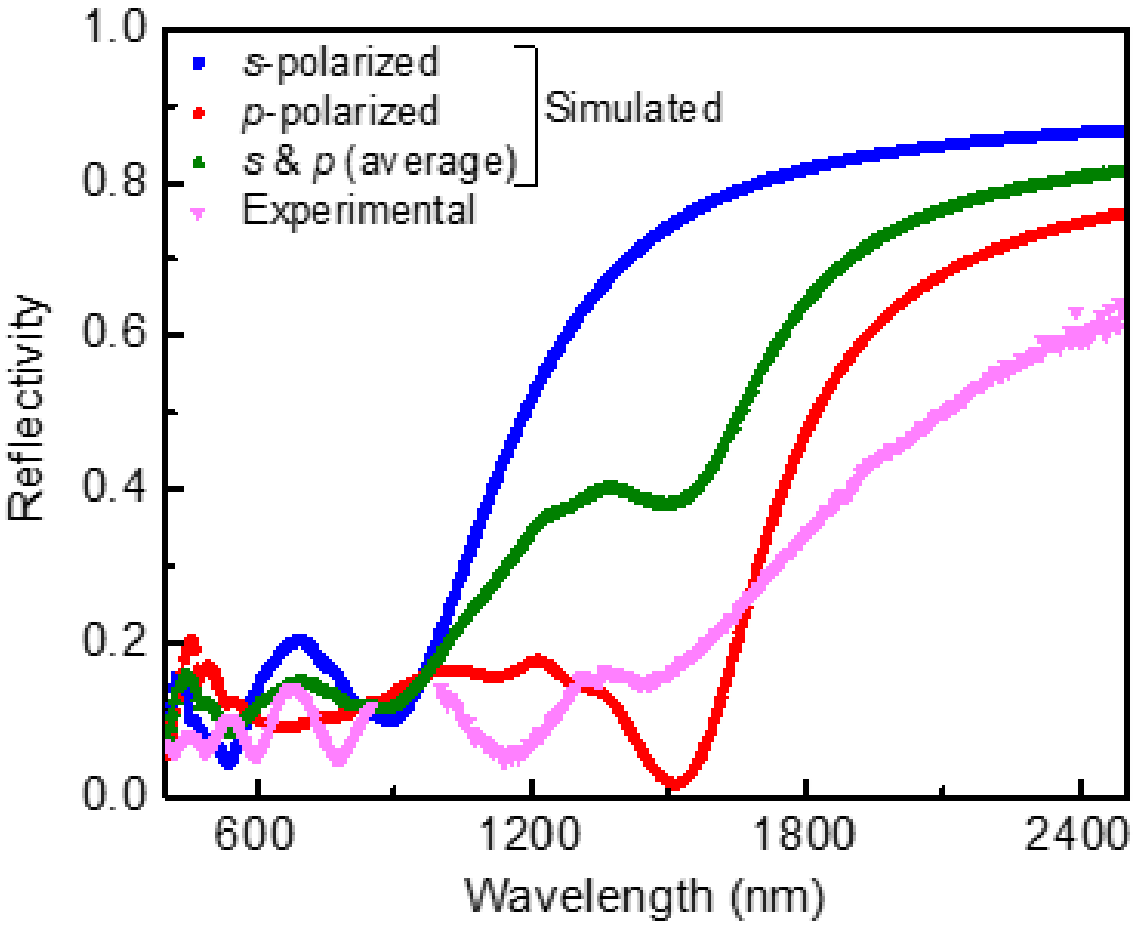

**Figure 3.11:** Simulated average reflectivity (0°-60 °) for the NS ITO coating on SS under different polarization of light, along with the experimental data.

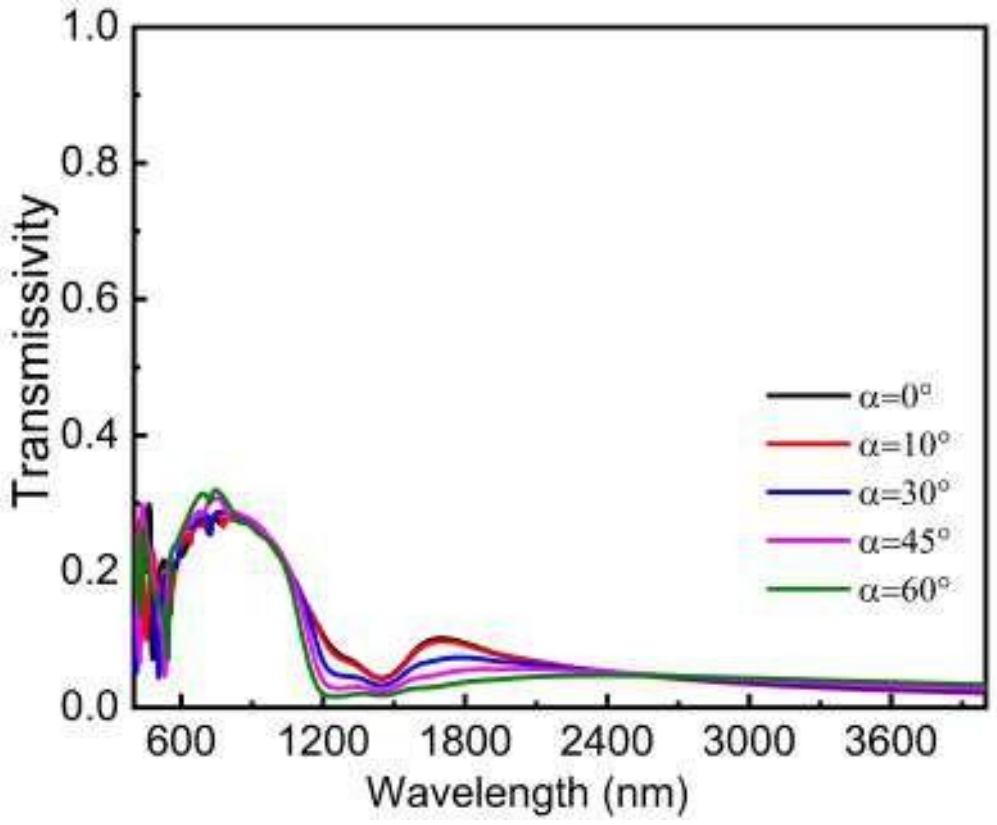

**Figure 3.12:** Transmissivity of the multilayer coating on glass substrate

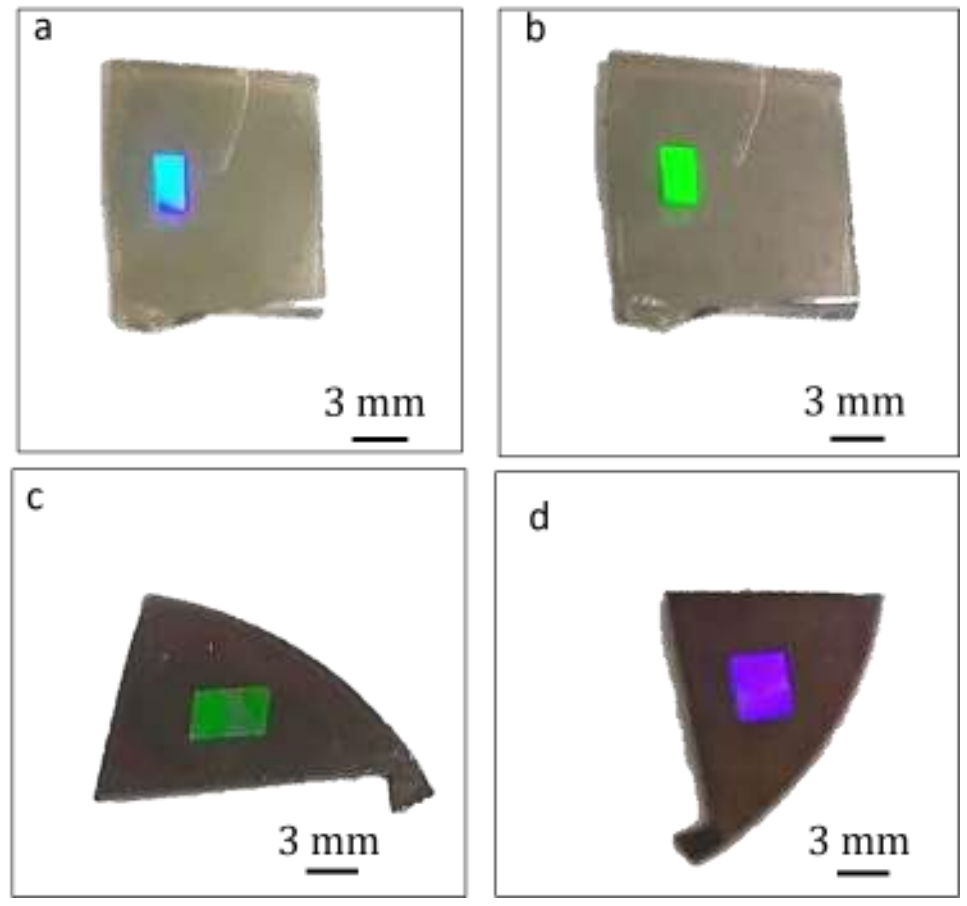

**Figure 3.13:** Optical images of the spectrally selective reflector on (a, b) glass substrate captured from different angles; (c, d) SS substrate captured from different angles. Different colours on the samples demarcate the nanostructured portion.

### 3.3.2 Role of each material and relevance of nanostructures

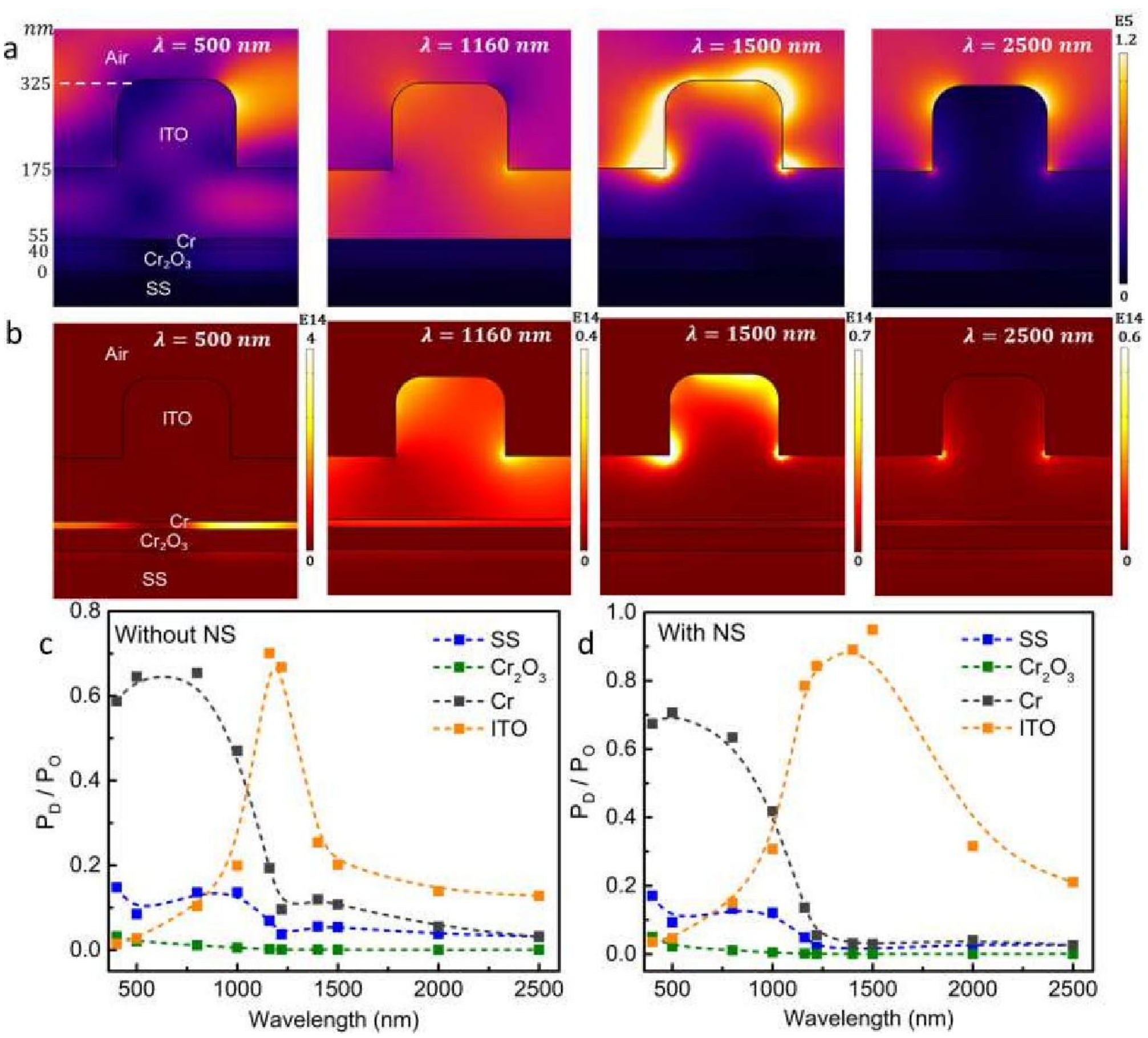

**Figure 3.14:** (a) Simulated electric field (V/m) and (b) dissipated power density (W/m$^3$) across the system at selected wavelengths; Spectral variation of power dissipated as a fraction of incident power, across the various layers (c) without nanostructures and (d) with nanostructures.

The simulations offer further insights into the role of each component of the multi-layer coating in eliciting the overall response. Figs. 3.14(a-b) plot the calculated electric field ($|\mathbf{E}|$) and power dissipation density ($P_D = 1/2(\omega\varepsilon^{''}|\mathbf{E}|^2)$) along the cross-section of the NS coating on SS, at selected wavelengths. Figs. 3.14(c-d) show the spectral variation of net power dissipated ($\int P_D dv$), in the various layers for coatings with and without NS. For $\lambda < \lambda_{ENZ}$, $P_D$ is highest ($\sim$70%) in the Cr layer, the remaining radiation being absorbed in SS. Dissipation in ITO increases as $\lambda \to \lambda_{ENZ}$ and is highest around $\lambda_{ENZ}$ since $\varepsilon^{'} \to 0$, which maximizes the **E** field in ITO and thus $P_D$.

Dissipation decreases at longer $\lambda$ due to the increasing reflectivity of ITO, thus decreasing field penetration. For the coating with the NS $\lambda_o$ red-shifts to 1500 nm, around 300 nm beyond $\lambda_{ENZ}$ and dissipation remains high in ITO between $\lambda_{ENZ}$-$\lambda_o$. The **E** field

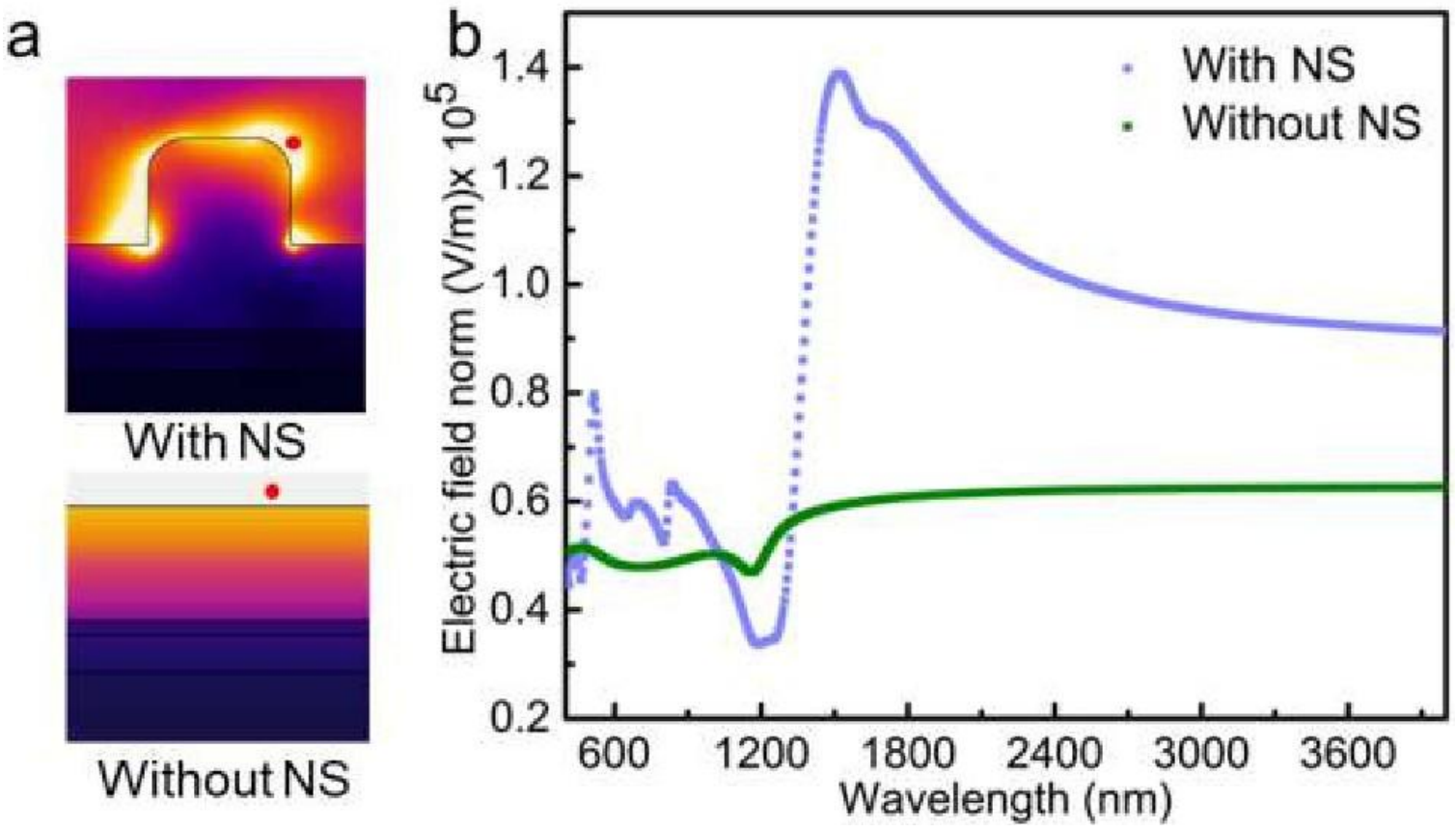

**Figure 3.15:** (a) Electric field plot of the coating with and without NS, (b) Simulated electric field magnitude (V/m) in the vicinity of NS and flat ITO without NS. Red dots indicate the vicinity for the coatings (with and without NS)

plot at 1500 nm shows signatures of dipolar resonances originating from the plasmonic response of the ITO NS, dimensions of which are comparable to the $\lambda$. Also, the calculated **E** field is enhanced in the vicinity of NS compared to the flat coating as shown in Fig. 3.15, indicative of the resonances in NS. They cumulatively impart high absorbance to the NS ITO coating between $\lambda_{ENZ}$-$\lambda_o$ and are responsible for red-shifting $\lambda_o$ showing that the narrow $\Delta\lambda \sim 400$ nm transition width across the wide incidence angle is achievable only with the NS. Note that the reflectivity is also reduced in the visible with the introduction of the NS, as validated by the experimental results in Fig.3.6d and Fig.3.6f. The above features demonstrate the dominant role played by the properties of ITO in achieving the step-function-like reflectivity response and contextualize the relevance of the NS.

### 3.3.3 Absorptivity and emissivity of the multilayer

The calculated emissivities for both the flat ITO and NS ITO coatings on SS, obtained by integrating their $R(\lambda)$ response show a low value of 0.2 in the IR as shown in Fig.3.16 a. The low emissivity is indicative of the coating's efficiency in minimizing the radiative loss for $\lambda > \lambda_o$. In the visible to NIR ($\lambda < \lambda_o$) the flat ITO coating shows an absorptivity of 0.82, which rises to 0.87 for the nanostructured coating. The absorptivity(A) and emissivity (E) of the coating on SS substrate were calculated using the equations 3.1 and 3.2, respectively.

$$A = \frac{\int_{0.4\mu m}^{1.5\mu m}(1 - R(\theta, \lambda))I_S(\lambda, T)d\lambda}{\int_{0.4\mu m}^{1.5\mu m} I_S(\lambda, T)d\lambda} \tag{3.1}$$

$$E = \frac{\int_{2\mu m}^{4\mu m}(1 - R(\theta, \lambda))I_b(\lambda, T)d\lambda}{\int_{2\mu m}^{4\mu m} I_b(\lambda, T)d\lambda} \tag{3.2}$$

$I_S$ is the solar radiation spectrum, $I_b$ is the blackbody radiation spectrum at 300 K and $R$ is the calculated reflectivity at a particular angle $\theta$ and wavelength $\lambda$. Fig. 3.16a plots the angle-dependent absorptivity and emissivity for the NS ITO coating and Fig.3.16b plots the spectral variation of absorptivity averaged over 0° - 60° angles for both the flat and NS ITO coatings.

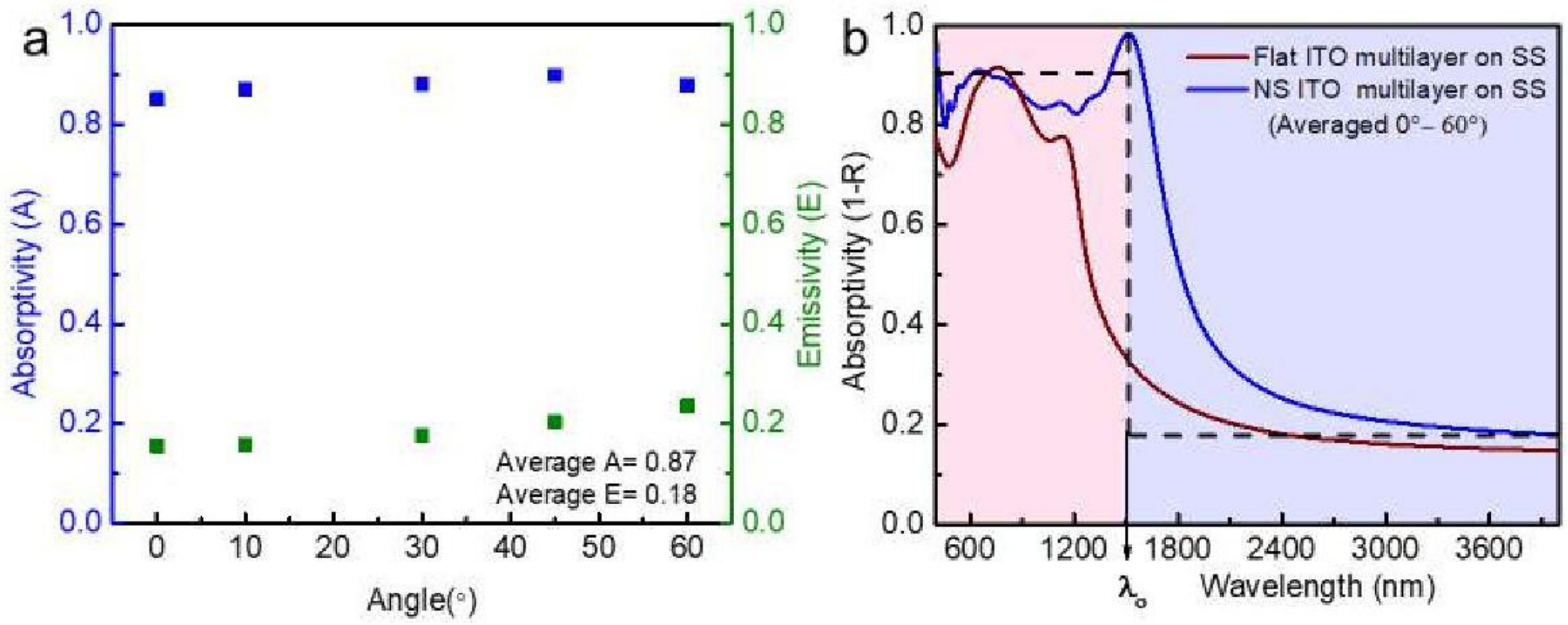

**Figure 3.16:** (a)Variation of Calculated absorptivity and emissivity of multilayer on SS with angle, (b) Spectral variation of absorptivity averaged over 0 ° - 60 ° angles for both the flat and NS ITO coatings.

### 3.3.4 Tunability and role of substrate

The $\lambda_{ENZ}$ of ITO which crucially controls the $\lambda_o$ of the coatings is determined primarily by the electron density ($N_e$) of ITO, thus varying $N_e$ systematically changes $\lambda_{ENZ}$ [169–172] and $\lambda_o$. is readily tuned in ITO by altering the density of oxygen vacancies ($N_{VO}$) via annealing in various $O_2$ partial pressures [169]. Annealing in the ambient decreases $N_{VO}$ and consequently, $N_e$ which red-shifts the $\lambda_{ENZ}$ and $\lambda_o$ as shown in Fig.3.17a and Fig.3.17c.

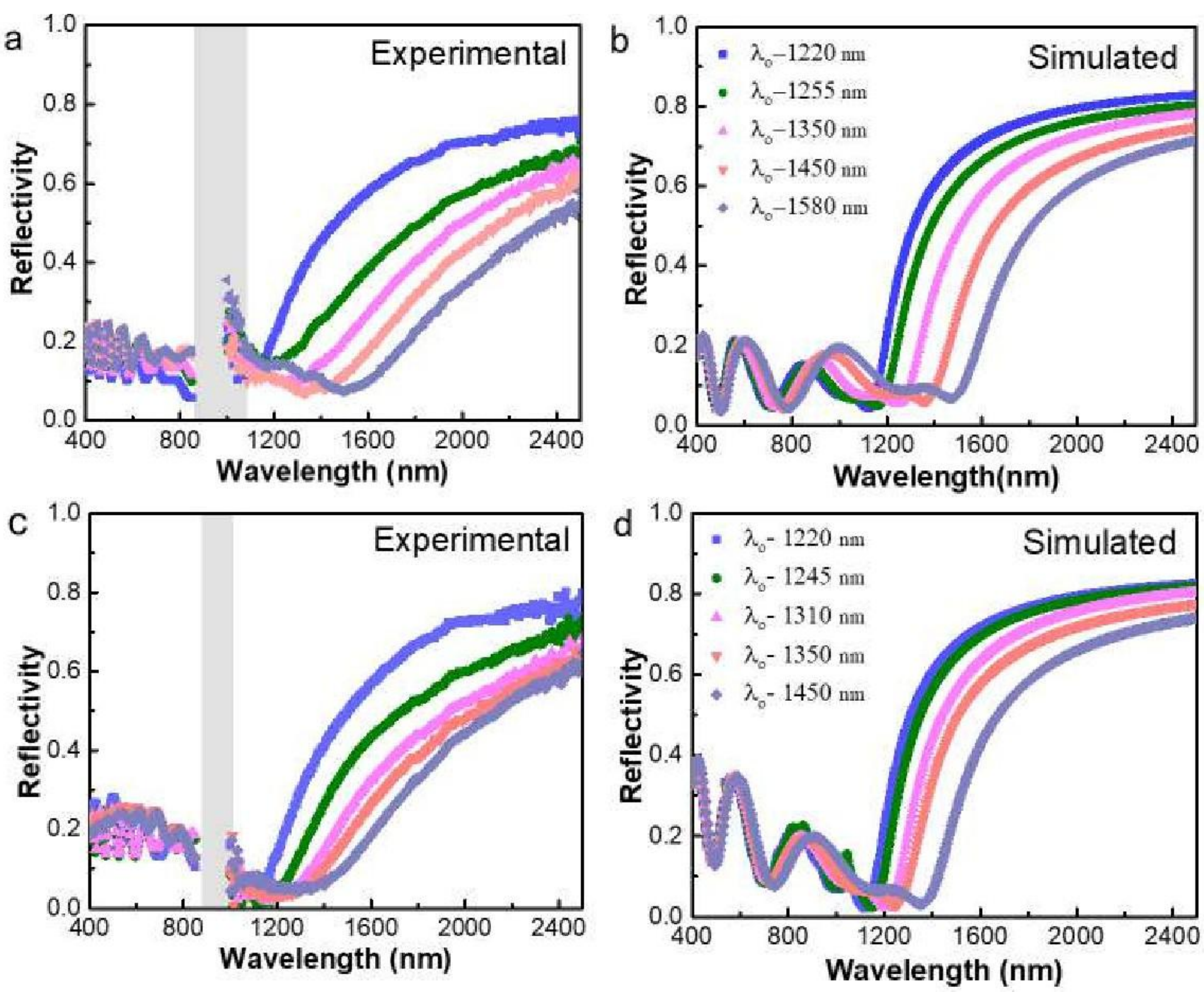

**Figure 3.17:** (a) Experimental and Simulated reflectivity spectra of (a, b) ITO/ Cr/ $Cr_2O_3$/SS; (c, d) ITO/ Cr/ $Cr_2O_3$/Glass on annealing over certain time periods at 350 °C. The shaded region in experimental plots-noise from the spectrometer.

| Annealing period (minutes) | 0 | 10 | 20 | 30 | 40 |
|---|---|---|---|---|---|
| $\lambda_{ENZ}$ (nm) | 1220 | 1255 | 1350 | 1450 | 1580 |

**Table 3.2:** Annealing of multilayer on SS in ambient pressure conditions

| Annealing period (minutes) | 0 | 10 | 20 | 30 | 40 |
|---|---|---|---|---|---|
| $\lambda_{ENZ}$ (nm) | 1220 | 1245 | 1310 | 1350 | 1450 |

**Table 3.3:** Annealing of multilayer on glass in ambient pressure conditions

Here, both for SS and glass substrates the reflectance spectra of the as deposited coating exhibit $\lambda_o \sim 1220nm$, which red-shifts from 1220-1580 nm and 1220-1450 nm upon annealing at 350 °C for 10 mins, as tabulated Tables 3.2 and 3.3 respectively. Fig.3.17b and Fig. 3.17d show the corresponding simulated spectra keeping all material parameters unchanged except decreasing $N_e$, which decreases from $1.05 \times 10^{21}$/cc to $6.62 \times 10^{20}$/cc. The results show that the cut-in wavelength may be variously tailored by thermal annealing in a conducive atmosphere and also by controlling the geometry of the grating pattern, which together provide a high degree of control over the step function.

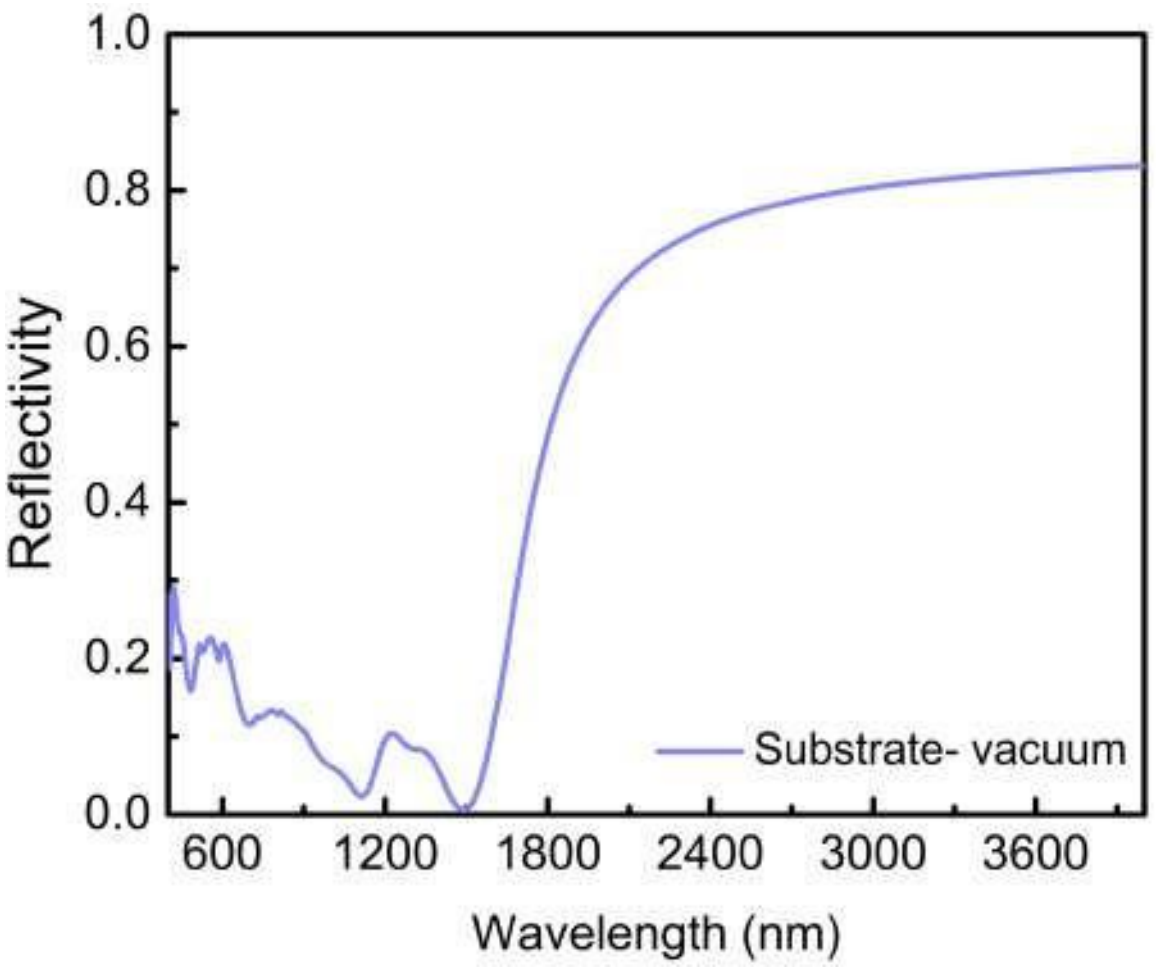

**Figure 3.18:** Simulated reflectance of the spectrally selective reflective coating at 45° with vacuum as the background substrate.

The novelty of the coating's design and applicability in imparting the step-function-like reflectivity onto diverse surfaces is elucidated by the simulated $R(\lambda)$ plot for an unsupported coating, i.e., with vacuum as the substrate, as shown in Fig.3.18. It demonstrates that the optical response is attributed solely to the coating and is minimally dependent on the optical properties of the substrate, though the latter's roughness needs to be substantially lower than the NS grating dimensions to ensure its performance in the IR.

## 3.4 Conclusions

In summary, we have developed a wide-angle, spectrally selective reflector coating that imparts step-function-like reflectivity on diverse substrates like SS and glass under unpolarized light. Based on a straightforward design consisting of three subwavelength-thick layers of $Cr_2O_3$, Cr, and ITO, adorned with a nanostructured grating, the average reflectivity $\sim$ 15% in the visible to NIR and becomes highly reflecting ($\sim$ 80%) above the $\lambda_o$. A combination of optical characterization and numerical simulations have been used to examine the role of individual layers in the integrated coating. Spectral variation in the calculated power dissipation across the layers elucidates the role of each layer, demonstrating the confinement of the electric field in the SS-$Cr_2O_3$-Cr layers, which uniformly suppresses reflectivity and maximizes absorption in the visible to NIR regime. The epsilon-near-zero property of ITO is exploited not only to impart strong reflectivity in the IR, but its nanostructured morphology is shown to improve the step function-like reflectivity. Finally, the easy tunability of ITO's $\lambda_{ENZ}$ via thermal annealing is shown to systematically shift the overall spectral response of the coating in the IR and this is perhaps the first demonstration of spectral tunability of a coating utilizing an ENZ material. Such a mechanism of obtaining tunable spectral selectivity over optically diverse substrates (opaque SS and transparent glass) is promising in the field of energy management and demonstrates that the design ideas may be translated to other ENZ materials across other spectral regimes. Moreover, the design strategy could be further enhanced using inverse design methods or machine learning algorithms to discover new material multilayers or geometries that optimize spectral selectivity, angular response, while keeping the "step-function" like reflectivity intact. Beyond energy management, the spectral control capabilities may be extended to applications in infrared camouflage, photodetectors, or selective thermal emitters with the selection of relevant ENZ materials where sharp transitions in reflectivity are important. Finally, this CMOS-compatible fabrication based on thin-film coatings and elemental nanostructuring is capable of achieving wide-angle, spectrally selective behavior offering a robust and cost-effective solution in optical engineering.

# Chapter 4

# Engineering band-selective absorption with epsilon-near-zero media in the infrared

Band-selective absorption and emission of thermal radiation in the infrared are of interest due to applications in emissivity coatings, infrared sensing, thermo-photovoltaics, and solar energy harvesting. The broadband nature of thermal radiation presents distinct challenges in achieving spectral and angular selectivity, which are difficult to address by prevalent optical strategies, often yielding restrictive responses. We explore a trilayer coating employing a nanostructured grating of epsilon-near-zero (ENZ) material, indium tin oxide (ITO), atop a dielectric ($SiO_2$) and metal (gold) underlayer, which shows wide-angle (0–60°) and band-selective (1.8 - 2.8 $\mu$m) high absorption ($> 0.8$). Numerical simulations and experimental results reveal that the ENZ response of ITO combined with its localized plasmon resonances define the high absorption bandwidth, aided by the sandwiched dielectric's optical properties, elucidating the tunability of the absorption bandwidth. Thermal imaging in the mid-infrared highlights the relevance of the ENZ grating, emphasizing the potential of this coating design as a thermal emitter. This study offers valuable insights into light-matter interactions and opens avenues for practical applications in thermal management and energy harvesting.

## 4.1 Introduction

Engineering absorption and emission properties of surfaces through optical coatings have garnered significant attention in recent years [173–175] due to their importance in applications like solar energy harvesting,[144, 176], thermal and optical imaging, [177, 178] photo-detection,[179] bio-sensing,[180, 181] and medical treatment,[182] radiative cooling systems,[159] etc. Various design strategies, using metal-insulator-metal (MIM) multilayers,[183] gratings,[184–186] meta-materials,[187–189] etc. have been explored to enhance absorption and emission properties of surfaces both over wide and selected spectral windows aiming to optimize directionality and efficiency of response[190]. Here, a nanostructured indium tin oxide (ITO) – $SiO_2$ – gold multilayer coating is investigated that shows wide angle, band-selective absorption and emission response in the near-infrared (NIR) with a bandwidth $\Delta\lambda$ across the central wavelength $\lambda_c$, as depicted graphically in Fig.4.1.

Band selective emitters and absorbers have been designed leveraging mechanisms like cavity absorption,[191, 192] plasmonic resonances,[193] and interferences supported by ultra-thin films of epsilon-near-zero (ENZ) materials[92]. ENZ media allow unprecedented light-matter interactions close to the ENZ wavelength ($\lambda_{ENZ}$), at which the material's optical properties show a cross-over from dielectric to metallic, with the real part of its dielectric constant ($\varepsilon(\omega) = \varepsilon' + i\varepsilon''$) going to zero. In this ENZ regime ultra-thin films of ENZ media (thickness $\sim$ 10 nm) exhibit angle and wavelength dependent perfect absorption, which is attributed to the excitation of ENZ modes along with extreme electric field enhancement and confinement in the thin film.[91, 92, 169, 194]. Structured ENZ media have been shown to impart broadband and band-selective absorption properties to various surfaces,[161, 190, 195–198] which leveraging Kirchhoff's law[101] show high emissivity within the same spectral band. Table 1 lists the properties of such band-selective coatings, with $\lambda_c$ denoting the central wavelength, extending from the visible to mid-IR and the bandwidth ($\Delta\lambda$) varying from 600 nm to over 5 $\mu$m. However, the experimental realization of such coatings suffers from two major challenges, the complexity of design and a strong directional dependence.

Here, we investigate strategies to address these shortcomings to demonstrate band-selective absorption and emission in the NIR, employing a tri-layer coating of a linear grating of an ENZ media (ITO) adorning a dielectric on metal ($SiO_2$/Au) underlayer on transparent BK7 glass. While glass has zero absorptivity in the NIR, the ITO/$SiO_2$/Au coating with ITO having $\lambda_{ENZ}$ of 1790 nm shows more than 85% absorption in the wavelength range 1800 - 2900 nm with $\lambda_c$ = 2.3 $\pm$ 0.5 $\mu$m and low absorption elsewhere. Notably, both absorption and radiation from the coated surface are omnidirectional up to 60° to the normal. Additionally, tunability of the spectral window is demonstrated by modifying $\lambda_{ENZ}$ through annealing ITO in oxygen-rich or oxygen-lean environments, which decreases and increases $n_e$, respectively. The universality of the design, including the relevance of the ENZ grating and the sandwiched dielectric, is discussed to show that the $\lambda_{ENZ}$ of ITO primarily control the low wavelength limit ($\lambda_i$) of the band with the high wavelength cut-off ($\lambda_f$) and thus the $\Delta\lambda$ determined by the dielectric spacer along with the grating periodicity and dimensions. Overall, the ENZ/dielectric/metal coating design showcases the inherent flexibility of the coating design, its ease of fabrication, and robustness, making it elemental in achieving band-selective absorption and emission. These findings encourage the integration of such coatings in applications like thermo-photovoltaics and devices for tailored IR emission.

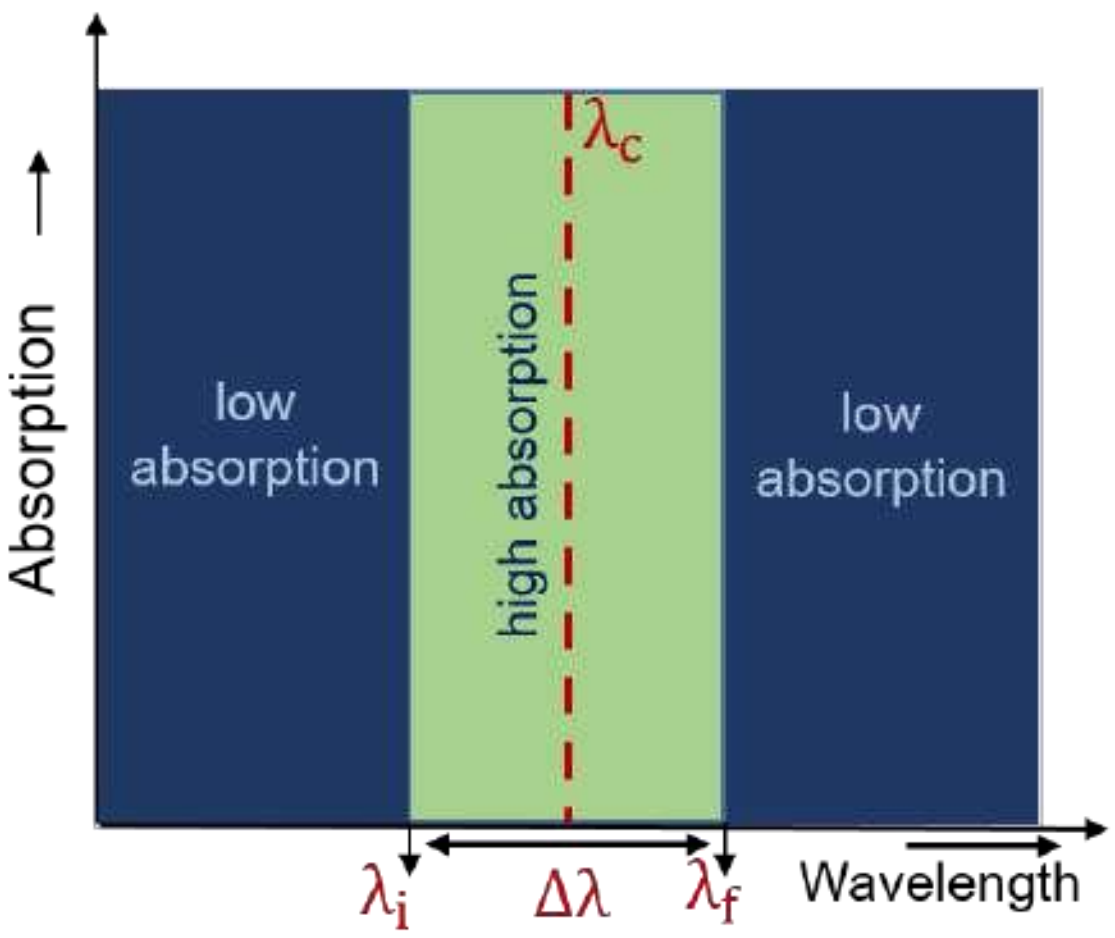

**Figure 4.1:** Graphical representation of a band-selective absorber

| Coating Material | $\lambda_c (\mu m)$ | $\Delta\lambda$ ($\mu$m) | Angular response (°) | Ref. |
|---|---|---|---|---|
| *[1]ITO NS | 4.25, 2.25 | 6.7,3.0 | 0-60 | [199] |
| *Au NS/ ENZ/$SiO_2$/ TiN/Si | 8.6 | 3.5 | 24 | [200] |
| *[2]Au NS/ITO | 2.1 | 1.6 | $< 50$ | [162] |
| *[3]Au NS/ITO/$SiO_2$ /Au | 1.5 | 0.6 | 0 | [201] |
| *MgO/$BaZr_{0.5}Hf_{0.5}O_3$/NiO | 14.9 | 2.5 | 50-70 | [198] |
| *Graphene meta-surface | 0.6 | 0.8 | 0-60 | [202] |
| Hyperbolic metamaterial | 4.0 | 2.0 | - | [203] |
| ITO/Au NS | 1.55 | 0.5 | $< 20$ | [204] |
| Graphene/$SiO_2$/Al | 2, 3, 4 | - | - | [205] |
| ITO nanostructures/ $SiO_2$/Au | 2.3 | 1.0 | 0-60 | This work |
| ITO/$SiO_2$ metamaterial | 1.75 | 1.45 | - | [206] |
| $SiO_2$/ SiO/ $Al_2O_3$/Al/Si MgO/ $Ta_2O_5$/ $TiO_2$/Al/Si | 9.5, 12 | 4.0 | 60-80 | [190] |

**Table 4.1:** Summary of band selective absorber coatings. * theoretical investigation, NS: nanostructures, [1]nanocylinder array, [2]split-ring array, [3]nanodisc array. Note : The numbers given in the table are approximate values.

## 4.2 Material and methods

### 4.2.1 Simulation

The finite element method is a widely used numerical simulation tool for determining electromagnetic properties of optical systems. A unit cell of the investigated system was designed in COMSOL 5.3a as shown in Fig.4.2 and was simulated using periodic boundary conditions. Two periodic ports were used where one was used to launch the incident wave and the other to collect the transmitted wave. Simulations were conducted using p-polarized light at selected angle of incidence and the reflectivity (R) was calculated. 100 nm gold on glass being a completely opaque substrate, the transmission through the system was zero and hence the absorption (A) is 1-R. The structure optimization was carried out to maximize absorption over a broad spectral range considering the feasibility of fabrication and experimental verification of the samples. It was used to minimize reflectivity over the IR over a selected spectral range with a straightforward fabrication protocol. The simulations were carried out in the range 1200 – 4000 nm though the experimental verification is limited to 2500 nm. The optical properties of the materials (Au, $SiO_2$) were taken from the literature[166, 168] and the Drude model was used to determine the optical properties of ITO as a function of three parameters, carrier concentration ($N_e$), scattering parameter ($\gamma$) and background permittivity ($\varepsilon_\infty$). The typical values of the parameters ($N_e$, $\gamma$ and $\varepsilon_\infty$) are taken from a previous publication[169] from the group.

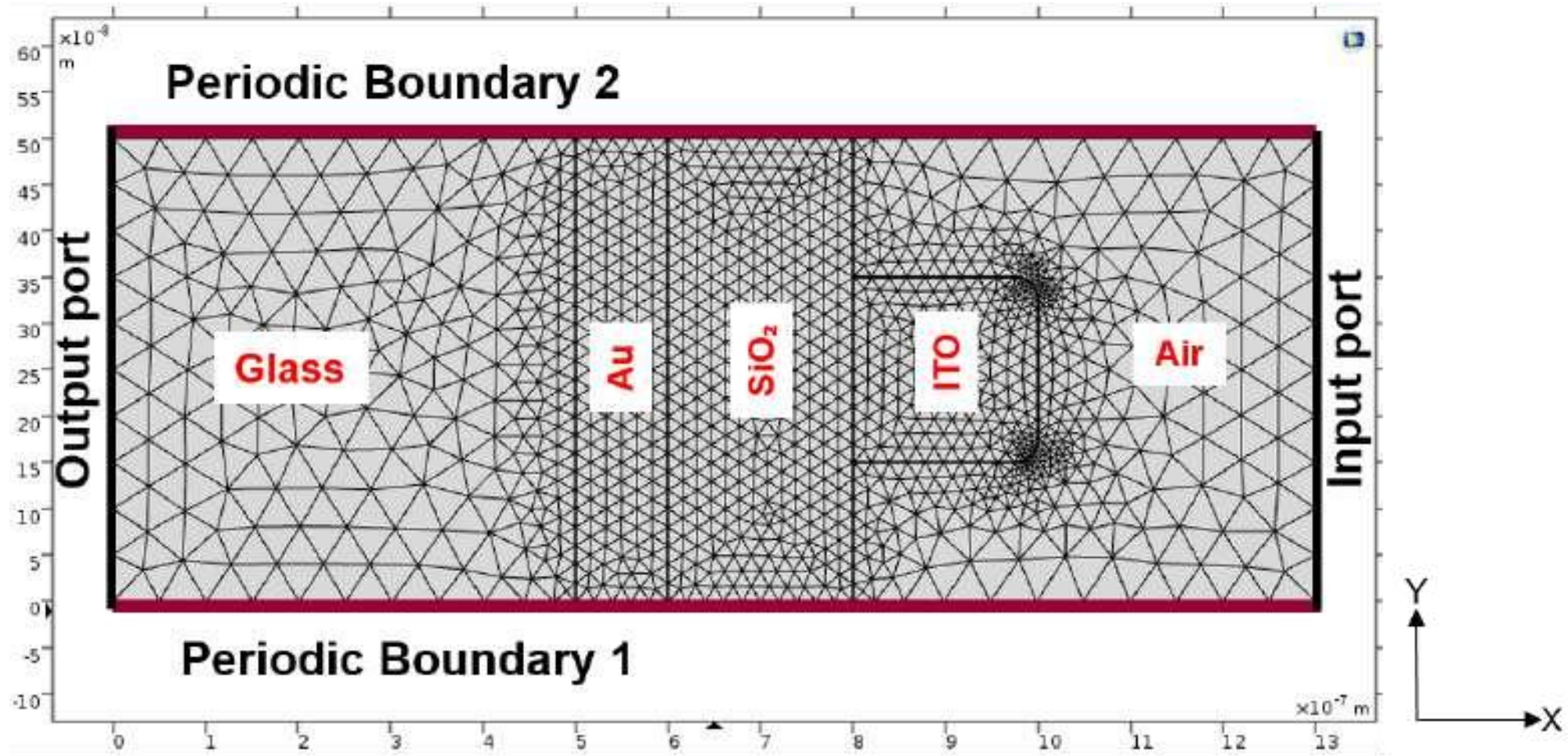

**Figure 4.2:** Cross-sectional view of the unit cell designed in the Wave optics module of COMSOL Multi-physics 5.3a.

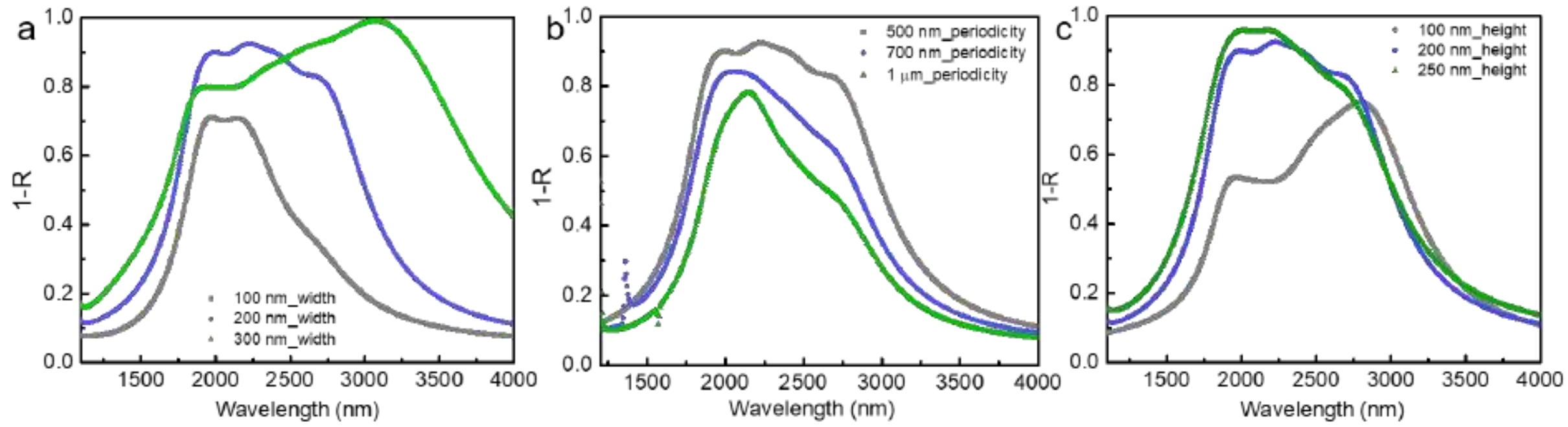

**Figure 4.3:** (a) Simulated 1-R plots for different widths (w) (100 nm, 200 nm, 300nm) of the NS, (b) for different periodicity (p) (500 nm, 700 nm, 1 $\mu$m) of the NS, (c) for different heights (h) (100 nm, 200 nm, 250 nm) of the NS array.

### 4.2.2 Sample fabrication

Cut pieces of glass were cleaned with acetone, isopropyl alcohol, and de-ionized water. 100 nm gold (Au) was evaporated on the substrates using thermal evaporation at a rate of 1 Å$/s$. $\sim$ 200 nm $SiO_2$ was sputtered on the Au/ Glass substrate using RF sputtering. Further, the $SiO_2$/ Au deposited substrates were spin-coated with 10 wt% PMMA resist and patterned using electron beam lithography (Raith Pioneer 2) to create the array of ITO nanostructures of $\sim$ 200 nm width and $\sim$ 500 nm (centre to centre) periodicity. Lithographed samples were developed in a solution of methyl isobutyl ketone (MIBK) and isopropyl alcohol (IPA) in $1:3$ volumetric ratios for 10 s followed by IPA wash for 10 s. Finally, 200 nm ITO was deposited onto the patterned substrates using RF sputtering with lift-off using warm acetone to yield the final grating on $SiO_2$/Au substrate. Details of RF sputtering of ITO and $SiO_2$ are given in Chapter 2.

### 4.2.3 Material and Optical Characterization

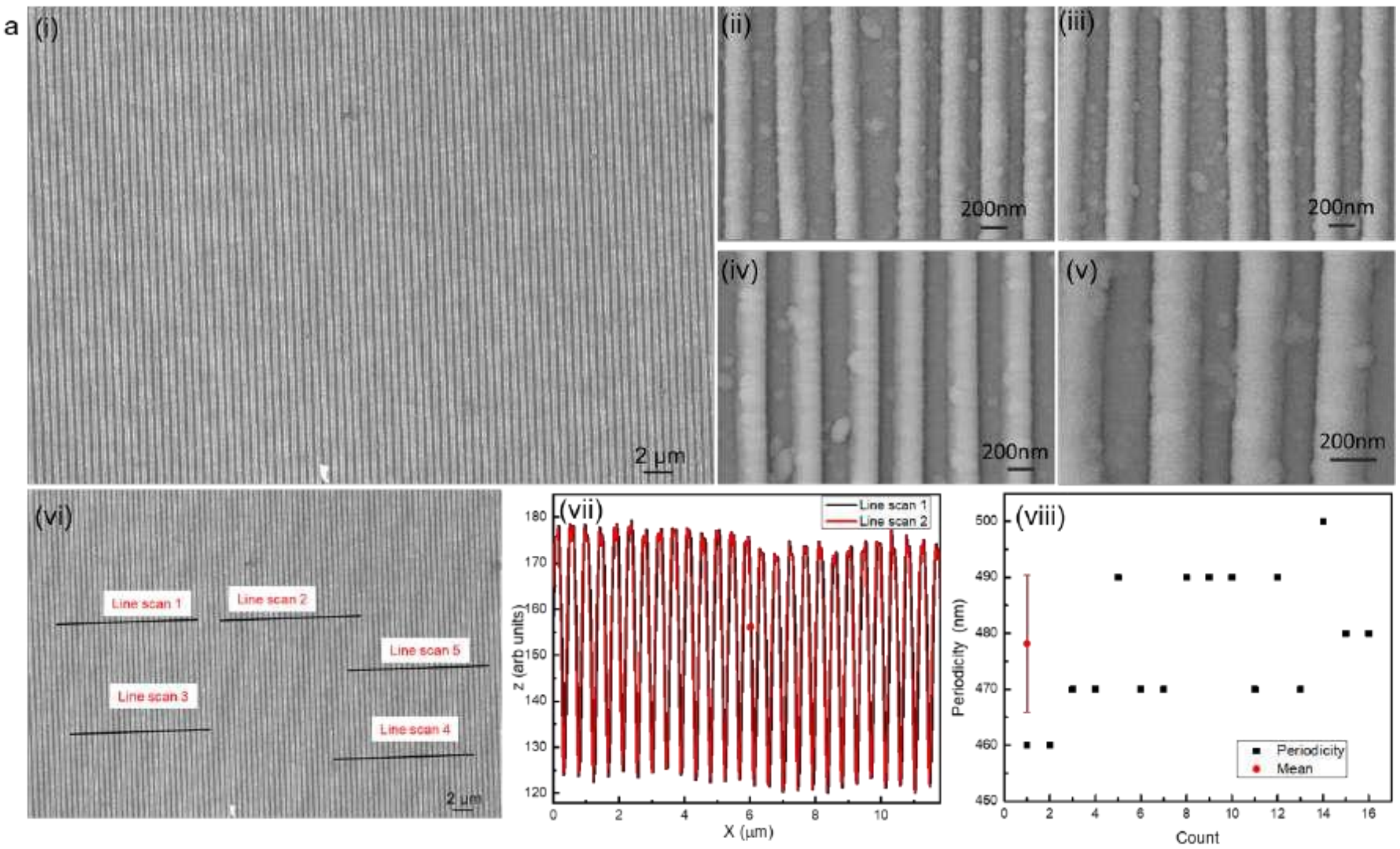

**Figure 4.4:** a. (i-vi) SEM images of the ITO grating/$SiO_2$/Au showing the line scans, (vii) Line scans showing periodicity of the grating, (viii) Periodicity vs number of nanostructures plot showing the average periodicity and the standard deviation.

Morphological characterization at each stage of sample development was performed using Nova Nano SEM 450 field emission scanning electron microscope (SEM). The angle-dependent specular reflectivity measurements were conducted using the universal reflectance accessory module of Perkin Elmer 950 spectrophotometer and the scattering data was obtained using the Integrating Sphere module of the Perkin Elmer 950 spectrophotometer after coating each layer in the 1200 nm-2500 nm wavelength range. Thermal emissivity measurements were performed using an IR camera (Fluke Ti480 Pro).

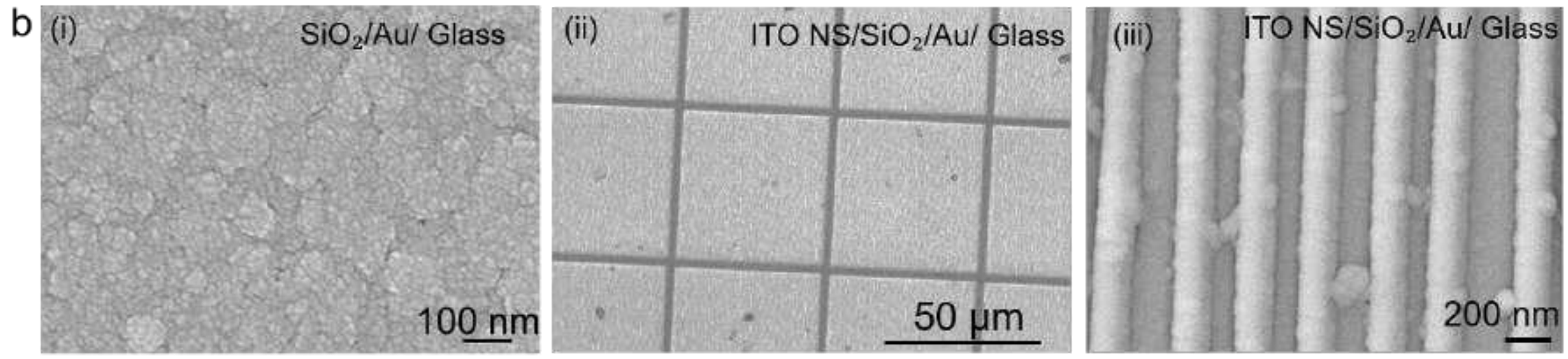

**Figure 4.5:** a.(i) SEM image of $SiO_2$/Au coating,(ii) Large area SEM image of ITO grating /$SiO_2$ /Au (iii) Magnified view of ITO grating/$SiO_2$/Au

## 4.3 Results and Discussion

### 4.3.1 Development of band-selective absorber and its aspects

Fig. 4.6a shows a schematic of the trilayer coating of ITO grating (periodicity $p \sim 500$ nm, width and height $w = h \sim 200$ nm ) on top of a 200 nm thick layer of $SiO_2$, on a 100 nm thick Au back-reflector. While the 100 nm Au film is opaque in the visible to NIR regime, the 200 nm $SiO_2$ thin film is transparent, which renders the coating opaque as shown in Fig. 4.7. However, inclusion of the ITO grating imparts strong spectral selectivity to the simulated absorption spectrum, as shown in Fig.4.6b. Under $p$-polarized illumination, the absorptivity of the coating, $A > 0.85$ within the spectral band 1800 - 2900 nm, for all angles of incidence up to $\approx 60º$. All simulations were performed with incident light having $|\mathbf{E_{in}}|$ = $10^4$ V/m, and details regarding the simulation model and optimization of the grating dimensions are available under the Materials and methods sections. Details of sample fabrication and characterization are given under the experimental methods section. The band-selective response is brought about under the dual action of the ENZ and the plasmonic response of ITO in conjunction with the refractive index contrast provided by the underlying layers. Fig.4.9a and 4.9b show the spectral variation of the simulated and experimental values of $A$ of the coating on glass for various angles of incidence, with the inset showing the secondary electron (SE) image of a

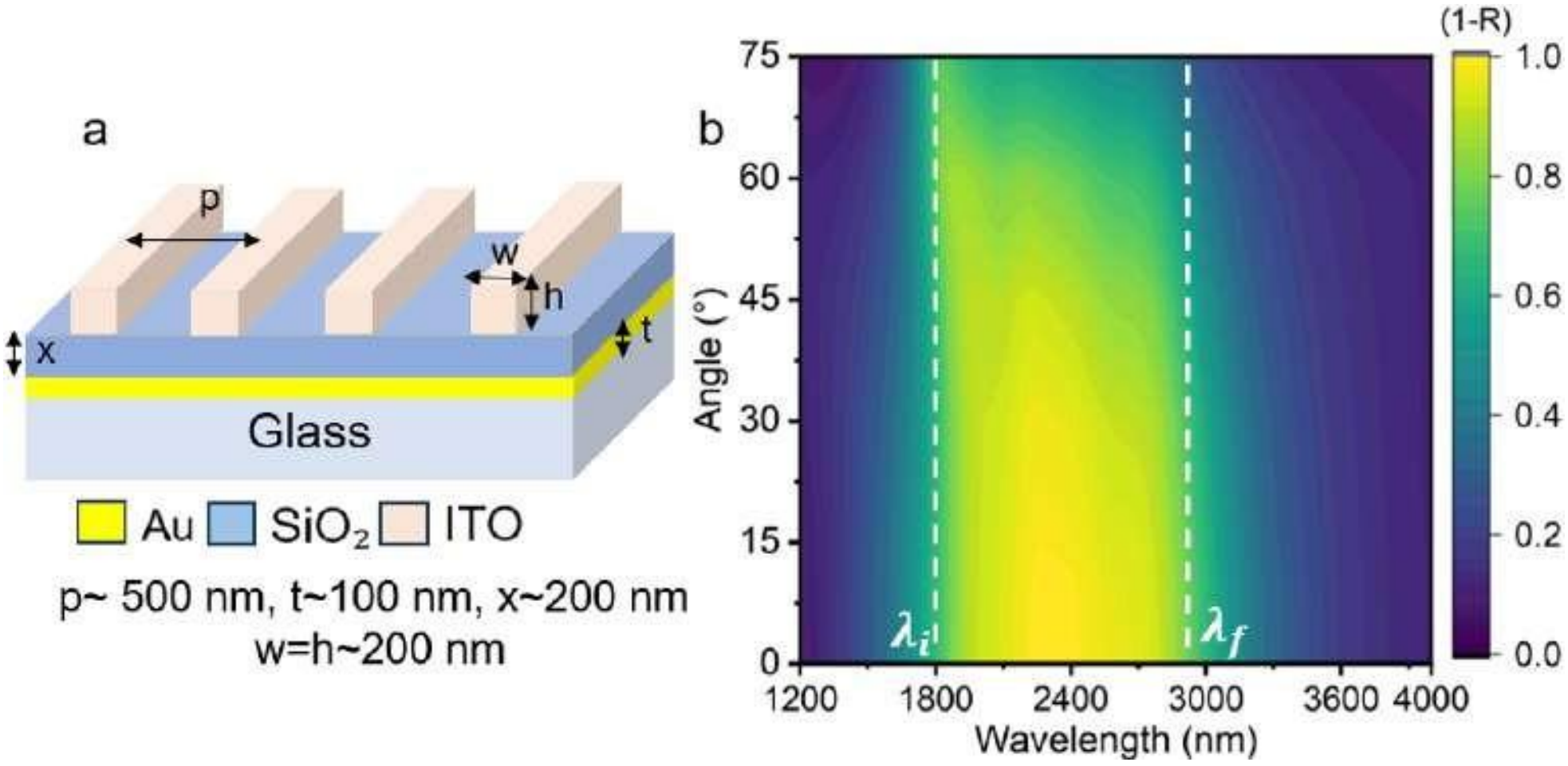

**Figure 4.6:** (a) Schematic of the band-selective absorber coating with optimized dimensions, (b) Simulated absorption of the coating for angles 0-75º for ITO with $\lambda_{ENZ}$ = 1790 nm under p-polarized light.

magnified view of the ITO grating. Fig.4.4 under the material characterization section shows a large area SE image of the grating along with magnified images recorded at various regions. Line scans recorded across multiple regions of the grating evidence the uniformity of the nanostructures with the grating periodicity $p \simeq 480 \pm 12$ nm (starting edge of one NS to the other). The low standard deviation benchmarks the quality of the grating necessary to elicit the band-selective absorption response from the coating. SE images across the developmental stages of the coating are further shown in Fig.4.5. Optical characterization of the ITO yields a $\lambda_{ENZ} \sim 1790$ nm, as shown in the spectral variation of $\varepsilon$ and refractive index ($\tilde{n}$) in Fig.4.8, which have been used to simulate the absorption spectrum in Fig.4.9a. Both results reproduce the high absorption above 1800 nm and at all angles up to 50º. Note that while the simulations were performed for $\lambda$ = 1200 - 4000 nm, instrumental limitations restricted experimental results up to $\lambda$ = 2500 nm. Hence, the falling edge of the band-selective response remained experimentally unexplored in the sample shown in Fig.4.9. The band-selective response is quantified in terms of the central wavelength ($\lambda_c$) of the band as stated earlier and its absorption bandwidth, $\Delta\lambda = \lambda_f - \lambda_i$, where $\lambda_f$ and $\lambda_i$ are the band-edge wavelengths between which $A$ reaches 70% of the maximum value as shown in Fig.4.10. Dashed lines in Fig.4.6 denote $\lambda_i$ and $\lambda_f$ at 1800 nm and 2890 nm, yielding $\Delta\lambda \sim 1.0\ \mu$m and $\lambda_c \sim 2300$ nm.

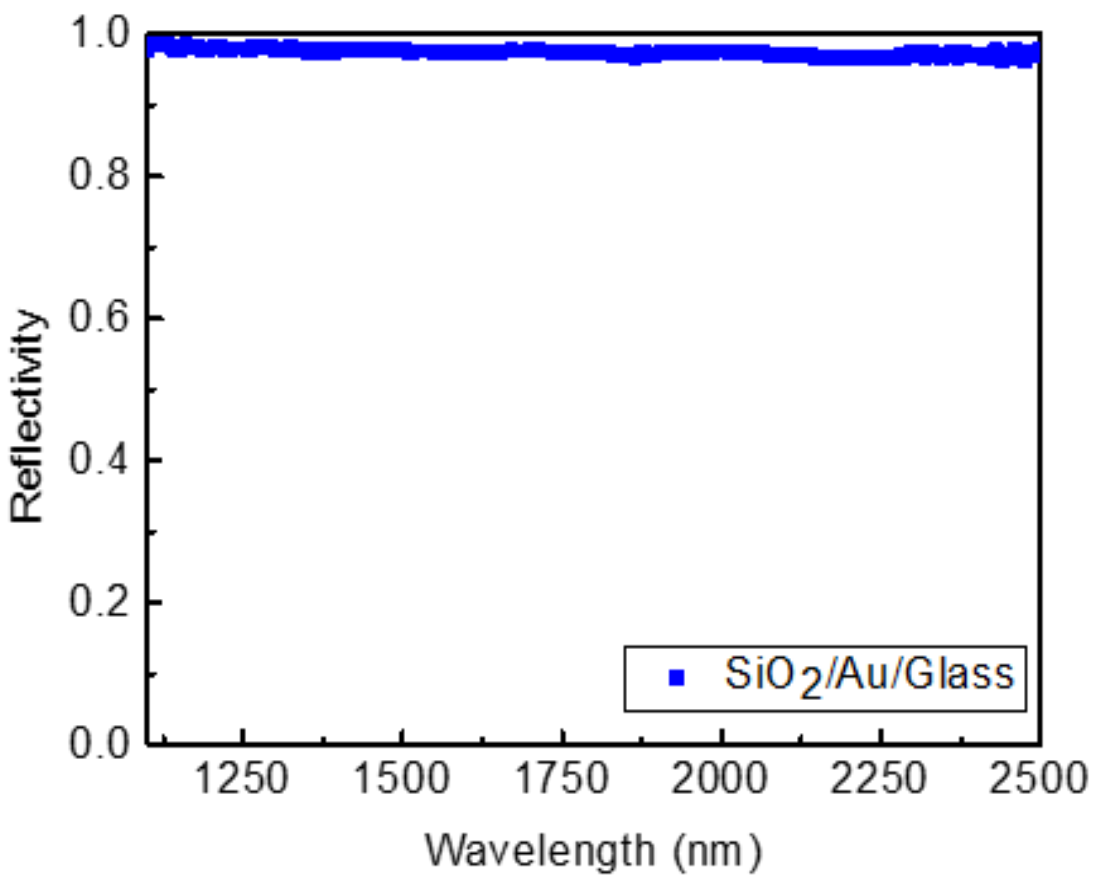

**Figure 4.7:** Measured reflectivity for $SiO_2$/Au/Glass.

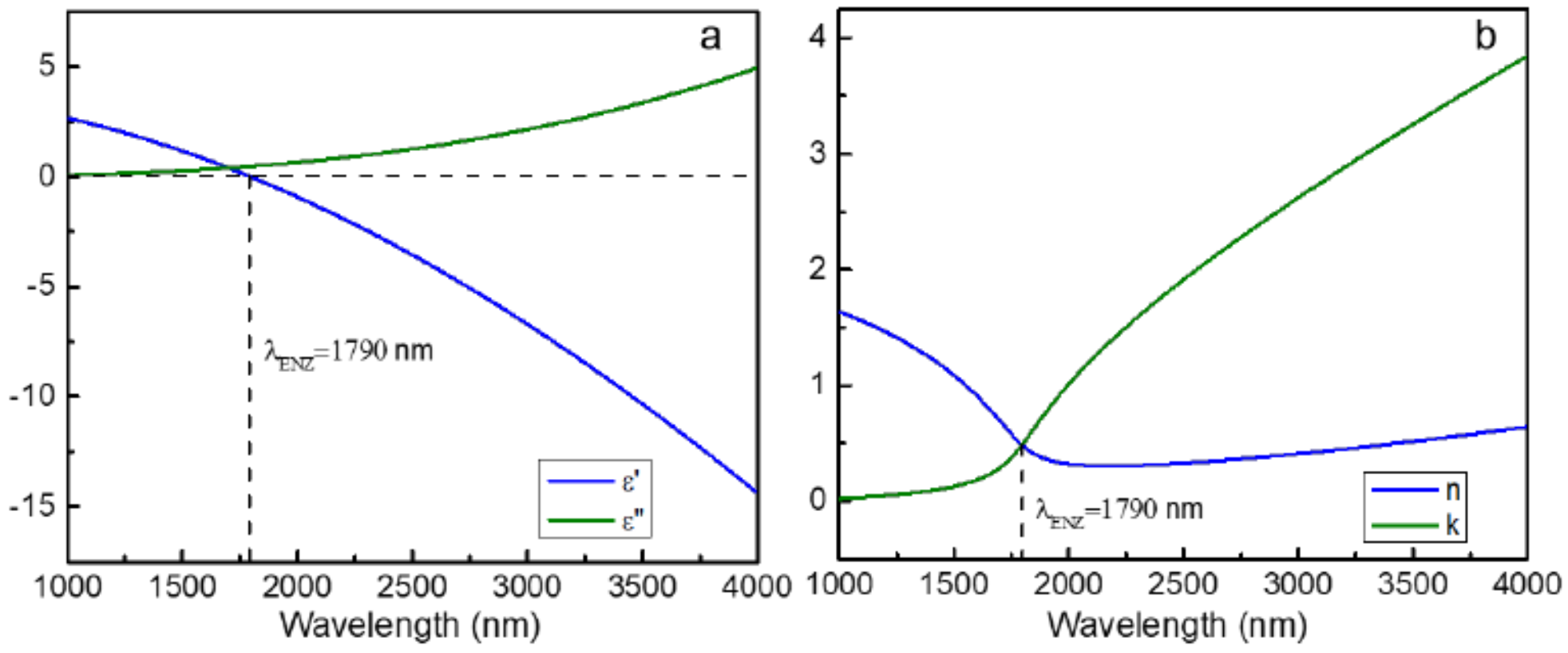

**Figure 4.8:** (a) Real ($\varepsilon'$) and Imaginary ($\varepsilon''$) plots of relative permittivity variation with wavelength, (b) Refractive index (n) and extinction coefficient (k) of ITO for $\lambda_{ENZ}$ = 1790 nm.

### 4.3.2 Investigating the developed band-selective absorption response

Its worth noting that $\lambda_i \sim 1800$ nm lies just beyond the $\lambda_{ENZ}$, where ITO enters its metallic regime i.e. $\varepsilon' < 0$. Fig 4.11a shows a series of simulated electric field magnitude ($|\mathbf{E}|$) plots across a vertical cross-section of the coating at selected wavelengths. ITO behaves as a dielectric ($\varepsilon' > 0$) for $\lambda$ = 1200 and 1500 nm with the crucial difference that $n > 1$ at 1200 nm and $n < 1$ at 1500 nm which decreases further at $\lambda = \lambda_{ENZ}$. Consequently, **E** field amplification and confinement within the ITO grating is the highest in the ENZ regime, which arises from the vanishing $\varepsilon'$ at $\lambda_{ENZ}$, and the resultant boundary conditions. Spatial distribution of dissipated power density ($P_D = \frac{1}{2}\omega\varepsilon''|E|^2$), across the coating cross-section is shown in Fig.4.11b at selected wavelengths. Here, $\omega$ is the angular frequency, and $\varepsilon''$ is the imaginary part of the relative permittivity. At the ENZ regime, as the **E** within ITO is amplified, the strengthening $\varepsilon''$ induces strong dissipation within the entire ITO NS, thus increasing absorption and tying $\lambda_i$ to $\lambda_{ENZ}$, as investigated later. Beyond $\lambda_{ENZ}$, as ITO enters its metallic regime, the $\varepsilon''$ increases further that limits **E** field penetration as shown in the Fig.4.11a plot for $\lambda$ = 2300 nm. In this metallic regime, the ITO nanostructures support localised surface plasmon resonances (LSPR), as evidenced by the local **E** field enhancement plots for $\lambda$ = 2300, 2890 and 3600 nm, at the corners of the ITO nanostructure. A decrease in loss in ITO due to a decrease in field penetration is offset by the plasmonic **E** field enhancement and increasing $\varepsilon''$. Importantly, the local plasmonic enhancement of **E** at the ITO NS is influenced by its dielectric environment, which accentuates the role of the underlying dielectric layer. Our investigations show that the dielectric underlayer crucially determines the spectral distribution of the LSPR modes, field enhancement and thus realizing the absorption band. Fig.4.12 shows the plasmonic response of the ITO grating on $SiO_2$, evidencing multiple plasmonic resonances that determine the $\Delta\lambda$ of the absorption band (see SI section 6 for details). Consequently, high absorption beyond $\lambda_{ENZ}$ is sustained by the LSPR amplification. The same grating doesn't show the third resonance when air replaces $SiO_2$ depicting the utility of the $SiO_2$ layer. SI Fig.4.13 plots the simulated **E** at two points on the ITO nanostructure and its spectral variation elucidating the role of the dielectric. The falling edge of the absorption band ($\lambda_f$) is primarily determined by the dimensions and periodicity of the ITO grating along with the refractive index of the dielectric spacer. While increasing $w$ of the NS (Fig.4.3a) redshifts $\lambda_f$ thereby widening $\Delta\lambda$ , decreasing grating density i.e. increasing $p$ narrows $\Delta\lambda$ along with a decline in overall value of $A$. The thickness ($h$) of the grating nanostructures primarily controls the strength of absorption as evidenced in (Fig.4.3a) . Hence, the ENZ response of ITO together with it's plasmonic resonances and the refractive index contrast provided by the back dielectric $SiO_2$ determines the absorption band. Spectral variation of power dissipation $P_D$, across the three components of the tri-layer coating, shows that the maximum energy dissipation within the high absorption band occurs in the ITO nanostructures, as shown in Fig.4.14. The ITO grating thus plays the dominant role in determining the spectral variation of absorption. The thickness of $SiO_2$ is so chosen that the incident light undergoes multiple reflections in the cavity formed between the metallic ITO grating and back Au reflector to enhance absorption at the ITO NS. The relevance of the dielectric layer is further elucidated via the simulated absorption plots for various dielectrics with different refractive indices shown in Fig.4.15. Here, $SiO_2$ is replaced with other infrared transmitting oxides like $HfO_2$ or $Cr_2O_3$, which have a constant refractive index ($n$) with low loss over the entire spectral regime investigated. While $SiO_2$ has $n \sim 1.46$, $HfO_2$ has $n \sim 1.9$ and $Cr_2O_3$ has $n \sim 2.2$ with a negligible imaginary components[166]. The results in Fig.4.15 show that $\Delta\lambda$ increases to 1.7 $\mu$m and 2.0 $\mu$m for $HfO_2$ and $Cr_2O_3$ with the $\lambda_i$ remaining fixed at 1790 nm. Since the spectral position of the plasmonic resonances of the ITO NS depends on the optical properties of the dielectric spacer, the $\Delta\lambda$ approximately doubles upon replacement of $SiO_2$ with $Cr_2O_3$. Thus, the spectral width of the absorption band is determined by the ITO grating in conjunction with the properties of the dielectric underlayer.

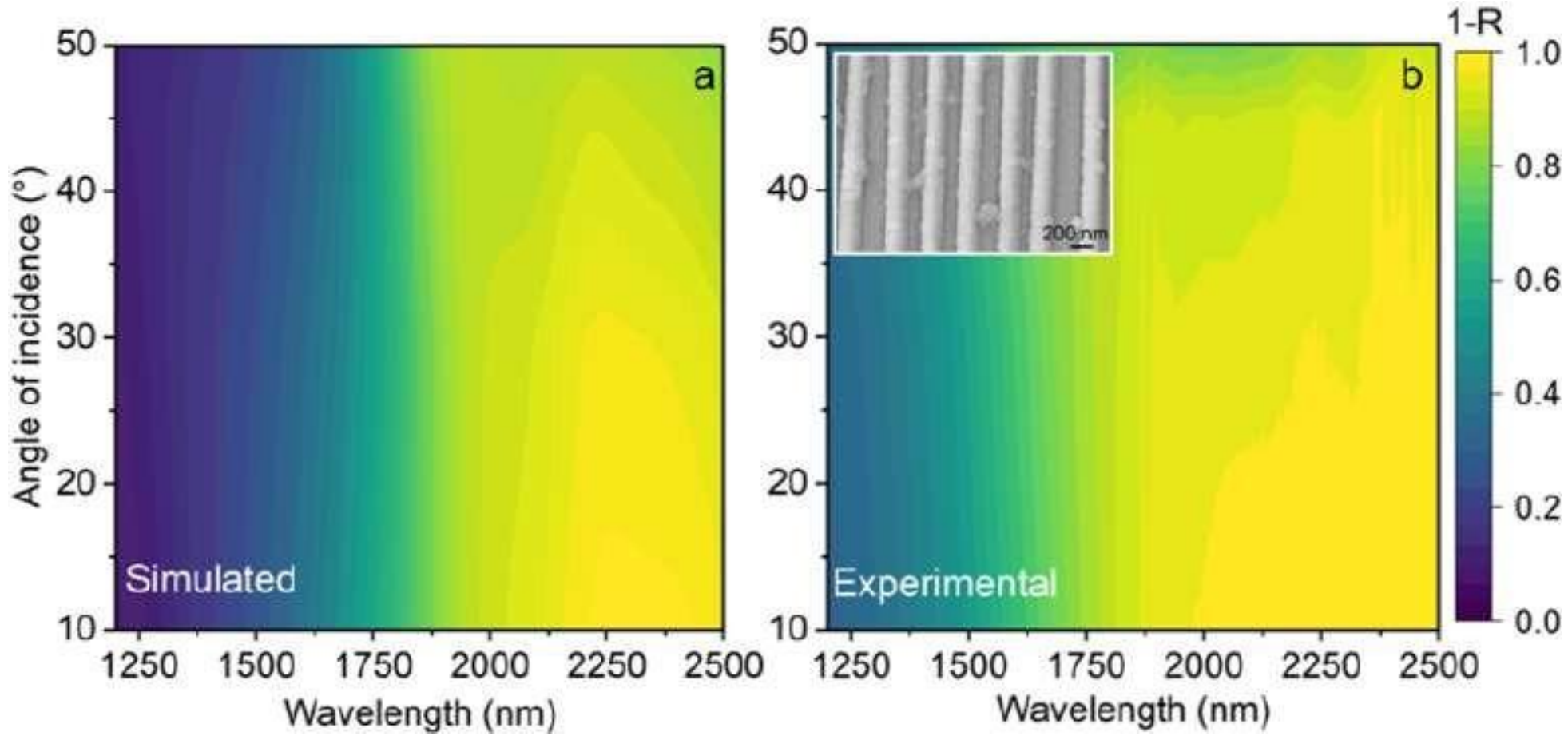

**Figure 4.9:** (a) Simulated and (b)Experimental response of the coating for 10º, 30º,45º and 50º angles. Inset shows the SEM image of the ITO grating on $SiO_2$/Au coated substrate.

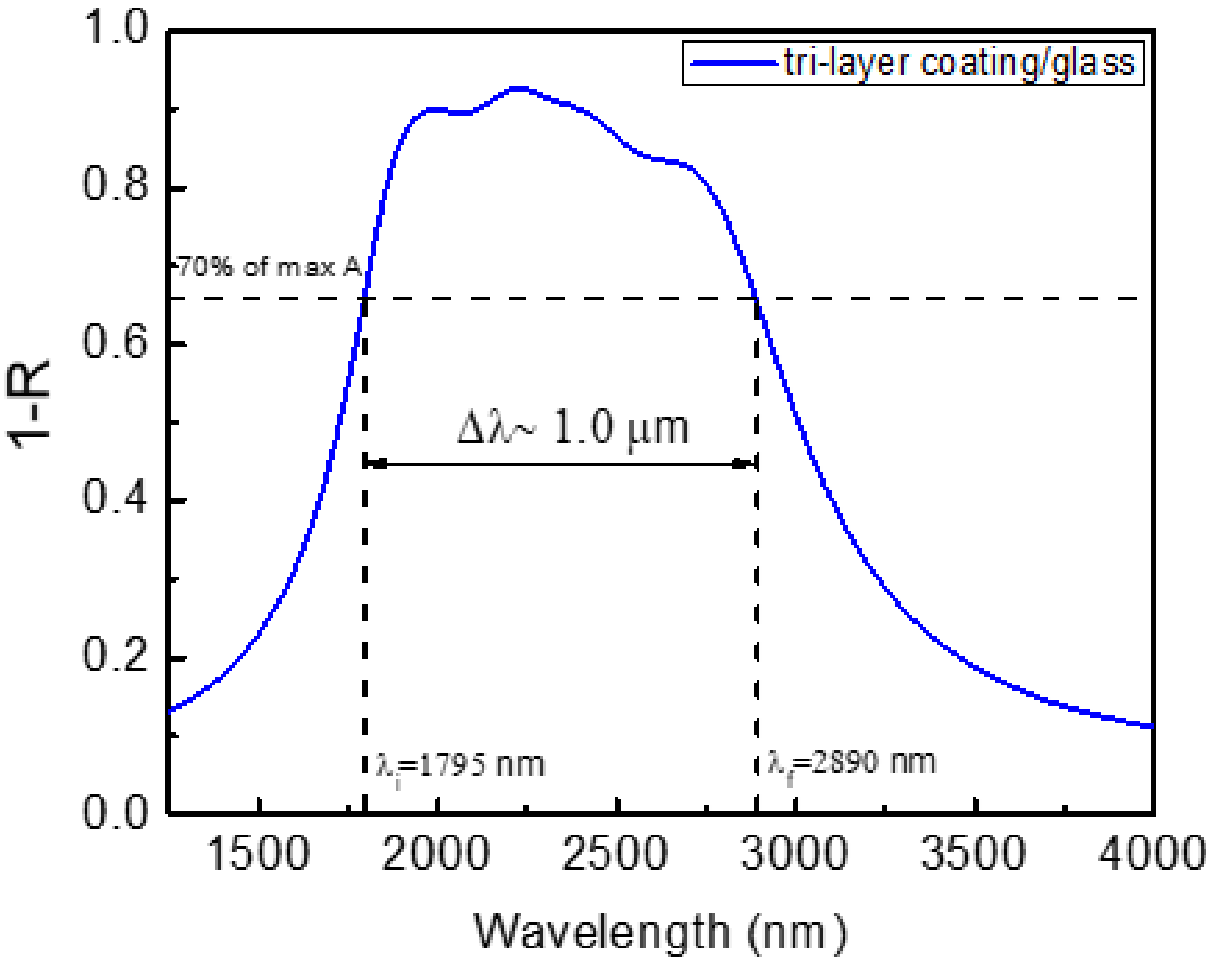

**Figure 4.10:** Optical response of the trilayer showing the bandwidth.

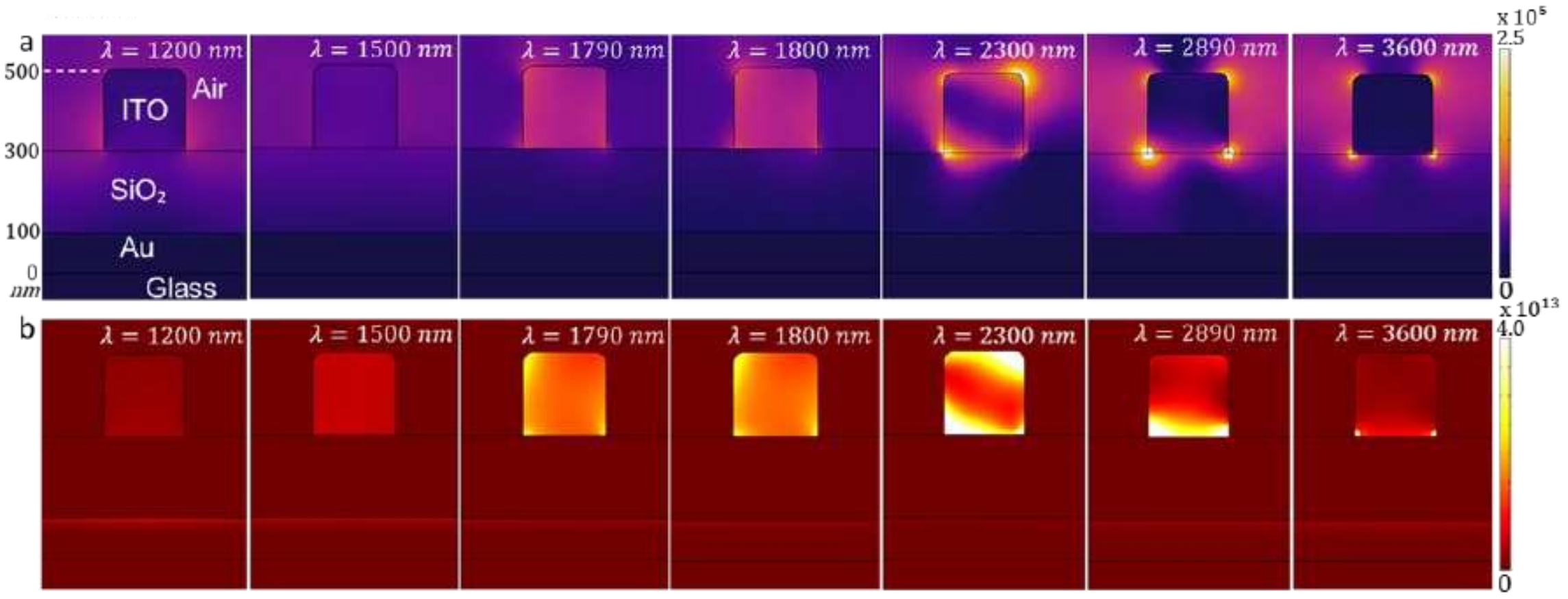

**Figure 4.11:** (a) Simulated electric field magnitude (V/m) and (b) dissipated power density (W/m$^3$) across the cross-section of the coating at selected wavelengths.

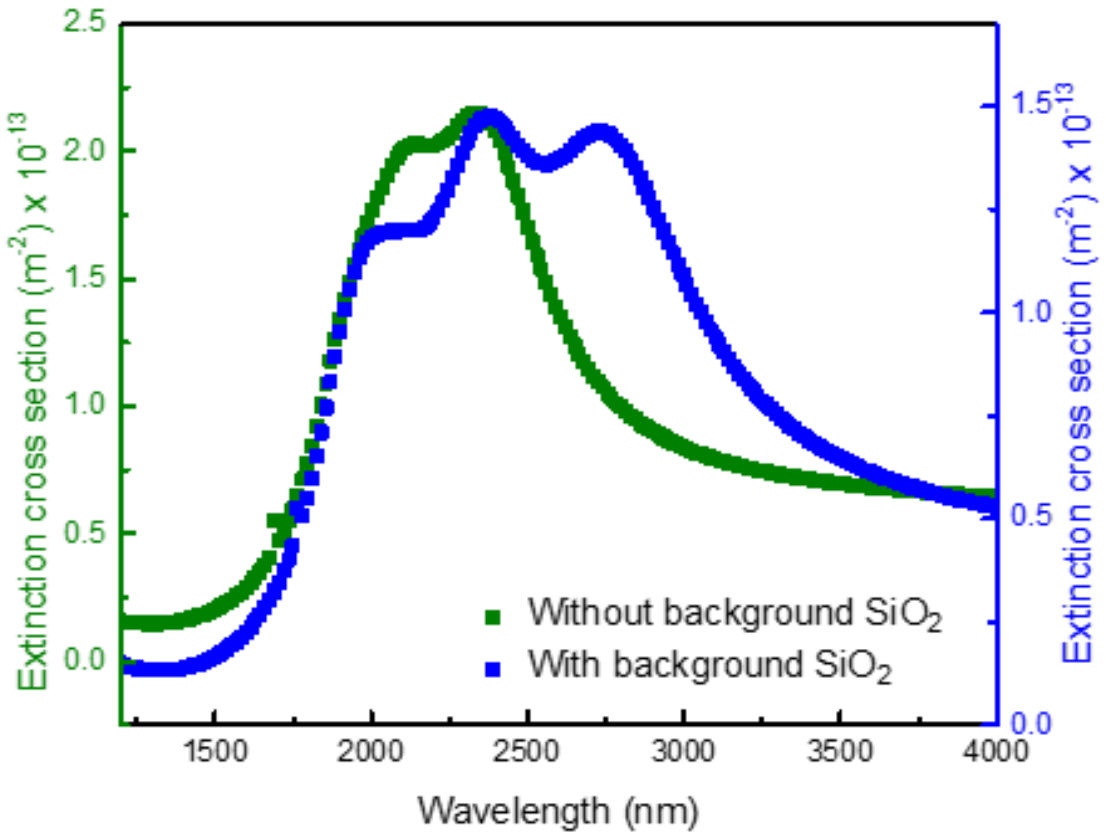

**Figure 4.12:** Extinction cross-section of ITO grating on $SiO_2$ and on air.

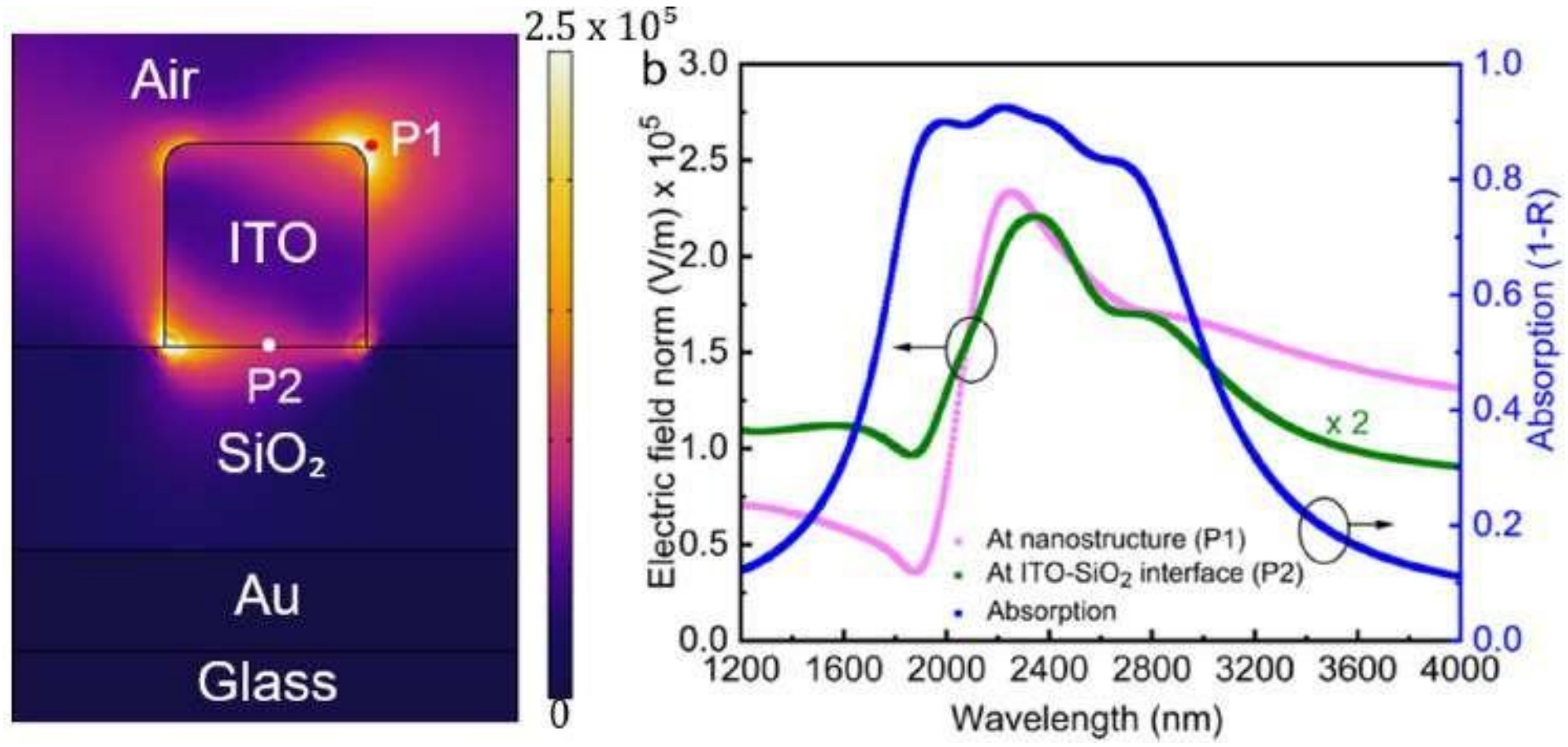

**Figure 4.13:** (a)Cross-sectional view of the simulated of electric field(V/m) plot near the nanostructures,(b) Simulated electric field at the NS and the ITO NS-$SiO_2$ interface along with total absorption.

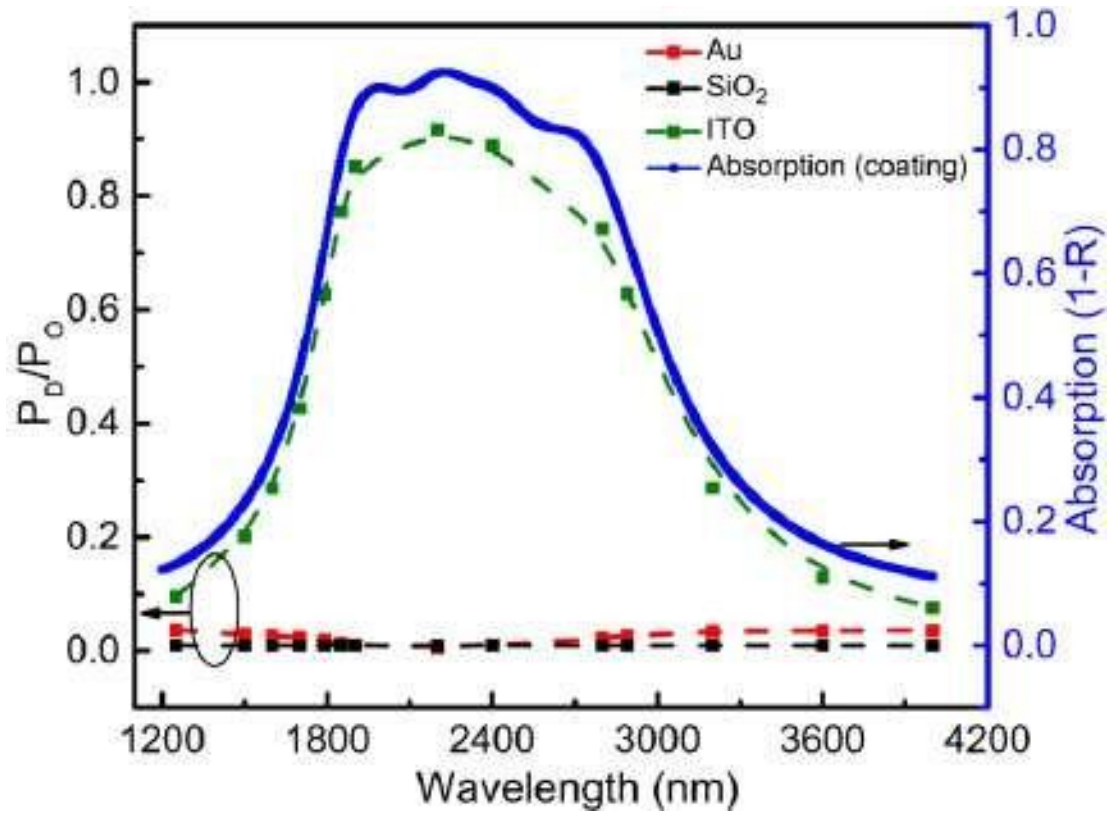

**Figure 4.14:** Simulated spectral variation of fractional power dissipated across the various layers of the coating and the overall absorption spectrum of the coating.

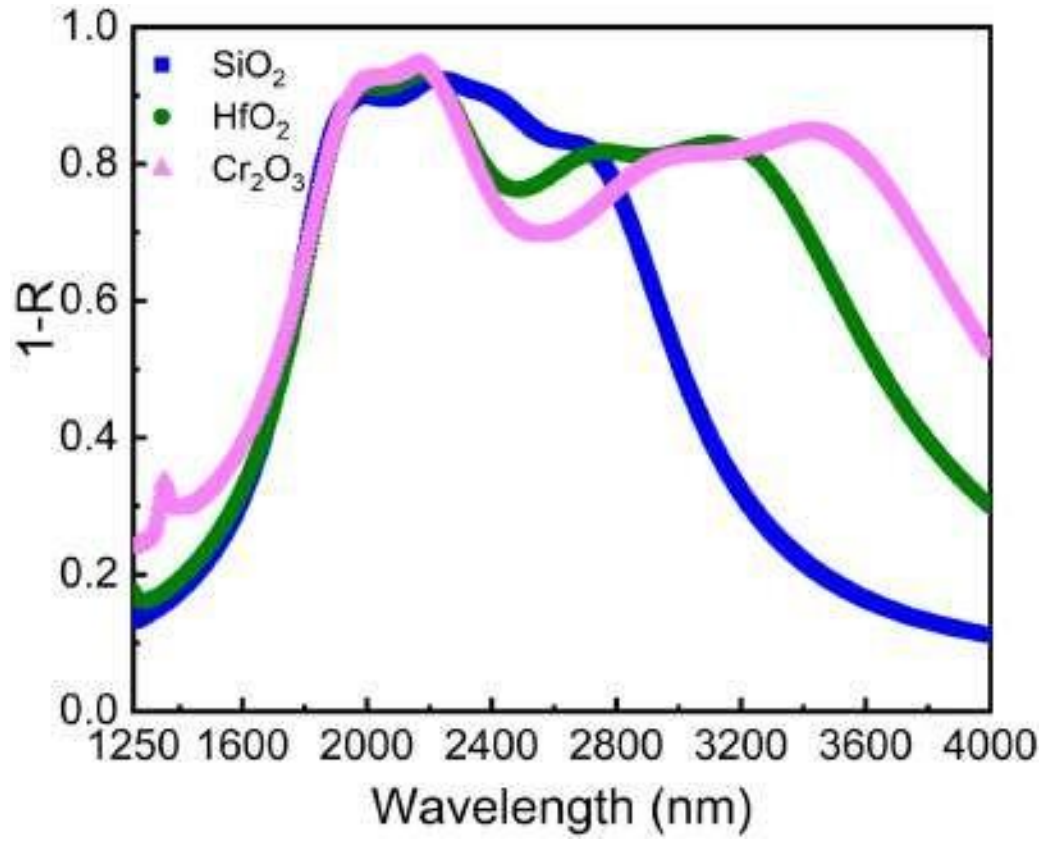

**Figure 4.15:** Simulated absorptivity plots with $SiO_2$, $HfO_2$, and $Cr_2O_3$ as the dielectric.

### 4.3.3 Tunability via varying $\lambda_{ENZ}$

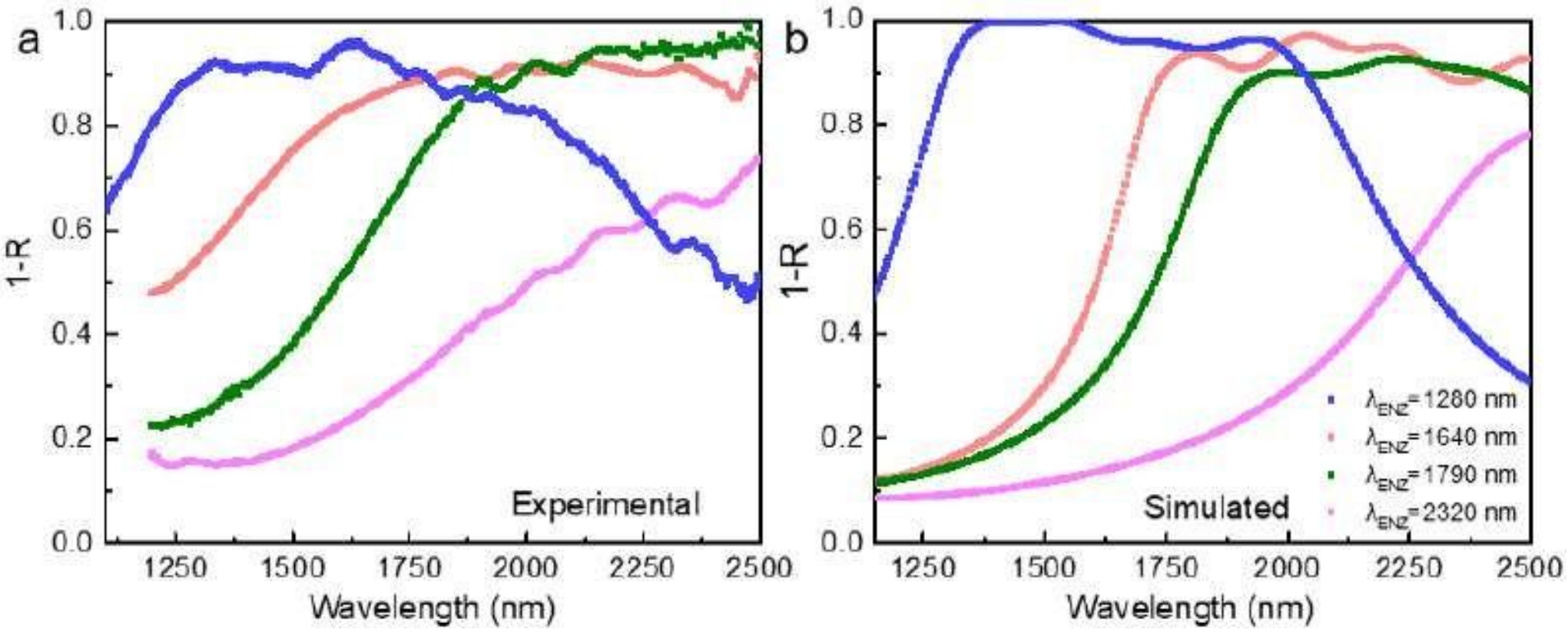

**Figure 4.16:** (a) Experimental and (b) Simulated absorption spectra showing tunability of the response.

As discussed earlier, $\lambda_i$ that is determined by the $\lambda_{ENZ}$ of ITO can be tuned by changing the number density ($N_e$) of ITO. Fig.4.16a shows the experimental absorption spectra after the coated substrate is annealed in oxygen-rich and -deficient environment, demonstrating spectral tuning of the absorption band. The as-coated substrate with ITO having $\lambda_{ENZ} \simeq$ 1790 nm with $N_e \sim$ 4.8 x $10^{20}$ $cm^{-3}$ was annealed in the ambient (oxygen-rich) at $\sim$ 350 $^\circ$ C for 15 minutes, which quenches oxygen vacancies in ITO, decreasing its $N_e$ to $\sim$ 2.9 x $10^{20}$ $cm^{-3}$ and red-shift the $\lambda_{ENZ}$ to $\sim$ 2320 nm, as per the standardized calibration reported earlier[169]. Samples with $\lambda_{ENZ}$= 2320 nm were subsequently annealed in an oxygen-deficient environment (vacuum at $\sim 10^{-6}$ mbar) at 350ºC in two consecutive steps of 30 minutes and 15 minutes. Annealing in an oxygen-lean atmosphere creates oxygen vacancies that increase $N_e$ and blue-shifts $\lambda_{ENZ}$ from $\simeq$ 2320 nm to $\simeq$ 1640 nm and then to $\simeq$ 1280 nm, across the two annealing stages. Following the change in $\lambda_{ENZ}$, the high absorption band shifts in the near-IR as shown in Fig.4.16a and replicated in the simulated spectra in Fig.4.16b. Table 4.2 lists the best-fit values of $\lambda_{ENZ}$ and $N_e$ obtained from the reflectivity data of ITO films processed as per the annealing protocol discussed above.

| Annealing period (minutes) | 0 | 30 | 45 |
|---|---|---|---|
| $\lambda_{ENZ}$ (nm) | 2320 | 1640 | 1280 |
| $N_e \times 10^{20}$ (/cc) | 2.9 | 5.7 | 9.45 |

**Table 4.2:** Annealing of trilayer coating in vacuum

### 4.3.4 Emissivity and thermal imaging

It is non-trivial that a 200 nm thick grating of an ENZ material (ITO) can impart high absorptivity to the dominantly reflecting underlayers of ($SiO_2$/Au) of the coating, substantially modifying the optical properties of a substrate. The emissivity of the coating within the high absorption band is calculated to be $\sim$ 0.8 as shown in Fig. 4.17a, which diminishes beyond the spectral band as shown in Fig.4.17b. The emissivity (E) of the coating on glass was calculated using the equation given below in the high absorption

band (1800-2800 nm).

$$E = \frac{\int_{1.8\mu m}^{2.8\mu m}(1 - R(\theta, \lambda))I_b(\lambda, T)d\lambda}{\int_{1.8\mu m}^{2.8\mu m} I_b(\lambda, T)d\lambda} \tag{4.1}$$

$I_b$ is the blackbody radiation spectrum at 300K and R is the calculated reflectivity at a particular angle $\theta$ and wavelength $\lambda$.

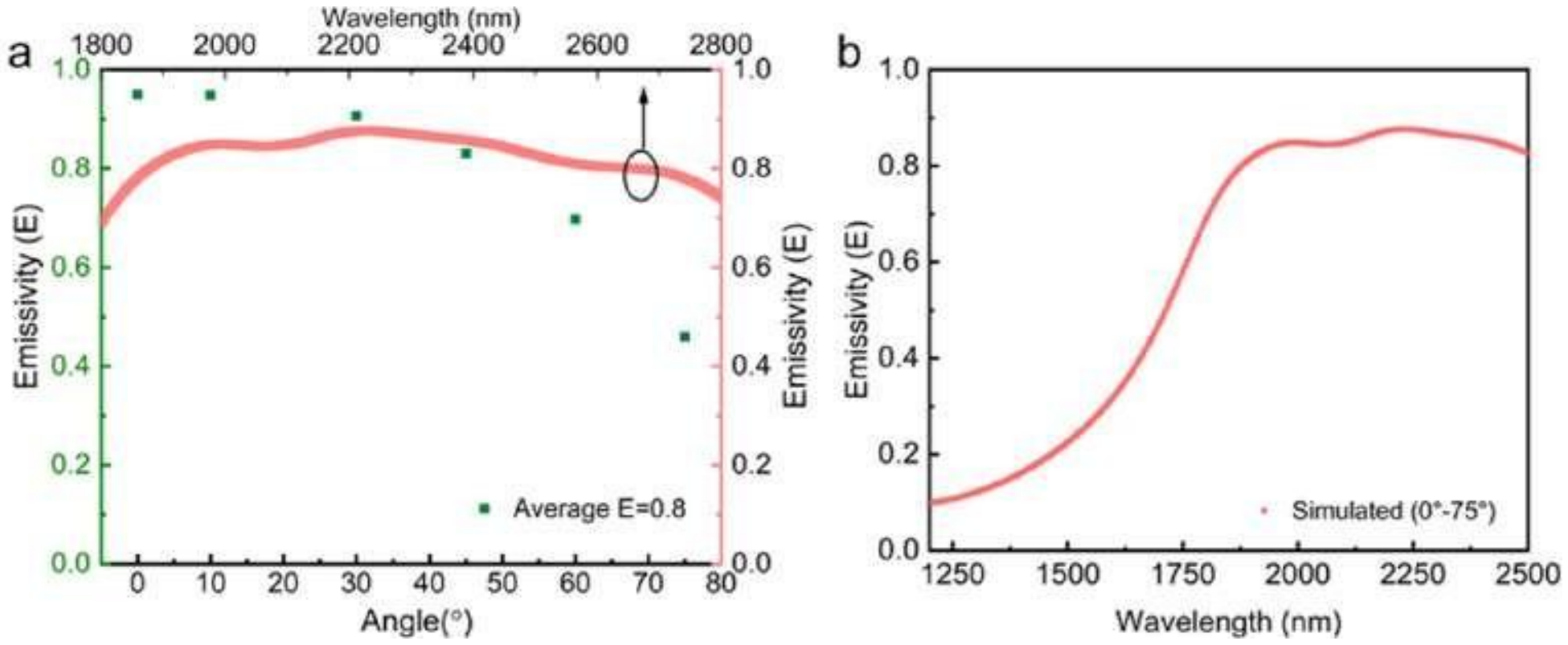

**Figure 4.17:** (a) Calculated emissivity of the coating on glass for different angles along with its spectral variation over the absorption bandwidth (averaged for 0-75°), (b) Simulated emissivity in 1200 -2500 nm wavelength range.

Importantly, the relevance of the ITO grating in increasing the emissivity of the coating is evidenced even in the mid-IR. Thermal images of a glass substrate fully coated with $SiO_2$/Au and a $\sim$ 3 $\times$ 3 mm square area coated with the ITO grating are shown in Fig.4.18, for four substrate temperatures. The thermal images are acquired with a thermal camera (Fluke Ti480 Pro), detecting in the wavelength range 8 $\mu$m - 14 $\mu$m. A forest of carbon nanotubes (CNT) served as the emissivity (E) standard, which showed $E \sim$ 0.98 as shown in Fig.4.19. The measured temperatures ($T_m$) of carbon nano tube sample is 39.2, 45, 52.5 and 63.6°C when heated at set temperatures ($T_s$) of 40, 46, 54 and 65 °C respectively. The emissivity of carbon nanotube forest on silicon is $\sim$ 0.98. In Fig.4.18, the temperature detected by the thermal camera, standardized with the CNT sample, on the $SiO_2$/Au coated region is far below the actual temperature due to the low emissivity of the Au film. However, the ITO grating coated region records a higher temperature due to the higher emissivity imparted by the ENZ grating compared to the uncoated background. Uniformity in emission across a wide angle is evidenced through thermal images acquired at different angles (0 - 45°) as shown in Fig.4.20. The thermal images were captured from a distance (x) of around 30 cm and the camera's aperture is around 2 cm wide. The thermal images (a, d) were acquired at an angle 15 ±5 degrees, (b, e) around 30 ±5 degrees, and (c, f) around 45 ±5 degrees to the vertical, as shown in the figure below. The substrate was heated to $\sim 40.0^{\circ}C$, for the images (a-c) and to $\sim 54.0^{\circ}C$ for images (d-f). For all cases the thermal maps show comparable temperature for the coated area.

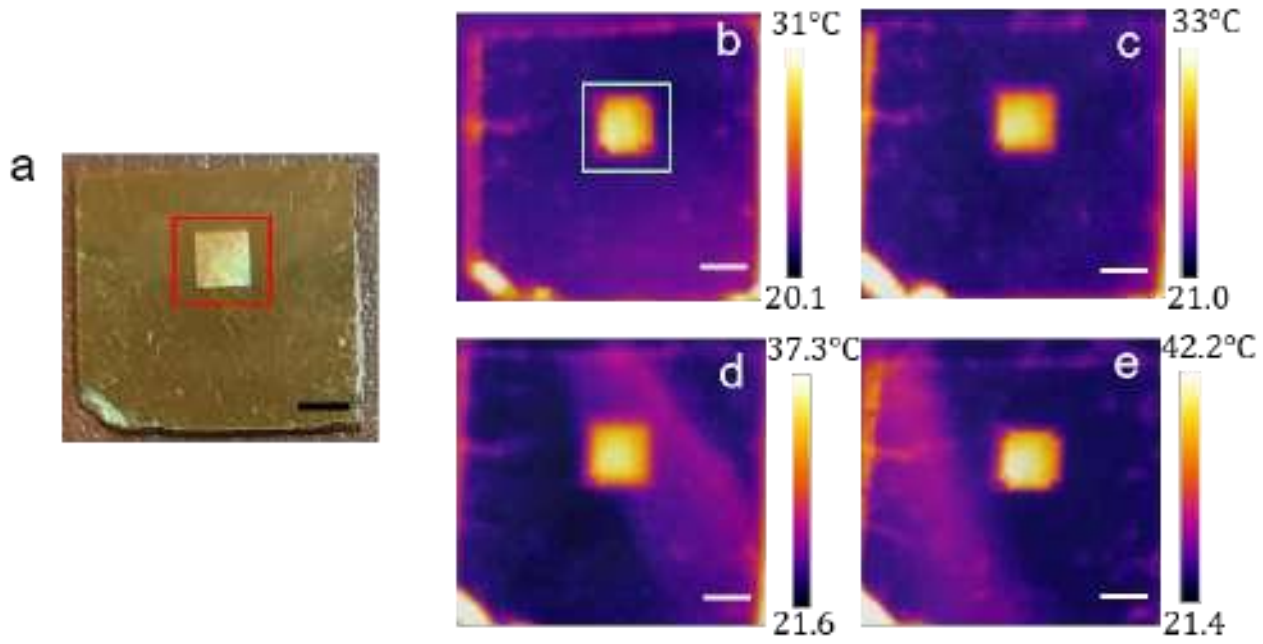

**Figure 4.18:** (a) Optical image of a glass substrate coated with $SiO_2$/Au, with the nanostructured grating over a selected region. Corresponding thermal images of the substrate heated to (b) 40°C, (c) 46°C, (d) 54°C, (e) 65°C. The red box in the optical image and the light blue box in the thermal image indicate the nanostructured region. Scale bar - 3 mm.

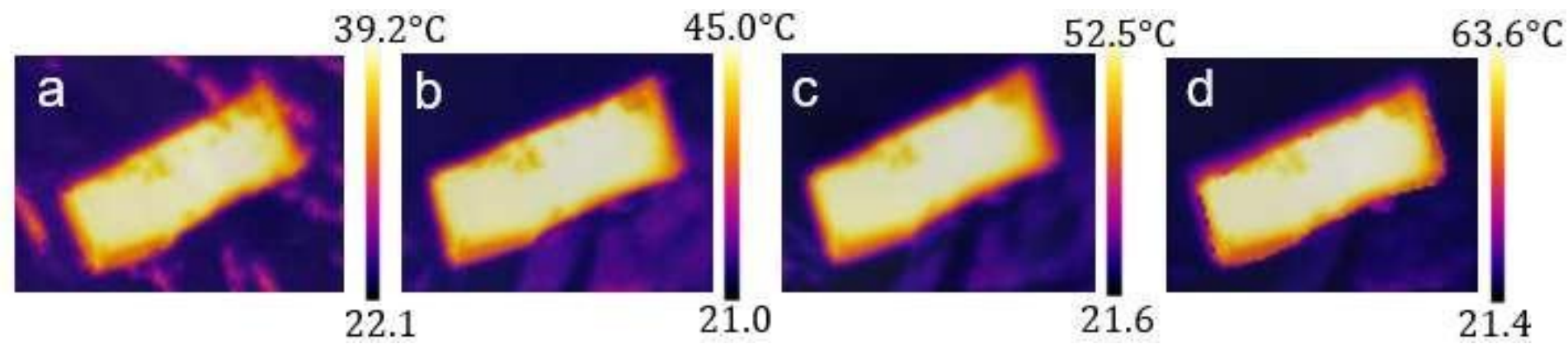

**Figure 4.19:** (a) Thermal images of the carbon nanotube forest on silicon when heated to (a) $\sim 40.0^{\circ}C$, (b)$\sim 46.0^{\circ}C$, (c)$\sim 54.0^{\circ}C$, (d)$\sim 65.0^{\circ}C$.

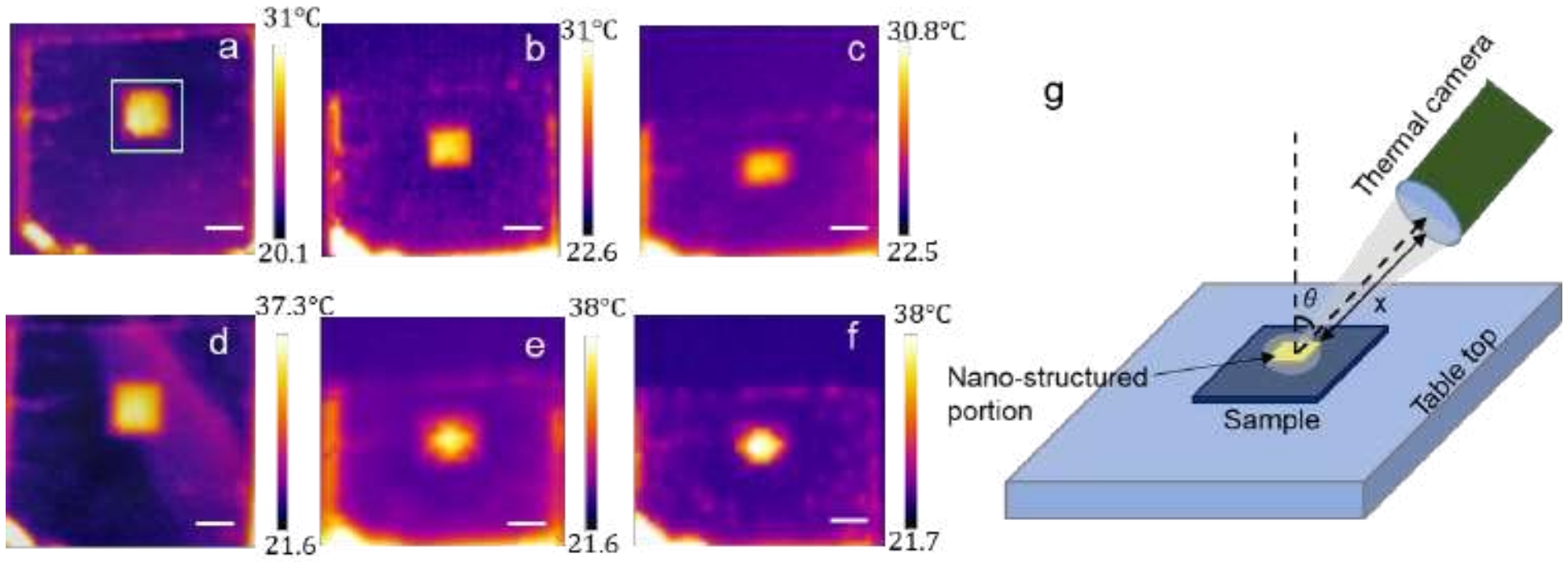

**Figure 4.20:** (a) Optical image of a glass substrate coated with $SiO_2$/Au, with the nanostructured grating over a selected region. Corresponding thermal images of the substrate heated to (b) 40°C, (c) 46°C, (d) 54°C, (e) 65°C. The red box in the optical image and the light blue box in the thermal image indicate the nanostructured region. Scale bar - 3 mm.

## 4.4 Conclusions

To conclude, we have investigated a coating with band-selective absorption, employing a nanostructured grating of ENZ media atop a highly reflecting dielectric-metal underlayer. Over 85% absorption, in a selected band (1800 nm - 2800 nm) over a wide angle (0° - 60°) is demonstrated, which arises primarily from a combination of the ENZ properties and plasmonic response of the nanostructured ITO grating, in conjunction with the optical properties of the dielectric underlayer. Experimental results, along with numerical calculations, delineate the roles played by the ENZ properties that enable electric field amplification and localization in the ITO grating, increasing absorption and the role of localized plasmon modes in determining the central wavelength and bandwidth of the high-absorption band. The tunable nature of the absorption band, by controlling the electron density of ITO and permittivity of the dielectric underlayer, displays the coating's customizability. Further, direct thermal emissivity measurements demonstrate the high emissivity of the nanostructured ENZ surface even at longer wavelengths, compared to its thin film counterpart. Further, the use of other infrared-transmitting oxides like $HfO_2$ or $Cr_2O_3$ is shown to broaden the tunability of the high absorption band for tailoring to specific applications. The investigation provides new directions and design principles in coating-based thermal energy management for potential thermo-photovoltaic applications. Finally, this study contributes to a broader framework connecting ENZ-based field confinement and plasmonic mode coupling in nanostructured geometries, offering design principles for future meta-surface absorbers. Altogether, this work bridges ENZ photonics and plasmonic nanostructuring design, depicting a path toward the next generation of high-performance, customizable thermal photonic surfaces.

# Chapter 5

# Enhancing light absorption and emission in monolayer $MoS_2$ using an epsilon-near-zero underlayer

Monolayer direct band gap semiconductors such as molybdenum disulfide are one of the most promising two-dimensional transition metal dichalcogenides for their attractive properties that cater to a wide variety of applications from photonics to optoelectronics. Even though their ultrathin nature and presence of defects make them applicable in field-effect transistors for various electronic applications. However, the same ultrathin nature becomes a drawback for photonics applications as the effective interaction of light with the material is lower and often affected by the presence of defects. This results in quenched light absorption and emission behavior. In this work, we explore a new strategy to enhance absorption as well as emission in CVD-grown monolayer $MoS_2$ using a visible plasmonic and epsilon-near-zero material- titanium nitride. This strategy doesn't involve any sophisticated nanostructuring, and the absorption and emission in $MoS_2$ are enhanced by simply using a TiN thin film. The optical properties of TiN thin film underneath $MoS_2$ boost the excitation and thereby enhance the light–matter interaction at the nanoscale. Hence, this work allows to open new avenues to engineer light–matter interactions in two-dimensional materials, thereby enabling their applications in photodetection, sensing and photovoltaic devices.

## 5.1 Introduction

Transition metal di-chalcogenides have attracted significant attention due to their fascinating multi-faceted physical properties. Their excitonic properties have been an interesting arena of research as they play a crucial role in determining potential applications in opto-electronic devices [207, 208]. Molybdenum di-sulphide ($MoS_2$) is one of the most well-investigated TMDCs for its robust optical properties. Monolayer $MoS_2$ has a direct band-gap and its A exciton emission lies at around 675 nm, while its B exciton lies at around 620 nm and there has been a substantial demand to enhance its absorption and emission [209]. The A exciton emission is substantially stronger in $MoS_2$ as it lies at a lower energy ($\sim$ 1.85 eV) compared to the B exciton ($\sim$ 1.98 eV) which allows more favorable transition probability[210, 211]. However, it is significantly affected by the substrate and the dielectric environment[212].

The most widely used approach to synthesize TMDCs, especially $MoS_2$, is chemical vapour deposition (CVD) as it gives a large area coverage of monolayers over the substrate. But often it suffers from drawbacks in terms of the presence of defects like sulfur vacancies etc.[199] Moreover, its extremely low thickness($\sim$ 0.7 nm) reduces its effective interaction with light and hence compromises light absorption. This, in turn, leads to weak excitonic emission in $MoS_2$. In this regard, various strategies have been explored to enhance the absorption and thereafter emission in $MoS_2$ through the introduction of strain[213], doping [174], plasmonic gratings[214], photonic crystals [215], etc., especially for CVD-grown ones. In an earlier communication, we demonstrated strain-induced emission enhancement in CVD-grown monolayer $MoS_2$ via transferring them on a periodic array of Au nanostructures on a silicon (Si) substrate [216]. However, most of the studies reported have shown rigorous fabrication processes involving nanostructuring that limits the practical applications. Here, we introduce a thin film of a material- titanium nitride (TiN) that exhibits plasmonic [217] and epsilon-near-zero (ENZ) response [218] in the visible regime (400 nm-800 nm) due to its high number density ($\sim 10^{22}$)/cc to enhance the absorption and thereby emission in $MoS_2$.
Titanium nitride (TiN) provides an industrially relevant material platform for various applications due to its complementary metal–oxide semiconductor (CMOS) compatibility, high laser- and thermal tolerance, etc [219, 220]. These attractive features have led to its extensive utilization in high-temperature photovoltaics[221]. Apart from these aspects, it is an emerging material for nano-photonics as it possesses plasmonic and user-tailorable optical properties, low loss compared to noble metals etc.[222] These have resulted its uasge in optical circuitry,[223, 224] nonlinear optical devices [225] and many other practical applications. Apart from these, it exhibits the unique ENZ response in the visible regime where its undergoes an optical transition from a dielectric to metal-like system with the real part of relative permittivity ($\varepsilon(\omega) = \varepsilon' + i\varepsilon''$) going to zero at the ENZ wavelength ($\lambda_{ENZ}$) witnessing certain interesting phenomena like extreme-field enhancement and confinement, and perfect absorption in the ENZ regime[91, 169, 195, 196]. Hence, this encourages checking how the optical properties of a monolayer $MoS_2$ would be modified by bringing a monolayer $MoS_2$ with excitonic properties in the same visible regime in proximity to it, expecting the possibility of extreme light-matter interactions in the two systems.The ENZ wavelength ($\lambda_{\mathrm{ENZ}}$) in TiN can be tuned by controlling the carrier concentration through the introduction and suppression of nitrogen vacancies [226]. Additionally, $\lambda_{\mathrm{ENZ}}$ in Tin is highly sensitive to the fabrication method and processing parameters [227]. Hence, this tunability offers a key advantage, enabling precise alignment of $\lambda_{ENZ}$ with the excitonic emission wavelength of $MoS_2$. TiN usually has a $\lambda_{ENZ}$ ranging between 550-700 nm, depending on the carrier density in the system and as mentioned A excitonic emission in $MoS_2$ lies at around 675 nm. Hence, the method of preparation and the deposition parameters of TiN can be optimized such that the A exciton emission in $MoS_2$ lies in the ENZ regime of TiN.

In this work, we investigate a novel approach to absorption and emission enhancement in monolayer $MoS_2$ leveraging the epsilon-near-zero (ENZ) properties of titanium nitride (TiN). 150 nm thick TiN films deposited for this study show an ENZ wavelength ( $\lambda_{ENZ}$) $\sim$ 690 nm. CVD-grown $MoS_2$ monolayers are transferred onto a 150 nm thick TiN-coated Si substrate to investigate the change in the absorption and the emission profile of $MoS_2$ in close proximity to TiN. The close spectral proximity of the $\lambda_{ENZ}$ of TiN with the emission line of $MoS_2$ brings about a cross-talk between the systems, which results in a significant enhancement of light-matter interaction in $MoS_2$. Finite element method simulations show that $MoS_2$ on TiN/Si together exhibits around 8 times enhanced wide-angle (0-60º) absorption in the visible regime compared to $MoS_2$ on bare Si. Further, spatially resolved photoluminescence (PL) and Raman spectroscopy measurements show a $\sim$ 6-fold increase in PL intensity and a $\sim$5-fold enhancement in Raman mode intensities for $MoS_2$ flakes partially lying on TiN-coated Si versus bare Si. A thin $SiO_x$ layer is introduced between TiN and $MoS_2$ to prevent charge transfer between the materials. This substrate-induced modification of optical response in $MoS_2$ is attributed to the typical optical properties of TiN derived from its ENZ behaviour, which amplify light absorption in $MoS_2$, thereby boosting emission. This approach opens new pathways for light modulation in 2D materials, thereby emphasizing ENZ materials as an effective material platform for advanced photonic applications.

## 5.2 Materials and Methods

### 5.2.1 Simulations and Numerical calculations

The absorption calculations were performed using finite element method modelling conducted in the Wave Optics Module of COMSOL® Multiphysics 5.3a. Fig. 5.1 shows the model geometry for which the optical response was simulated using Floquet periodic boundary conditions. The optical properties of Si was taken from the literature [228], with the Drude model used to determine that of TiN as a function of three parameters, carrier concentration $N_e$, scattering parameter $\gamma$, and background permittivity $\varepsilon_\infty$. The typical value of $\varepsilon_\infty$ of TiN is taken from the literature[229], whereas the $N_e$ and $\gamma$ are obtained by fitting the experimental reflectivity spectra through a custom-written code in MATLAB.

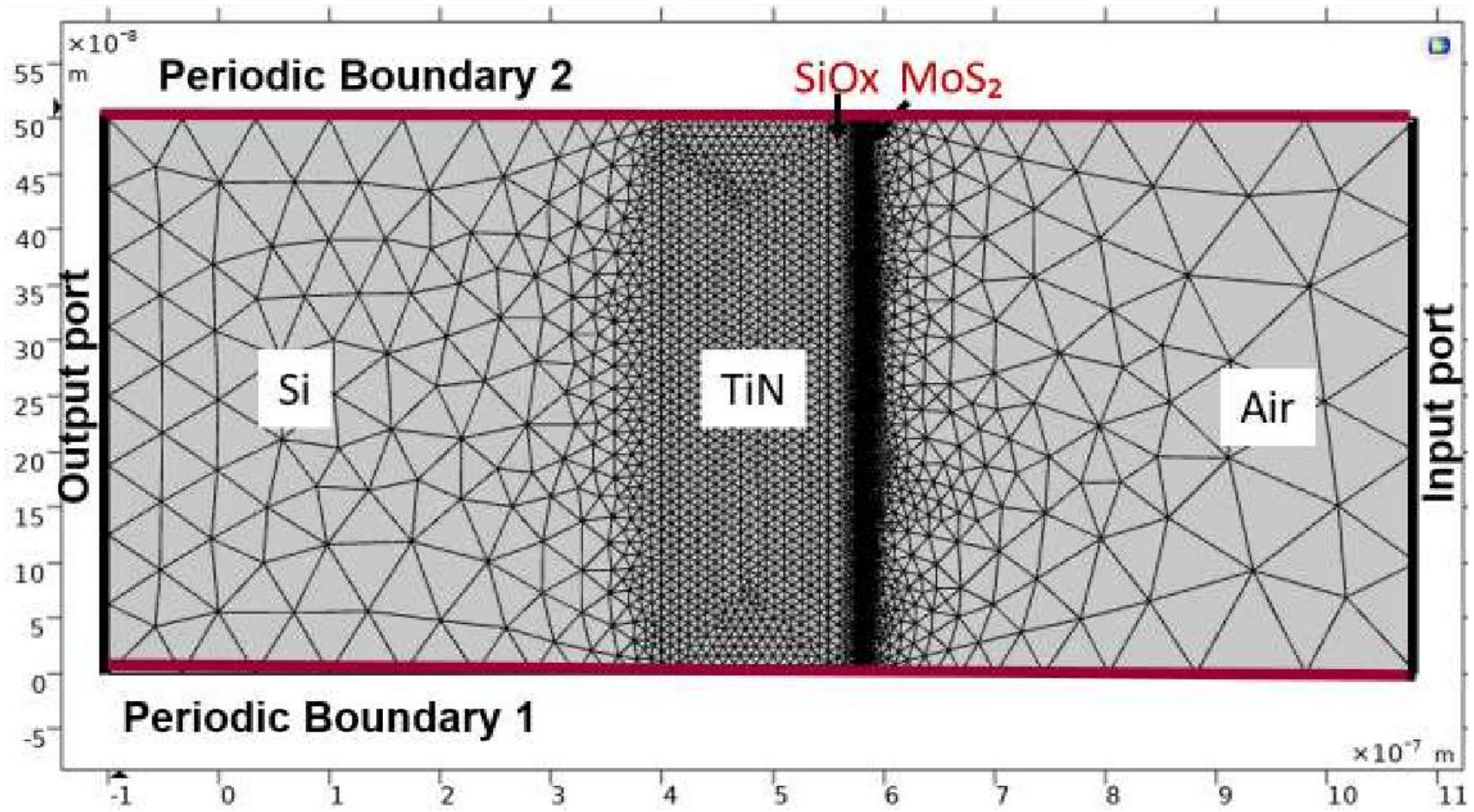

**Figure 5.1:** Image of the 2D model designed in COMSOL Multiphysics 5.3a for the absorption calculations.

The permittivity and refractive index of $MoS_2$ are obtained through fitting of a function obtained by superposition of three Lorentzian oscillators as given below:

$$\varepsilon(\omega) = 1 + \Sigma \frac{f_k}{\omega_k^2 - \omega^2 - i\omega\gamma_k} \tag{5.1}$$

| Material parameters | TiN | $MoS_2$ |
|---|---|---|
| $N_e$ ($cm^{-3}$) | $1.3 \times 10^{22}$ | $4 \times 10^{12}$ |
| $\gamma$ (eV) | 0.6 | 0.059, 0.12, 0.37 [118] |
| $m^*$ | 1.09 [230] | 0.54 [120] |
| $\varepsilon_\infty$ | 4.564 [122] | - |
| $\omega_k$ (eV) | - | 1.85, 1.98, 2.877 [118] |

**Table 5.1:** The material parameters used for TiN and $MoS_2$ for the fitting of the dielectric function.

where

$$f_k = a_k \frac{Ne^2}{m^* \varepsilon_0} \tag{5.2}$$

where k is 1, 2 and 3 corresponding to the three excitonic peaks in $MoS_2$ and $a_k$ is the fitting parameter, $N_e$ is the carrier concentration and $m^*$ is the effective mass of TiN and $MoS_2$ respectively. The typical values of all the material parameters for TiN and $MoS_2$ are given in the table 5.1.

### 5.2.2 Sample fabrication

Cut pieces of bare silicon substrates were cleaned with deionized (DI) water, acetone and isopropyl alcohol (IPA) and followed by mild wash in 1% HF solution for 10s to remove the native oxide layer from the Si substrates. 150 nm of TiN was coated on to the etched Si substrates using standard reactive RF sputtering from a Ti target in the partial pressures of Argon and nitrogen gas (4:1 ratio of Ar:$N_2$ flow) ( under a constant substrate annealing at 500 °C. The TiN/Si substrates were spin coated with 10% PMMA resist. 40 $\mu$m x 40 $\mu$m square windows were opened on the TiN/Si substrates using standard electron beam lithography (Raith Pioneer 2). The patterned substrates were developed in a solution of methyl isobutyl ketone (MIBK) +IPA (1:3 ratio by volume) for 10 seconds followed by IPA wash for a further 10 seconds to expose the TiN films through the patterned windows. The exposed TiN was etched in a solution of hydrogen peroxide ($H_2O_2$) diluted with DI water (1:5 ratio by volume). After the complete etching of the exposed TiN, the PMMA layer was dissolved in acetone and the patterned TiN substrate was obtained. Further, CVD-grown $MoS_2$ was transferred onto TiN substrates using the wet etch transfer method in SiOx-coated TiN substrates[231].

### 5.2.3 Sample Characterization

The reflectivity measurements of thin films of TiN/Si samples were conducted using a Perkin Elmer Lambda 950 spectrophotometer at specific angle of incidence. SEM images were acquired with Nova Nano SEM 450 field emission scanning electron microscope coupled with an Apollo X energy dispersive x-ray analysis (EDS) system which was used to collect the EDS spectra of the TiN film on Si. X-ray diffraction (XRD) studies were performed with CuK$\alpha$ = 1.540 Å at 45 kV (Empyrean, PANanalytical). Atomic force microscope (AFM) images were acquired with Bruker Multimode 8 under ambient conditions. Spatially resolved Photoluminescence (PL) and Raman measurements were conducted with a Horiba Xplora plus Raman setup based on a confocal microscope. PL, Raman mapping, and spectra acquisition were performed using 532 nm laser excitation through a 100x objective with numerical aperture (N.A.) - 0.9. All the spectra for PL and Raman were recorded with gratings with a ruling density of 600 gr/mm and 2400 gr/mm, respectively, in ambient conditions.

## 5.3 Results and Discussions

### 5.3.1 Fabrication and characterization of TiN thin film on Si

Fig.5.2a shows the AFM topography of a titanium nitride (TiN) thin film deposited on a Si substrate via reactive sputtering from a titanium target in the presence nitrogen and argon gas introduced into the chamber in the ratio of 4:1 . The film exhibit typical an average ($S_a$) and RMS roughness ($S_q$) of 4 nm, as quantified by AFM topography scans over a 3 × 3 $\mu m^2$ area (see Fig.5.2a(ii)), the film attesting to a good quality film consistent with previously reported results on films prepared by similar techniques [232, 233]. The formation of TiN on Si was verified using multiple material characterization techniques. First, energy-dispersive X-ray spectroscopy (EDS), shown in Fig.5.2b, con-

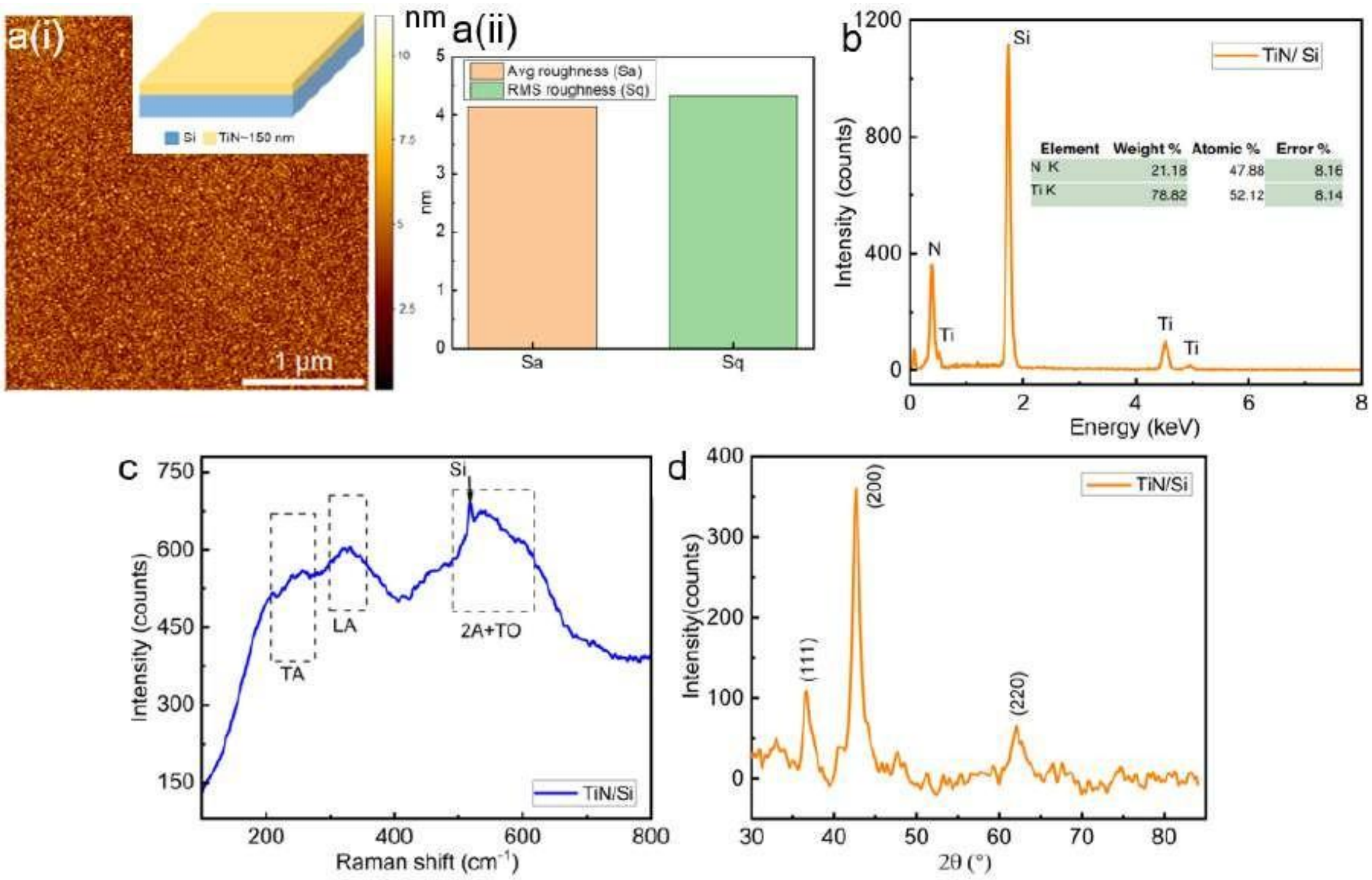

**Figure 5.2:** a(i) AFM topography image of ∼150 nm TiN film on Si substrate with the inset showing the schematic of the sample and a(ii) plot showing the average and rms roughness of the film for the topography shown in a(i), (b) EDS spectra with inset showing the atomic %, (c) Ramanspectra (d) XRD of the TiN film on Si.

firms the presence of Ti and N on the Si substrate through their characteristic peaks and with the atomic % indicating Ti:N ratio ∼ 1.1. Raman spectroscopy further identifies the TiN thin film as it shows the characteristic Raman modes: TA, LA, and a broad peak associated with the 2A and TO modes (Figure 5.2c), consistent with previously reported results [234]. Finally, X-ray diffraction (XRD) shows three distinct peaks characteristic of polycrystalline TiN, further confirming its formation (Fig.5.2d) [235]. Fig. 5.3a shows the experimental and simulated reflectivity of bare Si substrate exhibiting a flat reflectivity of around 30% from 400 to 2500 nm wavelength range while Fig.5.3b shows that of the 150 nm TiN film (thickness measured using AFM as shown in Fig.5.9c) deposited on the Si substrate. Fig.5.3b shows that the film exhibits low reflectivity in the visible below 600 nm with the minima at $\lambda = 555$nm revealing TiN is a dielectric in the visible below 600 nm, beyond which the reflectivity increases and saturates to $\sim$ 60% at around 1200 nm signifying that it is a good reflector in the IR and have a optical dielectric to metal transition beyond 600 nm. Even though TiN is a wide band gap semiconductor, it has a relatively high carrier density ($\sim 10^{22}$/cc), which renders metal-like characteristics with optical properties that are well described using the Drude model. The experimental reflectivity data were fitted with Drude parameters - carrier concentration ($N_e$), high frequency permittivity ($\varepsilon_\infty$ ) and scattering parameter ($\gamma$) for the 150 nm TiN film on Si, and the corresponding reflectivity was simulated as discussed in the simulation section under the material and methods and yielded $\lambda_{ENZ}$ of the deposited TiN film to be $\sim$690 nm, where the real part of relative permittivity goes to zero and the real part of refractive index exhibits the same value as that of the imaginary part, as shown in Fig.5.3c as indicated by the vertical black dashed line. The deviation of the experimental reflectivity from the simulated reflectivity can be due to the fact that experiments were conducted under unpolarized light, while the simulations were conducted with p-polarization. Also, the inability to exactly quantify the loss might lead to an overestimated reflectivity in the simulations at longer wavelengths.

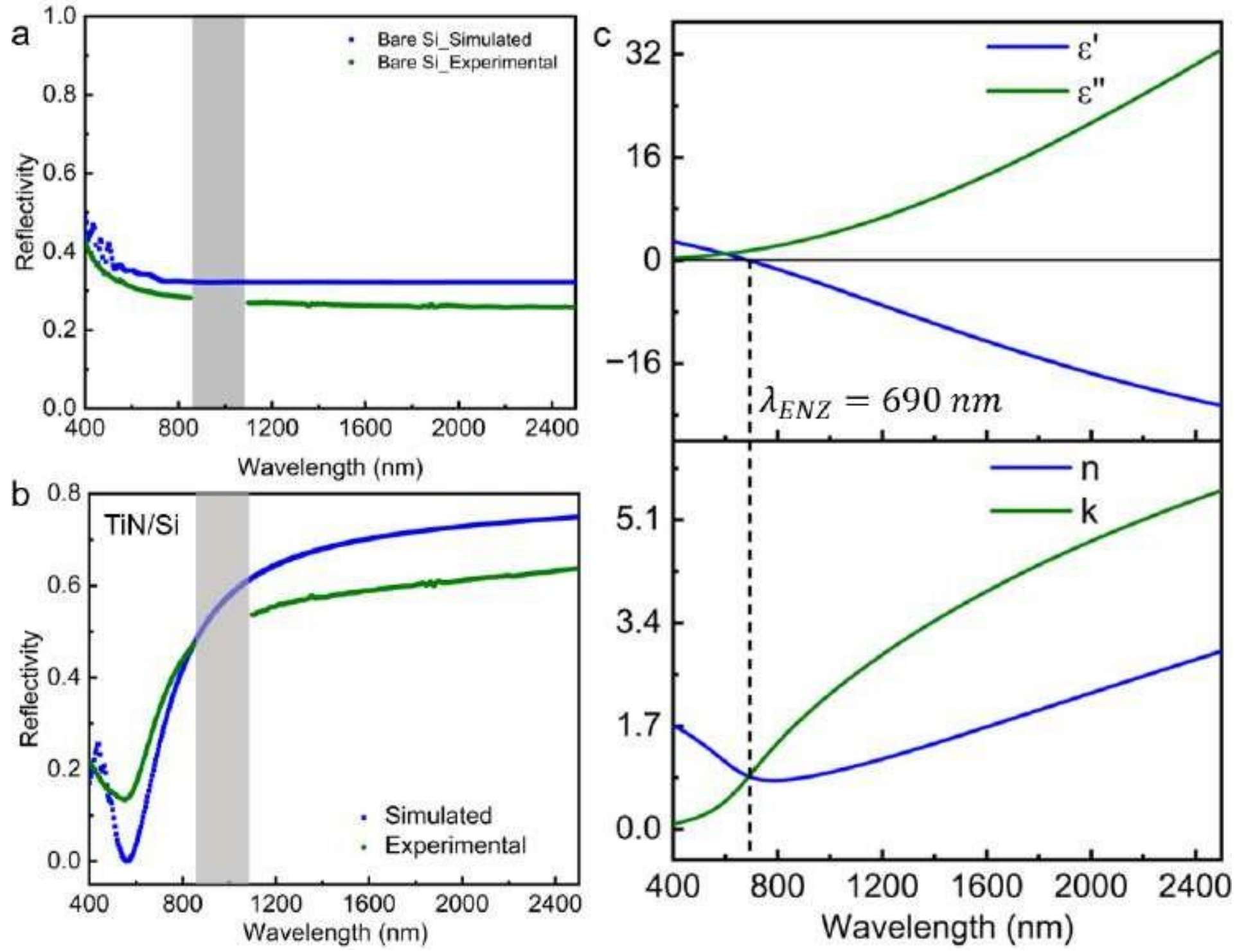

**Figure 5.3:** (a) Simulated and experimental reflectivity spectra of (a) bare Si substrate (b) the TiN film on Si at 10° angle of incidence, (c) Real and imaginary components of relative permittivity, refractive index and extinction coefficient of the TiN film showing $\lambda_{ENZ} \sim 690$ nm as indicated by the black vertical dashed line. The shaded regions in (a, b) are excluded to mask spectrometer noise while acquiring the experimental reflectivity.

### 5.3.2 Absorption enhancement in $MoS_2$

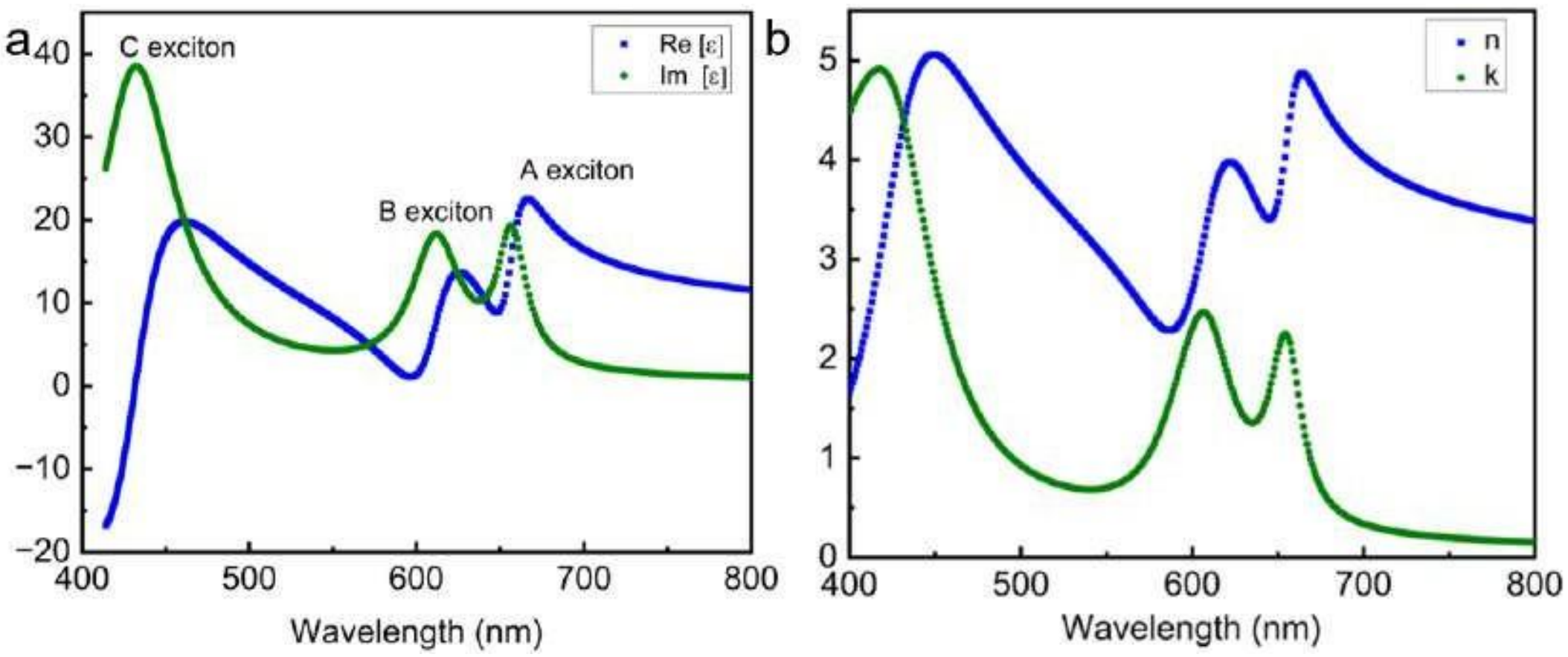

**Figure 5.4:** (a) Real and imaginary parts of relative permittivity plots for monolayer $MoS_2$, (b) Real and imaginary parts of refractive index plots for monolayer $MoS_2$

Fig. 5.4a and b show the permittivity and refractive index plots of $MoS_2$ in the visible regime obtained by fitting a function to the three excitonic peaks of $MoS_2$. The function is a superposition of three Lorentzian oscillators given in equation 5.1. It shows the three characteristic peaks for the excitons- A, B and C as demarcated, exhibited by monolayer $MoS_2$. The two distinct A and B excitons arise from direct band gap transitions near the K-point in the Brillouin zone. However, they are split due to spin-orbit coupling in the valence band. The A and the B exciton, both are associated with the ground state, while the B exciton is associated with higher spin-orbit split state [236]. Fig.5.5 shows the simulated absorption for a free-standing monolayer $MoS_2$ in air, illustrating the weak absorption in these systems. Fig. 5.6a shows the schematic of the TiN and the $MoS_2$

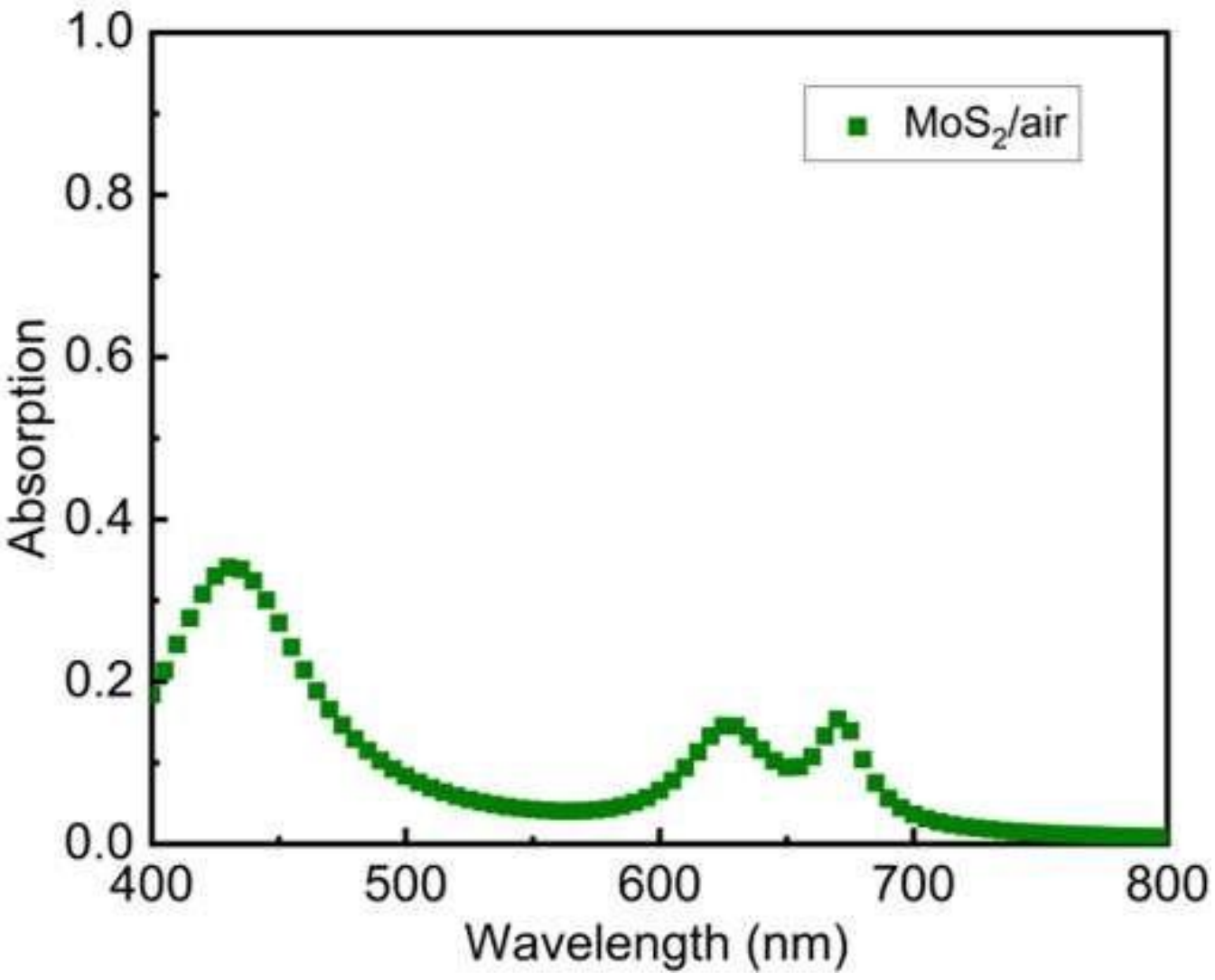

**Figure 5.5:** (a) Simulated absorption for a free standing monolayer $MoS_2$

system explored for absorption enhancement. Initially, we conducted finite element calculations for simulating the absorption of a monolayer $MoS_2$ lying flat on a Si substrate, then we introduce a TiN layer of $\lambda_{ENZ}$ around 690 nm between $MoS_2$ and Si where we observe, the total absorption enhances upto 80% with the two excitonic peaks becoming prominent as shown in Fig.5.6b. Further, the introduction of a SiOx layer between the $MoS_2$ and the back TiN layer boosts the total absorption to 90%. The SiOx layer was introduced with the intention to prevent any charge transfer between the $MoS_2$ and the TiN layer for the emission measurements, as discussed later. However, we observe that it helps to enhance the absorption further by increasing the effective interaction of light with the system. Also, in order to check the limit of maximum absorption enhancement, we systematically varied the thickness of TiN from 10 to 300 nm, and we observed that the maximum absorption is obtained with a 300 nm film of TiN, as shown in Fig.5.6c and saturates beyond that. This proves that the thickness of TiN plays an integral role in the absorption enhancement. Fig. 5.6d shows the total absorption plot for a 150 nm thin film of TiN with the variation in $\lambda_{ENZ}$ of TiN. We observe the maximum absorption happening at the A exciton wavelength of $MoS_2$ when the $\lambda_{ENZ}$ of TiN is fixed at 690 nm. When the $\lambda_{ENZ}$ goes below or well above 690 nm, the absorption at the A exciton wavelength decreases. This indicates that the optical properties of the underlying TiN, quantified by the $\lambda_{ENZ}$, play a dominant role in determining the absorption of the entire system. Further, we checked the absorption happening only in the ultrathin $MoS_2$ layer as shown in Figs.5.7a and 5.7b. We observe that the total absorption in $MoS_2$ is enhanced up to 20% at the A exciton and B exciton wavelengths at all angles from 0-60° due to the underlying TiN layer (Fig.5.7a) with $\lambda_{ENZ}$ fixed at 690 nm indicating wide-angle absorption enhancement in $MoS_2$ due to the underlying TiN. Also, on variation of $\lambda_{ENZ}$ of TiN, we find that the maximum absorption at the A exciton wavelength of $MoS_2$ is observed when the $\lambda_{ENZ}$ of TiN is closely aligned with the A exciton emission wavelength (Fig.5.6b). The absorption in $MoS_2$ is slightly increased when $\lambda_{ENZ}$ = 500 nm and the TiN becomes essentially metallic for $MoS_2$, but it drops when TiN has $\lambda_{ENZ}$ = 400 nm as TiN becomes a better metal and most of the light is reflected, thereby reducing the effective interaction of light with $MoS_2$. Hence, there is no further improvement in absorption at the A exciton wavelength is observed. This indicates how crucial is the $\lambda_{ENZ}$ alignment with the excitonic wavelengths for effective light absorption in $MoS_2$. This is further validated as the absorption enhancement is not observed when we simulate the absorption replacing the TiN layer with an Au film of same thickness of 150 nm as seen in Fig.5.8. Inspite of having plasmonic properties in the visible, a simple Au film cannot enhance the near field unless engineered in the form of a grating, etc. At ENZ, TiN exhibits lower reflectivity than Au, resulting in higher optical path lengths in the $MoS_2$ layer and thereby improving absorption. Moreover, Au is highly reflective at 690 nm ($\sim$ 80%) so the effective light interaction with $MoS_2$ is reduced.

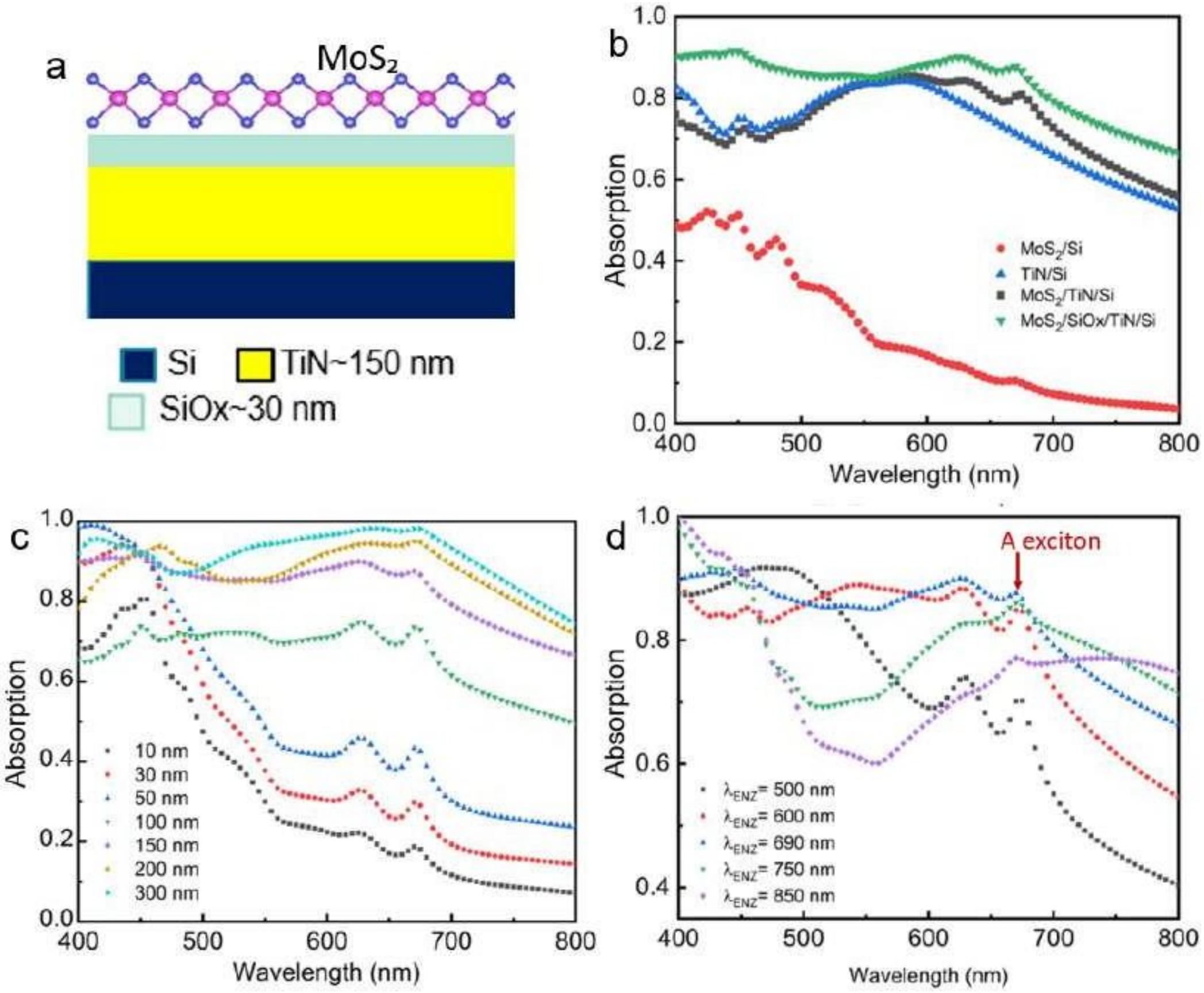

**Figure 5.6:** (a) Schematic of the structure for studying the optical response of $MoS_2$ on TiN substrate, (b) Simulated absorption plots for bare $MoS_2$ lying on flat Si substrate, $MoS_2$/TiN/Si, TiN/Si $MoS_2$/SiOx/TiN/Si, Simulated absorption for the entire $MoS_2$/SiOx/TiN/Si system (c) with variation in thickness of TiN and $\lambda_{ENZ}$ fixed at 690 nm, (d) with variation in $\lambda_{ENZ}$ of TiN and thickness fixed at 150 nm

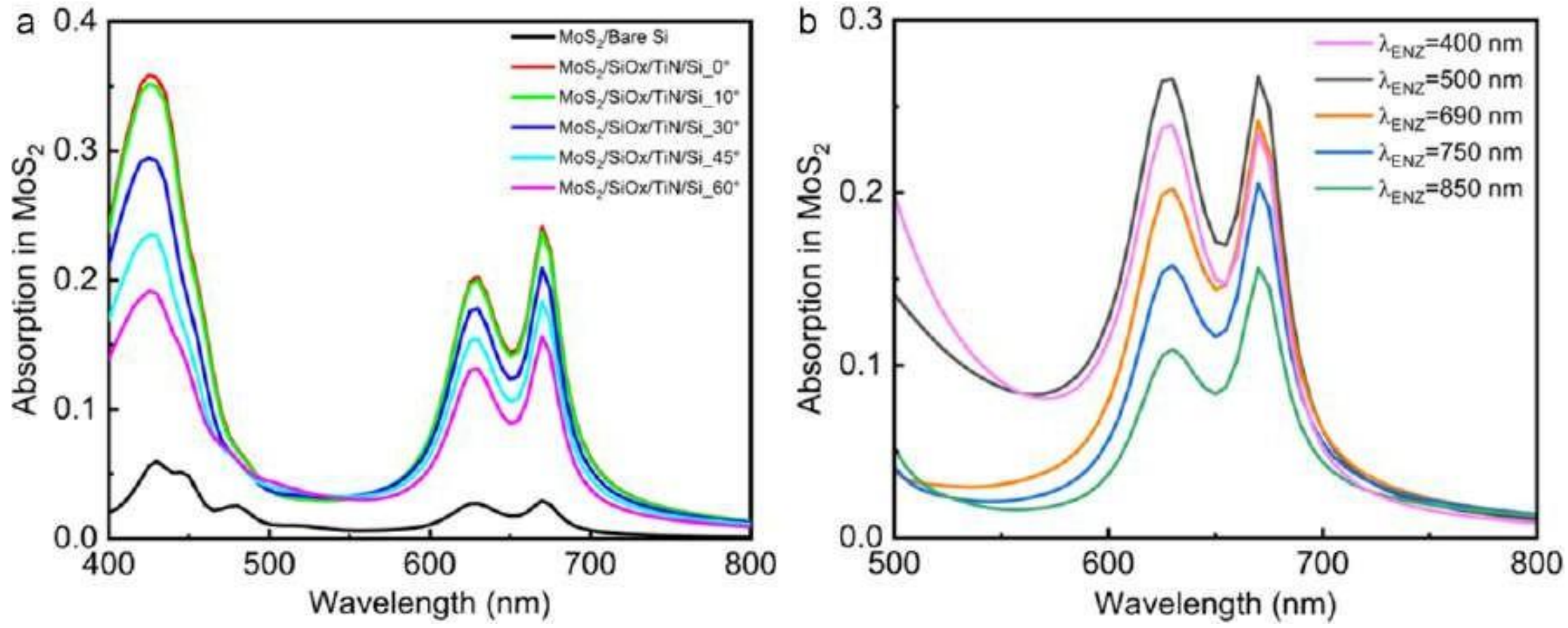

**Figure 5.7:** (a) Simulated absorption occuring in $MoS_2$ lying on SiOx/TiN/Si (a) with variation in angle of excitation from 0°-60° with $\lambda_{ENZ}$ of TiN fixed at 690 nm, (b) with $\lambda_{ENZ}$ variation of the background TiN through varying the carrier density in TiN.

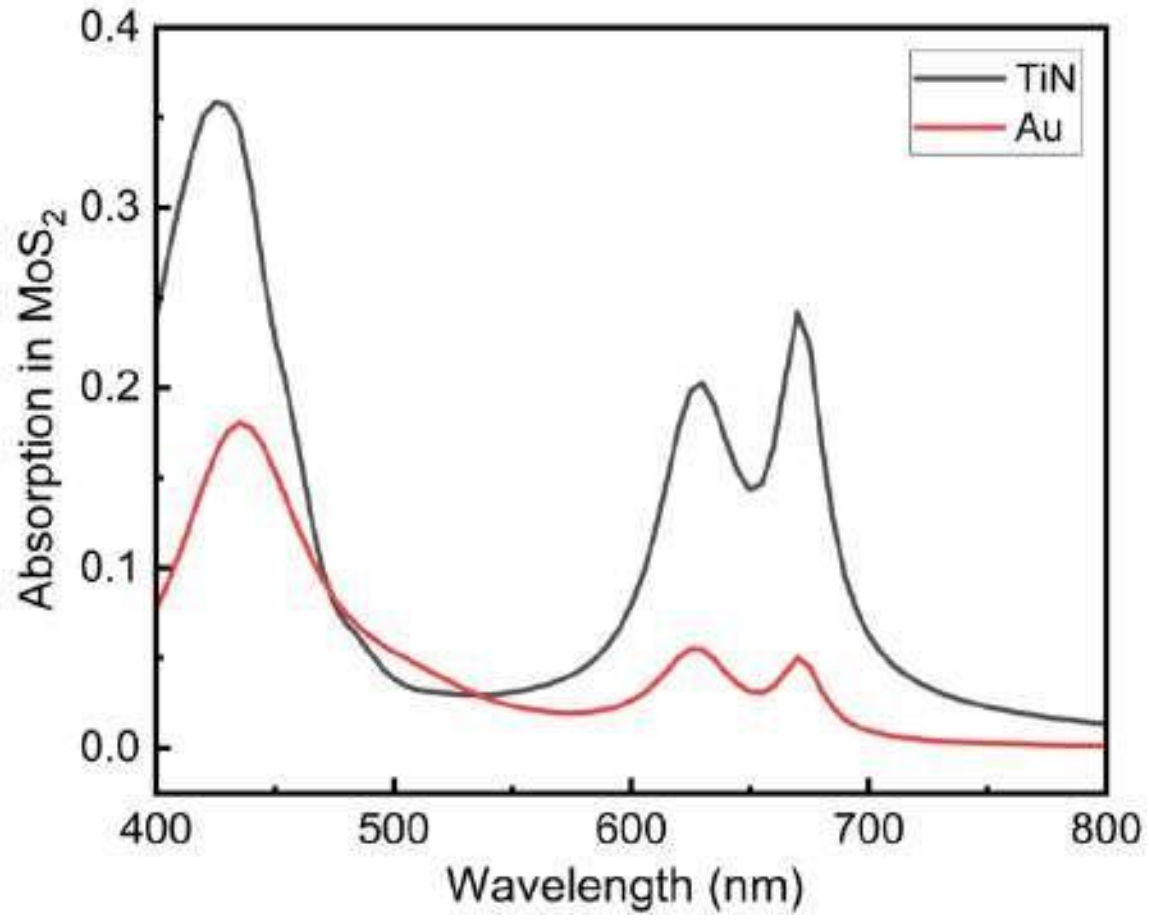

**Figure 5.8:** Simulated absorption in monolayer $MoS_2$ with underlying 150 nm TiN and Au film on Si.

### 5.3.3 Emission enhancement in $MoS_2$

It is expected that any enhancement in absorption will be reflected in enhanced emission properties following Kirchoff's law. Hence, we probed the emission in this system through photoluminescence (PL) measurements. To check the emission enhancement from a single flake, we patterned square windows of width $\sim$40 $\mu$m using e-beam lithography on the TiN film on Si and completely etched the TiN so that the CVD grown flakes $MoS_2$ flakes can be transferred in such a way that they partially lie on the TiN as well as on Si as shown in the schematic Fig.5.9b. This scheme was adopted to compare the PL and Raman signals from the same $MoS_2$ flake on TiN with that of bare Si. In ad-dition, $\sim$30 nm of hydroxy silses-quioxane (HSQ) polymer was coated on the patterned TiN/Si substrate prior to $MoS_2$ transfer and annealed at 500$^{\circ}$C for 1 hour to convert it to SiOx and thereby isolating the $MoS_2$ flakes from TiN. CVD-grown $MoS_2$ was transferred on the patterned SiOx/TiN/Si substrate using the wet-etch transfer method (for details, refer to chapter 2). SEM image of the transferred $MoS_2$ flakes on patterned SiOx/TiN film is shown in Fig.5.9c. Fig. 5.9d shows the AFM topography of the $MoS_2$ flake lying partially on SiOx/TiN/Si and SiOx/Si (green dashed line shows the outline of the flake). The better contrast is provided by the AFM adhesion image recorded simultaneously with the AFM topography as shown in Fig. 5.9e.

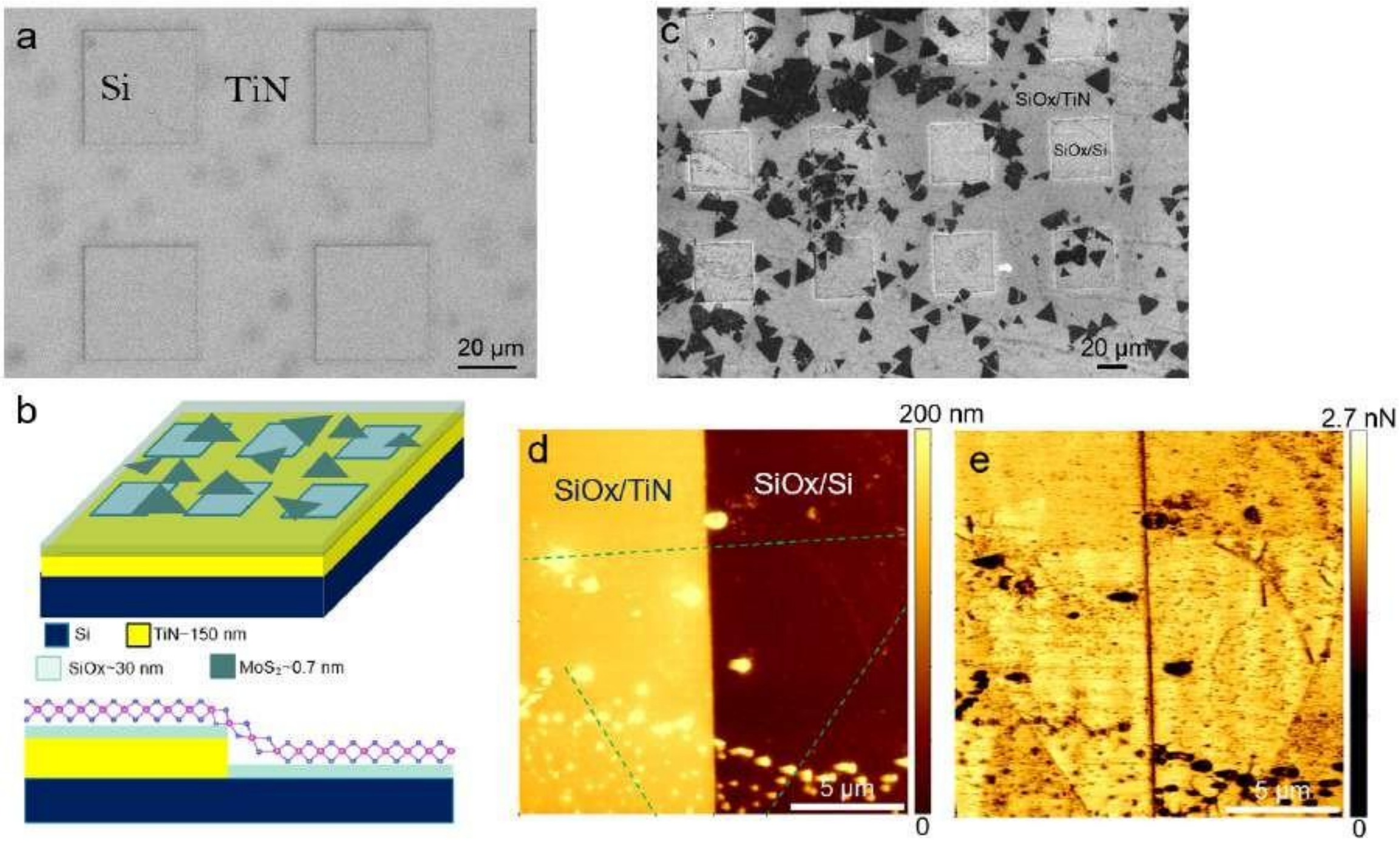

**Figure 5.9:** (a) SEM image of Si windows opened in TiN film via lithography and selective etching (b) Schematic of the structure for studying the optical response of $MoS_2$ on TiN substrate, (c)SEM image of $MoS_2$ flakes transferred on the patterned SiOx/TiN/Si substrate, (d) AFM topography showing the flake lying partially on HSQ coated TiN and Si, (e) AFM adhesion image showing the contrast of the flake with respect to the substrate.

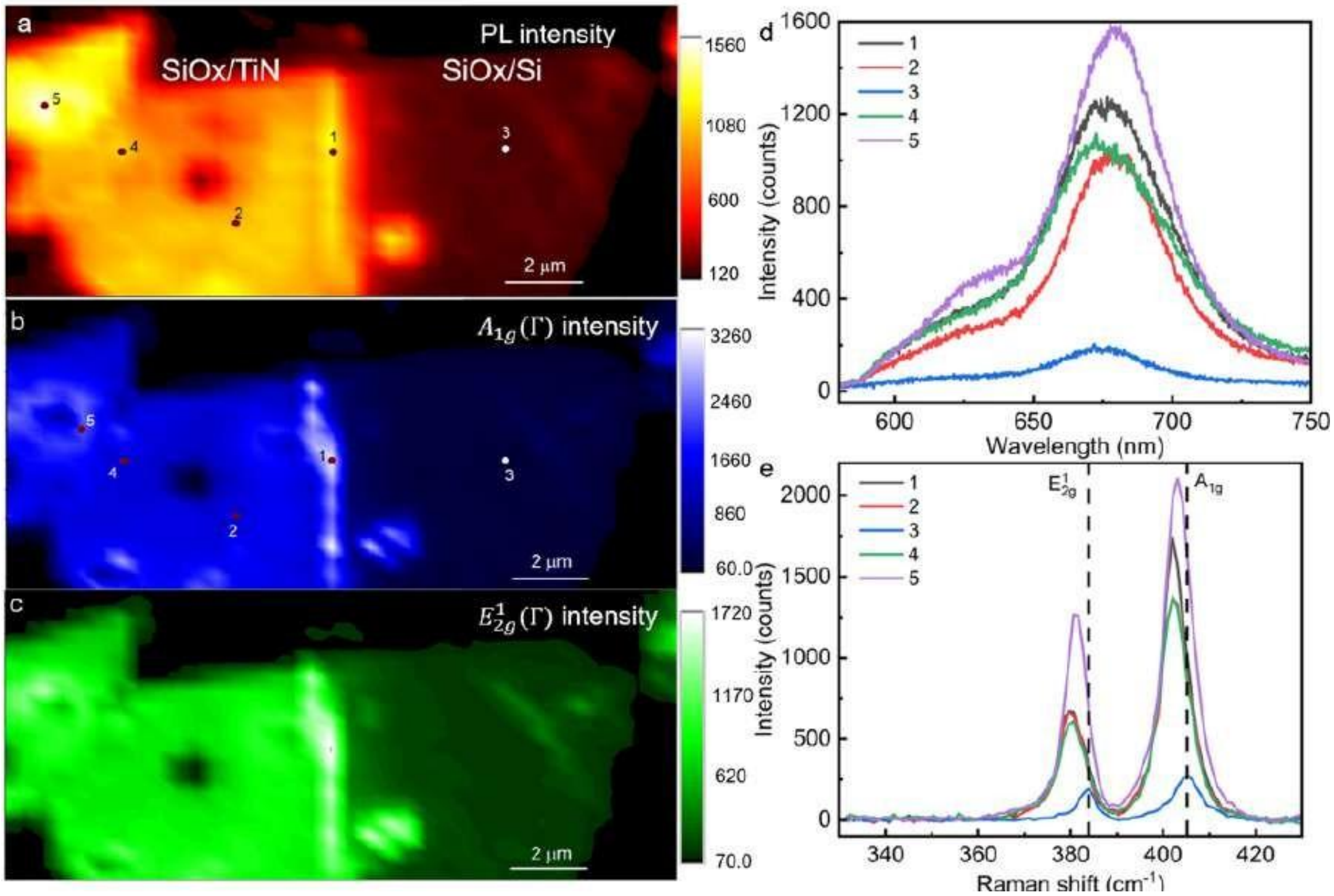

**Figure 5.10:** Spatial maps of (a) PL intensity; (b, c) Intensity maps for $A_{1g}(\Gamma)$ and $E^1_{2g}(\Gamma)$ raman modes of the $MoS_2$ flake lying partially on SiOx/TiN and bare SiOx/ Si; (d) PL spectra plotted at selected locations as shown in the intensity map (e) Raman spectra plotted for the same locations for the $MoS_2$ flake on and off TiN.

Thereafter, we characterized the PL and Raman response of the flake lying partially on $SiO_x$-coated TiN and Si as shown in Fig. 5.9d. Fig. 5.10a shows the spatial map of the PL intensity at the A exciton wavelength ($\sim$ 675 nm) of the $MoS_2$ flake partially lying on SiOx/TiN/ Si and bare SiOx/Si, recorded under 532 nm laser excitation. The locations denoted as 1, 2, 4 and 5 are positions of the flake lying on SiOx/TiN and 3 is on bare SiOx/Si. It is evident that locations 1, 2, 4, and 5, which are on TiN record enhanced PL intensity compared to location 3 (bare SiOx/Si). Moreover, location 1 and throughout the edge of the TiN film record a consistent elevated PL intensity as the flake suffers from strain at this location due to the height difference between the TiN film and the Si substrate and $MoS_2$ has to effectively bend. The strain reduces the band-gap locally, leading to exciton funelling [237–239], thereby increasing the PL intensity further. The PL spectra show that the PL intensity is essentially ∼6 times enhanced for $MoS_2$/SiOx/TiN in comparison to $MoS_2$/ SiOx/Si as shown in Fig. 5.10d. However, due to certain local topographic undulations, the PL is further enhanced occasionally, as seen in the spectrum for position 5, where both the TiN and strain have a combined effect on the PL enhancement. Moreover, the presence of the insulating $SiO_X$ layer between $MoS_2$ and TiN isolates the two materials and hence, PL is expected not to be affected due to charge transfer between the two materials. TiN has a work function of ∼4.8–5.0 eV [240] and monolayer $MoS_2$ has band gap of ∼ 1.8 eV with conduction band minimum (CBM) at ∼4.0 eV and valence band maximum (VBM) at ∼5.8 eV (all relative to vacuum level)[241]. Hence, TiN tends to favor electron injection into $MoS_2$ which is forbidden by the SiOx layer. The PL peak for the transferred $MoS_2$ flake corresponding to A exciton is observed at ∼ 675 nm and it can vary slightly depending on the sulfur vacancies. As mentioned earlier, the TiN film prepared here has $\lambda_{ENZ}$∼ 690 nm. Note that we define the ENZ regime to be the wavelength range ($\Delta\lambda$) where the real part of the permittivity varies between +1 to -1. Hence, here the A emission wavelength for $MoS_2$ lies well within the ENZ regime of TiN. Hence, the emission enhancement in $MoS_2$/SiOx/TiN/Si can be justified to be primarily resulting from the refractive index contrast provided by the back ENZ TiN layer on Si. This is even verified when we record the PL spectra for a monolayer $MoS_2$ lying directly on TiN without the SiOx layer showing enhanced intensity as compared to bare Si, as seen in Fig.5.12. However, in this case, the relative enhancement is low as compared to $MoS_2$/SiOx/TiN/Si proving that the 30 nm thick SiOx layer increases the effective interaction of light with $MoS_2$, thereby improving absorption in $MoS_2$ as discussed earlier.

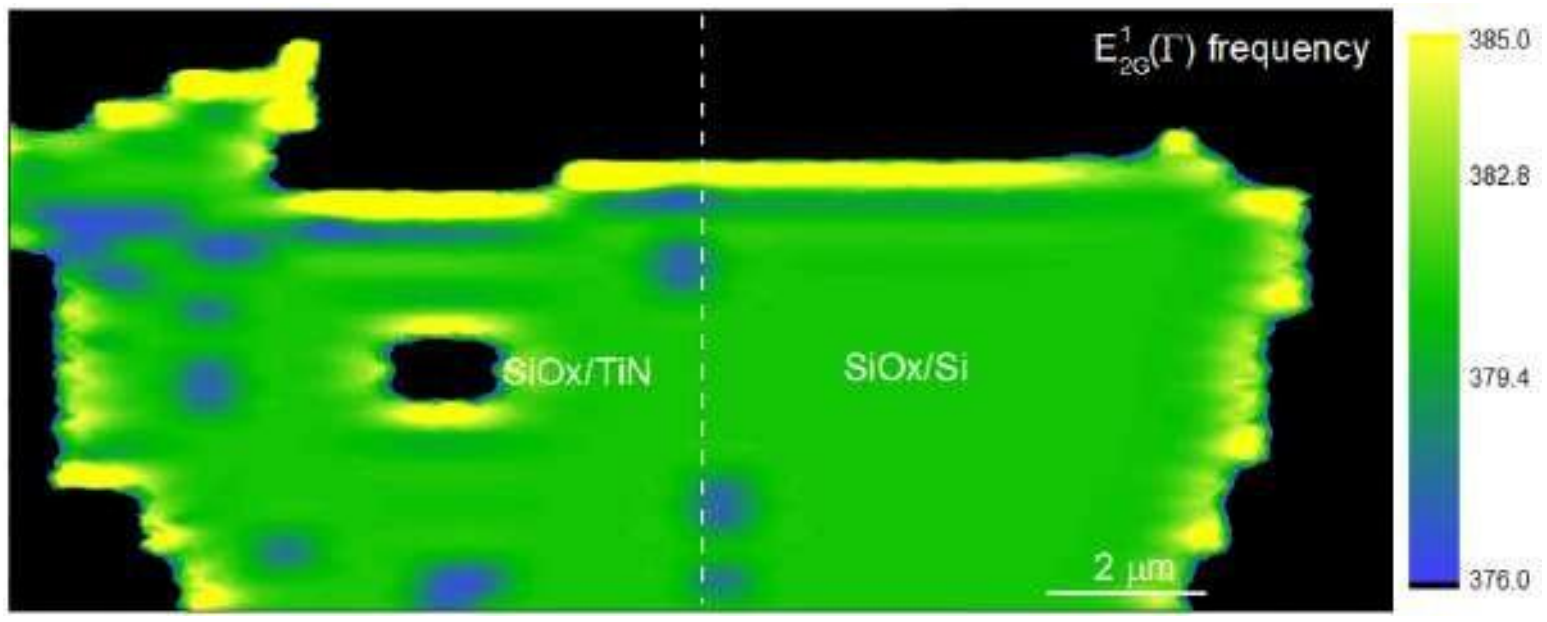

**Figure 5.11:** Peak shift map of $E^1_{2g}(\Gamma)$ raman mode.

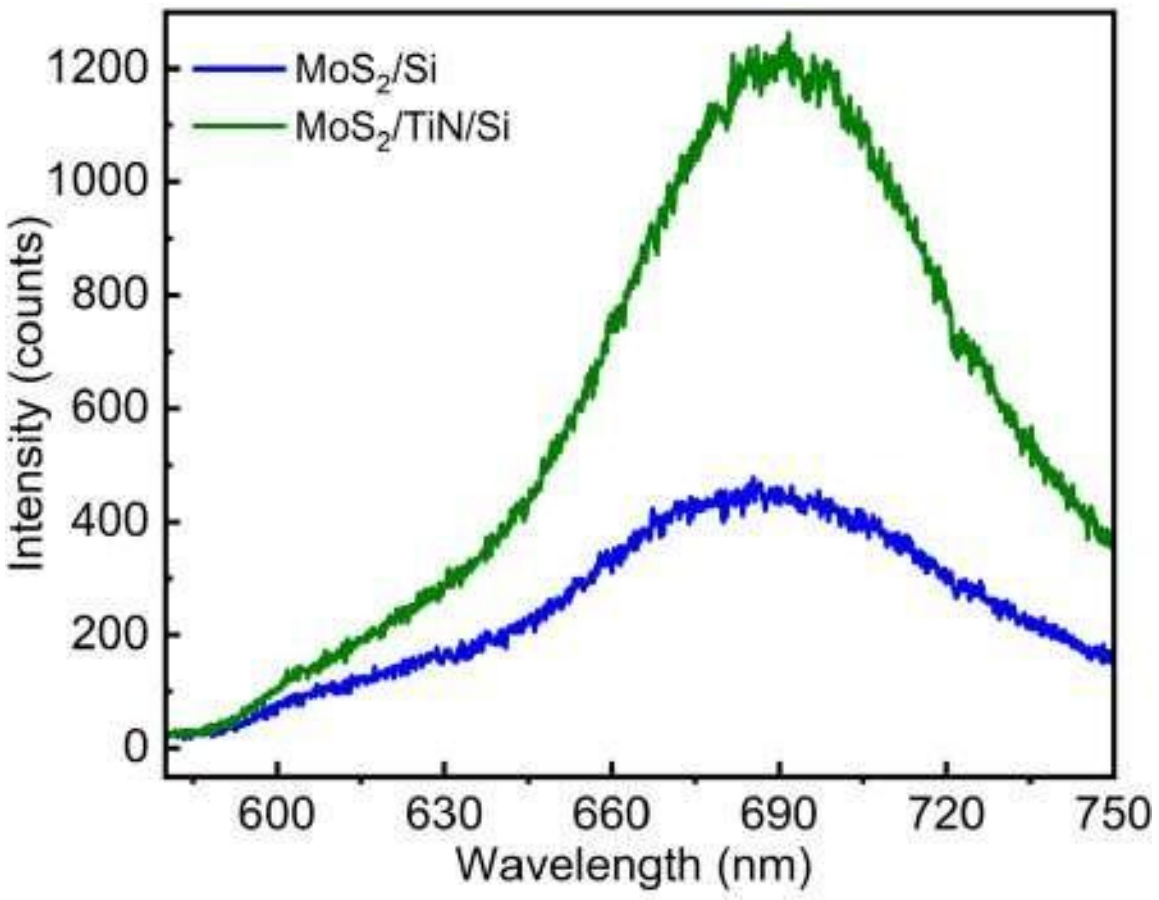

**Figure 5.12:** PL spectra recorded on flake monolayer $MoS_2$ lying partially on TiN/Si and bare Si

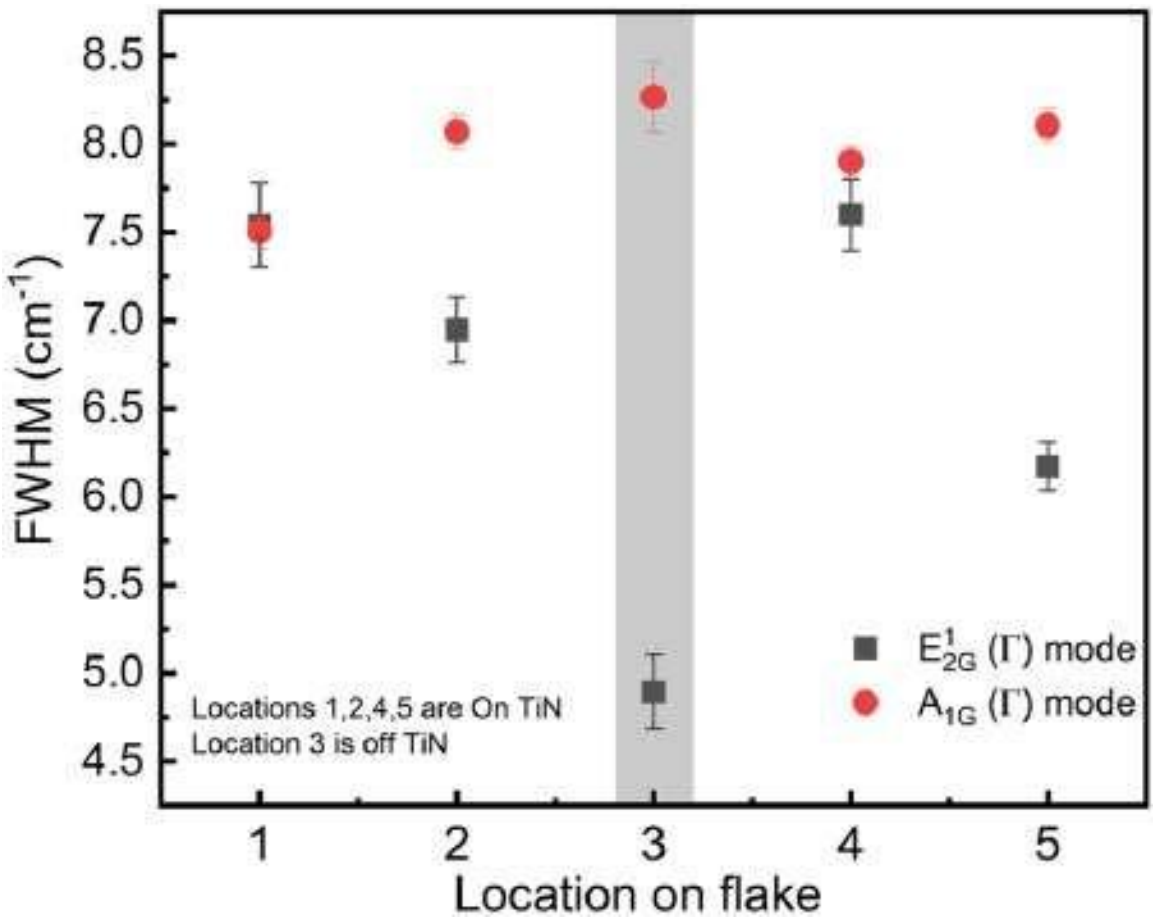

**Figure 5.13:** FWHM plot for the in-plane and out of plane $E^1_{2g}(\Gamma)$ and $A_{1g}(\Gamma)$ raman modes in $MoS_2$ on and off SiOx/TiN.

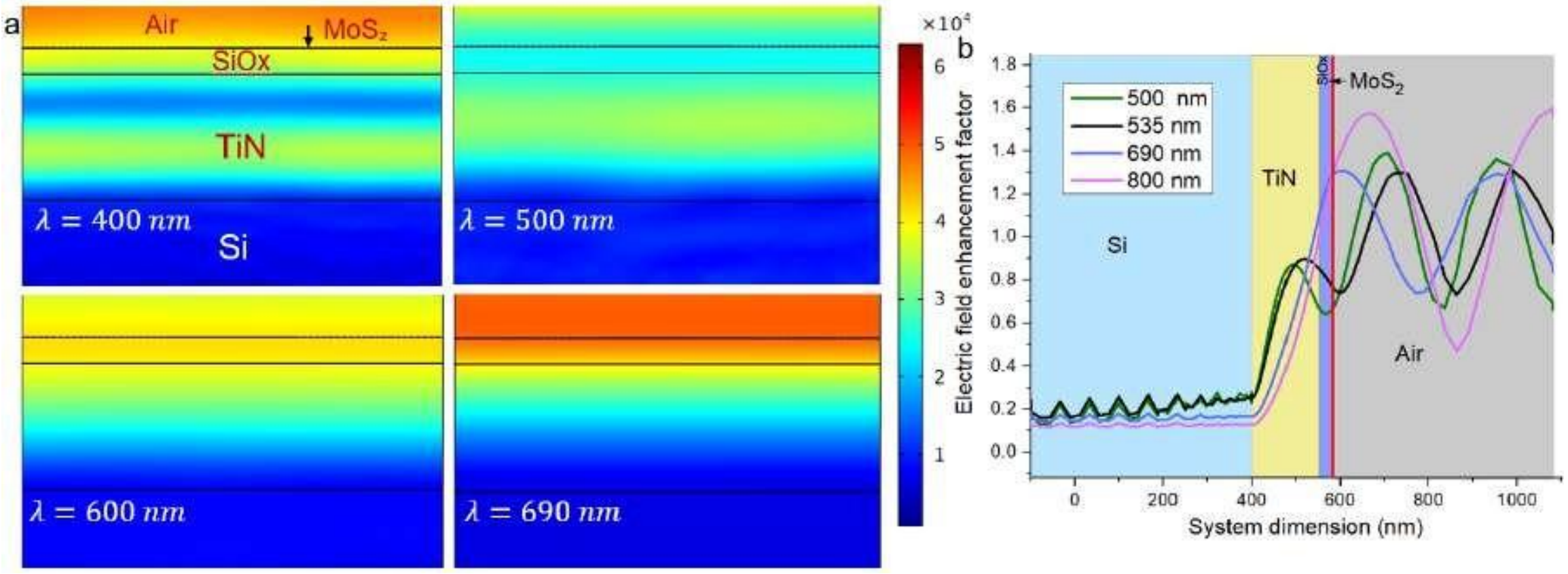

**Figure 5.14:** (a) Electric field profile (V/m) plotted at various wavelengths for a TiN film with $\lambda_{ENZ}$ = 690nm, (b) Variation of Electric field enhancement plotted across the different domains of the TiN+$MoS_2$ system.

Figs. 5.10 b and 5.10c show the spatial maps of the Raman intensity of the two characteristic modes in $MoS_2$- $E^1_{2g}(\Gamma)$ and $A_{1g}(\Gamma)$ at ~385 $cm^{-1}$ and ~ 405 $cm^{-1}$ respectively for a flake lying partially on SiOx/TiN/Si and bare SiOx/Si substrate. $E^1_{2g}(\Gamma)$ mode denotes the in-plane and $A_{1g}(\Gamma)$ mode denotes the out-of-plane Raman modes in $MoS_2$ and the intensity maps are plotted at the frequency values showing the maximum intensity. Similar to the PL intensity map, the Raman intensity maps plotted for the frequencies exhibiting the highest intensity also record enhanced intensity of the flake on TiN compared to Si, with location 1 and the flake lying on the TiN edge exhibiting the highest intensities consistently on both modes. The $E^1_{2g}(\Gamma)$ mode is red-shifted by ~ 2 $cm^{-1}$ and the $A_{1g}(\Gamma)$ mode is red-shifted by ~ 4 $cm^{-1}$ at selected locations as indicated in the flake. The shift in both modes uniformly can be due to heating effects as discussed later, as the measurements were performed using a higher laser intensity (10% of maximum power), which might lead to a red shift in the spectra[242]. Here, the contribution of strain can be neglected as the average roughness of the bare TiN film is less than ~4 nm, even though it may have a minimal contribution due to local topographic undulations. The peak position map of the $E^1_{2g}(\Gamma)$ mode plotted, shown in Fig.5.11, supports this claim, as it shows no significant peak shift for the flake on the TiN region compared to the Si region. However, there is a broadening in the in-plane $E^1_{2g}$(Γ) peak when the flake lies on SiOx/TiN (locations 1,2,4,5) as seen from the FWHM plot for both the modes shown in Fig.5.13, indicating that the SiOx/TiN increases the phonon-phonon scattering in $MoS_2$. SiOx and TiN, both being materials with low thermal conductivity [243, 244] dissipate heat poorly during laser exposure, which causes local heating of $MoS_2$, thereby leading to peak broadening.

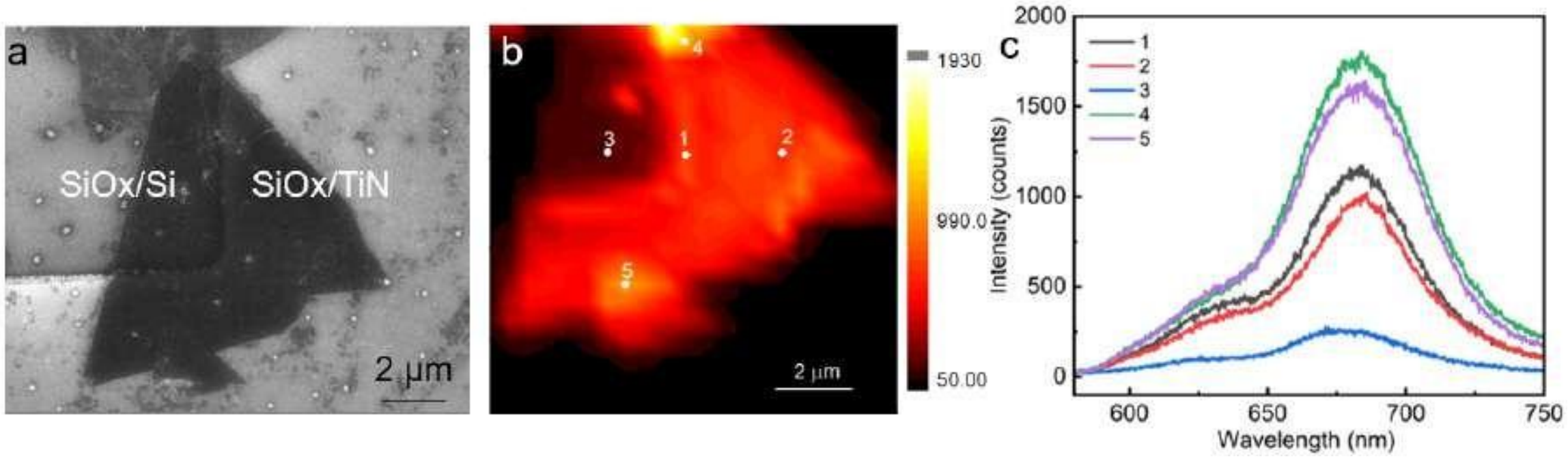

**Figure 5.15:** (a) SEM image of a bilayer $MoS_2$ flake lying partially on SiOx/TiN/Si and SiOx/Si, (b) Spatial map of PL intensity of the $MoS_2$ flake, (c) PL spectra plotted for specific loactions as shown in (b).

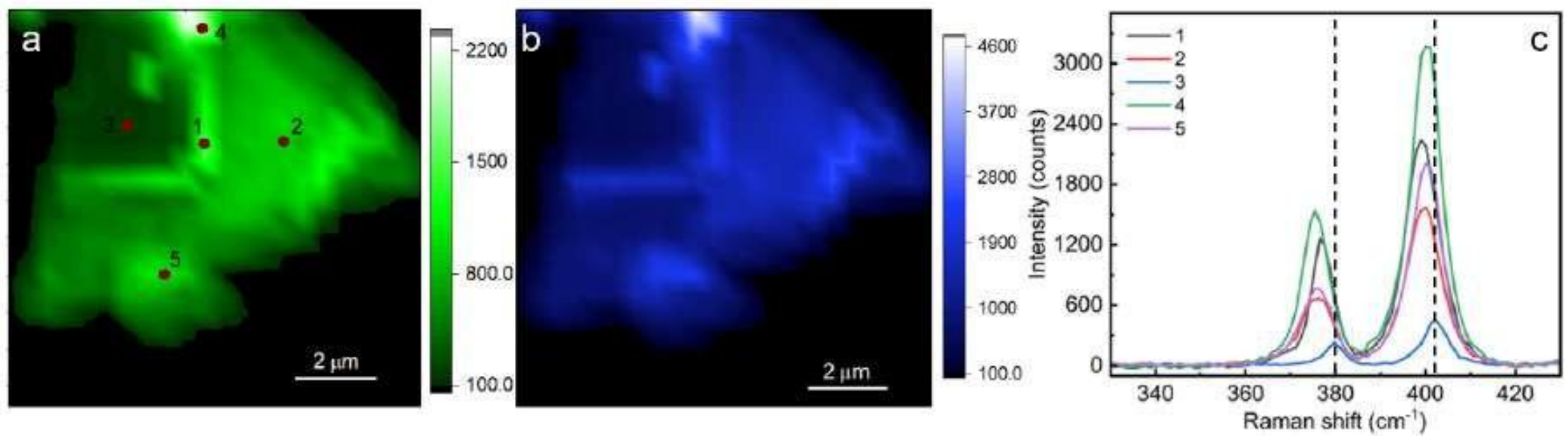

**Figure 5.16:** Spatial maps intensity for (a) $E^1_{2g}$ (Γ), (b) $A_{1g}$(Γ) raman modes of the bi-layer $MoS_2$ flake lying partially on SiOx/TiN/Si and bare SiOx/Si, (c) Raman spectra plotted for different locations on the flake as marked in (a).

The intensity enhancement in both Raman modes indicates the local field enhancement due to the presence of the background TiN/Si substrate. To confirm local field enhancement in the flake, we performed FEM calculations as shown in Fig.5.14. The enhancement factor was calculated by dividing the simulated electric field value by the input electric field ($E_{in} = 3.8 \times 10^4$ V/m). It shows that at $\lambda = 690$ nm and beyond, the electric field is enhanced ($\mathbf{E} > 1$) in $MoS_2$. The PL peak intensity in $MoS_2$ due to A exciton emission lies at $\lambda = 675$ nm, which falls within the ENZ regime of the background ENZ substrate. Therefore, it can be primarily inferred that the enhancement in both PL intensity and Raman signal arises from the electric field enhancement due to the refractive index contrast provided by the underlying TiN/Si substrate, which enhances absorption in $MoS_2$, as previously discussed. Moreover, the experiments were performed using a high numerical aperture (N.A. = 0.9) objective, allowing light to be incident over a broad range of angles rather than only at normal incidence ($0^\circ$). This indicates at the possibility of both s- and p-polarized excitation, suggesting that the ENZ properties of the substrate contribute to the observed effects. However, further investigations are required to confirm this as well as to explore the possible role of Fresnel reflections in enhancing the PL and Raman signals.

To confirm the reproducibility of our results, we performed spatially resolved PL and Raman measurements on a separate flake, as shown in Fig.5.15 and Fig.5.16. This flake was a bilayer; nevertheless, a six fold enhancement in PL intensity was observed. The Raman mode intensities also showed enhancement, consistent with the findings from the monolayer flake. Also, the edge exhibited higher intensity because of the additional local strain induced by the height difference. Therefore, combined theoretical and experimental results establish the overall optical property modification in $MoS_2$ due to the underlying substrate.

## 5.4 Conclusion

In summary, we have demonstrated ENZ substrate-based absorption and emission enhancement in $MoS_2$ devoid of any sophisticated nanostructuring. The near field is enhanced due to the typical optical properties of the underlying TiN layer, facilitated by the ENZ response. So, placing $MoS_2$ in proximity to TiN enhances wide-angle absorption in $MoS_2$ and thereby enhances the emission. Also, the enhancement in the Raman mode intensities typically indicates the enhanced electric field due to the underlying SiOx/TiN films. The 6 times enhancement in PL intensity through the mere introduction of a TiN layer denotes the efficiency of this strategy. Also, the presence of an ultrathin SiOx layer poses an advantage by not only restricting the charge transfer but also improving the absorption by increasing the effective interaction of light with the $MoS_2$ with the underlying TiN. Hence, combined theoretical and experimental results elucidate the important role played by TiN as a material platform in effectively modifying and enhancing the optical properties in ultrathin flakes of $MoS_2$, thereby encouraging the usage of epsilon-near-zero materials as versatile platforms in photonic applications. This strategy also provides insights into applications to other TMDCs or 2D semiconductors, enabling platform-level enhancement of light-matter interaction across a range of van-der Waals materials. Altogether, this study demonstrates that ENZ materials like TiN provide a powerful, fabrication-friendly route to controlling light-matter interactions in 2D materials, opening new avenues in scalable photonics and optoelectronics.

# Chapter 6

# Conclusions and Future Outlook

As understood from the previous chapters, the uniqueness and multifaceted properties of ENZ materials provides an elegant platform in recognition of novel optical technologies towards the development of practical photonic devices. The various works that have been described showcases direct applications resulting from integration of various materials ENZ materials in optimized dimensions and architectures. Moreover, the successful fabrications of large area ENZ patterns over diverse substrates and experimental demonstrations of spectrally selective optical responses supported by theoretical investigations straightaway addresses the challenges that the field had been facing. This chapter summarizes the key results and findings of this thesis and later provides an idea of the future studies that can be conducted motivated from the present work.

## 6.1 Conclusions

- First, we have successfully demonstrated a wide-angle, spectrally selective reflector coating with a "step-function"-like reflectivity response that is generated on a opaque-SS as well as transparent-glass substrate elucidating the relevance of the architecture as a substrate independent design. The design is straightforward and consists of easily fabricable CMOS compatible materials of $Cr_2O_3$, Cr, and ITO as the ENZ material with subwavelength thickness and structures. The average reflectivity obtained in the visible was $\sim$ 15% while the that in NIR ($\sim$ 80%). The transition from low to high reflectivity was demarcated by a cut-in wavelength $\lambda_o$ determined by the $\lambda_{ENZ}$ and plasmonic resonances of ITO which was employed both in thin film and nanostructured forms. We examined two forms of the coating: one with the top ITO nanostructures along with the thin film and other having only the ITO thin film. In both cases we observed the step-function like reflectivity is generated. However, for the flat ITO coating the visible reflectivity was higher than the nanostructured counterpart. Hence, indicating at the requirement of nanostructuring. Both experimental as well as numerical simulations concurrently points out the roles of individual material with ITO playing the key part in achieving the "step function"-like reflectivity. Further, the tunability of ITO's $\lambda_{ENZ}$ via a simple annealing protocol make it even more advantageous in comparison to the existing coatings as summarized in Chapter 3. Hence, this study provides a simple yet a remarkable design holding relevant applications in thermal energy management. Overall, this architecture offers a robust and cost-effective solution in optical engineering.

- In the second study, we constructed a coating design such that it relayed a band-selective absorption response, employing again a nanostructured grating of ENZ media (ITO) on top of highly reflecting dielectric-metal ($SiO_2$/Au) underlayer. The design imparted a wide-angle (0° - 60°) response with over 85% absorption, in a wavelength band of 1800 nm - 2800 nm which was understood to be arising primarily from a combination of the nanostructured ITO's ENZ properties and plasmonic response in conjunction with the optical properties of the dielectric under-layer which became evident when $SiO_2$ was replaced with other dielectrics such as $HfO_2$ and $Cr_2O_3$ which have higher refractive indices than that of $SiO_2$ . In this case too, experimental investigations together with numerical calculations delineated the roles played by the ENZ properties of ITO that enabled electric field enhancement and confinement in the ITO grating which increased the absorption and generated the band-selective response. In this study, the effect of the localized plasmon modes in ITO facililated by the surrounding dielectrics were crucial determining the central wavelength and bandwidth of the high-absorption band. Finally,the tunable nature of the entire absorption band was demonstrated by controlling the electron density of ITO and permittivity of the dielectric underlayer which relayed the customizability of the coating. Further, direct venture into thermal emissivity measurements to demonstrate the high emissivity of the nanostructured ENZ surface even at longer wavelengths explained the efficacy of the design. Overall, the investigation provided yet another direction towards realizing coating designs with ENZ systems which would be elemental in thermal energy management for potential thermo-photovoltaic applications.

- To push the boundaries beyond engineering the optical response of surfaces, we finally investigated the potential of a visible ENZ material- titanium nitride in enhancing light absorption and emission in a utrathin 2D material- molybdenum di-sulphide grown via chemical vapour deposition. We numerically showcased that the presence of a 150 nm thin layer of TiN underneath $MoS_2$ resting on a silicon substrate leverages the average absorption by 20% in $MoS_2$ over a wide angle ($0^\circ$ - $60^\circ$). This didn't involve any sophisticated nanostructuring unlike the previous two cases. Further, we also checked the emission from $MoS_2$ through photoluminescence(PL) measurements. To test the TiN's effect in PL enhancement, a simple structuring was conducted to open Si windows on the TiN film such that after transfers of $MoS_2$ flakes, it partially lies on TiN as well as Si. A thin layer of SiOx was introduced $MoS_2$ and TiN/Si to prevent any charge transfer between the two layers. PL showed a 6 times enhancement in the A exciton emission intensity while Raman also showed significant intensity enhancement in both the modes of $MoS_2$ indicating that the near-field is enhanced due to the typical refractive index constrast provided by the underlying TiN/Si substrate facilitated by the ENZ property of TiN with further investigation in ENZ based effects to be explored. Overall, this approach opens new pathways for light modulation in 2D materials, leveraging ENZ materials for advanced optoelectronic applications.

## 6.2 Future Outlook

The various findings presented in this thesis build the background for further research both fundamentally as well as from application perspective. A few of them are outlined as follows:

- Further research is required to better understand and control the interactions between electromagnetic fields and ENZ media, particularly based on the idea of nonlinear optics and metamaterials. The strong non-linearity aspect in ENZ media have made them a relevant candidate to look into integrated photonic applications. Hence, design of metasurfaces and metamaterials with these materials to check non-linear responses will be an important step in realizing practical nonlinear optical devices.

- As discussed, TMDCs are susceptible to various factors like substrate interaction, defects, strain etc. In this scenario, it would be interesting to check the combined effect of strain and ENZ phenomena on the optical and electronic properties of these materials using nano-patterned substrates of ENZ materials.

- In this thesis, mainly two ENZ systems have been explored which appears to be highly promising for practical applications. However,challenges still exist in terms of material losses, surface roughness, and realizing ENZ phenomena in specific wavelength ranges. Hence, further research is needed to improve the quality of the films and structures through rigorous optimization of the fabrication processes.

# Appendices

# Appendix A

# Enhancement of local optical, electronic and catalytic properties in strain-engineered $MoS_2$

## A.1 Introduction

As discussed in Chapter 5, transition metal dichalcogenides (TMDCs), have gained significant attention due to their attractive material properties. The ultrathin 2D nature and superior electronic, optoelectronic, and mechanical properties have made them strong contenders for emerging applications [111]. Originating from their incredible mechanical properties, the introduction of strain significantly modulates their band structure, and may be used to tailor their optoelectronic properties[245]. Additionally, the flakes often show formation of wrinkles and bubbles due to differential adhesion, surface adsorbates or may be artificially engineered through non-uniform straining. Wrinkles are high-strain features nucleated as mechanisms of strain relaxation and are susceptible to several factors such as substrate interaction, number of layers, environmental humidity and temperature etc [246, 247]. Thus, their structure provides information regarding various interactions vital to designing efficient devices. In this chapter we discuss, how we introduced non-uniform strain in monolayer $MoS_2$ through substrates with periodic patterns gold (Au) nanostructures of different geometries. Through multiphysics investigation we probe the modification in local optical and electronic properties and how it can be used in practical applications such as hydrogen evolution.

## A.2 Introduction of non-uniform strain and the modification of different properties in $MoS_2$

### A.2.1 Patterning of surfaces with varying periodicities and geometries of nanostructures

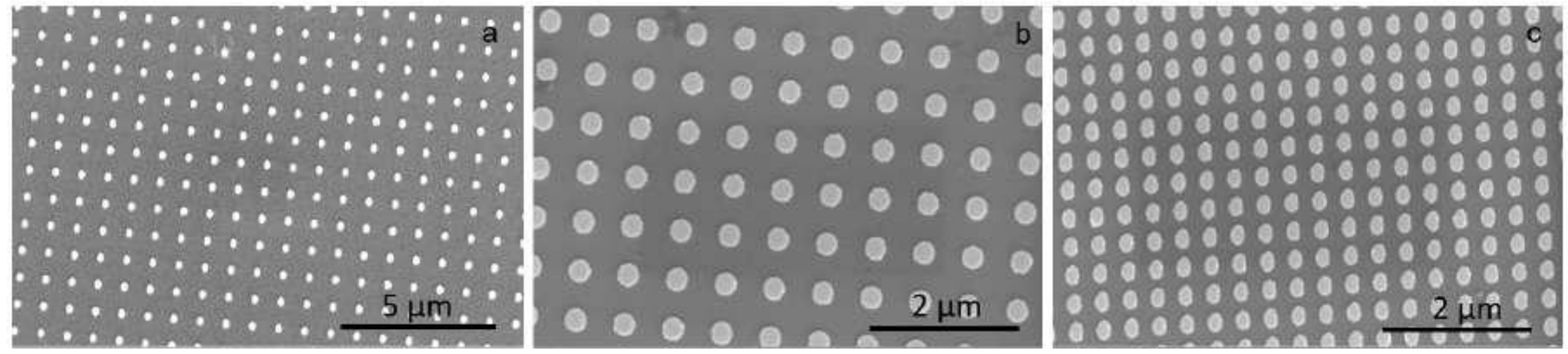

**Figure A.1:** SEM images of Au nanocylinders of (a) 2 $\mu$m,(b)1 $\mu$m and (c) 0.5 $\mu$m periodicity

Bare Si as well as $SiO_2$/Si substrates were patterned with metal nanostructures such as gold as well as dielectric nanostructures as shown in Figs.A.1 and A.2. As shown in Fig.A.1 cylindrical nanostructures of Au with 300 nm diameter and 50 nm thickness with varying periodicities from 2-0.5 $\mu$m were fabricated initially. The dimension and aspect ratio of the Au nano-cylinders were chosen such that the broad plasmonic resonance of the Au-nanostructure lies in the IR, beyond 1000 nm and as a result plasmonic enhancements do not interfere with the strain-induced effects on PL and Raman spectra. Moreover, the CVD-grown $MoS_2$ flakes typically have a typical dimension of 20- 30 $\mu$m. Thus, the periodicity and dimensions were so chosen such that a significant number of nanostructures lie within a monolayer area, to maximize the effective induced strain, both locally and across the entire monolayer flake. The nano-cylinder peridicity (centre to centre) distance was kept at 2 $\mu$m in the initial samples to enable spatially resolved PL and Raman spectroscopy measurements via an optical setup that typically has a spatial resolution $\sim$ 0.7 $\mu$m.

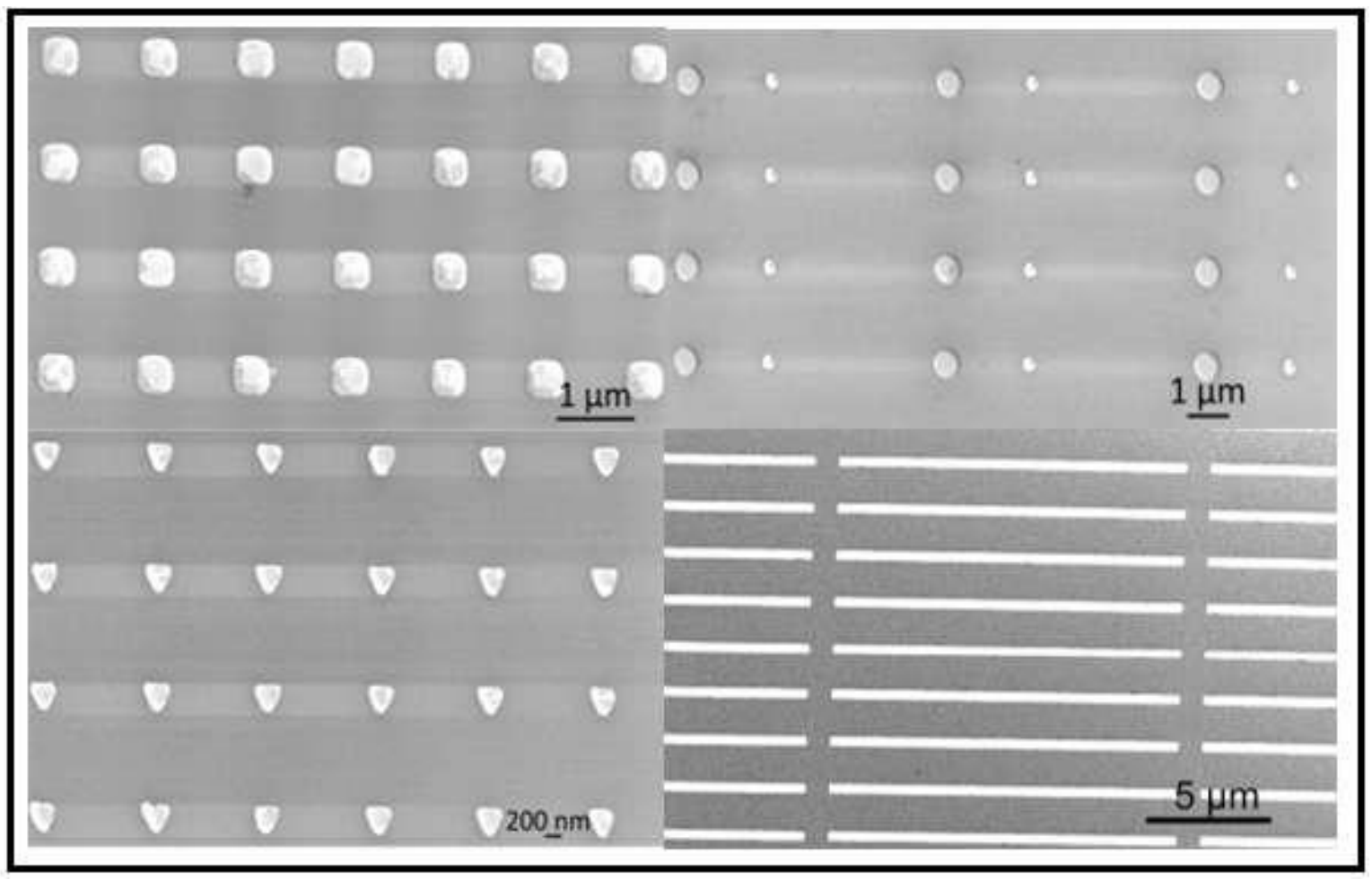

**Figure A.2:** SEM images of e-beam lithographed (a) square,(b) triangular, (c) circular and (d) ridge nanostructures of Au on $SiO_2$/Si substrate.

Later, to understand the origin of strain in the flakes in the form of wrinkles and occasional nanobubbles, fabrication of patterns of different geometries was conducted as shown in Fig.A.2. The edge dimension of the square and the triangular nanostructures was kept at 500 nm with a typical thickness of the structures being 50 nm. The circular nanostructures had varying diameters 200-500 nm with the same thickness. The ridges had typical width and length of 300 nm and 15 $\mu$m length, respectively. Such variation in the structure geometry and dimensions helped to understand the origin of wrinkles when monolayer $MoS_2$ drapes over these structures as discussed in section A.2.2.

### A.2.2 Strain maps

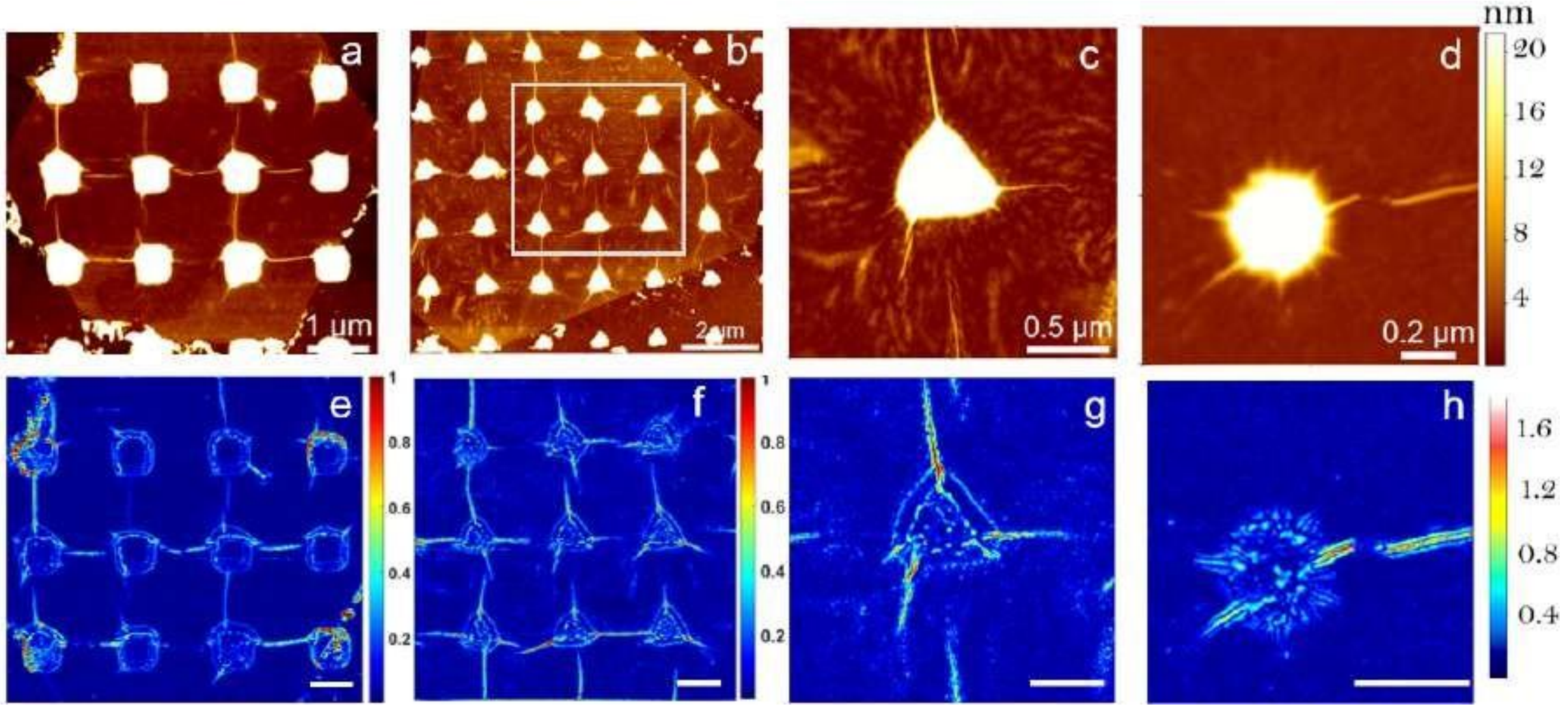

**Figure A.3:** Large area AFM topography images of monolayer $MoS_2$ transferred on (a) square nanostructures,(b) triangular nanostructures, Zoomed AFM topography images of $MoS_2$ on (c) single triangular, (d)circular structure, (e,f,g,h) Corresponding strain maps of the AFM topographies as shown in the top row. Color bar of (c,d) are are same. Color bars of (g,h) are same. Scale bar of (e,f,g,h) is 500 nm

Fig.A.3 show the large area as well as zoomed AFM topography images of $MoS_2$ flakes draping on Au nanostructures as well as the strain maps obtained from the AFM topographies using the continuum elasticity theory as shown in equation A.1 [248].

$$|\varepsilon_{zz}| = |\frac{\eta t}{1-\eta t}[\frac{\partial^2 h}{\partial x^2} + \frac{\partial^2 h}{\partial y^2}]| \tag{A.1}$$

where, $\eta(= 0.25)$ signifies the Poisson's ratio [249] and $t(= 0.8\ nm)$ is the thickness of monolayer $MoS_2$ and $h$ represents the local topographic variation in height for which the strain maps are calculated. From the AFM images it is evident that the local curvature of the structures decides the generation of of wrinkles as they mainly originate from the cornners of the nanostructures often forming an interconnecting network with occasional bubbles. The strain distribution at the periphery of the nanostructures and over wrinkles are clearly depicted in the corresponding strain maps as shown in Fig.A.3 (e,f,g,h). These strain features lead to local bandgap modulation with change in carrier density, having a direct effect on the optical and transport properties of $MoS_2$.

### A.2.3 Photoluminescence Enhancement

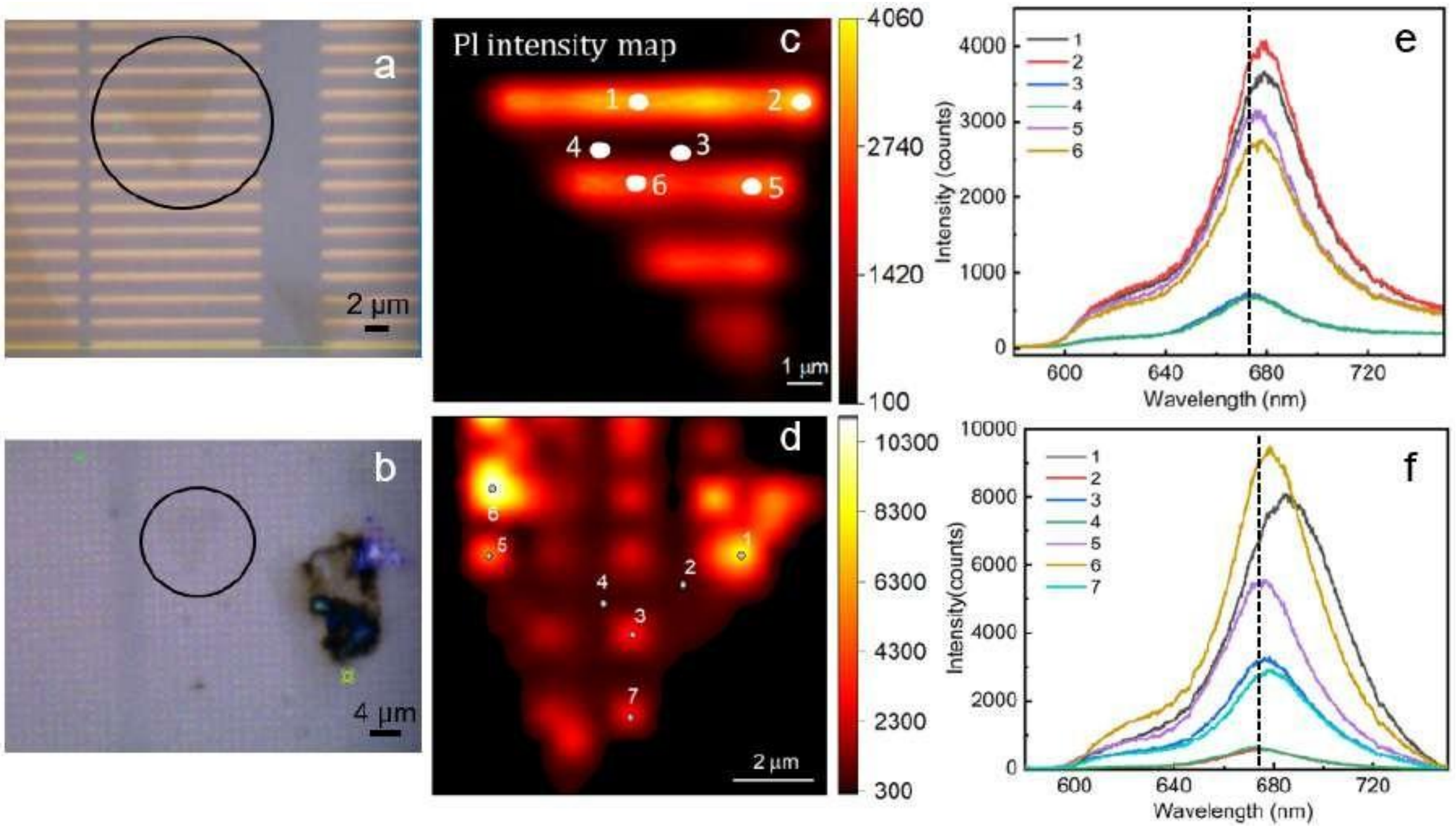

**Figure A.4:** (a,b) Optical microscope images of two flakes transferred on Au nanostructures on Si substrate. (Black circles show positions of the flakes) PL intensity map of monolayer $MoS_2$ on periodic (c) nano-ridges, (d) cylinders, (e,f) PL spectra plots for the selected locations as indicated in spatial maps of (c) and (d).

Fig.A.4 shows two instances of spatially resolved PL intensity maps conducted on monolayer $MoS_2$ transferred on the patterned ridges (width$\sim$ 300 nm and thickness 50 nm) cylinders ( diameter $\sim$ 300 nm and thickness 50 nm). Both the $MoS_2$ flakes are grown under identical conditions using CVD, where the intrinsic defect density in individual flakes is not expected to vary significantly. In both cases, we observe enhancement of PL intensities for the flake on top of the nanostructures which is also confirmed from the PL spectra plots at selected positions on the flake. For the locations '1, 2, 5 and 6' where the flake lies on the structures( Fig.A.4c), the intensity is enhanced

while for'3,4' which is off the nanostructures, it remains unchanged (Fig.A.4e). Similar observations are reproduced in another flake transferred on cylindrical nanostructures as shown in Fig.A.4d and the corresponding spectral plots in Fig.A.4f. Moreover, the spectral plots for both the flakes show that PL is red-shifted on the nanostructures. The strain induced leads to the local lowering of the conduction band minima at the strained regions, creating potential wells, which leads to "funnelling" of excitons toward local potential minima[237–239]. Hence, strained regions in $MoS_2$ exhibit enhanced PL intensity in comparison with unstrained regions. Also, the red shift in the PL peaks indicates the direct-to-indirect band gap transition due to strain. Now, in the first case (Fig.A.4c), the nano ridges are made of Au, while in the second case, the cylinders are made of a dielectric material, $Al_2O_3$. The first case induces an uniaxial strain, and we observe that the highest PL intensity obtained is $\sim$ 4000. In case of cylindrical nanostructures (Fig. A.4d), the strain induced is biaxial and hence the effective strain in the $MoS_2$ flake is increased, and we witness an overall enhanced PL with maximum intensity $\sim$ 10,000 as seen from the spectra plotted in Fig.A.4f. Moreover, in the second case, there is no chance of charge transfer as the structures are made of dielectric. Hence, this gives an idea that biaxial strain with dielectric nanostructures is more effective in resulting strain-induced PL enhancement in $MoS_2$.

### A.2.4 Raman Peak shift and enhancement

Spatially resolved Raman measurements were also conducted on the same flake (Fig. (A.4)) as shown in Fig.A.5. Here as well, the Raman intensities of both modes are enhanced. The in plane $E^1_{2g}(\Gamma)$ mode shows variable red-shift upto $\sim 5\ cm^{-1}$ while out of plane the $A_{1g}(\Gamma)$ mode is red-shifted upto $\sim 4\ cm^{-1}$ as seen in the frequency maps in Figs. A.5c and d. The red-shift in $E^1_{2g}(\Gamma)$ confirms the local strain being induced in the flake [245] commensurate with the PL measurements while that of $A_{1g}(\Gamma)$ is influenced by local number density [250] and provides a relative measure of the same. Thus, the corresponding spatial maps in Figs.A.5a and A.5b demarcates regions of high and low strain and $n_e$, respectively. Hence, the correlation between the PL and Raman maps is indicative of the common origin for the responses, i.e. strain.

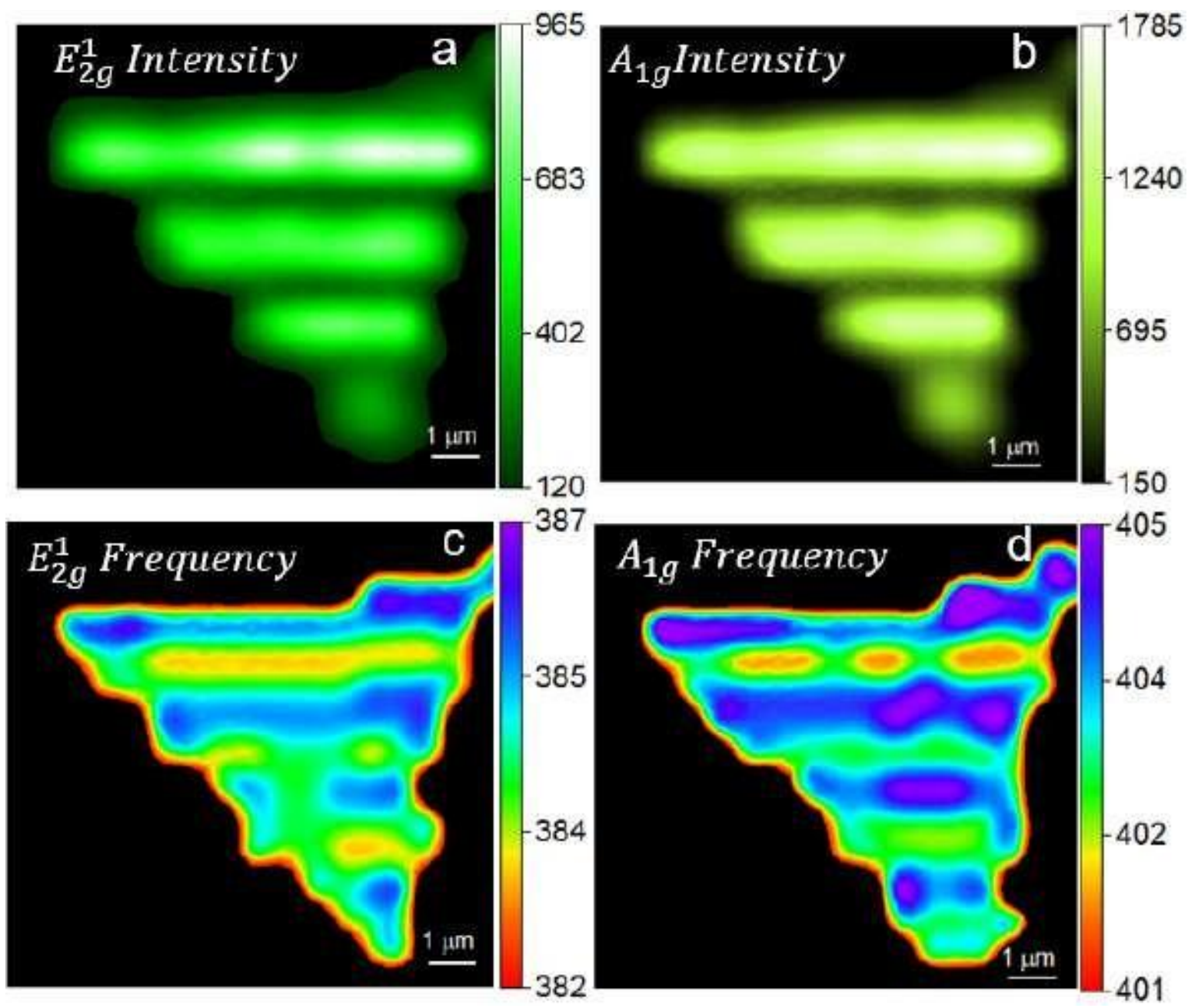

**Figure A.5:** (a) Raman intensity map of (a) $E^1_{2g}$ (Γ) mode, (b)$A_{1g}$(Γ) mode, Spatial maps of Raman mode(c) $E^1_{2g}$(Γ) frequency, (d) $A_{1g}$(Γ) frequency of the $MoS_2$ flake on patterned periodic ridge structure.

### A.2.5 Mobility enhancement

As shown in Fig.A.1, strain was induced in $MoS_2$ flakes by transferring them on Au nano-cylinders with varying periodicities in a controllable way. With conformal draping of $MoS_2$ over the nanostructures, the strain induced effects are not only limited to the periphery of the nanostructures but also to the intervening regions as wrinkles and nanobubbles, thereby the entire flake is non-uniformly strained. With an increase in local carrier density, non-uniform strain should have an impact on the electrical transport along the 2D flakes. To investigate that field effect transistor (FET) devices were fabricated on unstrained and all the strained samples with varied nanostructure density, in bottom gate configuration as shown in Fig.A.6a. Figs.A.6d and e show the source-drain characteristics and transfer characteristics for a strained $MoS_2$ device. Overall strain in the sample were varied and increased by increasing the nanostructure density by reducing their peridicity as $s : 2\ \mu m$, $1\ \mu m$ and $0.5\ \mu m$, with the corresponding FET devices labeled as $1\times$, $2\times$ and $4\times$ strained, respectively. The density of wrinkles and bubbles is expected to be governed by the nanostructure density and hence higher the nanostructure density, higher should be the wrinkle density and the overall strain.

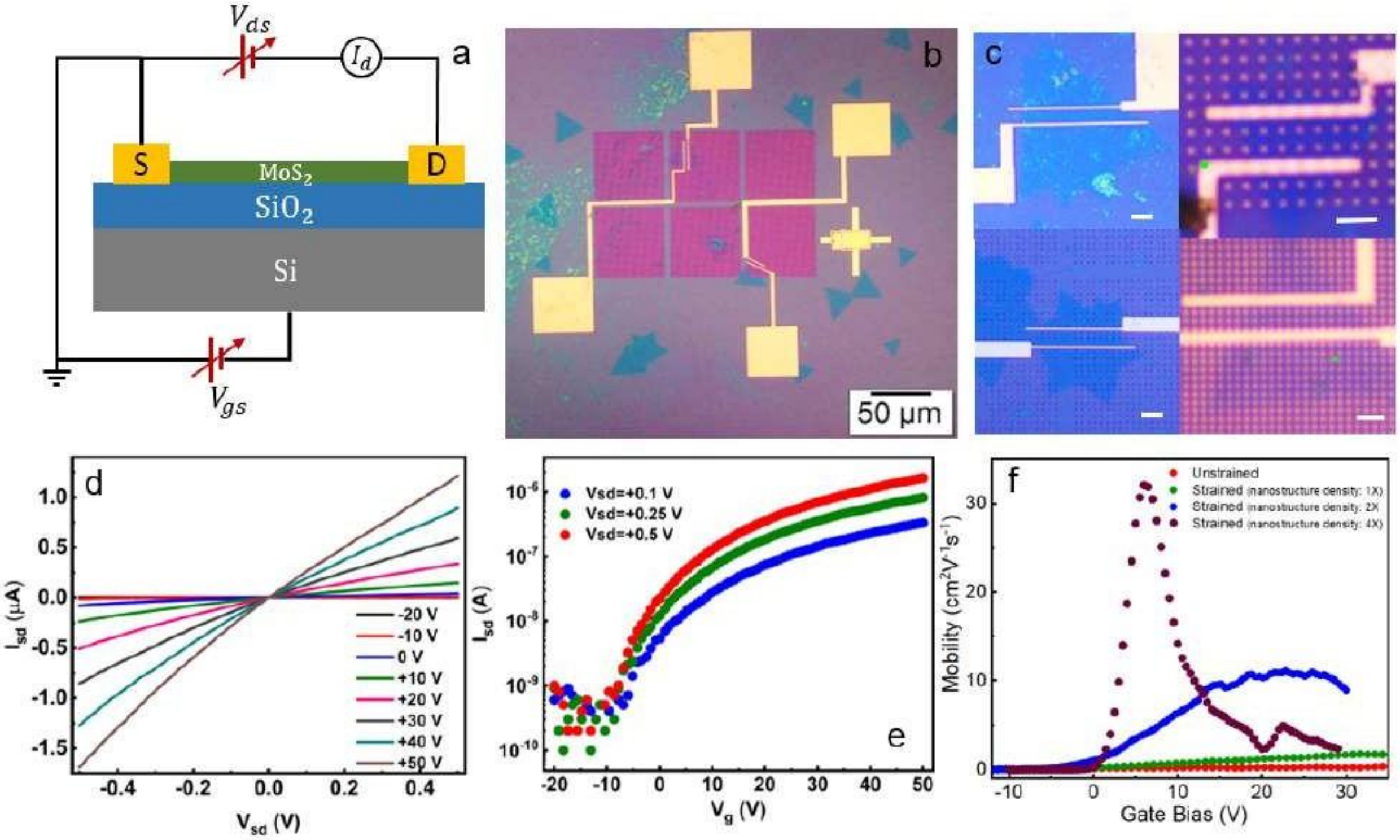

**Figure A.6:** (a) Schematic of the device configuration of field effect transistor, (b) Optical microscope image of the fabricated FETs on strained $MoS_2$ samples, (c) zoomed images of the FET devices of bare $MoS_2$ as well as with varying nanostructure periodicity, (d) Source-drain characteristics with varying gate bias of a typical strained $MoS_2$ device, (e) Transfer characteristics of the strained for different source drain bias, (f) Variation in field effect mobility as a function of nanostructure density. Scale bars in figure 'c' is 2 $\mu$m.

However, exact quantification of %strain across the samples remains non-trivial due to the non-uniform nature of strain distribution. Although, the local %strain variation does not change across the samples, ”overall strain” in the flake does increase with the nanostructure density which is demonstrated for the strained samples with progressive increase in nanostructure density which leads to a systematic increase in electron mobility, up to 60 times compared to an unstrained $MoS_2$ as shown in Fig.A.6f. The field effect mobility was calculated using equation A.2.

$$\mu_{FE} = \frac{L}{W}\frac{t_{ox}}{\varepsilon_0\varepsilon_r V_{sd}}\frac{dI_{sd}}{dV_g} \tag{A.2}$$

where, tox is oxide layer thickness, $\varepsilon_0$ denotes vacuum permittivity, $\varepsilon_r(= 3.9)$ is dielectric constant of $SiO_2$, and L, W are channel length and width, respectively. The $\mu_{FE}$ values were calculated from two-probe FET characteristics and are known to be limited by the nature of the electrical contacts. However, from the linear nature of the current vs voltage plots in the source-drain characteristics, it is understood that there is minimal contribution of contact resistance in the estimated mobility.

## A.3 Practical application of strained $MoS_2$ devices

### A.3.1 Enhancement of hydrogen evolution activity

The poor in-plane electrical conductivity and inert basal plane activity in monolayer $MoS_2$ often lead to major challenges in realizing its practical application. Here, a direct application of the strain-induced CVD-grown $MoS_2$ monolayers was realized to enhance their hydrogen evolution reaction (HER) activity [251]. Device configuration was modified for the electrochemical measurements. In this, a single contact was fabricated on the strained $MoS_2$ domains instead of the two-probe as was done in case of FETs. However, the method of introduction of strain was the same. Electrochemical measurements were performed on the strained $MoS_2$ domains using microcell devices fabricated by standard lithography techniques. The larger contact pads were fabricated using standard photolithography while the micron sized electrodes connecting bigger contact pads to the specific domains were fabricated through e-beam lithography, followed by Cr/Au metallization (2/50 nm) using thermal evaporation, followed by lift-off to obtain the devices as shown in the Fig.A.7a. Next, the devices were spin-coated with the PMMA polymer of thickness 390 nm, and a reaction window was opened on top of the strained devices through e-beam lithography. The electrochemical HER measurements were done using a standard three-electrode setup. For micro-reactor measurements, Au electrodes were used as the working electrode. The counter and reference electrodes were a sharp platinum rod and an Ag/AgCl reference micro-electrode. The electrochemical measurements were done in a small drop of 0.5M $H_2SO_4$ solution. The linear sweep voltammetry was used to measure polarization curves from 0 to -400 vs the reversible hydrogen electrode (RHE) at a scan rate of 5 mV/s. The current densities were calculated by normalizing the measured currents by the surface area of the window exposed to the electrolyte solution. In this case, the strain was varied in the samples by changing the nanostructure density as discussed in the previous section and the best HER activity was obtained in the maximum strained sample, as shown in the current density vs RHE plots and overpotential value as shown in Figs.A.7b and c.

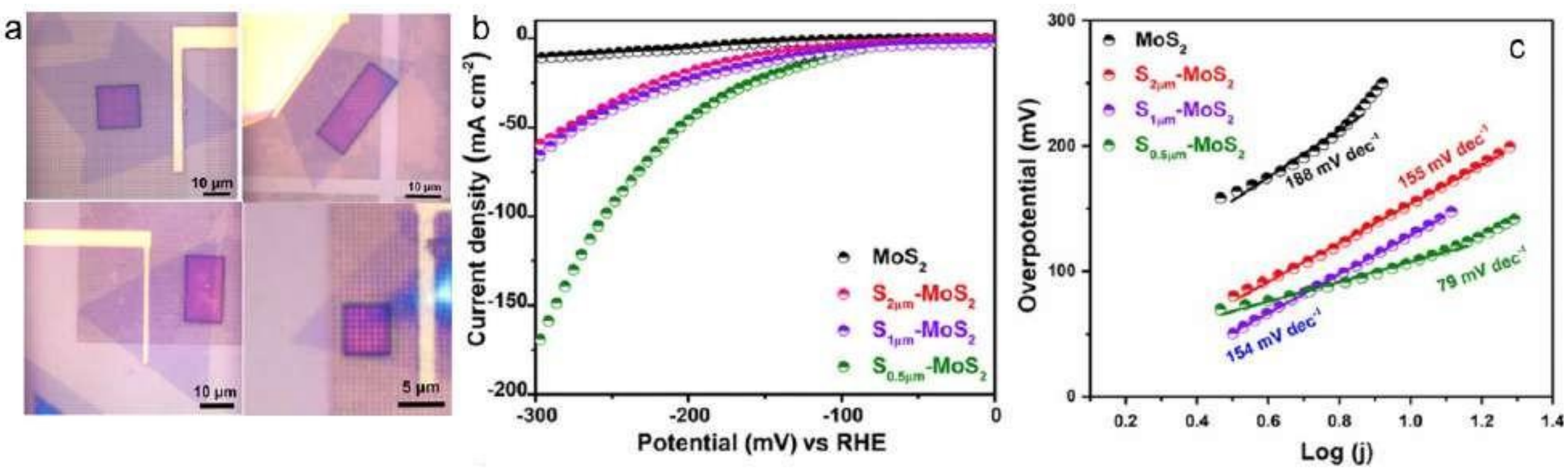

**Figure A.7:** (a) Optical images of various devices showing the windows opened for HER activity, (b) Linear sweep voltammetry (LSV) curves recorded for pristine $MoS_2$ and strained $MoS_2$ with varying inter-pillar spacings,(c) the corresponding Tafel plots.

## A.4 Conclusion

In summary, non-uniform strain was successfully introduced in the $MoS_2$ monolayers through lithographically patterned substrates of varying dimensions and geometries. Demonstration of modification of local optical properties was done through spatially resolved PL and Raman measurements which helped in gaining comprehension of the underlying phenomena. Direct practical applications has been realized through the performance of the strained devices in mobility and HER activity enhancement. Overall, the substrate-induced strain control scheme demonstrated here is scalable and provides a straightforward strain control route that is crucial for engineering device functionality, and compatible with existing on-chip processing technology for incorporation in standard electronic devices.

# Appendix B

# Defect induced modification of optical properties in exfoliated $MoS_2$ flakes via ion irradiation

## B.1 Introduction

Controlled defect creation in transition metal di-chalcogenides such as $MoS_2$ can be another strategy to modify their optical and electronic properties apart from strain as discussed in Appendix 1. As a result, diverse applications can be realized in form of single photon emitters [252], efficient electro-/photocatalysts [253], sensors [254] etc. Various approaches, such as vacuum annealing [255], electron beam irradiation [256], focused X-ray beam irradiation [257], and focused ion beam exposure [258] to induce defects in $MoS_2$ in a controllable manner, have been explored with the target to minimize effective damage to the material. This chapter discusses the creation of defects in thin exfoliated $MoS_2$ flakes on $SiO_2$/Si substrate via He+ and Ar ion irradiation with a typical optimized energy of 7 keV and varying ion fluence typically of the order $10^{14}$ $/cm^2$ and its impact primarily on optical properties to understand the role of defects in realizing efficient opto-electronic applications.

## B.2 Ion beam irradiation

Ion beam irradiation is a sophisticated technique of doping, functionalizing materials through bombardment with a focused or collimated beam of ions with varying values of energy or ion dose, depending on the requirements. It is a flexible method that can induce defects or particle deformation in a well-controlled manner[259]. The control parameters in the ion irradiation process are typically the angle of incidence of ions, energy, fluence/dose of the ion beam, type of ion, and temperature. Fig.B.1 shows the schematic of a ion implanter with the various components that facilitate ion irradiation. As shown in the schematic, the ions emitted from the ion source is extracted and passed through a analyzing magnet which gives the specific direction to the ion beam. Further it is acclerated to suitable velocity and focused through a set of magnetic lens before it is incident on a target/substrate.

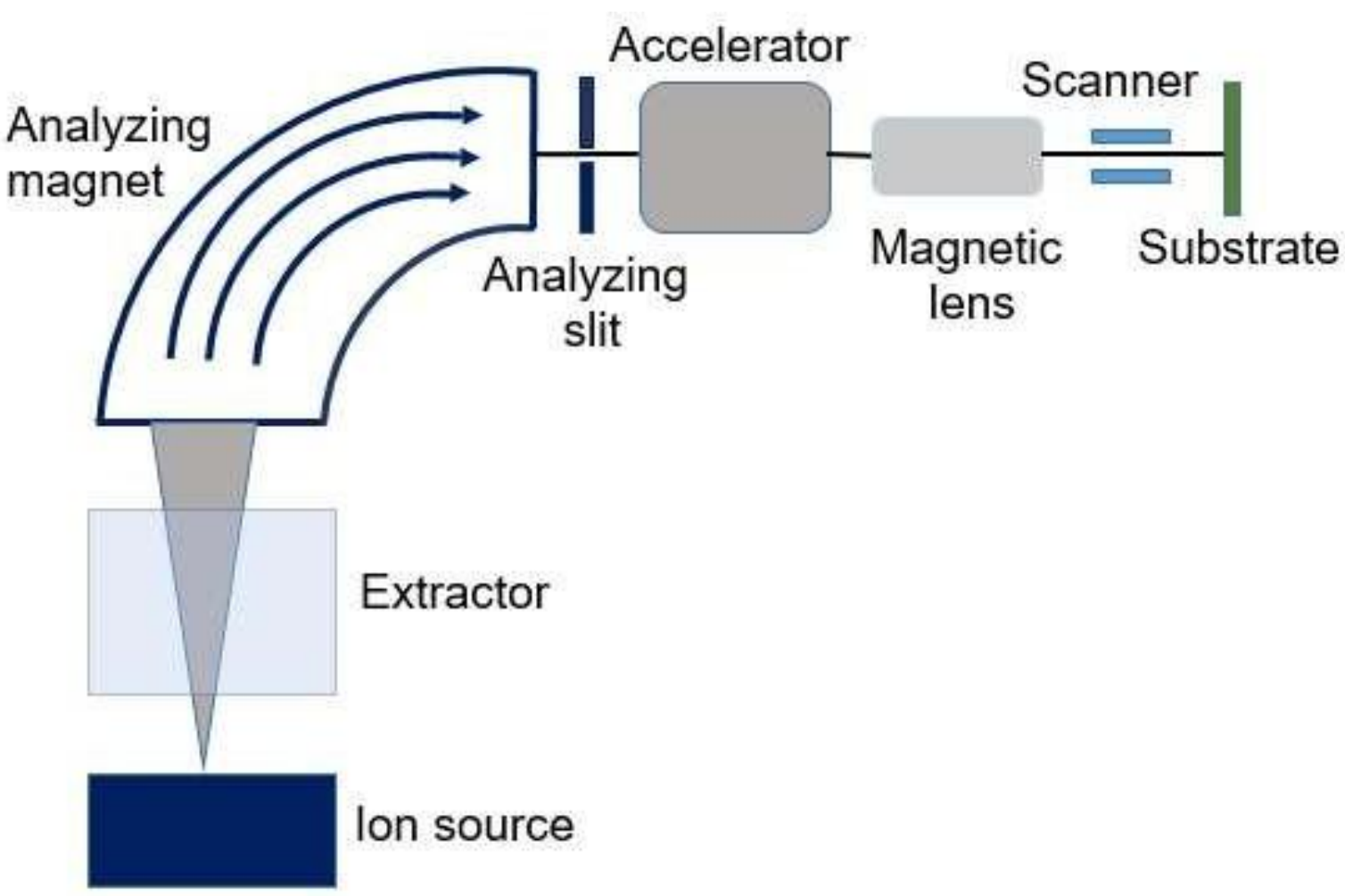

**Figure B.1:** Schematic of an ion irradiator.

## B.3 Photoluminescence, Raman and x-ray photo-electron spectroscopy on ion irradiated $MoS_2$

Thin flakes as well as monolayers of $MoS_2$ were exfoliated from bulk single crystals of $MoS_2$ purchased from "2D semiconductors". Fig.B.2 shows the optical image of an exfoliated flake on $SiO_2$/Si substrates where the layers vary from three to six as quantified from AFM topography line scans (Fig.B.3). These flakes were introduced to an ion irradiation chamber evacuated to $\sim 10^{-7}$ mbar pressure. The flake was irradiated with He+ion with an energy of 7 keV and ion fluence of $10^{14}$ $/cm^2$. This specific value of energy of ions was chosen on the basis of some primary hypothesis. First, higher en-ergy (> 10 keV) might lead to material damage due to their low thickness (∼0.7-10 nm).

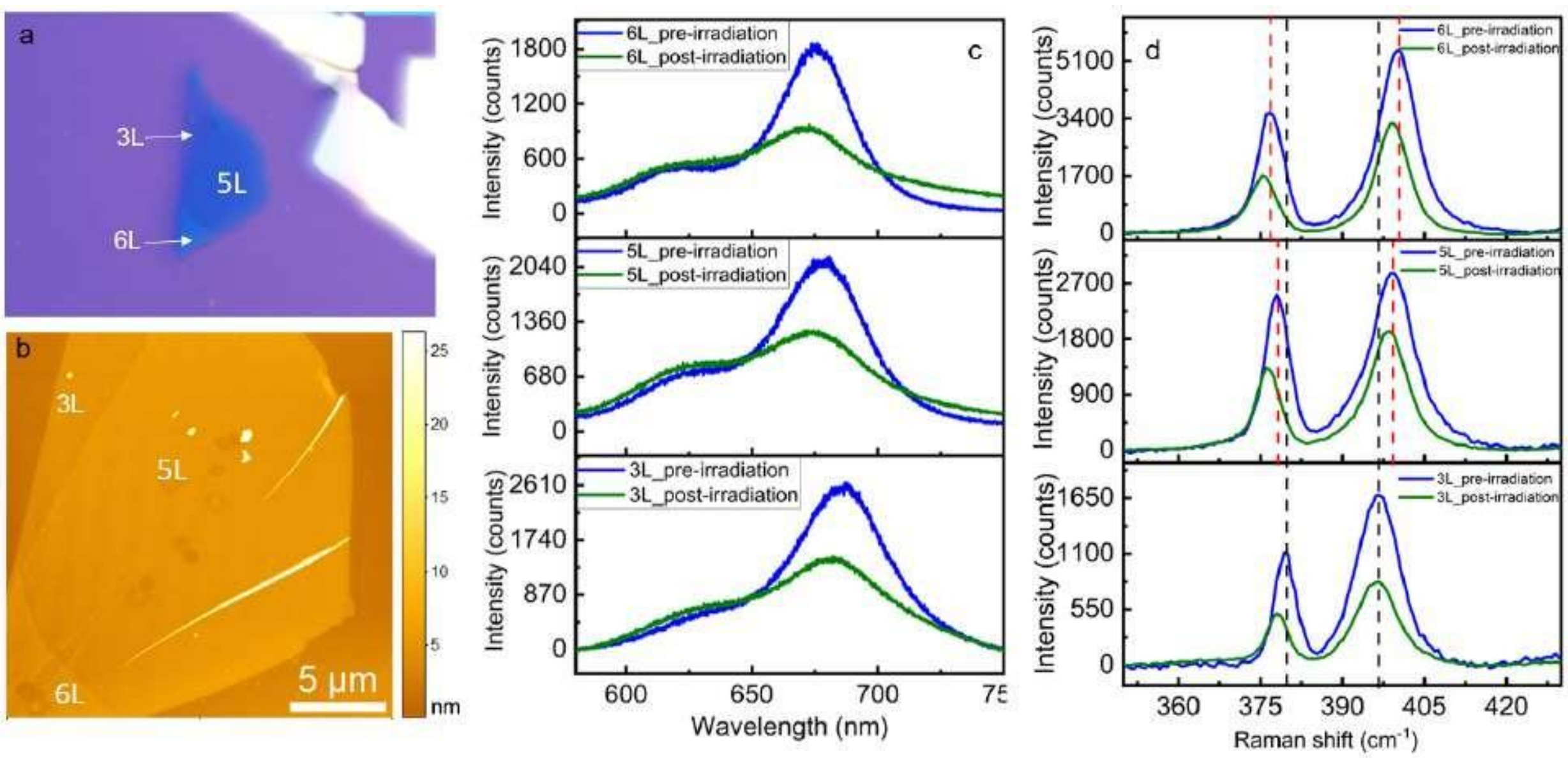

**Figure B.2:** (a) Optical image of an exfoliated $MoS_2$ flake with varied layers on $SiO_2$/Si substrate, (b) AFM topography image of the same flake, (c) PL, (d) Raman spectra recorded pre and post-irradiation across the different layers of the flake. Black-dashed lines in (d) indicate the positions $E^1_{2g}$ (Γ) and $A_{1g}$(Γ) modes in 3L $MoS_2$ pre-irradiation and provide an eye guide to the relative shift in peak positions when the layer number changes to 5L and 6L and the red-dashed line shows the positions $E^1_{2g}$(Γ) and $A_{1g}$(Γ) modes in 5L and 6L $MoS_2$ in pre-irradiation condition.

Secondly, very low energies (<2 KeV) might lead to non-uniform irradiation of ions on the sample. In addition, previous reports on ion irradiation on TMDCs directed us to the initial choice of the type of ion, the typical range of ion energy, fluence etc [253, 260]. PL and Raman spectroscopic measurements were conducted pre and post-irradiation as shown in Figs.B.2c and d, respectively.

Pre-irradiation, the PL shows a decrease in intensity with the increase in layer number (3 to 6), consistent with previous observations [261]. Post irradiation, PL shows systematic quenching of the A exciton emission intensity across all the layers with relatively less impact on B exciton emission, indicating that defects have been created in the samples (Fig.B.2c) [262]. He ion irradiation might induce defects like S vacancies as

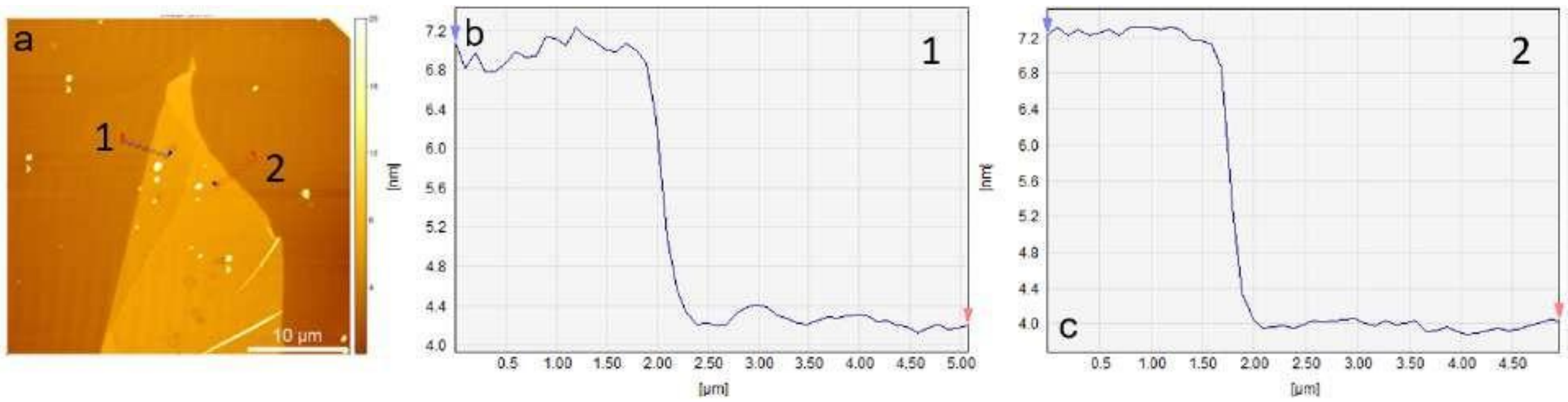

**Figure B.3:** (a,b,c) AFM topography of the exfoliated $MoS_2$ flake with varied layers on $SiO_2$/Si substrate with the line scans quantifying the exact number of layers in the flake.

well as interstitial defects in the lattice structure of $MoS_2$. These defect centers lead to the non-radiative recombination of excited electrons and holes and thereby quench the PL intensity due to a reduction in the radiative recombinations to emit photons. Raman measurements also indicate alteration of the crystal lattice of $MoS_2$ after ion irradiation as shown in Figure B.2d. The inplane $E^1_{2g}(\Gamma)$ mode around $380 cm^{-1}$ shows a red shift across all the layers, indicating the introduction of strain in the lattice leading to mode softening [247] due to the defects introduced. The out-of-plane $A_{1g}(\Gamma)$ also shows a slight red-shift upon irradiation, indicating doping into the material. In addition, the intensities of both modes are quenched, hinting at defect creation and the modification of the pristine lattice structure. Next, a large area uniform exfoliated flake of thickness

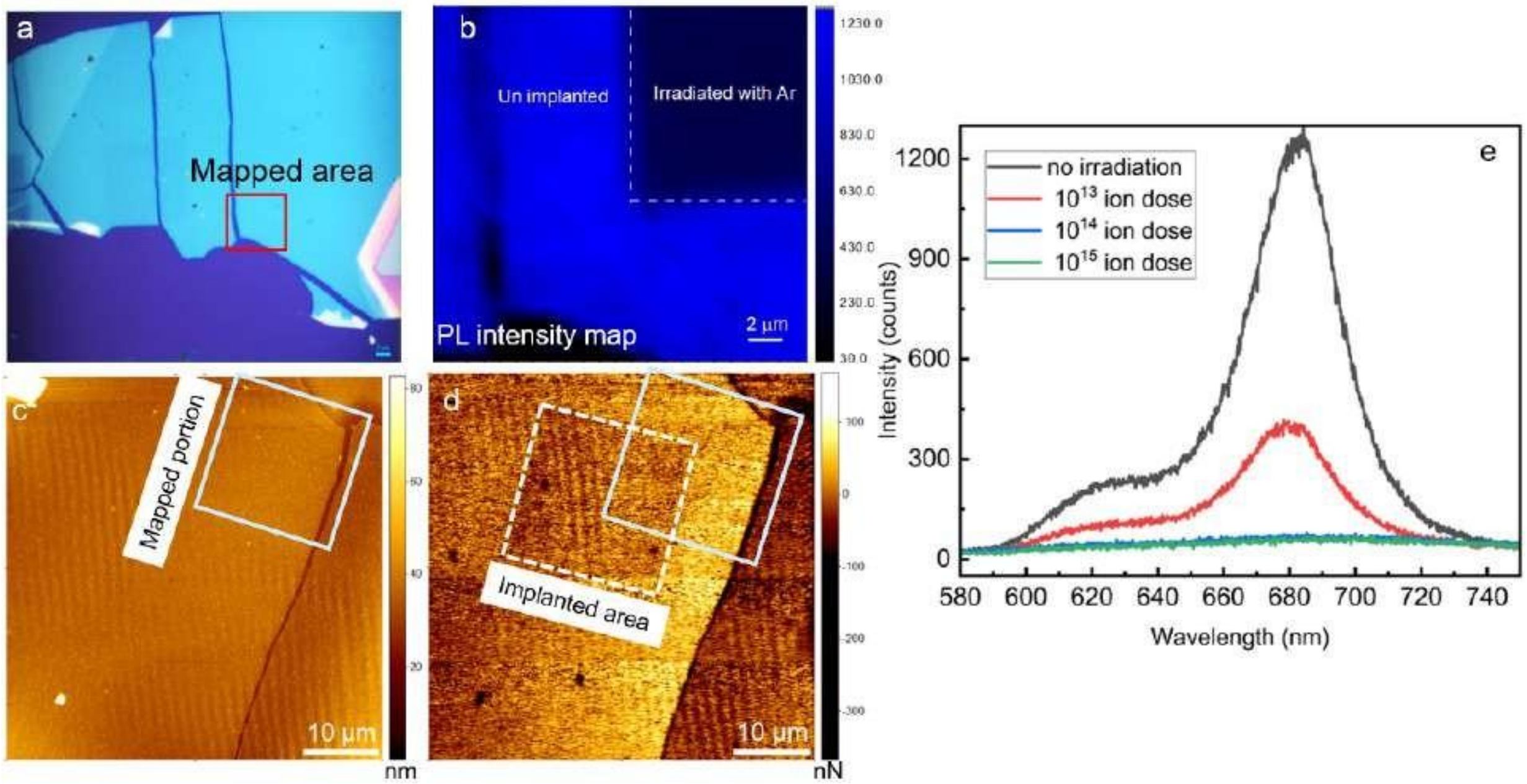

**Figure B.4:** (a)Optical image, (b) PL intensity map, (c) AFM topography, (d) AFM disipation image of the exfoliated flake in $SiO_2Si$ substrate, (e) PL spectra plotted for three irradiated regions with varying Ar ion doses with a spectra of un-irradiated location on the same flake.

~ 8 nm as shown in Fig. B.4a was irradiated selectively at three different positions with varying ion fluences of $10^{13}$, $10^{14}$ and $10^{15}$ $/cm^2$ keeping the energy fixed at 7 keV but this time the ion used was Argon. This was conducted in order to understand how an ion with a bigger size than He can induce defects in the $MoS_2$ lattice as a function of ion fluence. Fig.B.4b shows the PL intensity map of the flake with a region irradiated with $10^{13}$ $/cm^2$ Ar ion dose as demarcated by the white dashed box. Consistent with our previous observations, the irradiated area shows a quenched PL intensity compared to the un-irradiated part. Even though AFM topography (Fig.B.4c) doesn't show any noticeable change in morphology, AFM dissipation image (Fig.B.4d) shows a clear contrast difference between the irradiated and un-irradiated regions indicating the material's surface property has undergone some modification. However, the areas which were irradiated with higher ion fluences of $10^{14}$ and $10^{15}/cm^2$ have complete extinction of PL as shown in figure B.4e at room temperature (298 K). This confirms the significant damage to the material being caused due to the usage of higher ion fluence. X-ray pho-toelectron spectroscopy (XPS) was used to examine the surface elemental composition of the $MoS_2$ samples. The high-resolution Mo 3d and S 2p peaks convey important information about the stoichiometry and the valence state of the $MoS_2$ flakes. XPS measurements were conducted with the irradiated flakes and the results were compared to a pristine flake. Fig.B.5 shows the characteristic XPS peaks for the Mo $3d_{5/2}$ and $3d_{3/2}$ orbitals at 229.5 and 232.7 eV, respectively,and the S $2p_{3/2}$ and S $2p_{1/2}$ peaks at 162.3 and 163.5 eV, respectively. The XPS spectra of Mo 3d and S 2p peaks for the irradiated flakes show a shift towards lower binding energy compared to pristine samples, showcasing the formation of sulphur vacancies[251].

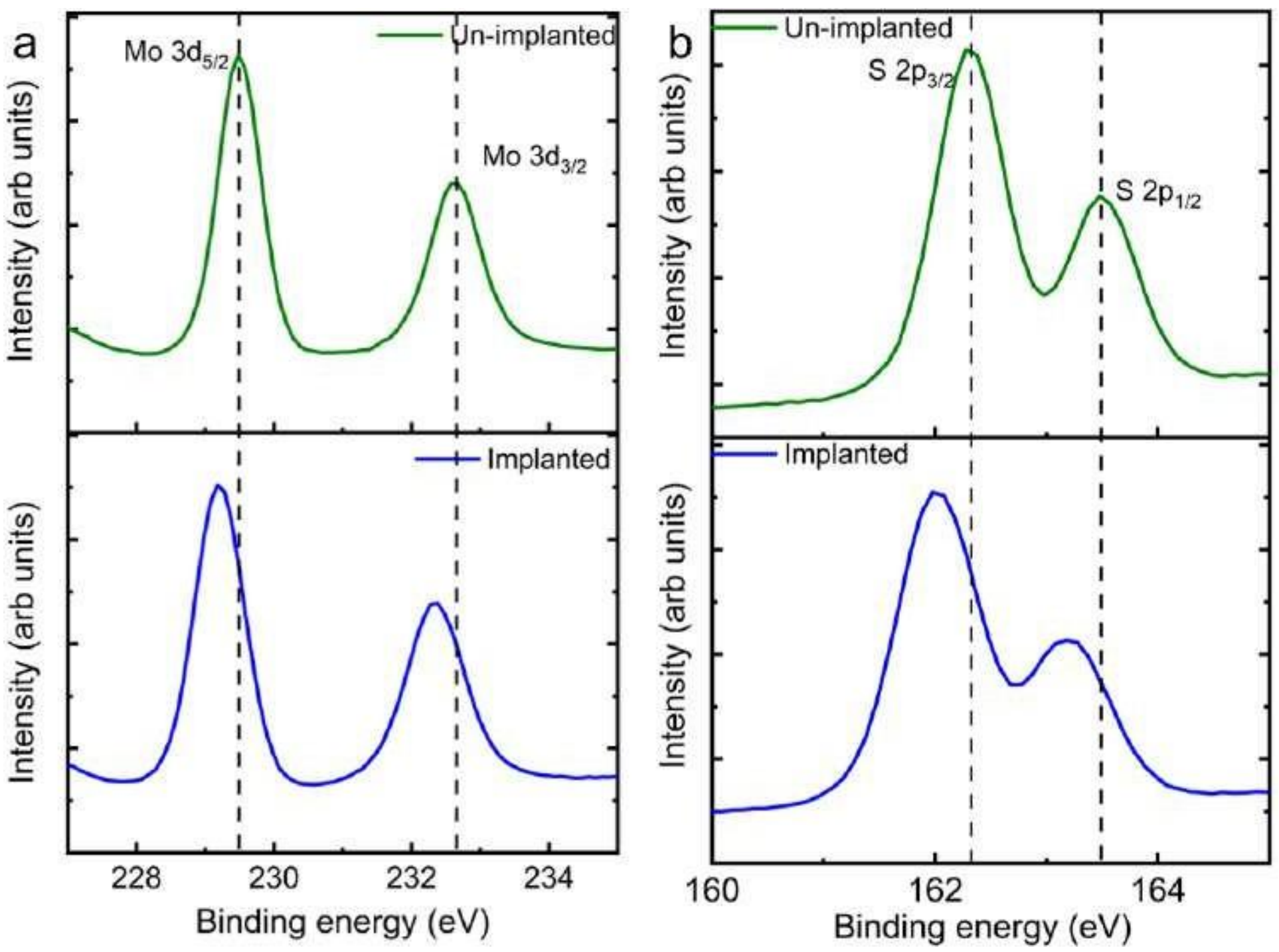

**Figure B.5:** XPS spectra of (a) Mo 3d and (b) S 2p of irradiated (implanted) and pristine (un-implanted) $MoS_2$ flakes.

## B.4 Conclusion

Hence, this work has primarily investigated another potential strategy of modification of optical properties in exfoliated crystals of $MoS_2$. Initial results show PL intensity quenching due to the generation of defects in the form of S vacancies, validated by XPS. Complete extinction of PL with high fluence Ar ion irradiation gives a clear depiction that the defects induced in the lattice of $MoS_2$ are highly susceptible to the size of the ion as well as the fluence with which it is irradiated. Raman measurements showcase the introduction of strain due to defect formation in the lattice. However, further investigations have to be conducted to better understand the underlying phenomena. Overall, this work provides a preliminary understanding of how defects play an effective role in deciding the optical response of TMDCs in the realization of practical applications.

# List of Abbreviations

**AFM** atomic force microscope

**CVD** chemical vapour deposition

**ENZ** Epsilon-near-zero

**FET** field-effect transistor

**ITO** Indium-tin-oxide

**SEM** Scanning electron microscope

**TiN** Titanium nitride

**TMDCs** transistion metal dichalcogenides

# List of Symbols

$N_e$ carrier concentration

$\varepsilon_\infty$ high frequency permittivity

$\sigma_{abs}$ absorption cross section

$\sigma_s$ scattering cross section

$E_{in}$ input electric field

$\gamma$ scattering parameter

$\lambda_{ENZ}$ epsilon-near-zero wavelength

$\sigma_{ex}$ extinction cross section

$\varepsilon'$ real part of relative permittivity

$\varepsilon''$ imaginary part of relative permittivity

$k$ imaginary part of refractive index

$n$ real part of refractive index

# List of Publications

- **Sraboni Dey**, Kirandas P S, Deepshikha Jaiswal Nagar and Joy Mitra, ”Epsilon-Near-Zero metal oxide based spectrally selective reflectors”, *ACS Applied Optical Materials*, **2 (7)**, 1360–1366 (2024).

- **Sraboni Dey**, Kirandas P S, Deepshikha Jaiswal Nagar and Joy Mitra, ”Engineering band-selective absorption with Epsilon-near-zero media in the Infrared”, *ACS Applied Energy Materials*, **8(4)**, 2328–2334 (2025).

- Arijit Kayal*, **Sraboni Dey***, Harikrishnan G., Renjith Nadarajan, Shashwata Chattopadhyay and Joy Mitra, ”Mobility enhancement in CVD-grown monolayer $MoS_2$ via patterned substrate induced non-uniform straining”, *Nano Letters*, **23(14)**, 6629–6636 (2023). **(*Equally Contributing Authors)**

- Renjith Nadarajan, **Sraboni Dey**, Arijit Kayal, Joy Mitra and M. M. Shaijumon, ”Enhancing hydrogen evolution reaction activity through defects and strain engineering in monolayer $MoS_2$”, *Chemical Science*, **15**, 18127-18134 (2024).

- **Sraboni Dey**, Renjith Nadarajan and Joy Mitra, ”Enhancing the Luminescence of $MoS_2$ Using an Epsilon-Near-Zero Underlayer”, *IEEE Xplore*, CLEO/Europe-EQEC 65582.2025.11111334, pp.1-1 (2025).